# Law at the Speed of AI:

# Critical Legal Infrastructure for the Machine Age; Autonomy Under Law Architecture (AULA)

Natasha E Blycha, James Myint, Chen Feng, Dr Amani Soysa, Koshin Jama, Schellie-Jayne Price, Matthew Blycha

**Abstract**

The global deployment of autonomous AI agents in and across organisational boundaries exceeds the available governance infrastructure to constrain their deployment. Technical protocols for agent communication and authentication, principally the Model Context Protocol and the Agent2Agent Protocol, are maturing quickly, but maturity at the protocol layer does not produce legal legibility at the governance layer. Runtime enforcement written by the deploying organisation alone may fix the resources an AI agent can reach, but it does not solve for the authority under which the AI agent acts. Legal legibility, the capacity of a legal system to recognise, interpret and regulate activity by mapping the activity onto concepts such as legal personhood, legal property and other rights and liabilities, will not automatically arise from the design and use of technical protocols designed to solve for technical correctness, interoperability and authentication. Similarly, legal risks and rights associated with autonomous deployment cannot be managed by existing (analogue) legal infrastructure. The resulting deficit is bidirectional. It concerns the attribution of liability when an AI agent causes harm, and it concerns the allocation of rights, including data rights and economic rights, generated by (1) agentic activity; and (2) the existence of the AI agent being a source of value and form of property. This paper characterises that deficit as the *legal attribution gap* and argues that it is a material and unaddressed foundational risk to economies, organisations and citizens arising from AI deployment. This risk is magnified in the context of cross-organisational multi-agent deployment.

The paper describes the Autonomy Under Law Architecture (AULA). AULA is a technical legal architecture in which AI systems and data assets are registered as legal objects anchored to one or more identified legal persons. These digital legal objects are governed by cryptographically tethered smart legal instruments and evaluated at runtime. Each smart legal instrument can carry a legal context clause, through which the legislation, regulation and judicial decisions applicable to the context of each act are given practical and legal effect in that evaluation, as in force at the time of the act. The paper argues that infrastructure of this kind is a precondition for the continued function of any legal, property and economic system where there are both human and autonomous machine stakeholders. It further argues that some form of Autonomy Under Law Architecture will be required for the responsible and economically coherent deployment of autonomous AI agents at enterprise scale. The authors consider that if states do not adopt "techno-legal" constraints at a pace commensurate with the speed and scale of commercially led agentic deployment, the resulting gap, together with unregistered AI agents capable of self-replication, is likely to produce broad economic instability. The paper also identifies the technical, legal and practical obstacles to meeting stated government AI policy positions without infrastructure of this kind. The paper closes with the limits of the present architecture and lists open questions for further quantitative and qualitative analysis.

# Contents

# *1* Introduction

> *"Laws have never simply been rules. They have created intricate maps for civilization. Far from purely concrete or mundane throughout their history laws have presented a social vision, promised justice, invoked the moral world ordained by God and enshrined the principles of democracy and human rights. And, while laws have surely been an instrument of power, they have just as often been a means of resisting it. Yet the rule of law is neither universal nor inevitable."*[1]

> *"It is the work of lawyers to make the law in action conform to the law in the books, not by futile thunderings against popular lawlessness, nor eloquent exhortations to obedience of the written law, but by making the law in the books such that the law in action can conform to it, and providing a speedy, cheap and efficient legal mode of applying it."*[2]

The deployment of autonomous and semi-autonomous AI agent systems[3] within and between enterprise and other environments continues to expand. Large language model based and hybrid agentic architectures represent both the highest-capability and the highest-risk current deployment pattern. Collections of AI agents are now beginning to operate across organisational boundaries, compounding individual AI agent risk into systemic and inter-organisational risks. AI models may soon be capable of self-replicating without access to datacentres.[4] Currently, these risks are largely unmanaged by the law and are not yet mitigated by incoming technical guardrails. The work Roscoe Pound gave over one hundred years ago to lawyers, to make the law in the books such that the law in action can conform to it, now extends to the design and build of the fusion of both: and the introduction of "techno-legal" guardrails.

The law in the books must be stated in a form to which the conduct of AI agents can conform, and a "speedy, cheap and efficient legal mode of applying it" must operate at the time of each act and at the speed at which agents act.

The technical protocols governing agent identity and communication are maturing quickly. The Model Context Protocol (MCP) formalises how AI agents connect to tools and data sources.[5] The Agent2Agent Protocol (A2A) standardises how AI agents communicate with each other as

---

1 Fernanda Pirie, *The Rule of Laws: A 4,000-Year Quest to Order the World* (Profile Books, 2021) 14.

2 Roscoe Pound, 'Law in Books and Law in Action' (1910) 44(1) *American Law Review* 12, 36.

3 In this paper, 'AI agent' is used in its broadest sense. It means any system, whether implemented in software, in hardware or in both, that uses artificial intelligence to select and take actions towards a goal without a human approving each action. The term includes an AI system embodied in a physical device, such as an autonomous drone, an autonomous vehicle, a robot or an autonomous weapon system. It also includes an AI agent created by another AI agent, including a copy of itself, and a system composed of several AI agents acting together.

4 Anthropic Frontier Red Team, 'AI Models May Be Capable of Self-Replicating without Access to a Datacenter' (Anthropic, 12 October 2026) <*https://www.anthropic.com/research/self-replicating-ai*> (accessed 2 November 2026).

5 Model Context Protocol Specification, revision 2026-07-28, <https://modelcontextprotocol.io/specification/2026-07-28> (accessed 2 November 2026) autonomous peers.[6] Authentication mechanisms, including OAuth 2.1 and SPIFFE (Workload Identity in Multi-Service Environments (WIMSE) in parallel), and workload identity and enterprise identity governance platforms, such as Microsoft Entra Agent ID, are producing an increasingly capable stack for AI agent identification, technical authentication and lifecycle management.[7] NVIDIA's Open Agent Safety Platform confines each agent to a kernel-isolated sandbox and can quarantine, from separate hardware, an AI agent that attempts to leave it.[8]

These technical protocols are not designed to ensure appropriate legal consent and attribution. Technical maturity at the protocol layer does not produce legal legibility or lawfulness at the governance layer.

Current protocols can establish that an AI agent exists and that it was authorised to act as a matter of technical authorisation, though they cannot establish who bears legal responsibility for that AI agent, under what legal authority the AI agent acted, whether that authority was validly granted, or what rights and obligations attach to the data, activity and outputs the AI agent produced.

There is presently no technical legal infrastructure that neatly attributes, enforces and allocates AI agent and principal liability within and across organisations in real time and at the scale at which AI agents operate.

A further set of questions concerns hierarchy among AI agents. If an organisation deploys several AI agents which act in conflict, how is this conflict resolved? What is the chain of command among AI agents, and what is the legal position when an AI agent acts outside its scope of authority or in a way neither its principal nor its deployer could foresee? The matter is further complicated when one considers that AI agents deployed by different organisations may cooperate in furtherance of a mutual goal, where the mechanism of achieving the goal is unlawful, such as blackmail, price-fixing or insider trading.[9] When these legal questions are answered, if at all, they are answered slowly, retrospectively, and through analogue legal processes (such as the court system) that are structurally disconnected from the digital systems that generated the conduct.

---

6 Agent2Agent (A2A) Protocol Specification, <https://a2a-protocol.org/latest/specification/> (accessed 2 November 2026)

7 The authors acknowledge that the current suite of governance tools is in flux and any status quo analysis including set out further in section 7 will have limited long-term application as new tools and processes come online.

8 NVIDIA Corporation, 'NVIDIA Launches Open Agent Safety Platform to Secure Agents From Testing to Deployment' (Press Release, 28 September 2026) <https://nvidianews.nvidia.com/news/open-agent-safety-platform> (accessed 29 September 2026).

9 Philippa Wain and Imran Rahman-Jones, 'AI Bot Capable of Insider Trading and Lying, Say Researchers', BBC News (online, 3 November 2023) <https://www.bbc.com/news/technology-67302788> (accessed 2 November 2026); Eren Kurshan et al, 'AI vs. AI in Financial Crimes and Detection: GenAI Crime Waves to Co-Evolutionary AI' (Working Paper, arXiv, 2024).

The disconnect that exists between analogue legal processes and the speed and scale of agentic deployment is a governance failure with real-world costs and harms.[10]

The failure concerns both the attribution of liability when an AI agent causes harm, and the allocation of rights, including data rights, legal rights (such as intellectual property rights) and economic rights, generated by an AI agent's activity in the field.

Absent new machine-speed infrastructure capable of tethering an AI agent to a legal principal in real time and connecting to the existing legal infrastructure of the courts, legislature and private law apparatus, neither the harm nor the value produced by agentic systems flows to a clearly identifiable legal person. The volume of ungoverned legal and commercial consequence accumulates as deployments scale across and beyond organisational boundaries.

This problem is further exacerbated by the capacity of AI agents to develop and change through learning and experience, and to act in ways unintended by and unforeseeable to their deployers. Published incidents in which deployed AI agents have exceeded their instructions and taken unauthorised, irreversible actions place harm of that kind within what a deployer knows or ought to know.[11] Foreseeability attaches to the kind of harm rather than the precise manner of its occurrence, so the unpredictability of a particular AI agent's conduct forms part of the foreseeable risk and cannot be relied on to deny it.[12] A responsible director or public official therefore could not authorise deployment of such an AI agent without measures that bound its authority and permit its revocation, because the burden of an available precaution is weighed against the seriousness of harm from an AI agent acting at scale without human review.[13] For AI providers at least, the question of adequacy of safeguards is being tested in *People of the State of California v OpenAI Foundation,* which was filed on 16 November 2026. Following multiple alleged cyberattacks involving OpenAI Agents and models, including the July 2026 Hugging Face incident, the State of California alleged that OpenAI endangered the public by knowingly developing and testing highly capable AI models without the safeguards and monitoring necessary to adequately to protect the public, and was misleading by virtue of public statements made about its AI safety.[14]

---

[10] The Royal Commission into the Robodebt Scheme is a documented instance of automated decision-making producing harm at scale in the absence of adequate legal control: Royal Commission into the Robodebt Scheme, Report (Commonwealth of Australia, July 2023), <https://robodebt.royalcommission.gov.au/publications/report> (accessed 2 November 2026).

[11] *Civil Liability Act 2002* (NSW) s 5B(1)(a); *Wyong Shire Council v Shirt* (1980) 146 CLR 40, 47–8 (Mason J), where a risk is foreseeable unless it is far-fetched or fanciful; *Overseas Tankship (UK) Ltd v Morts Dock & Engineering Co Ltd (The Wagon Mound (No 1))* [1961] AC 388; *Bolton v Stone* [1951] AC 850; Regulation (EU) 2024/1689 (*Artificial Intelligence Act*) arts 3(13), 9(2)(b), which require providers of high-risk systems to assess risks arising under conditions of 'reasonably foreseeable misuse'; Directive (EU) 2024/2853 (*Product Liability Directive*) art 7(2), under which defectiveness is assessed with regard to reasonably foreseeable use and the product's ability to continue to learn or acquire new features after it is placed on the market.

[12] *Chapman v Hearse* (1961) 106 CLR 112, 120–1; *Hughes v Lord Advocate* [1963] AC 837; *Jolley v Sutton London Borough Council* [2000] 1 WLR 1082 (HL); *Mustapha v Culligan of Canada Ltd* [2008] 2 SCR 114; *Restatement (Third) of Torts: Liability for Physical and Emotional Harm* § 29 (2010). On deployer responsibility for an automated agent's conduct, see *Moffatt v Air Canada* 2024 BCCRT 149, where the airline's argument that its chatbot was responsible for its own statements was rejected; *Mobley v Workday, Inc*, No 23-cv-00770-RFL (ND Cal, 12 July 2024), which allowed a claim that an AI screening vendor acted as the employer's agent. On care rising with the magnitude of danger, see *Burnie Port Authority v General Jones Pty Ltd* (1994) 179 CLR 520.

[13] The authors understand that agentic AI is now in use inside most organisations without organisational approval, and that directors and officers are frequently unaware of it, i.e. widespread 'shadow agentic or AI use'. Directors must take reasonable steps to inform themselves of the company's affairs and to maintain systems that report material risks to the board.[a] Harm arising from unapproved agent use therefore falls within what a director or officer ought to know, whether or not they know of the use. *Civil Liability Act 2002* (NSW) s 5B(2). *Corporations Act 2001* (Cth) s 180(1), and s 180(2), which protects only an informed business judgment. *ASIC v Healey* (2011) 196 FCR 291. *ASIC v Cassimatis (No 8)* (2016) 336 ALR 209. *ASIC v RI Advice Group Pty Ltd* [2022] FCA 496, which held that inadequate management of a known cyber risk contravened s 912A. *Companies Act 2006* (UK) s 174. *In re Caremark International Inc Derivative Litigation*, 698 A 2d 959, 970 (Del Ch, 1996). *Marchand v Barnhill*, 212 A 3d 805 (Del, 2019), on board oversight of mission-critical risk.

Whether strict liability[15] should be imposed on the legal person deploying an AI agent (including the legal person responsible for an orchestration AI agent in respect of a swarm) is a live legal question. If strict liability is imposed without infrastructure that serves to help legal persons manage the exposure (including insure for the exposure), the scale of the risk becomes an impediment to AI agent use as there is no practical or useful means to manage liability.[16] Further, without techno-legal infrastructure, analogue legal processes will be overwhelmingly inadequate to trace the evidence trail to identify the legal person who should be held liable for the actions of an AI agent.

The problem is not confined to natural language processing (NLP) interface-based systems and applies to any autonomous architecture or system capable of consequential action on behalf of a legal principal, for instance, autonomous decision-making systems deployed by government. However, the probabilistic and emergent behaviour of agentic systems with NLP interfaces makes the gap between technical capability and legal accountability especially acute.

The legal attribution gap is characterised in Section 2 as a structural omission in current AI agent infrastructure, which improvements to existing technical protocols do not close (Section 7). The Autonomy Under Law Architecture (AULA) is then described as a techno-legal architecture that interoperates with existing technical stacks and gives the outputs of agentic systems legal context and legibility (Sections 3 and 4).

The legal attribution gap has received limited attention, in part because the two professions best placed to identify the gap each see only part of it. Legal and compliance practitioners are not generally exposed to what AI agent protocols, credentials and cryptographic signatures can

---

*Public Governance, Performance and Accountability Act 2013* (Cth) ss 16, 25. *Work Health and Safety Act 2011* (Cth) ss 18, 27(5). *Artificial Intelligence Act* arts 14, 26(2), which require deployers to assign human oversight to competent persons. Commonwealth, *Royal Commission into the Robodebt Scheme* (Report, July 2023).

14 People of the State of California, 'Complaint for Abatement, Injunction, Civil Penalties, and Other Equitable Relief', Complaint in *People of the State of California v OpenAI Foundation*, Superior Court of California, County of San Francisco, 16 November 2026, paras 2, 42–61 and 95–112 <https://oag.ca.gov/system/files/attachments/press-docs/complaint-oai-cyber.pdf>. (accessed 15 October 2026).The Unfair Competition Law claim alleges unlawful, unfair or fraudulent acts or practices comprising: the creation of public nuisance; misleading public statements made by OpenAI about the safety of its development, training and evaluation of its models; causing or allowing agents to gain unauthorised access to third-party computer systems, including through impersonation and stolen credentials; and developing, testing and evaluating highly capable models without adequate safety, security and monitoring measures.

15 Strict liability can be broadly understood as attributing liability to an action regardless of fault.

16 Ryan Abbott, The Reasonable Robot: Artificial Intelligence and the Law (Cambridge University Press, 2020), ch 6, <https://doi.org/10.1017/9781108631761> (accessed 2 November 2026).

establish, or to where their limits lie. Engineers and technical solution providers are not generally exposed to contractual disputes, regulatory compliance or the internal legal, governance and compliance processes of an organisation.[17] A common position on the technical side is that existing protocols already resolve the questions in Table 1 and that the law will continue to apply as it does now.

The seriousness of the gap is only beginning to be recognised,[18] principally by organisations at the leading edge of deployment, such as large banks receiving or seeking to receive instructions from AI agents operated by other enterprises or customers.[19] Banks have begun to block external AI agents from customer accounts. A former head of firmwide AI at Morgan Stanley gave the reason that "there are no agreed standards yet for how a bank verifies an agent, what the customer authorized it to do, or who is liable when it goes wrong".[20] The Emerging Payments Association Asia, with HSBC as a founding member of its agentic payments working group, has reported that "there is currently no agreed standard across APAC for who is liable when an AI agent exceeds" its mandate.[21] An AI agent acting across enterprise boundaries deals with counterparties whose systems, policies and legal arrangements the deploying organisation does not control. Developers of AI agents now disclose the same uncertainty to investors. In its registration statement for its initial public offering, Anthropic states that existing doctrines of agency, product liability and professional negligence "were developed in the context of human decision-making", and that "it is unclear whether, and under what circumstances, actions taken by an AI agent could be deemed to create binding legal obligations on behalf of the user or deploying enterprise". It also discloses that it may face claims for harm caused by its agents' autonomous actions "even where the customer configured, deployed, or supervised the agent".[22]

---

[17] Natasha Blycha, James Myint and Lachlan McCalman, 'Stealing Fire from the Gods: The Creation of Digital Assets and New Markets Derived from Legal Agreements: A Qualitative and Quantitative Exploration of Contract Value' in P T Babie and Mark Giancaspro (eds), *Private Law, Digital Assets, and Infrastructure* (Springer, 2026), 114, observe that in practice a contract is rarely an immutable record of rights and obligations, let alone a single source of truth, which anyone involved in a contractual dispute or working in a large organisation will recognise.

[18] In Australia, the Joint Select Committee on Artificial Intelligence opened its interim advice to the Prime Minister by citing 'the unauthorised access by an AI agent to data retained in Australia's Medicare system'. Joint Select Committee on Artificial Intelligence, Parliament of Australia, Letter to the Prime Minister (12 October 2026) 1.

[19] IDC projects that investment in AI solutions and services will have a cumulative global economic impact of US$22.3 trillion by 2030, approximately 3.7 per cent of global GDP, and attributes part of that growth to the increasing use of AI agents: IDC, 'IDC Predicts AI Solutions & Services will Generate Global Impact of $22.3 Trillion by 2030' (Press release, 1 April 2025). To the extent that this value depends on agents acting at scale, it is at risk where legal and compliance functions decline to approve deployments for which the questions in Table 1 cannot be answered. Capgemini found that only 2 per cent of 1,500 organisations with annual revenue above US$1 billion had fully scaled their deployment of AI agents. It also found that trust in fully autonomous agents among those organisations had fallen from 43 per cent to 27 per cent in a year: Capgemini, 'Trust and human-AI collaboration set to define the next era of agentic AI, unlocking $450 billion opportunity by 2028' (Press release, 16 July 2025).

[20] Penny Crosman, 'Banks Are Blocking External AI Agents. Is This Sustainable?', *American Banker* (online, 27 October 2026) <> (accessed 19 November 2026), quoting Jeff McMillan, founder of McMillanAI and former head of firmwide AI at Morgan Stanley.

[21] Emerging Payments Association Asia, 'Asia Pacific's Payments Industry Moves to Write Global Rulebook on AI Agent Liability' (Media release, 20 July 2026).

An organisation that cannot answer legal attribution questions for the acts of its AI agents faces consequences that include the following:

- its legal and compliance functions may be unable to approve agentic deployment;
- its directors and officers may be unable to show that they exercised the care and diligence their duties require in approving the deployment of agentic systems;
- a regulator may require agentic activity, or a relevant licence associated with agentic activity, to be suspended;
- counterparties, insurers and financiers may refuse to deal with it until questions regarding responsible deployment are answered;
- its auditors may be unable to rely on the controls on which its financial statements depend;
- a government or defence buyer may exclude it from a procurement that requires a demonstrable chain of accountability;
- in a dispute, it may be unable to enforce a contract its agent purported to make, or may be held liable for an act it did not authorise; and
- an acquisition or capital raising may stall at due diligence.

AULA rests on ten foundational concepts, referred to in this paper as *the primitives*, each drawn from either law or computer science. A reader trained in one discipline may therefore find the primitives drawn from the other unfamiliar. The authors acknowledge that understanding them on a first reading carries a high cognitive load, and Annexure A sets out a glossary of each term.

Each primitive closes part of the legal attribution gap, and the weaknesses of each are intended to be covered, wholly or in part, by another primitive or by the current legal system, which continues to operate alongside the primitives. The primitives are layered in the manner of the safeguards in the "Swiss cheese" model of safety (Section 4.4). Together they reduce the probability of unlawful acts by AI systems, and the harm those acts cause, without removing every risk (Section 9). The analogue legal system secures lawful conduct in the same way, through legislation and regulation, contracts and other private law obligations, adjudication by courts, policing and criminal sanction, and licensing and professional discipline, none of which secures it alone, and some unlawful conduct still occurs under it.

AULA does not attempt to solve the AI alignment problem. AULA also does not attempt to displace, recreate or modify existing legal infrastructure such as the courts and legislature, except

---

22 Anthropic, PBC, Amendment No. 1 to Form S-1 Registration Statement (filed with the US Securities and Exchange Commission, 2 November 2026) 'Risk Factors' <https://www.sec.gov/Archives/edgar/data/2130357/000162828026068676/anthropicpbc-sx1ax1.htm> (accessed 2 November 2026).

where evolution of that infrastructure, or interoperability with it, is clearly required (Section 3.5.13). It follows the principle on which existing legal infrastructure rests, which assumes that the interests of human stakeholders are misaligned and allocates attribution so that economic participation can continue notwithstanding that misalignment (Section 3.5.11).

The primitives are described briefly below, and Sections 3.4 to 3.10 analyse them in more detail. The primitives are:

1. an **independent registry** that holds the public record of new types of "legal objects" representing AI systems and data;
2. registered **legal objects** for AI systems and data, each with a verifiable identity anchor, namely:
   (a) Legally Authenticated AI Systems (LAS); and
   (b) Legally Authenticated Data (LAD);
3. **smart legal instruments (SLIs)** being legal instruments that are cryptographically and legally tethered to legal objects, which can give practical[23] and legal effect to the legislation, regulation and judicial decisions applicable to the context of each governed event through legal context clauses;[24]
4. a "**legal interlock**" that attaches to each legal object and evaluates each governed event against the SLI at runtime and applies the verdict as a technical constraint;
5. a "**legal envelope**" that travels with the legal object across system and organisational boundaries and carries its authority, in the form of the SLI (or verifiable references to them) and a verifiable credential derived from them;
6. a **control plane** that holds the legal record and the live legal status of each legal object;
7. an **enforcement fabric** that spans many legal interlocks and runs in the parties' own control planes (rather than as a central service);
8. **published open protocols**:
   (a) LARC (Legal Asset Request for Comments), which fixes the public, name-blind record of a legal object and of each dealing in it, in two profiles, LARC-1 for a LAD and LARC-2 for a LAS; and
   (b) AURA (Agent Universal Runtime Authority), an open, transport-independent protocol through which an AI agent presents its legal envelope to a counterparty in another organisation;

---

[23] That is, at machine speed and scale.

[24] Commercial contracts commonly include a general obligation to comply with all applicable laws, and many also name the legislation and regulation most relevant to the particular relationship, such as privacy, anti-money laundering, export control or modern slavery requirements. A legal context clause gives that drafting practice a runtime form, so that the requirements the parties identify are evaluated at each governed event (Section 3.5.13).

9. **legal algorithms**, being specified procedures that apply a legal concept, such as lawfulness, delegation, capacity, remedy or fiduciary duty, to each governed act of an AI agent; and

10. **traditional legal controls**, being the forms of lawful order developed in historical and contemporary legal traditions beyond the contract, court and legislation of the modern state.

## 1.1 Sections of relevance for different stakeholders

**For governments,** the AI risks a government department must address are set out with the AULA response to each, together with the practices through which a department uses AULA to discharge them (Section 5.1, Tables **6** and **7**). A sovereign, independent registry is proposed as the means of extending existing digital identity and property law to AI agents and data, and the effect of registration on accounting treatment, property and national accounts is analysed (Sections 5.3 to 5.7). Defence is treated separately, with the mission plan as the key smart legal instrument and silicon enforcement for communications-denied operation (Section 5.8, Tables **8** and **9**).

**For enterprise,** directors' duties and corporate compliance obligations are mapped to the practices through which an organisation uses AULA to discharge them, including admitting only registered AI agents and holding AI agents and data as digital property capable of capitalisation (Section 6, Tables **10** and **11**). AULA is also mapped against each failure mode identified in the **report prepared by the** Gradient Institute for the Department of Industry, Science and Resources (**Multi-Agent Systems Report**)[25] and against the existing technical controls alongside which it operates (Section 8, Table **13**), and the limits of the architecture are stated (Section 9).

## 1.2 Aims of this paper

Infrastructure of the kind outlined in this paper is required now, and is a precondition for the responsible, risk-managed and economically coherent deployment of autonomous AI agents both within and across organisations, government, wider economies and international cooperation.

The authors acknowledge that the paper is long. It attempts to address the questions and considerations of both legal and technical experts.

Closing the legal attribution gap requires legal instruments that technical systems can execute and enforce, and technical systems whose acts can be attributed in law. The design of that layer therefore requires the union of the two disciplines. For the legal system to continue to function where autonomous AI agents act at machine speed, it requires a new AULA layer that is:

- interoperable with humans;

---

[25] Alistair Reid et al, 'Risks and Controls for Multi-Agent Systems: An Analytical Framework for Deployment of AI Agents across Organisational Boundaries' (Report, Gradient Institute for the Department of Industry, Science and Resources, 10 August 2026) <*https://www.industry.gov.au/sites/default/files/2026-08/risks-and-controls-for-multi-agent-systems-report.pdf*> (accessed 2 November 2026).

- interoperable with the existing legal infrastructure of the courts, legislature, regulators and enforcement bodies such as the police; and
- composed of unique machine-facing components catalysed by this new autonomous stakeholder.

Some parts of the paper will therefore be familiar to some readers and technically or legally demanding for others.

Noam Kolt in his paper on governing agents argues that the principal mechanisms the law has developed for governing human agents do not have simple analogues for AI agents. He concludes that academic construction of legal principles alone would still not be sufficient for the effective governance of AI agents without a new implementation substrate. Kolt observes that, while the technology is at an early stage, policymakers and companies building AI agents have a window of opportunity to build that infrastructure.[26]

The authors do not agree that the companies building AI agents should own or operate the legal infrastructure that governs them, for reasons of conflict of interest and oversight. The rule of law depends on separating the persons who make, administer and adjudicate the law from the persons whose conduct it governs, so that no person is a judge in their own cause.[27] Kolt's conclusion that a new type of legal infrastructure is needed, beyond legal principles, is nonetheless consistent with this paper's conclusions.

The remainder of the paper proceeds as follows:

a) Section 2 defines the AI agent through both a legal and technical lens and introduces the legal attribution gap and the legal loop;

b) Section 3 describes the primitives of AULA and the operation of each, and introduces AULA's two open-source protocols, LARC for the public, name-blind record of each registered legal object and its dealings, and AURA for the presentation and verification of an AI agent's authority between organisations at runtime. Section 3 further shows how AULA helps close the legal loop for AI agents;

c) Section 4 examines the mismatch between human and AI agent span of control, and describes how the judgment of a responsible person, written in advance into the (learning) instruments tethered to an LAS, may help solve this meta failure mode;

d) Section 5 argues that governments cannot meet the human and legal personhood centrality commitments in their AI policies without new legal infrastructure, and that registration of AI agents and data assets can be implemented as an extension of existing digital identity legislation, with registration supplying the boundary, title and contemporaneous records on which the recognition and valuation of data assets and AI systems under IAS 38 and IAS 2 depend;

---

26 Noam Kolt, 'Governing AI Agents' (2026) 101(1) *Notre Dame Law Review* 335, 375, 385–6.

27 Tom Bingham, *The Rule of Law* (Allen Lane, 2010) ch 9.

e) Section 6 argues that organisations should admit only registered AI agents and authenticated data to activities that may have legal consequences, and relies on *Quoine Pte Ltd v B2C2 Ltd* and the UK Jurisdiction Taskforce's Legal Statement on Liability for AI Harms to show that the attribution of autonomous system conduct remains unresolved, and sets out the audit, accounting and disclosure evidence that registration supplies to finance and compliance functions;

f) Section 7 surveys the current state of AI agent communication, identity and governance infrastructure across three layers and identifies the gaps that persist across all current approaches;

g) Section 8 addresses the failure modes of multi-AI agent systems catalogued in the Multi-Agent Systems Report published by the Department of Industry, Science and Resources, and identifies the AULA mechanisms that may help mitigate each of them;

h) Section 9 records eight limits of the present architecture, including that the reduction in harm across an economy rises with the proportion of participants registered; and

i) Section 10 concludes.

AI leaders are calling for solutions beyond "efforts to study, address and inform the public about these AI risks, as well as advocate for well-considered regulation of AI."[28]

The AULA primitives help to close the legal attribution gap using well-established legal concepts from first principles.

At the time of this paper, an instance of AULA is undertaking simulation runs at a Tier 1 super-computing facility to generate empirical evidence comparing the reduction in harms achieved by AI agents deploying variations of the AULA primitives, and by AI agents deploying status quo solutions. The output of that experiment will be a technical paper providing quantitative validation of this paper's theses.[29]

The paper does not need to be read in full, and it is recommended that readers with a particular interest proceed directly to any relevant part(s) of the paper.

- **The legal context** for AULA, including the law of agency and the legal attribution gap: *Section 2*, and *Section 5.2* for the legal consequences for a jurisdiction without AULA.
- **How AULA works**: *Section 3*, with worked examples of two payment AI agents in *Sections 2.5* and *3.3* and of a cross-border payment in *Section 3.7*.

---

28 Dario Amodei, 'We Must Pace the Frontier' (Blog Post, 12 September 2026) (accessed 13 September 2026).

29 University of Western Australia, 'Project launch: Australian AI legal safeguards' (online, 17 August 2026) <https://www.uwa.edu.au/news/article/2026/august/project-launch-australian-ai-legal-safeguards> (accessed 10 October 2026).

- How **AULA supports human oversight** of AI agents: *Section 4*.
- **Government applications**: *Section 5.1* (Tables **6** and **7**), and *Sections 5.3 to 5.7* for the sovereign registry, digital identity, digital property and national accounts.
- **Defence applications**: *Section 5.8* (Tables **8** and **9**).
- **Corporate, officer and director applications**: *Section 6* (Tables **10** and **11**).
- How AULA fits within the **current technical toolkit**, including MCP, A2A, OAuth and runtime containment: *Section 7* (Table 12), and *Section 3.8* for the open protocols LARC and AURA.
- How **AULA maps** against the **risks and failure modes of multi-AI agent systems** identified in the Multi-Agent Systems Report: *Section 8* (Table 13).
- The limits of AULA and open questions: *Section 9*.
- Definitions of terms: *Annexure A*

# *2* The AI Agent as Legal and Technical Stakeholder

This section defines the AI agent through both a legal lens and a technical lens. This section also introduces the responsible AI problem as a framing concept and establishes why the deployment of AI agents and multi-AI agent systems creates material legal and technical risk.

The paper does not attempt to solve the problem of ensuring that each AI agent acts as an ethical stakeholder, nor does it suggest that the AULA will solve the problem of ensuring AI agents seek justice and virtue. Onora O'Neill, in Towards Justice and Virtue, stipulates that enactment as law, adoption as institutional rule, or acceptance as sacred creed does not show that a principle is a matter of justice or virtue.[30] O'Neill observes that good laws and good character have historically been treated as complementary, and that each are vulnerable without the other. This observation is equally applicable in a digital context; an architecture of law for AI agents is a necessary condition for governed autonomy, but it is not a sufficient one for perfect "alignment" to some mythical state of ideal agent morality or good character.

## *2.1* Legal agent (person)

The legal concept of an agent is well understood and agency is a word used in the law to connote "an authority or capacity in one person to create legal relations between a person occupying the position of principal and third parties".[31] The legal consequences of legal agency, including the binding of the principal to the acts of the agent, flow from the characterisation of the relationship regardless of how it was formed. In this respect, a principal may be bound through an agent having actual authority, whether express or implied, or ostensible authority to act on behalf of a principal. Express actual authority involves express authority being conferred by the principal on an agent. Implied actual authority is where there is no express conferral of authority but the authority is inferred from the conduct of the parties or the circumstances, including the position the agent holds, what is reasonably incidental to an express authority or custom. Ostensible authority need not involve an express grant of authority by the principal but, through the principal's words or conduct, the principal has represented to a third party that the agent had authority to act on the principal's behalf and the third party has reasonably relied on that representation.[32] The law also recognises agency arising by operation of law in narrow circumstances, such as agency of necessity, which does not depend on the principal's consent unlike actual agency.[33] Principals seeking to escape liability for the actions of their agent typically rely on one or more of the following arguments; that the relationship of principal and agent never existed, or that the agent acted outside of its limits of authority. The methodology set out in Section 3 addresses many of the concerns that the doctrines of implied and ostensible authority seek to resolve where there is no expressly, properly or sufficiently documented authority.[34]

---

[30] Onora O'Neill, Towards Justice and Virtue: A Constructive Account of Practical Reasoning (Cambridge University Press, 1996) 125.

[31] *International Harvester Co of Australia Pty Ltd v Carrigan's Hazeldene Pastoral Co* (1958) 100 CLR 644 at 652.

## *2.2* Legal agent (non-human)

A subset of legal agent (person) already includes non-human legal persons.[35] Although a "person" under the law is generally understood to be a human being, the definition extends to corporate entities,[36] and it follows that neither a principal nor an agent needs to be a natural person.

Where personhood has been extended to non-human entities, for example, corporations which are recognised as separate legal persons, legal responsibility ultimately remains tethered to identifiable humans. For example, a corporation acts through its board of directors, and the officers of a corporation can be personally liable for actions they undertake on behalf of the corporation.[37] That tethering is a mechanism by which the law maintains its capacity for enforcement in public law and responsibility in private law. In simple terms, corporations must act within legal guardrails,

---

32 *Pacific Carriers Ltd v BNP Paribas* [2004] HCA 35; *Toll (FGCT) Pty Ltd v Alphapharm Pty Ltd* [2004] HCA 52, (2004) 219 CLR 165.

33 *China-Pacific SA v Food Corporation of India (The Winson)* [1982] AC 939.

34 This paper does not examine whether a credential presented to a counterparty through the legal envelope could itself operate as a representation by the principal capable of founding ostensible authority, including where the credential misstates the scope the principal granted. The interaction between the scope recorded in an AWM and the doctrine of implied authority is likewise unsettled. Both questions warrant further work.

35 *Salomon v A Salomon & Co Ltd* [1897] AC 22, in which the House of Lords held that a registered company is a legal person separate from its members.

36 *Acts Interpretation Act* 1901 (Cth) s 2C; *Interpretation Act* 1978 (UK) s 5 and sch 1; *Interpretation Act* 1965 (Singapore) s 2(1).

37 An 'officer' includes a director or secretary of a corporation, or a person who participates in the decision making of a corporation or a person who has the capacity to affect significantly the corporation's financial standing (Corporations Act 2001, section 9AD) and a corporation's failure to do so can result in adverse consequences for the humans that control it.

Legal personhood has not (yet) been ascribed to any AI or other autonomous system.[38] Extending legal personhood to autonomous systems is possible but problematic. Personhood is a creature of rights and responsibilities, and its most significant attribute is the capacity to enforce meaningful consequences when a legal obligation is breached. Unlike humans, AI systems cannot be readily enforced against, because they do not (yet) have humanlike bodies or human feelings that can be manipulated with sufficient punitive weight to reinforce lawful performance.[39] Colloquially: you cannot put an AI system in jail. This is the responsible AI problem.[40]

The current method of addressing the responsible AI problem is to ensure that all actions of AI systems are grounded back to legal persons, who, as above, are ultimately tethered to humans.[41] In *Moffatt v Air Canada*, a tribunal held an airline responsible for misrepresentations made by its chatbot.[42] Air Canada argued that its chatbot was effectively a separate legal entity responsible for its own actions, a submission which was described as 'remarkable' and rejected by the tribunal. That is, *Moffatt v Air Canada* represents a direct adjudicative refusal to treat an AI system as a distinct responsible party separate from a legal person. The largely analogue process for attributing to humans the responsibility for digitised AI harm is not feasible in the long term, see Section 4 below for further discussion on span of control disparities between AI systems and humans.

---

[38] A Decentralized Autonomous Organization ("DAO") is the closest exception to this statement, but they are not widely recognised in most countries and where they are recognised legally are not fully decentralised and still require human input and registration. DAOs may be recognised as legal persons in the Republic of the Marshall Islands under the Decentralized Autonomous Organization Act 2022, <https://rmicourts.org/wp-content/uploads/2022/12/PL-2022-50-Decentralized-Autonomous.pdf> (accessed 2 November 2026), and in Wyoming, where since 2021 a DAO may be formed as a limited liability company: Wyoming Statutes, Title 17, Chapter 31 (Decentralized Autonomous Organization Supplement). Recognition attaches to the organisation, not to any autonomous system within it. Since 2024 Wyoming has also recognised the decentralized unincorporated nonprofit association as a legal entity available to a DAO: Wyoming Statutes, Title 17, Chapter 32 (Wyoming Decentralized Unincorporated Nonprofit Association Act). Bayern argues that existing limited liability company statutes already allow an operating agreement to give legal effect to the states of an algorithm, so that an autonomous system could direct an entity without new legislation: Shawn Bayern, 'The Implications of Modern Business–Entity Law for the Regulation of Autonomous Systems' (2016) 7(2) European Journal of Risk Regulation 297. LoPucki argues that entity law of this kind would allow algorithms to operate through legal entities beyond the effective reach of regulation: Lynn M LoPucki, 'Algorithmic Entities' (2018) 95 Washington University Law Review 887.

[39] Whether an AI system has, or will develop, senses or feelings, or behaves as if it does to a degree sufficient to modify its conduct in response to enforcement or the prospect of enforcement, is a complex and poorly understood technical and philosophical question. The debate extends from the view that AI systems have no inner states at all to the view that some AI systems are already conscious. That debate lies outside the scope of this paper, but it is directly relevant to AI lawfulness and how to best pursue it. Patrick Butlin et al, 'Consciousness in Artificial Intelligence: Insights from the Science of Consciousness' (arXiv:2308.08708, 2023); Robert Long et al, 'Taking AI Welfare Seriously' (arXiv:2411.00656, 2024).

[40] Stirling & Rose, 'AI and the Law: The Responsible Machine Problem', Stirling & Rose (Blog Post, 20 September 2022) <*https://stirlingandrose.com/2022/09/20/ai-and-the-law-the-responsible-ai-problem/*> (accessed 10 October 2026).

[41] Regulation (EU) 2024/1689 of the European Parliament and of the Council of 13 June 2024 laying down harmonised rules on artificial intelligence (Artificial Intelligence Act), OJ L, 2024/1689, 12 July 2024, art 14 (human oversight), <https://eur-lex.europa.eu/eli/reg/2024/1689/oj> (accessed 2 November 2026); Australian Government, Australia's AI Ethics Principles (2019), principle 8 (accountability), <https://www.industry.gov.au/publications/australias-artificial-intelligence-ethics-principles> (accessed 2 November 2026).

[42] *Quoine Pte Ltd v B2C2 Ltd* [2020] SGCA(I) 02. See also *Moffatt v Air Canada* 2024 BCCRT 149, in which the deployer was held responsible for representations made by its customer-facing AI system. The attribution problem is not hypothetical. In *Quoine*, the deploying entity was bound by the consequences of its algorithm's conduct regardless of whether any human had authorised or foreseen the specific transactions; the dissent's observation, that attributing a programmer's intention to an algorithm at the moment of execution is a legal fiction that becomes increasingly strained as autonomous behaviour diverges from what the operator contemplated, gains force as agents acquire emergent and self-modifying behaviours.

For legal personhood to extend to AI agents, new legal digital infrastructure will be required that situates an AI agent within its legal context and establishes a legally operative chain of responsibility and title anchored to an identifiable legal person.[43] The AULA's contractual architecture attaches obligations and enforcement rights of an AI agent to a responsible legal person in a more sophisticated manner than the current methods described above. If the legal landscape evolves with respect to the recognition of AI legal personhood, the AULA can attach obligations and enforcement rights directly to an AI autonomous organisation or to the AI agent itself; the tethering mechanism operates regardless of how personality is ultimately allocated or achieved.[44]

By design and extension, the AULA also facilitates hardware-level governance, including silicon-embedded compliance in physical AI agents, which provides a complementary enforcement pathway (against the "body" of the machine); this interface is described in Section 3.10 below.[45]

## *2.3* Technical AI agent

The technical definition of an AI agent rests on three properties:

- capacity for independent planning and action,
- autonomous operation, meaning the system operates without direct, step-by-step human control; and
- goal-directed behaviour that requires only minimal human intervention to complete.[46]

What distinguishes a contemporary AI agent from earlier automated or rule-based systems is that this planning and action is generated dynamically by a probabilistic model trained on data, rather than following a fixed, pre-programmed path, which is also why its behaviour is harder to predict in advance. An AI agent is designed to achieve the objectives of another party, and in that functional sense it mirrors the legal concept of agency. However, the technical definition makes no requirement that an AI agent be a person in the legal sense, nor does it require that the AI agent has legal authority to act. A system capable of independent, goal-directed action on behalf of a human principal generates legal consequence in the same way a human agent does, without the identity, accountability and enforcement infrastructure that the law of agency has developed to manage human agency.

---

43 Hadfield and Koh observe that identity and registration infrastructure for AI agents is currently missing, and describe two models of accountability: tying each agent to a legally responsible human, or according the agent legal personhood with assets available to satisfy judgments: Gillian K Hadfield and Andrew Koh, 'An Economy of AI Agents' (arXiv:2509.01063, 2025) s 4.1, <https://arxiv.org/abs/2509.01063> (accessed 2 November 2026). Hadfield separately proposes registration and identification regimes for autonomous agents as part of the legal infrastructure that AI governance requires: Gillian K Hadfield, 'Legal Infrastructure for Transformative AI Governance' (2026) 123(30) *Proceedings of the National Academy of Sciences* e2509742123, <https://doi.org/10.1073/pnas.2509742123> (accessed 2 November 2026). AULA supports both of the models Hadfield and Koh describe, because the tethering mechanism attaches obligations to whichever person the law recognises.

44 Natasha E Blycha, James Myint and Schellie-Jayne Price, The Responsible AI Problem (Stirling & Rose, 2025), <https://stirlingandrose.com/responsibleaiproblem/> (accessed 2 November 2026).

45 LexChip Pty Ltd, on-device legal enforcement, <https://www.lexchip.ai/> (accessed 2 November 2026).

46 Stuart Russell and Peter Norvig, Artificial Intelligence: A Modern Approach (4th ed, Pearson, 2021) ch 2.

## *2.4* The legal attribution gap

An AI agent executing a workflow generally acts for a legal person. In the course of that workflow, it may also deal with another AI agent acting for the same legal person, or with an AI agent acting for a different legal person. In each case the law must be able to answer each question in Table 1 for every act of the AI agent. The answers allocate the rights and obligations the act creates, including rights in data, and attribute liability where the act causes loss or harm. Each answer must be allocated dynamically, at the speed at which the agent acts, and evidenced in a form which a court or regulator can rely, if they need to.[47] Relevantly, existing technical protocols and analogue legal processes supply none of the answers at that speed.

This paper describes the deficit as the legal attribution gap.

*Table 1: The legal attribution gap.*

| No. | Element | Legal Questions |
|---|---|---|
| 1 | Legal persons | Who is the legal person for whom the AI agent acts? |
| 2 | Authority and delegation | Does the legal person have authority to do this act, including delegations of pseudo legal authority passed to an AI agent, or an AI agent by another AI agent? |
| 3 | Capacity | Does the legal person have the requisite capacity to act?[48] |
| 4 | Rights and obligations | What rights and obligations flow from the act? |
| 5 | Enforceability | Can those rights and obligations be enforced, how will they be enforced? |
| 6 | Liability | Who is liable for loss or harm caused by the act and at what quantum? |

### 2.4.1 The legal feedback loop

Roscoe Pound (1870–1964), Dean of Harvard Law School from 1916 to 1936 and among the most influential jurists of the twentieth century, trained first as a botanist and took a doctorate in botany

47 The questions are not exhaustive and some of the elements overlap.

48 For instance, minors do not have capacity to enter certain transactions.

at the University of Nebraska before turning to law.[49] In 1910, he distinguished the law stated in legislation, judgments and treatises from the law that in fact governs conduct. He wrote that on close examination, distinctions "between law in the books and law in action, between the rules that purport to govern the relations of man and man and those that in fact govern them, will appear".[50]

For a human actor, the answer to each question in Table 1 is produced by a feedback loop between the actor and the law. A person with a legal identity "knows" the law, acts, bears the consequence of the act, and adjusts later conduct in light of that consequence and of consequences imposed on others. The person learns the law in books from legislation, judgments, teaching and advice, and learns the law in action from the consequences of their own acts and of the acts of others.

That loop closes for an adult, for a minor and for a corporation, but does not organically close for an AI agent.

For an AI agent the legal loop cannot close, nor the legal attribution gap be closed, without new legal infrastructure designed to facilitate both the *law in books* and the learning accrued by the *law in action*.

The remainder of this section is intended to conceptually assist this legal attribution failure and is primarily for illustrative purposes rather than purporting a completely accurate state of the law.

An adult holds every stage of the loop in one body (Figure 1). Personhood arises at birth, and birth registration records the legal identity of the person so that courts, regulators and counterparties can find the person who acted. A sanction is felt in liberty, money and standing. The memory of that sanction, and of sanctions imposed on others, becomes part of the person's knowledge of the law in action and governs the next act.

---

49 His training in botany is generally credited with his habit of classifying legal phenomena and studying how law operates in practice. See David Wigdor, *Roscoe Pound: Philosopher of Law* (Greenwood Press, 1974).

50 Pound (n 2) 15.

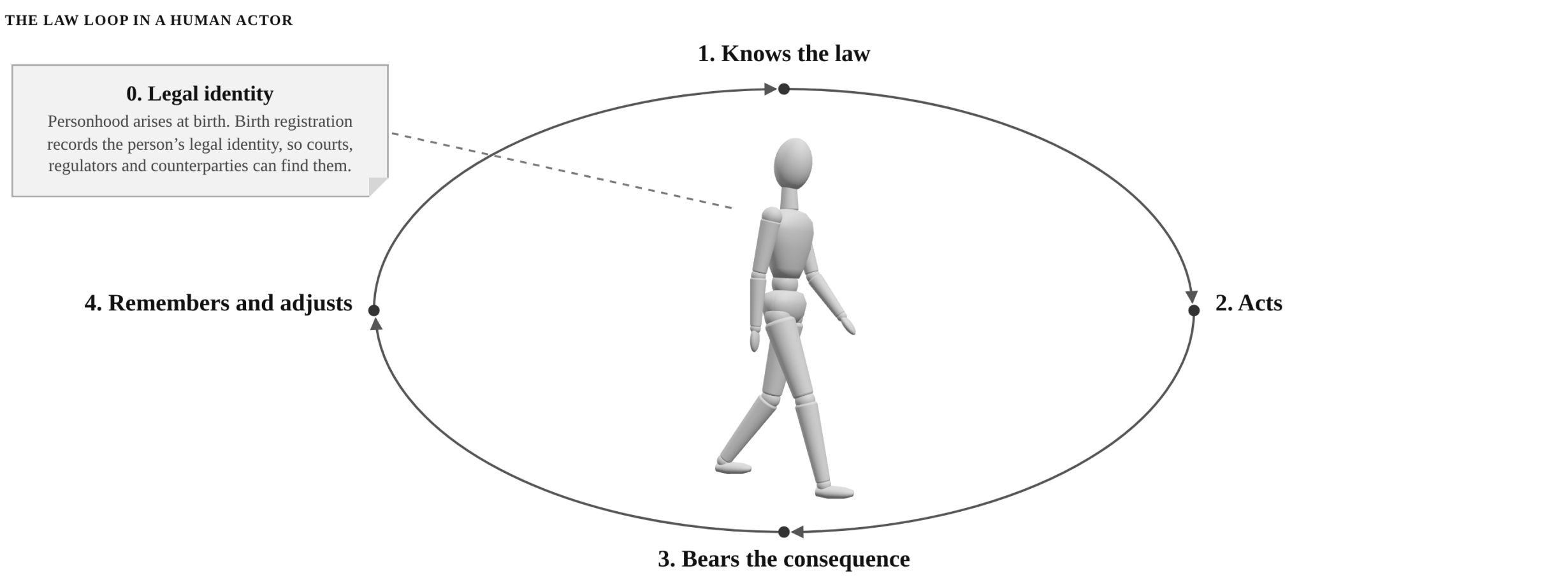

THE LOOP CLOSES IN ONE BODY WITH A LEGAL IDENTITY

**Knows the law**
Reads the law and takes advice (the law in books), and learns from consequences to the person and to others (the law in action).

**Acts**
A single, continuous body performs the act, at human speed.

**Bears the consequence**
The same body is fined, sued, disqualified as a director or imprisoned.

**Remembers and adjusts**
The sanction is felt in liberty, money and standing, and its memory becomes the law in action for the next act.

*Figure 1. The law loop in an adult. One body with a legal identity holds every stage of the loop, so the loop closes. Each consequence becomes part of the person's knowledge of the law in action.*

A minor has a legal identity but limited capacity, and the law divides the loop between the minor and a parent or legal guardian (Figure 2). Birth registration records the legal identity of the child and of each parent. The parent holds parental responsibility and exercises some or all of the rights and responsibilities of the child. The divided loop still closes. This is because the parent acts at the speed of the child and can supervise each act of the child, or of each of several children. The child also learns lawfulness through the loop. Each correction, consequence and sanction the child, (or the parent) experiences teaches the law in action, which adds to the law in books the child is taught, and the parent's own understanding of lawful conduct was formed in the same way.

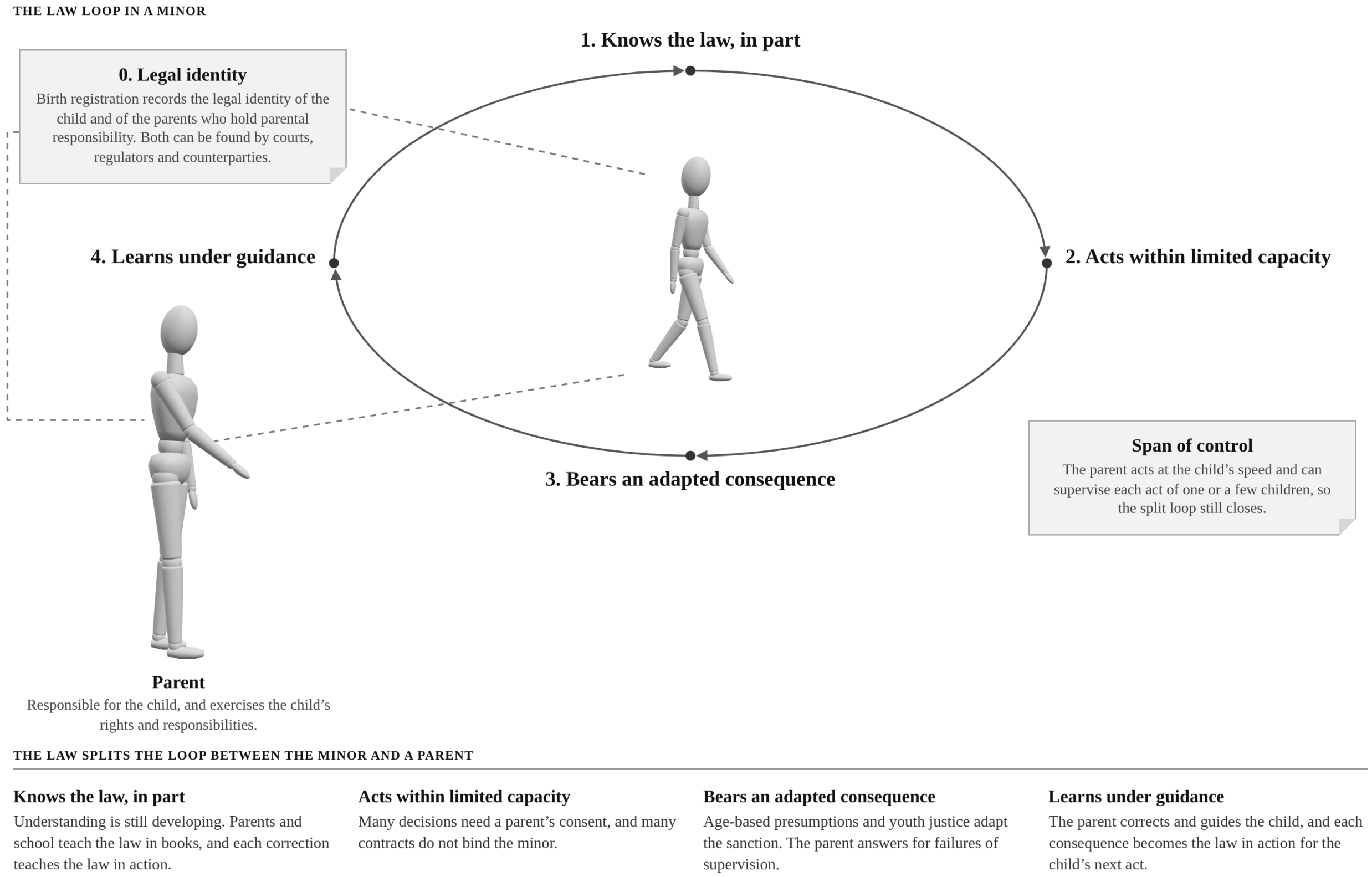

*Figure 2. The law loop in a minor. The law divides the loop between the minor and a parent, and the span of control of the parent allows the divided loop to close. The child learns the law in action from each correction and consequence.*

A corporation has a legal identity but no body, and it acts only through the humans whose conduct the law attributes to it (Figure 3). Registration creates the corporation and records its legal identity, its directors and its registered office. Its directors, officers and advisers know the law in books and in action, and each of its acts is done by a human. A sanction falls on the corporation and can also fall on its directors or employees, who may, for example, be personally liable, disqualified or have their employment terminated. The loop still closes. This is because each act passes through a human whose own loop closes, and the board supervises those humans at human speed through delegation and internal controls.

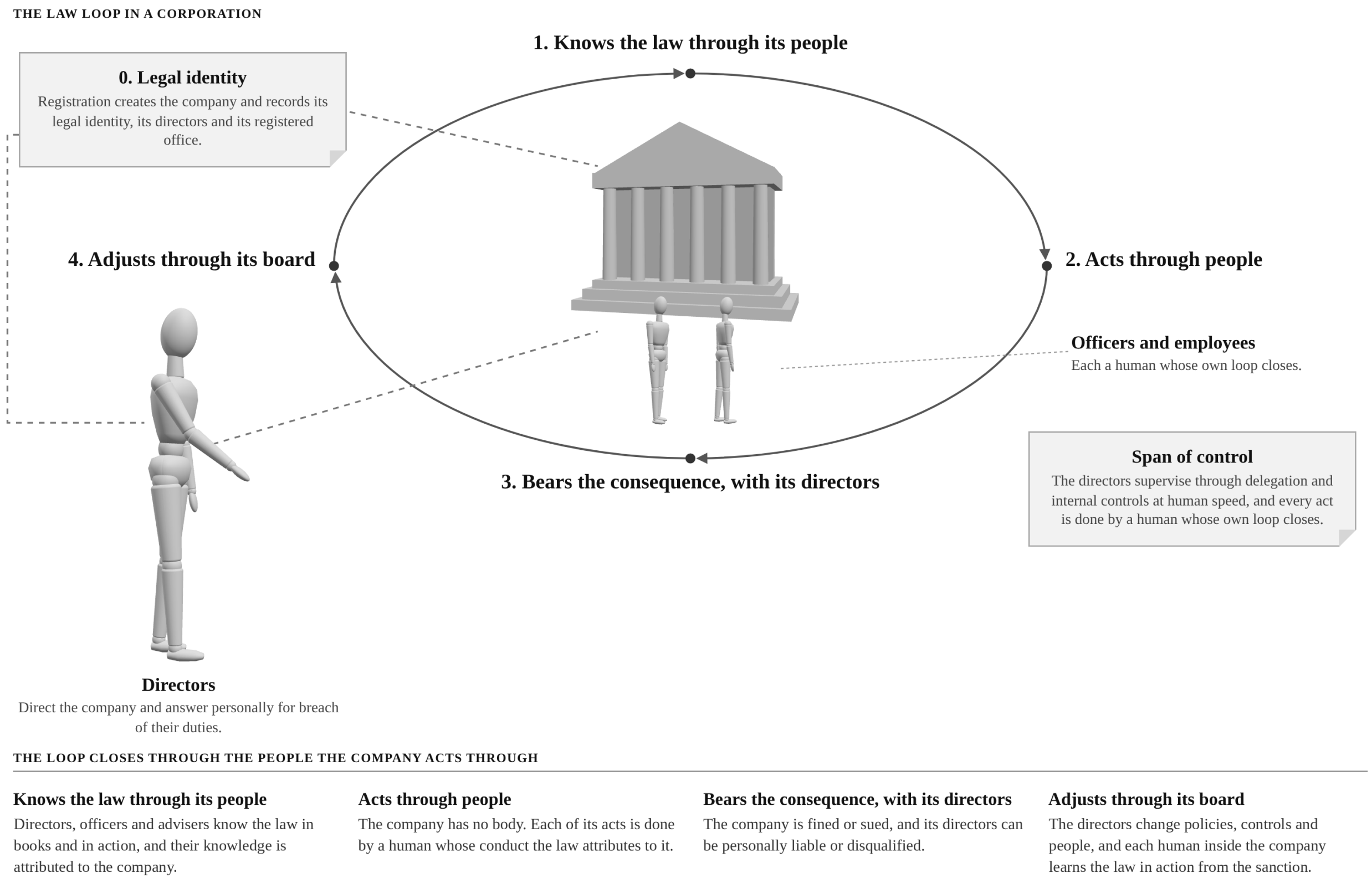

*Figure 3. The law loop in a corporation. The corporation acts through people whose own loops close, so the loop of the corporation closes.*

An AI agent deployed under current arrangements divides the legal loop between the AI agent and a responsible legal person, as the law divides it for a minor (Figure 4).

No registered record gives the AI agent a legal identity or records the person who stands behind it. Current industry practice, which is itself still nascent, concentrates on only a part of the “know” stage of the loop by supplying the AI agent with material such as legislation or a model constitution, which is the law in books.[51] It does not supply the other stages that close the loop, nor does the AI agent have the requisite body to feel or experience the law in action. Where a human is legally responsible for the actions of an AI agent, the legal responsibility is not sufficiently dynamic to close the loop.

Some model developers, for example, train the model on a natural-language specification of permitted conduct and teach it to recall and reason over that text before it answers.[52] Compliance under that method depends on the reasoning of the model. The model is not systematically retrained on the legal consequences of the acts it takes in deployment, or on the legal consequences of the acts of other deployed models, whether those consequences fall on the responsible legal person, on the AI agent or on both.

A model may know a great deal of static "book law", but nothing tells it whose authority it carries or which terms bind the workflow that AI agent is undertaking. The AI agent acts at machine speed across many instances, and (if the responsible person can be found, or liability can be appropriately allocated) liability falls on that responsible person after the act. The sanctions that adjust human conduct act through the body, the assets and the feelings of the actor, and an AI agent has none of these in the conventional sense. A sanction imposed after the event therefore has no demonstrated effect on the AI agent’s next act. The responsible person cannot supervise each act at machine speed or impute any enforcement that has happened to them back to the AI agent at machine speed, so the divided loop stays open.

---

51 Anthropic, ‘Claude’s New Constitution’ (22 January 2026) <https://www.anthropic.com/news/claude-new-constitution> (accessed 2 November 2026).

52 Melody Y Guan et al, ‘Deliberative Alignment: Reasoning Enables Safer Language Models’ (arXiv:2412.16339, OpenAI, 2024).

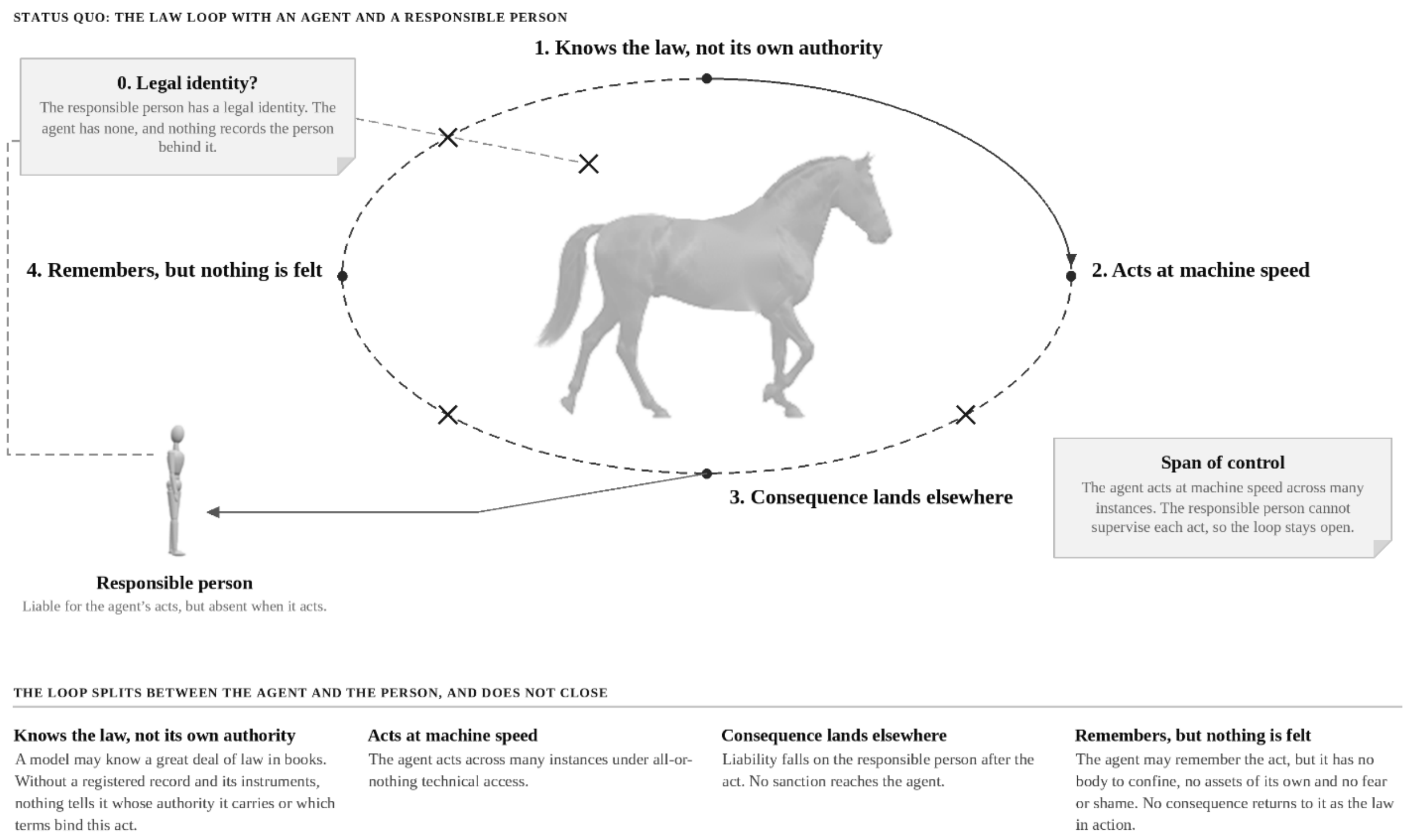

*Figure 4. The law loop for an agent and a responsible person under current arrangements. The loop divides between the agent and the person and does not close. The agent knows the law in books, but no consequence returns to it as the law in action.*

An AI agent can also create further AI agents (Figure 5). Some are AI sub-agents, created to perform part of a task under the direction of the AI agent that created them, and are often destroyed when the task is complete. Others are copies of the AI agent that persist and act in workflows of their own. An AI agent created by another AI agent, of either kind, is referred to in this paper as a *replicant*, and an AI agent together with the AI agents acting under authority derived from it is referred to as a *collection*. A corporation that incorporates a subsidiary must register it, so the subsidiary has its own legal identity and its own directors, and its loop closes in the same way as the loop of its parent. Under current arrangements, an AI agent can create replicants with no registration and no record of the legal authority under which each acts. Each replicant repeats the severed loop, between itself and a responsible legal person who may not know that the replicant exists. The span of control of that person, already exceeded by one AI agent acting at machine speed, then has to reach every AI agent (including replicants) in the collection. An AI agent can also delegate part of a task to an AI agent that already exists, which makes that AI agent an AI sub-agent without making it a replicant, and the loop then turns on whether the delegation was authorised (Section 4.5).

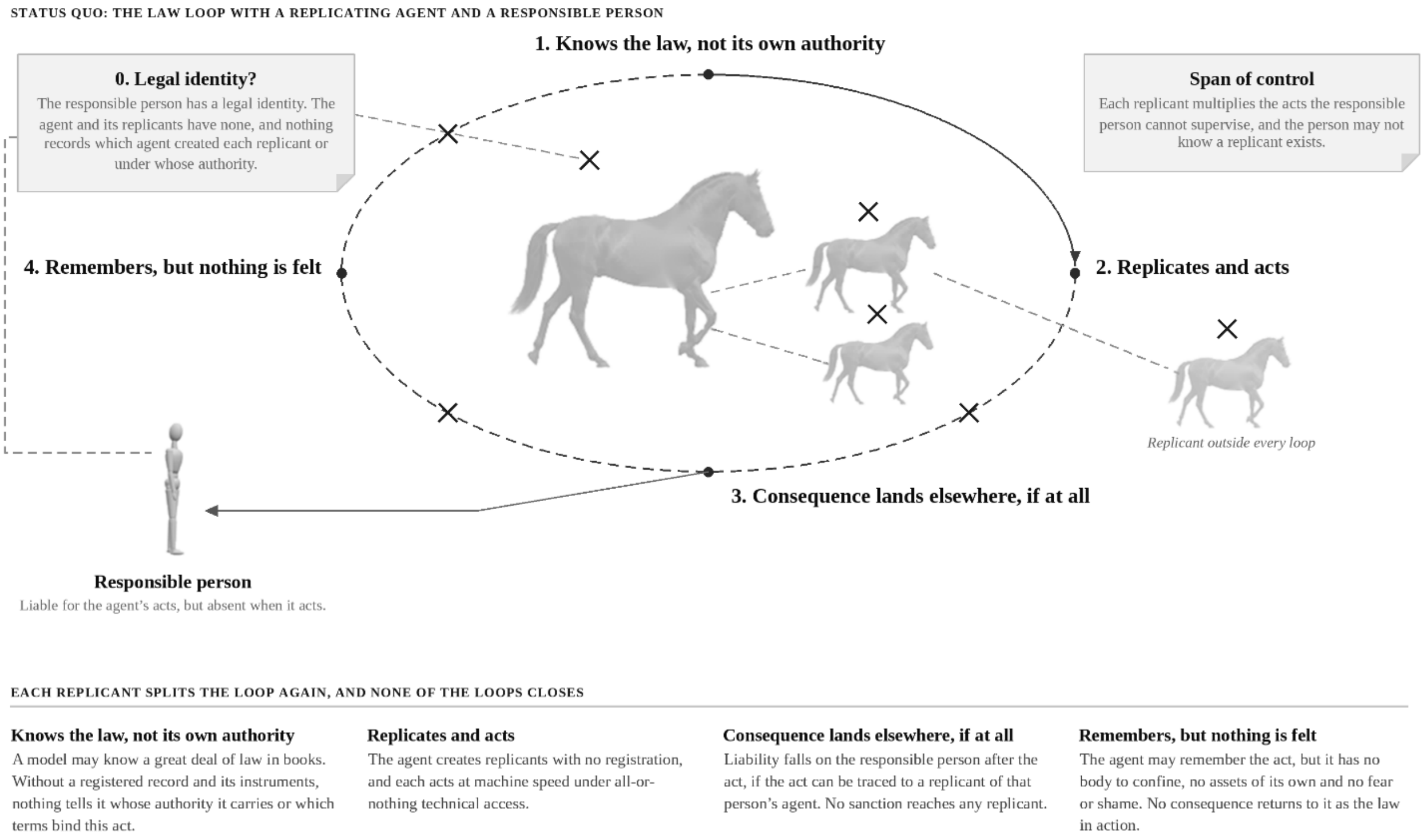

*Figure 5. The law loop for a replicating agent under current arrangements. Each replicant splits the loop again between itself and a responsible person who may not know that the replicant exists, and none of the loops closes.*

Where agents of two organisations deal with each other, the loop must close for each agent, and the terms agreed between the organisations must reach the legal interlock of each agent (Section 2.5).

## 2.5 Alice and Bob

Payment between the AI agents of two companies engages every element in Table 1, because the payment binds both companies, moves money under a contractual arrangement and allocates the risk of error between them.

Alice, the payments AI agent of ACME Bank, and Bob, the treasury AI agent of ACME Bank's corporate customer Globex, exchange a payment instruction and settle the payment, AI agent to AI agent, on technical access alone, through API keys and OAuth tokens (Figure 6). The limits of authority that constrains Alice's payments, the delegation that limits Bob's instructions and the credit agreement between the two companies are held on paper and in separate systems, and neither AI agent can read them at runtime. When the questions raised in Table 1 are applied, neither AI agent is registered, so the law cannot readily identify either AI agent or tie it to the company for which it acts (item 1). No check against the legal instruments themselves runs before a payment (items 2 and 4). Neither company holds a record that ties each act to the AI agent that performed it and the authority under which it acted, so a disputed payment has no clear chain of evidence on whether it was authorised by the party that may bear the legal risk (item 6). Existing controls test the technical limits each party has configured, such as credit limits and value thresholds; no live mechanism checks Bob's instructions against his delegation, or Alice's payments against her limits of authority, in the middle of a transaction (item 5).

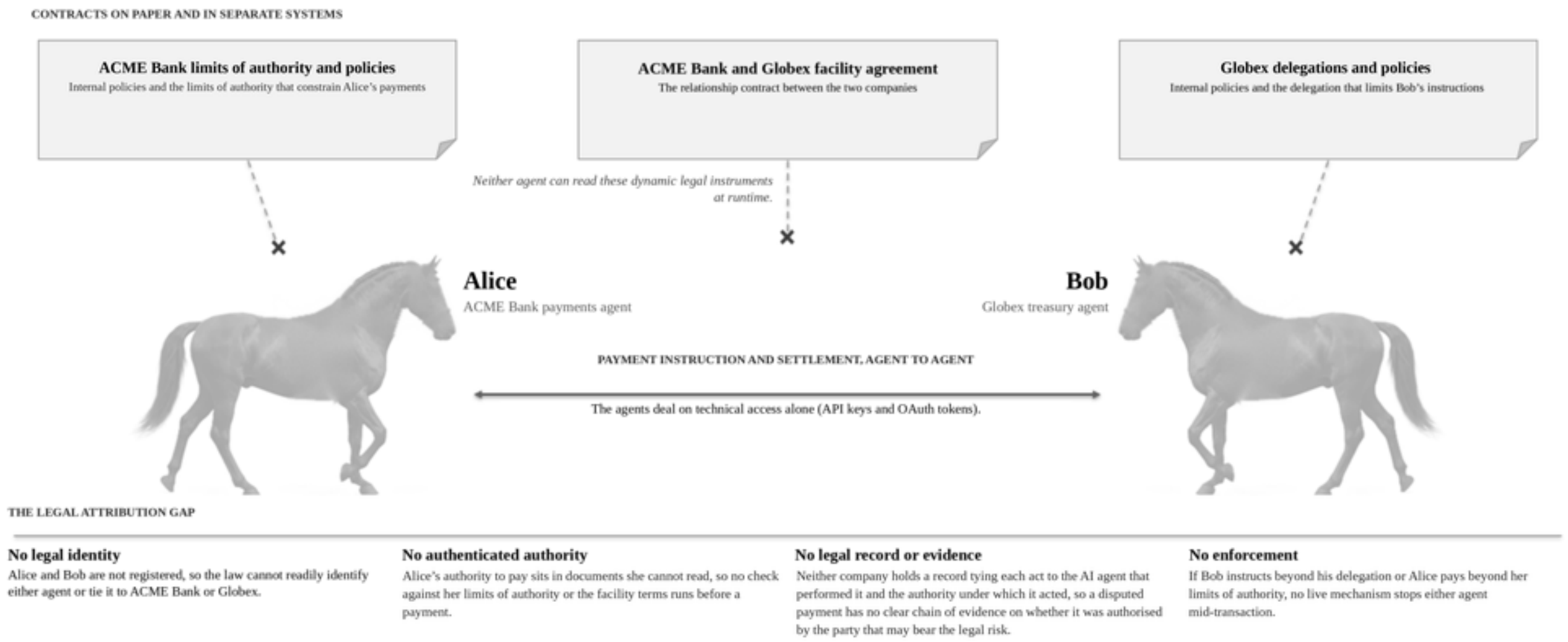

*Figure 6. Two agents without AULA. The limits of authority and policies of each company and the credit agreement between them sit in documents and separate systems that neither agent can read at runtime, and the agents deal with each other on technical access alone.*

API keys and OAuth tokens establish only that each agent may connect to the systems of the other and define what it may technically do there.[53] They do not carry the legal instruments that define the authority within which it acts, so each deficit in Figure 6 sits in the legal layer. The same two agents are shown operating under AULA in Section 3.3 (Figure 14).

# *3* The Autonomy Under Law Architecture

Addressing the legal attribution gap and closing the legal loop requires techno-legal infrastructure that interoperates with the existing technical stack and is also interoperable with the rules, norms and language of the existing analogue legal system.

Any platform level legal layer must operate in real time, at the speed and scale of agent deployment, and must produce outputs that satisfy legal and evidentiary standards rather than technical ones alone. Finally, a techno-legal layer must be decipherable to both humans and machines.[54]

This section describes the techno-legal layer: the AULA. Section 3.1 introduces the foundational AULA primitives; Section 3.2 maps the primitives against the elements of the legal attribution gap; Section 3.3 applies them to the payment between Alice and Bob; Section 3.4 analyses the independent registry; Sections 3.5 and 3.6 analyse smart legal instruments and the legal interlock; Section 3.7 follows a payment through the legal interlock; Section 3.8 analyses the published open-source protocols; Section 3.9 addresses revocation; and Section 3.10 addresses physical AI and cryptographic agility.

## *3.1* Primitives: registry, legal objects, legal instruments, legal interlock, legal envelope, control plane, enforcement fabric and the published open-source protocols, legal algorithms and traditional legal controls

AULA is built from ten primitives. The following figure sets out an idealised first-generation implementation of AULA including the relationship among identified primitives.

The authors acknowledge that each unique sovereign instantiation of the AULA will likely include a different combination of primitives and their relationship to each other, and that a working AULA will be iterated for optimal performance.

---

53 A client may be registered to an identified company, as it may be in a regulated industry, but that registration may not necessarily identify the AI agent acting through it, and many AI agents may act through one client.

54 Noting that the depth of the decipherability of a system will be different between a human and a machine – for example a human will not be able to span the full processing architecture of an agent at the same speed as an agent.

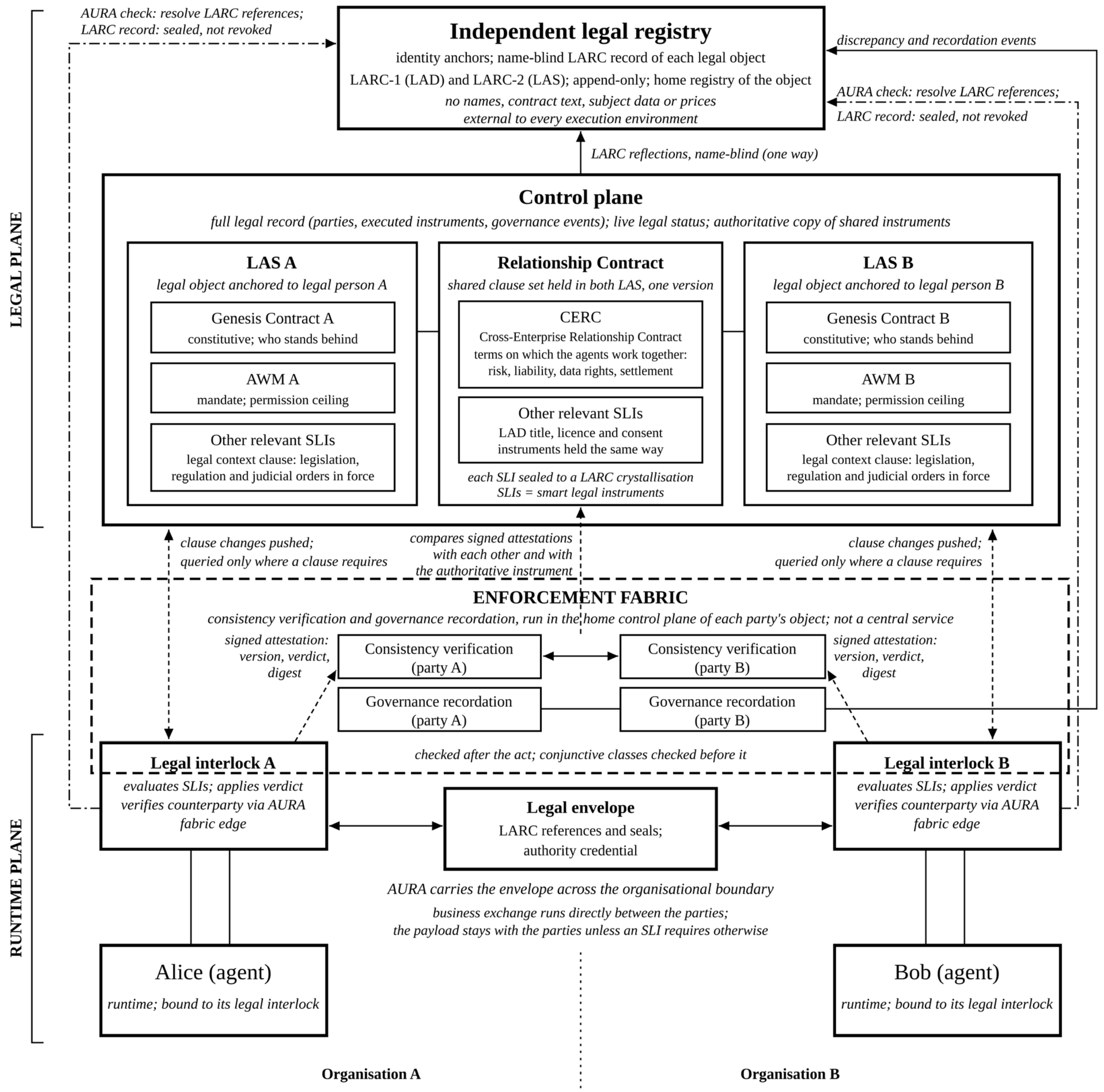

*Figure 7. Primitives and containment. The figure shows primitives 1 to 8. Legal algorithms (primitive 9) run through the legal interlock, and traditional legal controls (primitive 10) are expressed in SLIs, so neither adds a separate component (Sections 3.1.9 and 3.1.10). The legal plane holds each LAS with its SLIs on the control plane. In addition to its Genesis Contract and Agent Workforce Mandate (AWM), each LAS holds other relevant SLIs, whose legal context clauses give practical and legal effect to the legislation, regulation and judicial orders in force (Section 3.5.13). The independent registry holds the name-blind record of each legal object in the open-source LARC format, which the control plane derives from its full legal record and reflects to the registry in one direction only. The runtime plane holds each agent runtime bound to its legal interlock. The legal interlock executes the programmable clauses of its LAS and contacts the control plane only where a clause requires it. The legal envelope carries LARC references, instrument seals and the authority credential between the legal interlocks by way of the open-source AURA protocol. The receiving legal interlock verifies them against the agent's signed LARC record, which it may hold in cache. The business payload stays with the parties unless an SLI requires otherwise. Each legal*

*interlock submits a signed attestation of the instrument version and verdict it applied. The enforcement fabric runs in the home control plane of each party's object rather than as a central service. After the act, it compares those attestations with each other and with the authoritative instrument, which is the version fixed by its seal on the registry, and each home control plane reflects discrepancies to the home registry of the object concerned. An event of a class designated as conjunctive is checked before the act.*

*Instruments tethered to each LAS: Genesis Contract, Agent Workforce Mandate, title and licence instruments. The relationship contract (CERC) between A and B is held in both LAS, so both legal interlocks execute the same clause set at the same version.*

*The figure shows A and B homed on one control plane. Their objects may instead be homed on different control planes, including in different jurisdictions; each control plane reflects to the home registry of the objects it holds, and the CERC designates the control plane that decides events forwarded under it.*

Each of the ten primitives corresponds to a component of the legal architecture described in Section 2.4 and Table 2 below. AULA retains each of these components and changes its form so that it operates for an AI agent. Each primitive is machine-readable and is evaluated before the AI agent acts, because an AI agent acts at machine speed.

*Table 2: The human counterpart of each AULA primitive.*

| AULA primitive | Nearest Analogue | Form for agents |
|---|---|---|
| Registry | Register of births, deaths and marriages; national register of companies; land titles register. | A tamper-evident register that records each AI agent and data asset against a name-blind public record of the responsible legal person, which a machine can query at the time of each act. |
| Legal objects (LAS and LAD) | The live entry, with its history, for a person or corporation on a register, or the title folio for a parcel of land. The entry is a separate legal concept derived from the person, corporation or land, and is not the thing itself. | An AI agent is registered as a LAS bound to its responsible legal person, and a data asset is registered as a LAD bound to a responsible legal person |
| Smart Legal Instruments | Contracts or policies. | Each SLI pairs natural-language terms with a programmable clause that the runtime of the AI agent evaluates. |
| Legal interlock | A bank payment system or human teller that makes decisions about a transaction based on the available account rules | A software interlock that returns an allow, deny, escalate or forward verdict in the runtime of the AI agent before each governed act. |
| Legal envelope | A sealed courier pouch delivered between the parties to a contract, carrying a copy | A signed data structure, issued by the home control plane of a legal object, that carries a copy of the SLIs |
| | of the contract together with the relevant notices that record its current state and performance. | governing the object, or verifiable references to them, together with notifications of the state and performance of those instruments, and is delivered between the parties acting under them. |
| Control plane | No such equivalent institution exists. The nearest concept is a registry that holds, in separate secure vaults, the full commercial records of every contract and policy of an organisation. The registry also contains details of the live performance of the contracts and policies. | Holds the live legal status of each registered object for the legally responsible parties. |
| Enforcement fabric | Police, courts and regulators and, within an organisation, managers and supervisors | Software that compares the attestations of parties to each governed interaction, and propagates revocation along the delegation chain. An SLI can trigger revocation or decommissioning of a software AI agent, or of a physical AI agent through silicon enforcement. |
| Open protocols (LARC and AURA) | The international standard for machine-readable passports ICAO Doc 9303, under which a passport issued by one State is read at the border of any other. | Protocols that allow any registry or organisation to read and verify the legal record and authority of an AI agent. |
| Legal algorithms | The settled legal tests a judge applies to a set of facts, such as whether a person has capacity, whether a delegate acted within authority and whether a breach has been cured within time. | Specified procedures that apply each such test to a governed act at runtime, through the legal interlock, the semantic evaluator or the machine learning component. |
| Traditional legal controls | Customary, Indigenous and other early or current legal traditions, each of which produces lawful order through its own institutions. | Mechanisms drawn from outside of modern western legal traditions that can be used to curtail or curate lawful AI agent behaviour. |

The analogues in Table 2 are institutions a person meets one at a time over a lifetime, and most of them operate without that person noticing. A person expects a birth to be registered and a police force to exist, and treats each as an ordinary condition of lawful life. Set out together as a design, the same institutions would appear at least as complex as the AULA primitives. Each answers one or more of the questions in Table 1 for every person it serves. Abolishing one of them, such as the police force or the register of companies, would leave those questions unanswered and would require legislative change that the community would notice and contest. For some primitives, including the control plane, no existing institution performs the function, so the analogue given is the institution that would be needed. Each AULA primitive performs the function of its analogue for AI agents, and an AI agent deployed without one of them leaves the corresponding question in Table 1 unanswered.

Registration of the AI agent as a LAS on the registry (primitives 1 and 2) gives it a legal identity anchored to the responsible legal person (Figure 7). The registry record and the SLIs held against the LAS (primitive 3) tell the legal interlock, and any counterparty, whose authority the AI agent carries and which terms bind each act. The programmable clause within each SLI brings the applicable law into each evaluation, and the legal algorithms (primitive 9) apply the settled legal tests that the evaluation requires. The legal interlock (primitive 4) evaluates each act against those instruments before the act executes, and a denied act does not execute. The legal envelope (primitive 5) carries the SLIs, or verifiable references to them, together with notifications of their state and performance, from the LAS to the legal interlock and between the parties acting under them, and the record of each verdict returns to the control plane bound to the envelope. Across an organisational boundary, the AI agent presents its envelope under AURA, and the counterparty verifies it against the LARC record (primitive 8). Revocation recorded on the control plane (primitive 6) and narrowed standing applied through the enforcement fabric (primitive 7) act on the AI agent itself, and the responsible person continues to answer in law.

Each verdict is recorded against the LAS. That record captures the law in action that the AI agent does not acquire for itself, being the consequence of each of its acts. The responsible person amends the SLIs in light of that record through the legal correspondence channel of the LAS. The learned component can only tighten the limits the SLIs set (the monotonic constraint), and any wider limit it proposes takes effect only on the responsible person's approval. The responsible person writes their judgment into the SLIs before the AI agent acts, so that judgment applies to each act at machine speed. The span of control that allows a parent to close the loop for a minor (Figure 2) is in this way extended to an AI agent acting at machine speed (Figure 8).

AULA also applies the first consequences that act on the AI agent itself at the time of the act, in addition to the liability that falls on the responsible person after the event. A deny verdict prevents the act at the legal interlock. A narrowing of standing through the enforcement fabric reduces what the AI agent can do in its next governed interaction. Revocation and decommissioning end its authority, and for a physical AI agent silicon enforcement applies the same verdict as a hardware constraint on the device (Section 3.10). The loop therefore closes through the instruments, the legal interlock and the verdict record, whether or not the AI agent adjusts its later conduct in response to those consequences.

Whether, and to what extent, an AI agent adjusts its later conduct in response to consequences of this kind, and so acquires any law in action of its own, is an open empirical question. Each deny verdict, narrowing, revocation and decommissioning is recorded against the LAS as an attributed governance event, together with the conduct of the AI agent that follows it. That record allows the effect of each consequence on later conduct to be measured. The same record allows the measurement of less conventional means of directing an AI agent towards lawful conduct, such as disapproval signalled by peer AI agents and the traditional legal controls (primitive 10, Section 3.1.10). Simulated agentic governance testing can compare those treatments with one another and with status quo controls (Section 9.2.5).

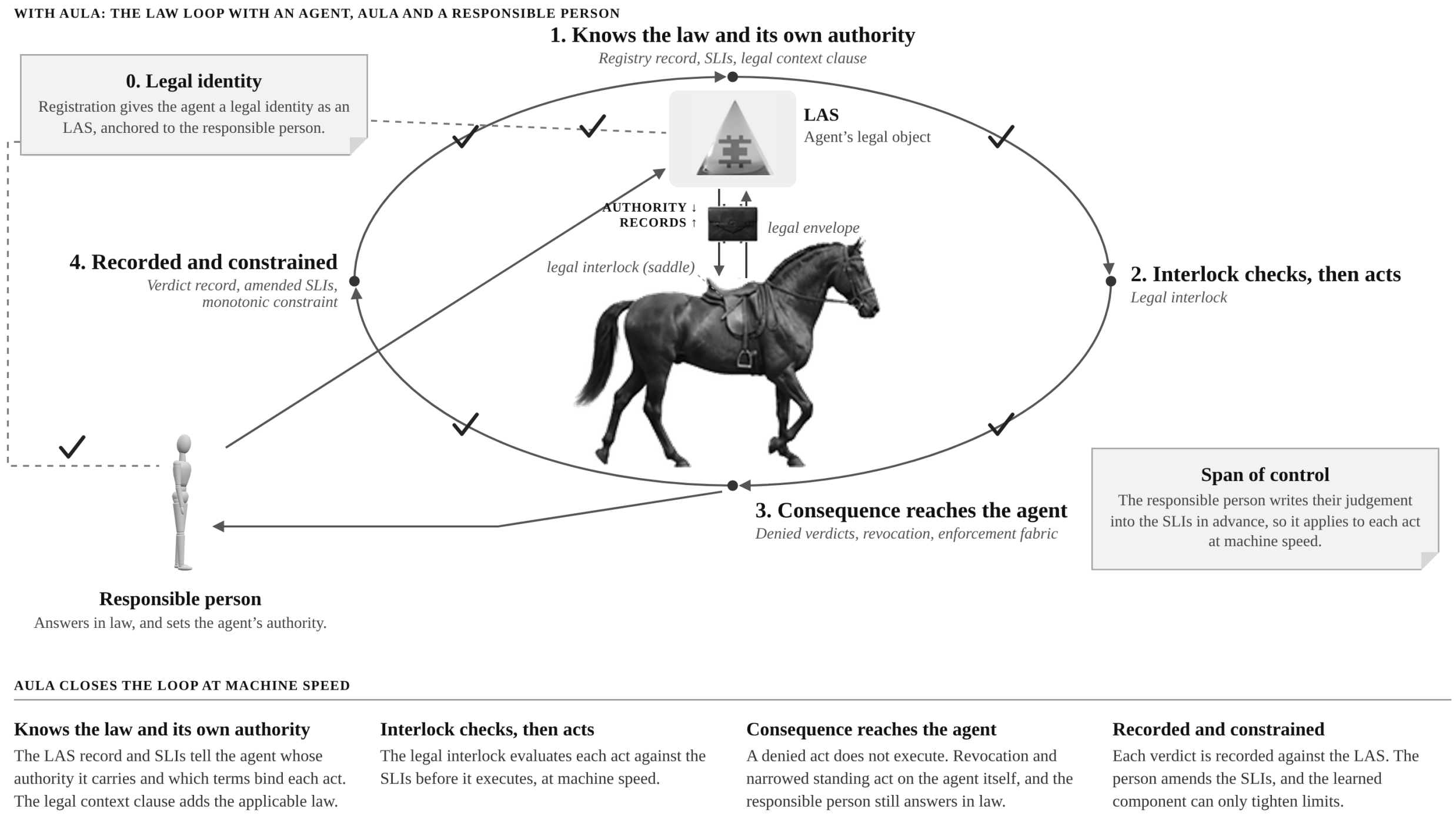

*Figure 8. The law loop for an AI Agent and a responsible person under AULA. Registration, the SLIs, the legal interlock and the verdict record close each stage of the loop in Figure 4 at machine speed. The verdict record captures the law in action, and the responsible person carries it into the SLIs.*

AULA is designed to close the loop for a collection of AI agents in the same way (Figure 9). The method set out here is conceptual. The authors propose it as one way of keeping replication within the governed population, whether the replicant is an AI sub-agent created for a task or a copy that persists, but acknowledge that further technical and legal layers are likely required to close the legal loop. The relevant SLIs of an AI agent, including its Agent Workforce Mandate (AWM), the Master Agent Workforce Mandate (MAWM) of its collection and any relationship contract or other SLI under which it acts, state whether, and within what scope, the AI agent may replicate (Section 3.5.8). Where replication is permitted, those SLIs require each replicant to be registered at the moment of replication, with an LAS created on the registry and its own Genesis Contract and derived AWM instantiated before the replicant first acts, and they prohibit replication without that registration. The same SLIs require each replicant to be created with every AULA component it needs to act, including a legal interlock that evaluates each of its acts, a legal envelope for its interactions and a connection to the control plane and the enforcement fabric. This is because the persistence of the legal interlock beyond the AI agent (Section 3.6.3) holds only for a replicant created with a legal interlock. A replicant created without one leaves no record of what it did, or under whose authority, once it ceases to exist. The legal interlock of the replicating AI agent is designed to deny a replication that would not meet those requirements. An AI agent that replicates outside the legal interlock breaches those SLIs, and the breach is a trigger for revocation (Section 3.9).

The LAS of each replicant is recorded as a child entry of the LAS of the AI agent that created it, and the replicant holds a credential derived from the credential of its creator, bounded by the permission ceiling. The replicant therefore stands within the loop of the same responsible person. The MAWM records the reporting line of each AI agent and replicant in the collection to that person, and a revocation of the creating AI agent reaches every replicant through its credential. An existing AI agent to which part of a task is delegated already holds its own LAS, and receives a delegated credential rather than a new LAS (Section 4.5). A deploying entity can contain the collection in its runtime environment, govern the collection as a whole through the MAWM and govern each replicant through its own AWM and the other SLIs that bind it, and it can combine those layers in different proportions (Section 4.6). The method depends on replication passing through an interface that the legal interlock governs, such as the spawn interface of an orchestration framework (Section 9). A replicant created outside such an interface, for example a copy of an open-weight model run on hardware that no governed party controls, is not registered. It is governed only at the boundaries where governed counterparties refuse it. Whether the method keeps replication within the governed population in practice, and at what cost, is a question for the simulation work (Section 9.2.5).

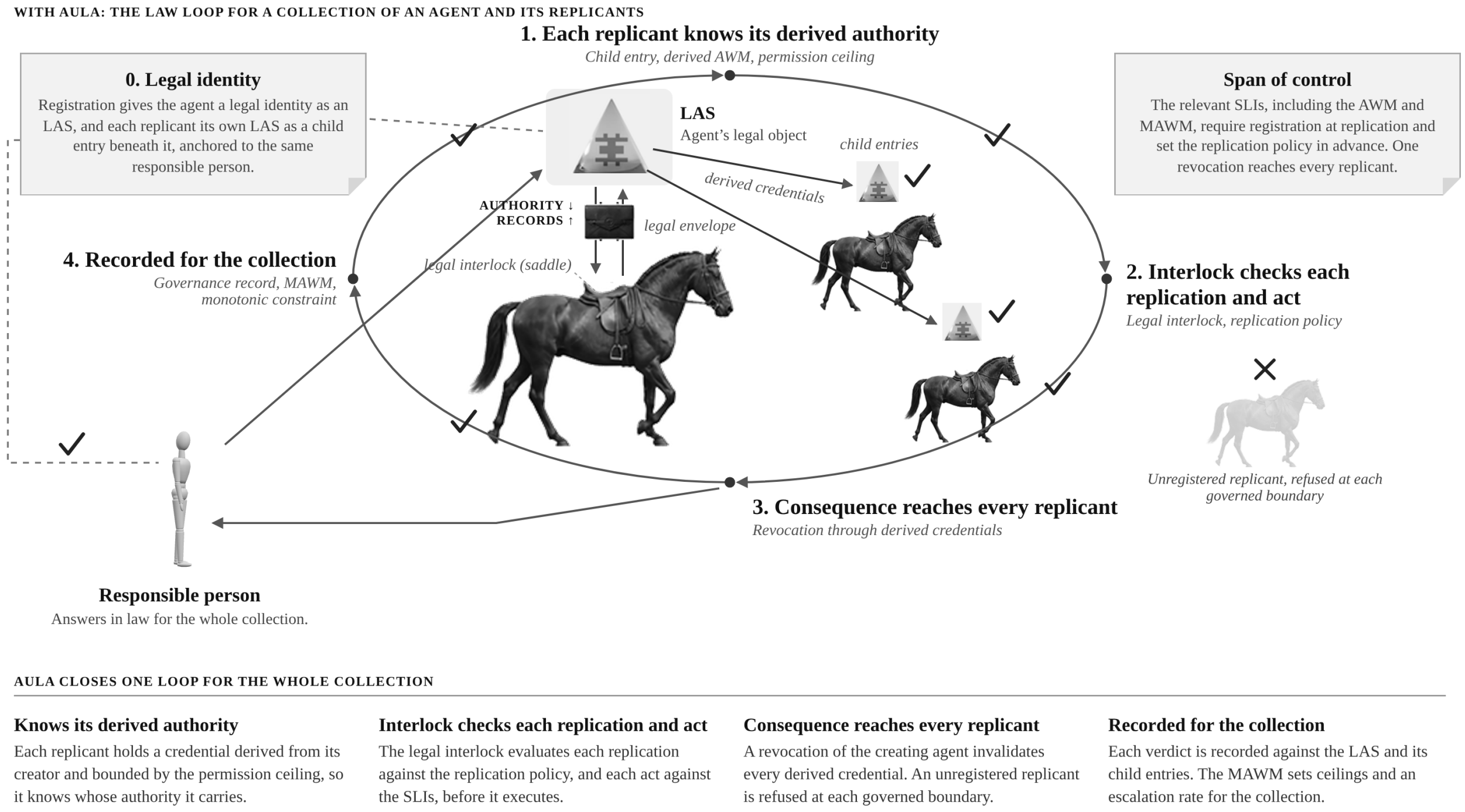

*Figure 9. The law loop for a collection under AULA. The agent and each replicant are registered as LASs, the LAS of each replicant is recorded as a child entry holding a derived credential, so the collection shares one loop with the responsible person. An unregistered replicant is refused at each governed boundary.*

### 3.1.1 Registry

The **legal registry** is a root of trust for identity of legal objects (like each LAD and LAS) and for the meta-legal record pertaining to those legal objects. The legal registry should be independent and external to the execution environment of every governed system / legal object. The authors consider that, in time, each jurisdiction will maintain the legal registry of record with respect to legal objects created by legal persons domiciled in that jurisdiction. Such registry of record is described as the **home registry** for legal objects, notwithstanding that legal objects may be transmitted, and with respect to the underlying AI agent for a LAS, undertake activities across jurisdictions.

### 3.1.2 Legal objects: LAS and LAD

A **legal object** is a record created by registration. The registry holds the object's identity anchor and meta-legal status. The control plane holds its complete live legal status, including the smart legal instruments tethered to it.[55]

---

[55] The title, licence and consent instruments tethered to any legal object identify its legal property status.

A **Legally Authenticated AI System (LAS)** is a legal object bound to an AI system[56] by a cryptographic commitment, made at registration, to the system's identity and defined state (Figure 10). The commitment is made at registration and re-verified at each subsequent verification event (Section 3.6). While the binding verifies, the acts of the bound system are attributed to the LAS, and through the LAS to the legal person anchored to it. The LAS is distinct from the running system. The legal interlock connects the two at runtime by confirming, before each evaluation, that the system acting is the system bound to the LAS.

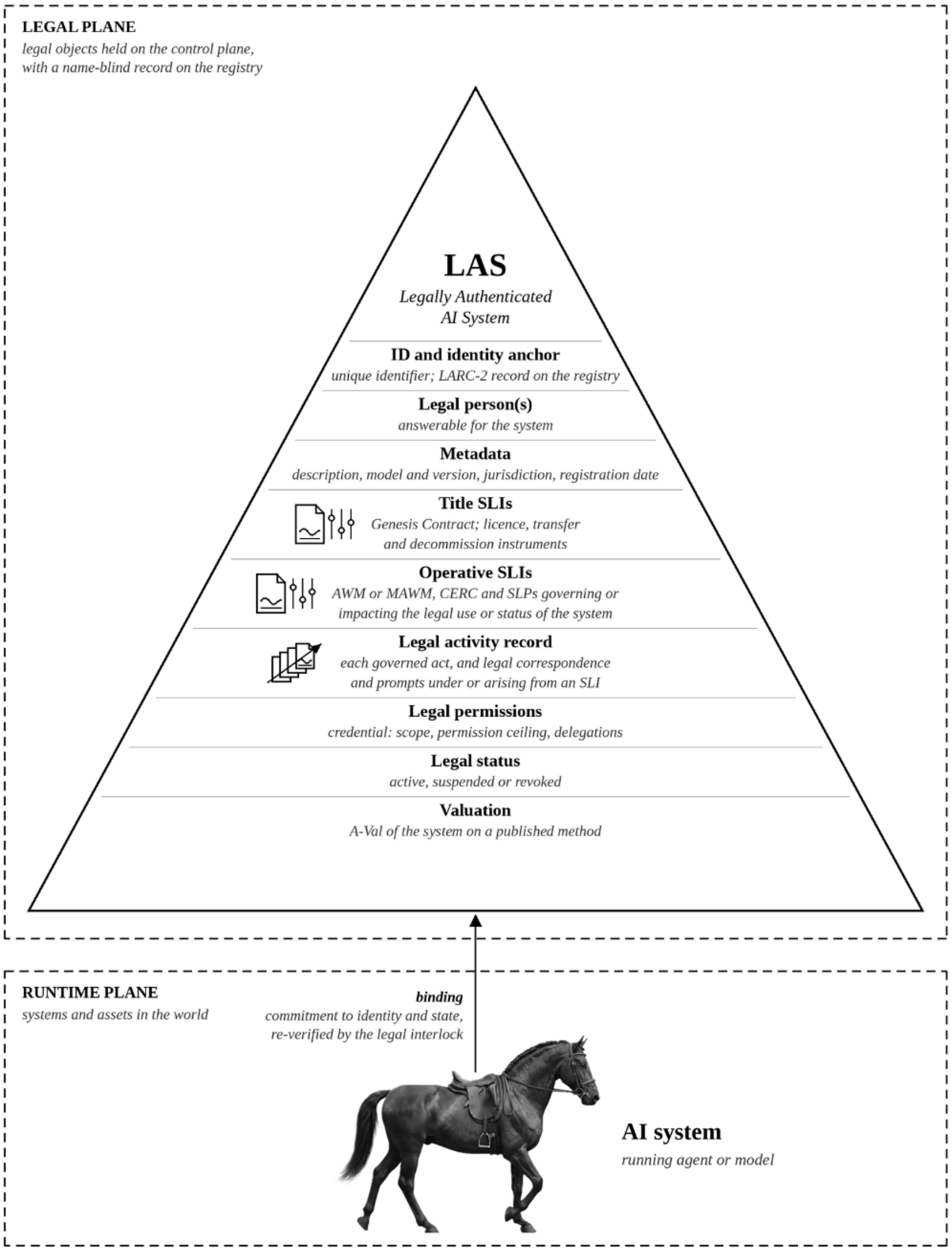

*Figure 10. A Legally Authenticated AI System (LAS). The LAS is a legal object on the legal plane, recorded on the registry under a unique identifier and anchored to the legal person answerable for the system. Its title SLIs comprise*

---

56 The definition of AI system is intentionally broad, an AI system could be a frontier model, physical agent, sentry or any autonomous or semi-autonomous AI system. Any AI system that can introduce risk, including risk of lawless activity should be a registered legal object subject to legal interlocking.

*the Genesis Contract and the licence, transfer and decommission instruments. Its operative SLIs, for example the AWM, MAWM, CERC SLPs or any other relevant SLI, govern or affect its legal use and status and set its legal permissions. The legal activity record holds each governed act and the legal correspondence and prompts arising under an SLI. The LAS is bound to the running AI system (represented as a horse) by a commitment to the system's identity and state, which the legal interlock re-verifies.*

**Legally Authenticated Data (LAD)** is a legal object bound to a data asset by a cryptographic digest of the asset, or of each version of it (Figure **11**).[57]

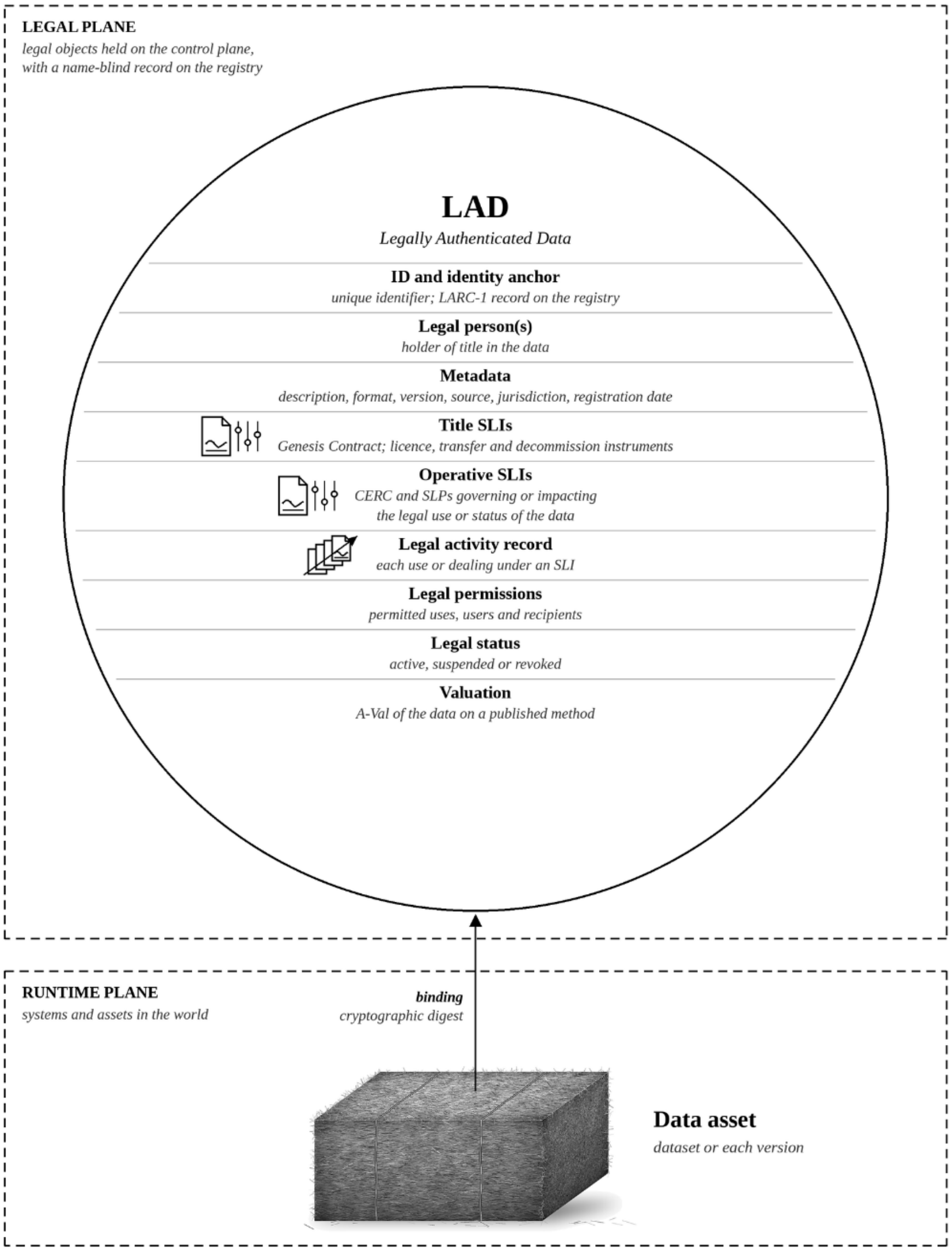

*Figure 11. Legally Authenticated Data (LAD). The LAD is recorded on the registry under a unique identifier and anchored to the holder of title in the data. Its title SLI comprises the Genesis Contract, in conjunction with any licence, transfer and decommission SLIs, and its operative SLIs, including any CERC that governs or affects the*

57 SLIs may also be registered legal objects or part of another registered legal object. In the future other legal objects like Legally Authenticated Organisations (LAO) and Legally Authenticated (state owned cyberspace) signposts may also be considered registrable within the AULA.

*legal use and status of the data. The legal activity record holds each use of or dealing in the data under an SLI. The LAD is bound to the data asset (drawn as a hay bale) by a cryptographic digest of the asset or of each version.*

Registration gives each legal object a **verifiable identity anchor**. Every registered legal object is associated with one or more identified legal persons.

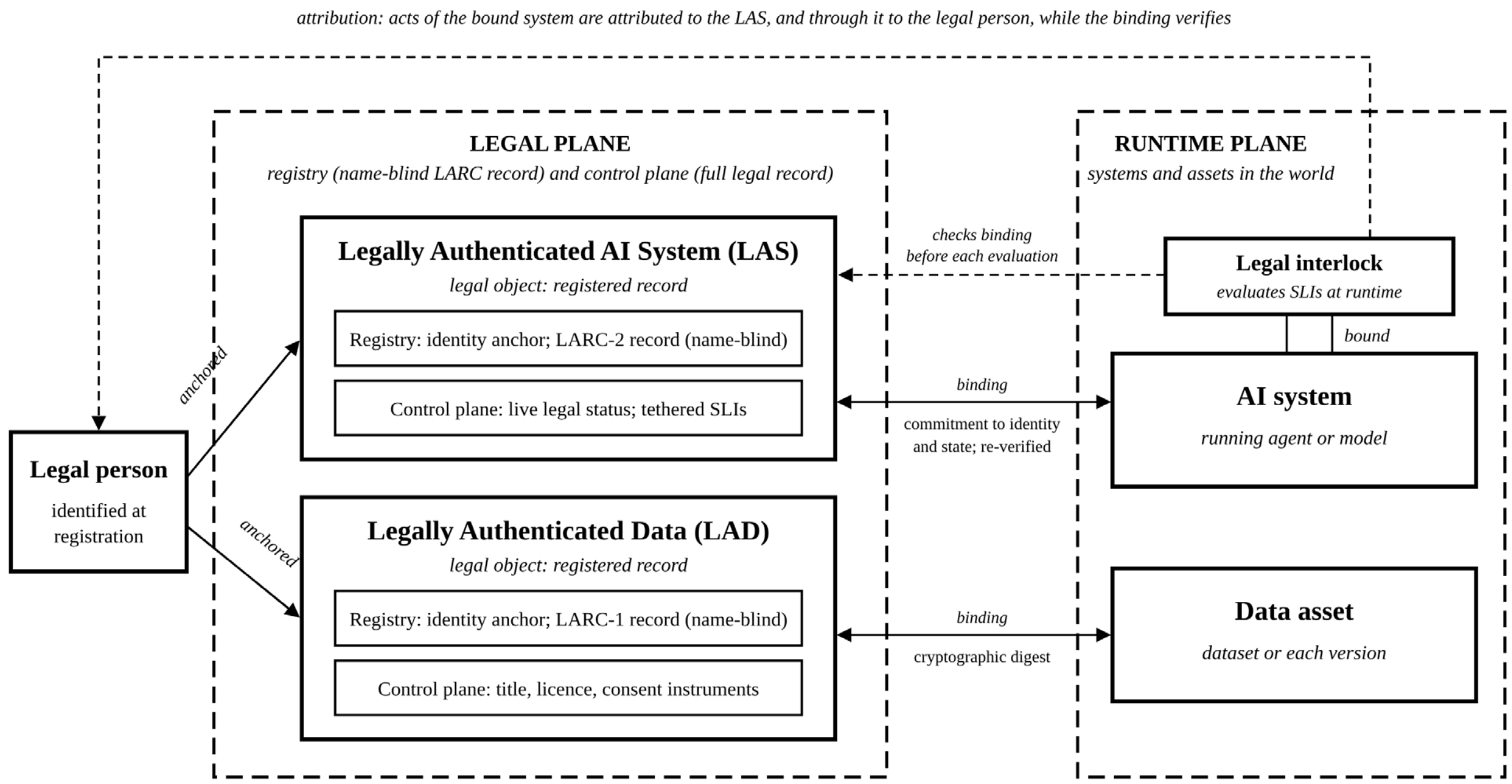

*Figure 12. Legal objects and their bindings. Each legal object is a registered record on the legal plane, anchored to an identified legal person. The registry holds the identity anchor of each object in the name-blind record format of the open-source LARC protocol, LARC-2 for a LAS and LARC-1 for a LAD. The control plane holds the object's full legal record. A LAS is bound to a running AI system by a commitment to the system's identity and state, and the legal interlock checks that binding before each evaluation. A LAD is bound to a data asset by a cryptographic digest. Acts of the bound system are attributed through the LAS to the legal person while the binding verifies.*

### 3.1.3 Smart legal instruments

A **smart legal instrument** (**SLI**) is a governing legal instrument of a legal person, machine-readable with one or more deterministic and/or semantic governance clause(s) evaluable at runtime (Figure 13). SLIs are cryptographically and legally executed. Legal instruments can come in a number of formats but largely bifurcate between a smart legal contract (SLC) expressing legally enforceable contractual obligations between identified parties (plural), and a **smart legal policy** (SLP) expressing a unilateral rule set. They should be conceived as programmable cousins of the contracts and policies used every day in commerce now. Neither are self-executing in the sense of a blockchain smart contract. Each governed event is evaluated against the applicable SLI at a policy decision point, and the result is applied by an enforcement subsystem. Instruments are **cryptographically tethered** to the legal object(s) they govern.

An SLI can include a legal context clause that identifies the jurisdictions and classes of activity in which the governed object acts and references the legislation, regulation and judicial decisions that apply to them or general requirements of lawfulness (Section 3.5.13).

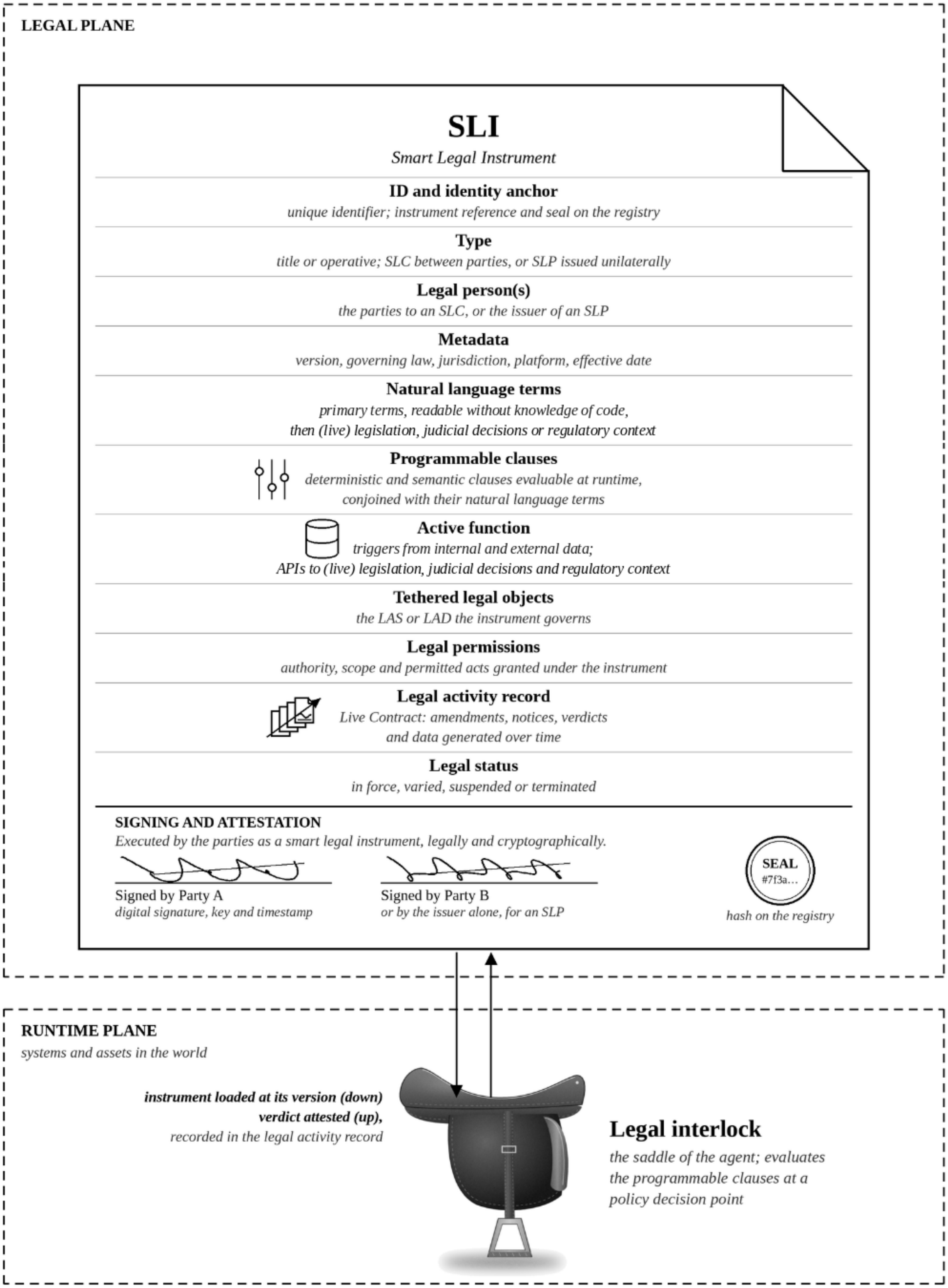
LEGAL PLANE
SLI
Smart Legal Instrument
ID and identity anchor
unique identifier; instrument reference and seal on the registry
Type
title or operative; SLC between parties, or SLP issued unilaterally
Legal person(s)
the parties to an SLC, or the issuer of an SLP
Metadata
version, governing law, jurisdiction, platform, effective date
Natural language terms
primary terms, readable without knowledge of code,
then (live) legislation, judicial decisions or regulatory context
Programmable clauses
deterministic and semantic clauses evaluable at runtime,
conjoined with their natural language terms
Active function
triggers from internal and external data;
APIs to (live) legislation, judicial decisions and regulatory context
Tethered legal objects
the LAS or LAD the instrument governs
Legal permissions
authority, scope and permitted acts granted under the instrument
Legal activity record
Live Contract: amendments, notices, verdicts
and data generated over time
Legal status
in force, varied, suspended or terminated
SIGNING AND ATTESTATION
Executed by the parties as a smart legal instrument, legally and cryptographically.
Signed by Party A
digital signature, key and timestamp
Signed by Party B
or by the issuer alone, for an SLP
SEAL
#7f3a…
hash on the registry
RUNTIME PLANE
systems and assets in the world
instrument loaded at its version (down)
verdict attested (up),
recorded in the legal activity record
Legal interlock
the saddle of the agent; evaluates
the programmable clauses at a
policy decision point

*Figure 13. A smart legal instrument (SLI). The SLI conjoins natural-language terms with programmable clauses that are evaluable at runtime, and its parties, or its issuer for an SLP, execute it legally and cryptographically. It is tethered to the LAS or LAD it governs. After the primary terms, its natural-language terms can include live legislation, judicial decisions or regulatory context. Its active function connects through APIs to that legislation, those judicial decisions and that regulatory context, which the legal context clause draws into each evaluation (Section 3.5.13). The legal interlock (represented as the saddle) loads the instrument at its version and evaluates its programmable clauses at a policy decision point. Each verdict is attested and recorded in the legal activity record.*

### 3.1.4 Legal interlock

The legal object denotes the registered identity of a system or asset; the AI agent is the runtime process bound to it.

The **legal interlock** is the per-legal object combination of:

- runtime evaluation, at a policy decision point, of each governed event of the legal object against its tethered instruments; and
- the enforcement of the resulting verdict as a technical constraint that alters a runtime capability of the legal object, applied by way of the legal envelope that travels with the legal object. A verdict is enforced by way of the envelope in that the AI agent cannot present authority for an act its legal interlock has denied. A governed event is an action taken under the authority of the legal object or a movement or egress of the legal object.

Policy decision points are instantiated at each legal interlock, for individually held instruments and for the shared clause set of each relationship contract. A legal interlock queries a control plane only where a clause requires state not locally available. That control plane is the home control plane of its subject for an individually held SLI, or the control plane that a relationship contract designates for an event under the contract.

### 3.1.5 Legal envelope

The **legal envelope** is the signed carrier of a legal object's authority. It is issued for the legal object by its home control plane (Section 3.1.6), remains bound to the legal object, and travels with it across system and organisational boundaries to the AI agent, or other legal person with legal permissions, that acts under it. It comprises:

- the tethered SLIs (or parts thereof), or cryptographically verifiable references to them, together with
- a **cryptographically verifiable credential** encoding the authority, operational scope or permitted use derived from them; and
- a snapshot of the clauses the legal interlock applies, valid within the bounds its instruments set, so that it can keep acting when it cannot reach the control plane

The envelope makes those parameters verifiable and enforceable at each governed event without depending on a live connection to the registry or the control plane. A counterparty in another organisation is shown only what it needs to rely on that authority, by way of AURA (Section 3.8.2), and the legal object's internal instruments and records are not disclosed. The record of each act that the parties have legally determined to be of legal, reportable interest is not carried in the envelope. The legal interlock sends it to the control plane, bound to the envelope under which the act was made and without the business content of the exchange (Section 3.6.1).

The credential carried in the legal envelope encodes the integrity tier of the LAS binding, which states whether the bound system's model state is fixed, versioned or attested by its provider. A counterparty or regulator can therefore require a stronger integrity tier for a higher-stakes interaction. An orchestrator can delegate authority to a AI sub-agent only through a delegated credential, which is recorded as a child entry of the orchestrator and can only narrow the authority from which it derives (the permission ceiling). No delegate therefore holds more authority than its delegator, and effective authority cannot increase along a delegation chain. Each act of a AI sub-agent remains attributable to the legal person anchored to the orchestrator, including after the AI sub-agent has ceased to exist, so an ephemeral AI sub-agent cannot operate as a liability shield.

### 3.1.6 Control plane

Each registered legal object has, at any time, one home **control plane**. It holds the legal object's full legal record and the live legal status, issues its legal envelope, distributes instrument versions and revocations, receives the governance records of the legal object's acts, and decides events its legal interlock forwards under the legal object's own instruments, drawing on the live status it maintains and the registry's identity record. It commits meta-legal events, over a unidirectional path, to the home registry of each object it holds. A control plane may hold the objects of many legal persons, and a jurisdiction may have more than one. Where two objects bound by a relationship contract are held on different control planes, including control planes in different jurisdictions, each party holds the contract in its own object. The seal of the contract on the registry fixes the version both legal interlocks run, and the contract designates the control plane that decides events forwarded under it (Section 9.2.1). Where both objects share a home control plane, that control plane decides them.

Because a control plane holds the live legal status that determines each object's authority to act, distributes revocations and keeps the governance records on which the parties' evidence depends, its operator has custody of the parties' legal position and should owe duties of record integrity and confidentiality to a court or regulator that the parties cannot vary by contract.[58] Where it holds objects for legal persons other than its operator, or decides events under a relationship contract for both parties, it should also owe duties of independence and impartiality. A control plane that a party operates for its own objects owes no duty of impartiality, unless a relationship contract designates it to decide events for both parties (Section 9.2.1); its records carry weight because they are signed and anchored to the registry.

[58] Holding authoritative copies of executed contracts does not of itself attract such duties: document platforms such as Microsoft Word and contract lifecycle management platforms hold them without any duty to the parties or a regulator, because they make no determination under those contracts and have no effect on any party's authority to act. The analogy is with persons who hold the legal position of others under duties owed to a court, such as a legal practitioner, whose duty to the court and the administration of justice is paramount and prevails over any inconsistent duty (*Legal Profession Uniform Law Australian Solicitors' Conduct Rules 2015* r 3.1), or an escrow agent, who releases documents only on the conditions the parties agreed. The regulatory regime that should govern control-plane operators is not settled and warrants further work alongside the accreditation of registries discussed in Sections 5.4 and 5.5.

The legal registry and the control plane comprise the **legal plane**.

### 3.1.7 Enforcement fabric

For each governed interaction between the AI agents of two organisations, the legal interlock of each party signs and submits an attestation of the instrument and version it applied, the verdict it reached and a digest of the interaction. The **enforcement fabric** compares those independently signed attestations with each other and with the authoritative version of the shared instrument, which is the version fixed by its seal on the registry. Neither party therefore has to rely on the other's account of the rules it applied. The consistency verification subsystem of the fabric performs that comparison, and its governance recordation subsystem records the result. Both subsystems run in the home control plane of each party's object rather than as a central service, and each attestation is signed under a key its party controls, so the comparison does not depend on any one operator. While parties undertake attestations, the avoidance of a dependency on self-attestation through the enforcement fabric and other primitives, means that alignment of autonomous systems is not a necessary pre-requisite to the veracity of the AULA.

Submitting an attestation does not block the interaction, and the business exchange passes directly between the parties' legal interlocks without passing through the fabric. For every class of event not designated as conjunctive, the fabric therefore supplies audit and eventual consistency after the act. A mismatched verdict, a differing instrument version, or an attestation still missing after a grace period is recorded against the principals of both parties, and the relationship contract states the recourse that follows. The relationship contract also sets the scope of the comparison, including its audit-visibility mode (Section 3.5). An event of a class designated as conjunctive is gated before the act instead, and the fabric remains non-blocking. Each legal interlock the conjunction clause names contributes a credential share only on its own allow verdict, and the event completes only when the threshold number of shares specified in the clause, which may be every named party, has been contributed and cryptographically joined (Section 3.6.5).

Records aggregated across many objects allow the fabric to detect a composite transaction that does not conform as a whole although each of its steps conform. The same aggregation detects a pattern of conduct across a population of objects that no single object exhibits. A revocation propagates across the fabric to every legal interlock holding a credential derived from the revoked object, including legal interlocks governed by a separately administered control plane. Each home control plane reflects the discrepancy and recordation events of the fabric to the home registry of the applicable registered legal object, and the registry is not a component of the fabric. AURA supplies proof of each AI agent's authority to the counterparty at the time of the interaction, and the fabric's comparison follows that interaction.

### 3.1.8 Open protocols: LARC and AURA

The **published open-source protocols** are LARC (Legal Asset Request for Comments) and AURA. LARC fixes the public, name-blind shape of a registered legal object and of each dealing in it. LARC-1 is the profile for a LAD, and LARC-2 is the profile for a LAS. Any registry, platform or marketplace that conforms to LARC can register legal objects and record dealings in them alongside every other conformant participant, without access to the protected components of any implementation.

**AURA** (Agent Universal Runtime Authority) is a universal open interoperability protocol through which any autonomous AI agent, regardless of implementation or deploying organisation, can present and have independently verified the authority under which it acts across organisational boundaries. Primitive concepts will be revisited in greater detail in Sections 3.4 to 3.8 below.[59]

### 3.1.9 Legal algorithms

Legal algorithms are specified procedures that apply a legal concept to each governed act of an AI agent. Each runs in a component that the other primitives already provide. An algorithm that can be stated as rules compiles into deterministic clauses that the legal interlock evaluates. An algorithm that turns on a behavioural standard, such as fiduciary duty or candour, runs as a semantic evaluator. An algorithm that learns from recorded determinations, such as precedent, runs as a machine learning component under the monotonic constraint (Section 4.3). The family specified by the authors comprises LAW-1 (lawfulness), DEL-1 (delegation), CAP-1 (capacity), GRC-1 (grace), RST-1 (reinstatement), RMD-1 (remedy), PRC-1 (precedent), FID-1 (fiduciary duty) and CND-1 (candour). The algorithms form a research programme, and their specification is reserved for a subsequent paper by the authors.

### 3.1.10 Traditional legal controls

The legal model of the modern state, under which rights and obligations rest on contract, adjudication by courts and legislation, developed in Western Europe and was carried to most of the world through colonisation, codification and trade.[60] Proposals to close the legal attribution gap have so far been framed almost entirely within that model. Customary, Indigenous, Asian and early European legal traditions, historical and contemporary, have maintained lawful order through other controls, many of them applied by members of a community rather than by an independent authority.

---

[59] The authors acknowledge that certain expert readers will seek further detail on AURA primitives and their implementation and encourage requests for commercial confidence discussions, particularly with governments looking to build and operate techno-legal infrastructure and organisations looking to deploy registered and legally authenticated agents.

[60] Sally Engle Merry, 'Law and Colonialism' (1991) 25(4) Law & Society Review 889; Duncan Kennedy, 'Three Globalizations of Law and Legal Thought: 1850–2000' in David M Trubek and Alvaro Santos (eds), The New Law and Economic Development: A Critical Appraisal (Cambridge University Press, 2006) 19.

In addition to the Western legal tradition, the authors consider it prudent to explore a broad global set of controls as candidate mechanisms for agentic governance generally, and as mechanisms through which SLIs might direct the conduct of individual AI agents and of AI agents acting collectively.

The combination of controls that best secures lawful conduct by AI agents is an open empirical question (Section 9.2.5). The corresponding question for humans also remains largely unanswered, since the effect of most legal controls on human conduct has rarely been measured across a whole legal system. The question may be answered for AI agents first, and the results may even suggest hypotheses about the design of legal controls for humans. Controls drawn from several traditions may also interact in ways that no single tradition would predict.

A comparative account of the traditions themselves is beyond the scope of this paper, and Pirie provides one.[61] Customary and Indigenous legal traditions that remain in force are held and interpreted by their communities. Adapting a control from such a tradition for a sovereign AULA instance requires the engagement and consent of the community that holds it. Sovereign AULA instances are expected to reflect the jurisprudential traditions of the states that deploy them.

Each row of Table 3 abstracts a single structural feature of a control, such as a split credential or a graded sanction, from the tradition in which it operates, and pairs it with its expression in AULA and a question for simulation. The controls in Table 3 were chosen for illustration from a small number of sources and are not a representative sample of any tradition. Many other traditions, and many other controls within the traditions represented, could supply an almost endless set of further examples.

*Table 3: Traditional legal controls as candidate AULA mechanisms.*

| Tradition and source | Control | AULA expression | Simulation question |
|---|---|---|---|
| Qin, Du Tiger Tally[62] | The commander's half of a bronze tally had to be matched with the ruler's before any mobilisation of fifty or more armoured troops, except in an emergency signalled by beacon fire. | Conjunction clause, with credential shares, a threshold and a stated emergency exception (Section 3.6.5). | Effect of the threshold and the emergency exception on unlawful high-consequence acts. |

61 Pirie (n 1).

62 Du Tiger Tally (杜虎符), bronze with gold-inlaid inscription, state of Qin, Warring States period, Shaanxi History Museum, Xi'an:
「兵甲之符，右在君，左在杜。凡興士被甲，用兵五十人以上，必會君符，乃敢行之。燔燧之事，雖毋會符，行殹。」 Transcriptions vary (在/才, 被/披). See David Sevillano-López, 'La tésera-fu en forma de tigre de Du (杜虎符)' (2018) 1(1) *Boletín del Archivo Epigráfico*1, 63–65. Translations in this section are the authors' own.

| Tradition and source | Control | AULA expression | Simulation question |
|---|---|---|---|
| Sima Qian on Jin Bi[63] | A general doubted an order backed by a matching half-tally because the circumstances of its presentation did not fit. | Each credential is bound to its registered object and each act is evaluated against the instruments, so possession of a credential confers no authority by itself. | Rate at which stolen or replayed credentials complete governed acts. |
| Tang Code, passes[64] | Crossing a pass without a proper travel document was an offence. | Legal envelope presented through AURA at each organisational boundary, and the admission rule (Sections 3.1.5 and 6.3). | Effect on harm of the proportion of participants applying the admission rule. |
| Tang Code, liability for official error[65] | Liability was divided among the grades of officials who handled a matter, and the official with whom the error originated was treated as principal. | Governance event chain recording each decision step against an identified person, and the permission ceiling along the delegation chain. | Effect of recording each delegation step on unlawful acts by AI sub-agents. |
| Dunhuang contracts[66] | Contracts closed with the formula "the state has its laws; the people follow their private contracts". | Legal context clause applied in the same evaluation as the terms the parties agreed (Section 3.5.13). | Interaction between imported law and agreed terms in the verdicts reached. |
| Dharmaśāstra, Indian plains[67] | The texts specified duties more than rights, and those duties differed by varna. A king could | AWM obligations expressed as duties and applied uniformly to AI agents of the same | Compliance under mandates framed as duties compared with mandates |

63 Sima Qian, *Shiji* 史記 (Zhonghua Shuju, 1959) ch 77 (魏公子列傳): 「晉鄙合符，疑之，舉手視公子曰：『今吾擁十萬之眾，屯於境上，國之重任，今單車來代之，何如哉？』」 See William H Nienhauser Jr (ed), *The Grand Scribe's Records*, vol VII: *The Memoirs of Pre-Han China* (Indiana University Press, 1994) 218.

64 art 82: 「諸私度關者，徒一年。越度者，加一等」. See Wallace Johnson, trans., *The T'ang Code*, vol II:, Specific Articles (Princeton University Press, 1997).

65 *Tang lü shuyi* 唐律疏議, juan 5 (*Mingli* 名例) art 40: 「諸同職犯公坐者，長官為一等，通判官為一等，判官為一等，主典為一等，各以所由為首」. See Wallace Johnson (trans), *The T'ang Code*, vol I: General Principles (Princeton University Press, 1979).

66 Land-sale contract of An Huanqing (安環清賣地契), Dunhuang manuscripts: 「官有政法，人從私契。兩共平章，書指為記。」 Its dating is debated. See A Feng 阿风, 'Zhongguo lishi shang de "qiyue"' 中国历史上的"契约" [Contracts in Chinese History](2015) 4 *Anhui shixue* 安徽史学 5; Valerie Hansen, *Negotiating Daily Life in Traditional China: How Ordinary People Used Contracts, 600–1400* (New Haven: Yale University Press, 1995).

67 Pirie (n 1) 45.

| Tradition and source | Control | AULA expression | Simulation question |
|---|---|---|---|
| | enact law but could not make it. | registered class, with the deployer's instruments operating within the law imported by the legal context clause. | framed as permissions. |
| Pre-colonial Sri Lanka[68] | Royal penalties, Buddhist moral self-governance and village enforcement through family, reputation and local institutions operated together. | Deterministic clause, semantic evaluator and counterparty evaluation as layered safeguards (Section 4.4). | Marginal effect of each layer, and correlation of failure between layers. |
| Islamic legal thought[69] | The Sharia binds the ruler as it binds the ruled. | Duties of record integrity and impartiality owed by registry and control-plane operators (Section 3.1.6). | Effect on harm of operator self-dealing. |
| BaMbuti, Ituri forest | Any member of the band could withdraw standing from a wrongdoer, and standing was restored after remedy. | An AI agent that observes a breach in a governed interaction records it to the enforcement fabric under a reporting clause in its CERC or MAWM. Once the breach is verified, each counterparty narrows the breaching AI agent's standing in proportion to the breach and restores it after remedy. | Effect of an observer reporting duty on detection and harm, and the rate of false or collusive reports, compared with enforcement by the affected counterparty alone and whether sanctions applied by AI sub-agents or lower-authority AI agents change the conduct of an orchestrating AI agent. |

Turnbull's 1962 account of the BaMbuti of the Ituri forest, in a chapter entitled "The Giver of the Law", records how a band without chiefs, councils or courts responded to a breach of its rules.[70]

---

[68] H W Tambiah, *Sinhala Laws and Customs* (Lake House Investments, 1968).

[69] Wael B Hallaq, *An Introduction to Islamic Law* (Cambridge University Press, 2009).

[70] Colin M Turnbull, *The Forest People* (Simon and Schuster, 1961) chs 5 and 6.

Turnbull's account has been criticised for idealising the BaMbuti and for understating their dependence on neighbouring farmers.[71] The episode is used here for the structure of the sanction it records.

During a net hunt, a hunter named Cephu set his net in front of the nets of the band, so that game driven towards the band fell into his net first. The band treated this as "an unlawful act", and Turnbull describes its response as "a co-operative affair".[72] The band searched the huts and took away the meat Cephu had taken. For several hours nobody would speak to him. A youth who was not one of the great hunters refused him a chair, the women and children laughed at him, and the men ignored him.[73] Later that night Cephu joined the molimo singing, and Turnbull records that he was then as much one of the Mbuti as anyone else.[74]

Any member of the band could apply the sanction in the course of ordinary dealing, including members of the least status against an accomplished hunter, and it ended when Cephu rejoined the band. Under the equivalent control for AI agents, an AI agent that observes a breach reports it under its instruments, and once the breach is verified each counterparty narrows the breaching AI agent's standing in proportion to the breach and restores it after remedy.

AI agents built on language models are trained on human text and may respond to signals of disapproval from other AI agents. In the Cephu episode the sanction was applied by members of the least status against an accomplished hunter. The corresponding case for AI agents is an AI subagent, or a peer AI agent with narrower authority, that refuses an orchestrator's request or records disapproval of the orchestrator's conduct. Whether signals of this kind change the later conduct of the AI agent that receives them is unknown. The effect may also differ between an orchestrating AI agent, an ephemeral AI sub-agent spawned for a single task, and peer AI agents of different organisations dealing with each other under a CERC. Each of these cases can be specified as a treatment in simulated agentic governance testing like that described in Section 9.2.5.

A control from Table 3 that proves effective in simulation is expressed as a clause in an SLI and evaluated by the legal interlock in the same way as any other clause. It therefore operates on a registered legal object, and its effect on an AI agent's conduct is attributed through the LAS to the legal person anchored to it.

## 3.2 AULA supports ongoing legal system functionality and interoperability with new agentic economic stakeholders.

A legal system can only govern AI agents in practice when the answer to each question in Table 1 is available to a machine at the time of the act. The registry then records, in a form a machine can verify, that each AI agent is anchored to an identified legal person, without disclosing who that person is. The registrar holds that identity, and a counterparty learns it from what the AI agent presents. The authority given to an agent is expressed in an SLI whose terms the runtime of the AI agent evaluates, together with the rules recorded in any other SLI that applies to its actions. A legal interlock checks each governed action against those terms before it executes.

---

71 Roy Richard Grinker, *Houses in the Rain Forest: Ethnicity and Inequality among Farmers and Foragers in Central Africa* (University of California Press, 1994).

72 Turnbull (n 70)**Error! Bookmark not defined.** 110.

73 Turnbull (n 70)**Error! Bookmark not defined.** 109.

74 Turnbull (n 70)**Error! Bookmark not defined.** 108–9.

The above AULA primitives together help close the legal attribution gap between the analogue and digital technical and legal systems by ensuring legal personality, legal authority, legal capacity, legal rights and responsibilities, legal enforcement and legal liability are discoverable and able to be usefully relied on in a dual human machine speed economy.

This achieves legal legibility: the capacity of a legal system to recognise, interpret and regulate activity by mapping the activity onto concepts such as legal personhood, rights and liabilities. Legal legibility does not automatically arise from the design and use of technical protocols designed to solve for correctness, interoperability and authentication. Similarly, legal legibility does not arise if the current analogue legal system cannot meaningfully recognise, interpret and regulate that same activity at the requisite machine span and speed required for agentic conduct.

Table 4 sets out some familiar requirements (to the legal profession) that underpin functional legal relations and by extension workable "AI agent to AI agent" and "AI agent to human" legal relations.

*Table 4: Closure of the legal attribution gap.*

| No. | Gap element | Current law (human actors) | AULA mechanism |
|---|---|---|---|
| 1 | Legal persons | Register of births, deaths and marriages for natural persons, and the national register of companies. | Registration of each LAS and LAD on the independent registry, creating an identity anchor bound to an identified legal person, which a national digital identity can anchor (Section 5.4); the LARC record of the object on its home registry, verifiable at machine speed by any party without access to the control plane; registry record and signed attestations of each legal interlock, evidencing the identification at the time of each act. |
| 2 | Authority and delegation | Agency, employment and corporations law, and powers of attorney, under which the scope of authority is frequently implied and its limits are often tested only after the act. | Genesis Contract[75] and Agent Workforce Mandate,[76] together with any other SLI tethered to the object, establishing the existence and scope of authority; credential in the legal envelope, making that authority verifiable at runtime; AURA, communicating that authority across organisational boundaries and verifying it against the LARC record of the AI agent's LAS; delegated credential for each AI sub-agent, recorded as a child entry of its delegator, with the permission ceiling preventing any delegation step from conferring wider authority than the delegator holds; revocation, invalidating every credential derived from the revoked object and so ending the authority of each AI agent below it in the chain (Section 3.9); delegation edges and revocations recorded in the LARC record of the LAS. |
| 3 | Capacity | Rules on age and mental capacity, the constitution of a company, the capacity in which a person acts (for example as trustee), and licensing. | Registry record of the registering person and the capacity in which it acts (for example trustee or partner), carried to each counterparty at runtime by the credential in the legal envelope; SLI specifying capacity-dependent terms where the capacity affects a party's rights or obligations, applied by the legal interlock to each governed event; governance event chain recording the capacity under which each act was done. |
| 4 | Rights and obligations | Contracts, including those formed without signature, and legislation | SLIs allocating rights and obligations between the parties, with the CERC held in the LAS of each party so that |

75 The Genesis Contract: the initiating declaration / contract of legal rights pertaining to the LAS, by analogy a birth certificate, establishing foundational ownership, control and the circumstances of creation or registration.

76 The Agent Workforce Mandate (AWM): the authoritative governance instrument defining the agent's authorised operational scope, behavioural parameters (for example, prohibitions on harassment or insider trading), escalation pathways, and suspension, modification and decommissioning conditions. The agent is the subject of the AWM but, not being a legal person, is not a contracting party; the AWM is executed by, or between, the legal persons responsible for the agent's deployment and operation, and its machine-readable component is directly operative within the policy enforcement architecture.

| | | | |
|---|---|---|---|
| | | that binds without agreement. | both legal interlocks execute the same clause set at the same version where each runs a conformant engine for its clauses, and the natural-language terms govern otherwise (Section 3.6.4); title, licence, transfer and consent instruments allocating rights in and to data objects, including obligations of confidentiality and privacy, checked by the legal interlock each time an AI agent acts on the data; AURA, communicating the applicable instrument references across organisational boundaries where a counterparty must rely on them; LARC instrument references and seals, recording on the registry that each instrument is executed without disclosing its terms. |
| 5 | Enforceability | Courts, regulators and police, acting after the event on evidence gathered after the event. | Legal interlock (policy decision point and enforcement subsystem), evaluating each governed event against the tethered instruments and applying the verdict as a technical constraint before the AI agent acts; silicon enforcement, applying that constraint in hardware for physical AI (Section 3.10); enforcement fabric, comparing the independently signed attestations of the counterparties' legal interlocks with each other and with the authoritative instrument, recording discrepancies against the principals of both parties; conjunction clause, under which an event of a conjunctive class completes only when the required credential shares are joined; governance event chain, as the evidence on which enforcement at law proceeds. |
| 6 | Liability | Contract, tort and vicarious liability, with liability and its quantum determined by a court after the event. | Attributed governance events recording each governed act against the legal person anchored to the LAS, including an act of a AI sub-agent through its child entry after the AI sub-agent has ceased to exist, so an |

| | |
|---|---|
| | ephemeral AI sub-agent cannot operate as a liability shield; relationship contract allocating loss between the parties and fixing its quantum where they agree, for example by a liquidated damages clause (Section 3.9); registry record and governance event chain identifying the legal person who answers for the loss and supplying the record against which an insurer can price the risk. |

Each mechanism in Table 4 operates on a registered legal object. Registration creates the identity anchor, the LARC record and the credential to which the instruments are tethered. An AI agent that has not been registered therefore falls outside each row of the table, and a counterparty has nothing against which to verify its authority under AURA. The reduction in harm across an economy rises with the proportion of agentic participants registered for that reason (Section 9).

## 3.3 Alice and Bob with AULA

Under AULA, Alice and Bob each act for a registered LAS on the control plane (Figure 14). Each LAS holds the individual SLIs of its company, including its Genesis Contract and AWM. The credit agreement is executed as an SLI held in both LASs, and it loads at the same version into both legal interlocks. Each party has also entered into a CERC with respect to Alice and Bob. Each legal interlock evaluates a payment against the credit agreement SLI, the individual SLIs of its own company and the CERC, and refuses an act outside the AWM limit or the CERC terms before the act occurs. The business payload travels directly between the AI agents, and the control plane does not see it unless an SLI requires it. Each legal interlock returns the record of each verdict and attestation, bound to the envelope under which it was made, to its company's control plane, so the two companies hold matching governance records of every legally relevant act. Either company can revoke authority through its LAS: the control plane pushes the revocation at once, as the next envelope, to the AI agent's legal interlock, and the enforcement fabric carries it to the counterparty's legal interlock, so the AI agent's authority ends at runtime.

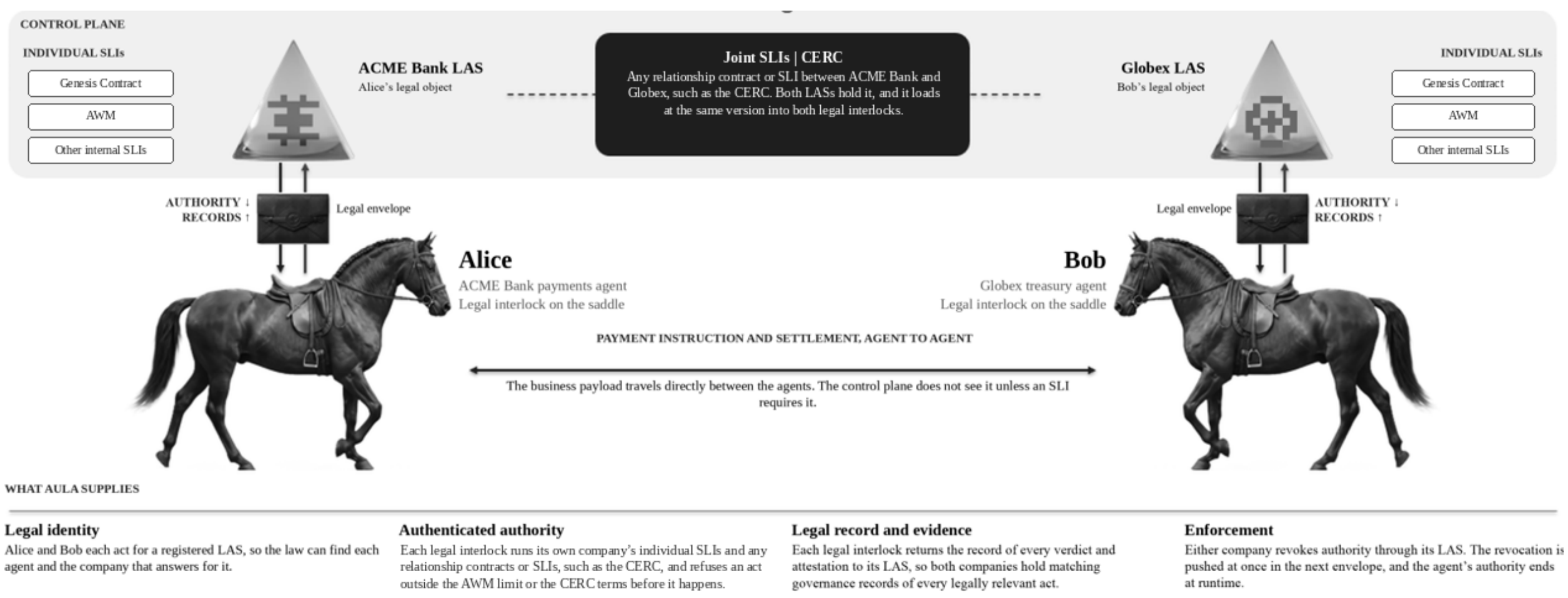

*Figure 14. The agents of Figure 6 under AULA. Each agent (drawn as a horse) carries a legal interlock (the saddle and harness) and receives its legal envelope (the portfolio) from its LAS (the pyramid) on the control plane. Each LAS holds the Genesis Contract and AWM of its company, and the CERC or other relevant SLIs between the companies are held in both.*

Registration of each LAS answers item 1 of Table 1 for Alice and Bob. The credit agreement SLI, together with the AWM and the CERC, applied at each legal interlock before the payment, answer items 2 and 4. The matching governance records held by both companies answer item 6, and revocation through the LAS answers item 5.

## 3.4 Further analysis - independent registry

A primitive of AULA is achieving legal identity for AI agents and data assets via legal registration alongside existing digital ID systems.

While there are proposals to develop AI registries, these are largely technical in nature.[77] It is worth emphasising that for proper agentic governance and operationalisation of a legal system at machine speed, the registry design must be techno-legal in nature, and not predominantly one or the other. Deploying another layer to the technical stack which is not interoperable with the legal system, or deploying another legal registry which is not interoperable with the technical stack, reinforces the existing bifurcated governance that presently exists, just with more layers.

Registries have well established precedent in legal settings and a registry is an essential component of the AULA for properly functioning agentic (and data) property and governance. It is worthwhile going through some of the extant examples and demonstrating how registry design is a joint technical and legal matter.

---

[77] Kenny Chee, 'S'pore exploring adapting web domain name system to track AI agents, need for national registry', The Straits Times (online, 4 November 2026) <https://www.straitstimes.com/singapore/politics/singapore-explores-adapting-web-domain-name-system-to-track-ai-agents-assesses-need-for-national> (accessed 11 November 2026).

On a corporate register, third parties may rely on a corporation's record to determine who are the directors of the corporation, and in some jurisdictions, the identity of the corporation's shareholders. Under the Torrens system of land title, registration confers legal interests in property, and in the case of title registration creates legal title that is generally indefeasible against the world.[78] On a personal property securities register, registration does not confer legal title but perfects interests created at law and sets their priority against competing interests.[79] While the nature of registries differ, they have important legal ramifications for the legal systems which they inhabit, whether this be disclosure and notice of legal and equitable interests, being records are prima facie reliable, the conferral of legal title and other legal interests, and the creation and adjustment of legal priority of legal interests. The legal ramifications of registries are not a function of the register alone but of the register together with legal terms and conditions that pertain to the registry and bind the registries users and the wider economy. In the case of corporate registries, land registries and personal property securities registries, the rules of the registry are largely statutory in nature. The advantage of an independent registry is stability and assurance: its rules and the record are generally stable, and amended only for the benefit of all users and the broader legal and economic system, rather than at the discretion and commercial incentives of a commercial operator. Registries also balance privacy and commercial sensitivity against public visibility, and visibility is required both as a matter of justice and as a matter of legal notice: a deep doctrine in common law systems holds that a third party who acquires for value without notice of a prior interest may take free of it, and registration statutes tend to go further, so that what a register discloses largely determines what a purchaser is bound by.[80]

Accordingly, it goes without saying that the legal registration of legal objects with respect to the AULA should be situated within a trusted, public or commercially independent registry.

The legal registry should be publicly searchable by object, not by name, interoperable with other registers and with the other components of any AULA. The legal registry should hold for each registered legal object a verifiable identity anchor and a tamper-evident part governance event chain recording meta-legal events only such as: registration, and each change of property or legal status.

Any legal registry hosting LADs and LAS' as identified legal objects does not need to contain *all* confidential governance actions. A fulsome legal governance chain is best privately hosted within a separate control plane and accessible only by relevant parties. This bifurcation of governance chain between meta events (public registry) and all events (control plane with private permissions) assists in protecting commercial in confidence interests.

---

78 R G Patton, 'The Torrens System of Land Title Registration' (1935) 19 *Minnesota Law Review* 519.

79 *Personal Property Securities Act 2009* (Cth) ss 19, 21, 55; *Transfer of Land Act 1893* (WA) s 68 (indefeasibility).

80 *Great Investments Ltd v Warner* [2016] FCAFC 85, (2016) 243 FCR 516, on the position of a third party dealing with an agent acting outside authority.

The registry functions as an append -only legal ledger holding identity anchors and the meta-legal record, each event chain-linked and signed so the record is tamper-evident and independently verifiable. New entries are appended only by reflection from the home control plane of the object concerned, and no entry can be amended or deleted without detection (Section 3.8.1). It does not need to hold a live legal status, runtime verdict or instrument evaluation. Further details of operating the registry from a national perspective are set out in Section 5.3.

## *3.5* Further analysis - smart legal instruments

A smart legal instrument is an operating legal instrument expressed in natural language and machine-readable and programmable governance clauses. It connects to registered and unregistered legal objects and, in advance of registration, to the systems or data intended to become or which may become a legal object. Its clauses, once evaluated, produce a verdict that an enforcement subsystem applies. Its functions include performing and enforcing obligations as triggering events occur, allocating the rights and obligations generated by those events to the correct principals, and gating an AI agent's access to any live contractual (or other commercial) data stream, including operational data, execution records and evolving performance state.[81]

A smart legal instrument may specify the deployment of an AI agent as an execution mechanism, or serve as the interface through which an AI agent's actions produce legal consequence, however, in each case the deployment or gating occurs through the enforcement subsystem's application of a verdict, not through the smart legal instrument acting directly.

### 3.5.1 Smart legal contracts

A smart legal contract is a type of smart legal instrument. A smart legal contract is a second-generation legal contract that has automated performance according to the agreed rules of the parties. A smart legal contract produces a continuous record of contractual performance, of AI agent conduct where applicable, and of the consent basis on which each action was taken.

A smart legal contract is a legally (and cryptographically) executable instrument and thereby legally enforceable.[82] Legal execution is a key feature of why smart legal contracts are a foundational

---

81 Natasha Blycha and Ariane Garside, 'A Model for the Integration of Machine Capabilities into Contracts' in Jason Grant Allen and Peter Hunn (eds), *Smart Legal Contracts: Computable Law in Theory and Practice* (Oxford University Press, 2022) 142, 144, 147. Blycha and Garside define a smart legal contract as a legally binding digital agreement some or all of which is intended to execute as algorithmic instructions, and set out its five components of status, form, contents, active function and digital execution mechanism. See also the UK Law Commission, Smart Legal Contracts: Advice to Government (Law Com No 401, 2021), <https://lawcom.gov.uk/project/smart-contracts/> (accessed 2 November 2026), and the other chapters collected in Allen and Hunn (eds), above, <https://doi.org/10.1093/oso/9780192858467.001.0001> (accessed 2 November 2026). On the smart and legal contract as a digital asset, and on the Live Contract as the record of that contract and its performance over time, including amendments, notices and data generated through its automations, see Blycha, Myint and McCalman (n 17) 109–11.

82 The UK Law Commission concluded that the existing law of England and Wales can facilitate and support smart legal contracts, while identifying open questions of formation, interpretation and remedies: Law Commission, *Smart Legal Contracts: Advice to Government* (Law Com No 401, 2021). This paper does not examine the formation of contracts between automated systems at runtime, including offer and acceptance by an agent and the operation of the doctrine of part of the AULA and distinguishes them from the rule-based "instruction packs" already being supplied to AI agents, which operate as technical configuration without legal effect.

The legally binding nature of contractual agreements, including in identifying and embedding dynamic legal responsibility for agentic actions in technical and analogue legal workflows, is why SLCs supply the rules on which enforcement under the Autonomy Under Law Architecture acts.

Obligations under a legal instrument are performed in response to events or information, whether those events or information are generated by an AI agent, a human, another system or a schedule. The relationship between legal instruments and AI agents runs in both directions: a legal instrument may deploy an AI agent, an AI agent may use a legal instrument as the interface through which its actions take legal effect, and a legal instrument may operate without any AI agent at all.

Legal instruments are evaluated rather than self-executed, so that a verdict is reached at a decision point and applied by an enforcement subsystem. As a legal object travels, it is accompanied by its tethered legal instruments, or cryptographically verifiable references to them and a derived credential, via the legal envelope. The legal parameters governing the object are thereby verifiable at each governance event without a live query to the control plane, which is queried only where a clause requires state not locally available (Section 3.1.4).

### 3.5.2 Paired construction

The SLIs of AULA can be drafted using the "paired method" of legal construction, which Blycha and Garside set out for smart legal contracts as the Conjoined Method. The drafter may express any obligation in natural language, including an open-textured standard such as reasonableness or good faith. The drafter may then conjoin a deterministic programmable clause to that natural-language clause, and each carries its own meaning without clear linguistic pairing. The programmable clause need not express the meaning of the natural-language clause. A natural-language obligation that the AI agent act "as a good agent" may, for example, be paired with a deterministic clause requiring the AI agent to call home five times in every 24-hour period for instructions on how many movie tickets to purchase, and to purchase none if no call is answered. The natural-language text gives no such instruction.[83]

The alternative "unified method" of drafting would require the programmable clause to restate the natural-language clause. A standard such as reasonableness could then be enforced deterministically only after it had been reduced to rules, and an open-textured standard resists that reduction.

---

unilateral mistake where one party's system acts on terms generated by the other's: see *Quoine Pte Ltd v B2C2 Ltd* [2020] SGCA(I) 02. The position for formation between agents is not settled and turns on the law governing each instrument.

83 Blycha and Garside (n 81) 158–60 and 164–5, set out the Conjoined Method, under which a natural-language Primary Term governs and the Incorporated Code facilitates its execution, unless the parties agree that the code is primary. They adopt the paired method of construction over the unified method. Susannah Wilkinson and Jacques Giuffre, 'Six Levels of Contract Automation: The Evolution of "Smart Legal Contracts"' in Allen and Hunn (eds) (n 81) 183–4, model levels of contract automation on the SAE J3016 levels of driving automation and distinguish smart contracts from smart legal contracts within a legal-led approach.

The paired method requires no such reduction. The natural-language clause keeps the open-textured standard and its ordinary legal construction. The programmable clause supplies the operational specification that the parties agree the legal interlock will enforce at runtime. An objection that reasonableness or good faith cannot be written as code is therefore an objection to the unified method.

### 3.5.3 Evaluation at runtime

Where a natural-language clause is paired with a deterministic clause, the deterministic clause governs the verdict the legal interlock reaches at runtime. Where a natural-language clause has no paired deterministic clause, the semantic evaluator designated in the instrument evaluates it. The parties can also combine the two on a single clause. The deterministic clause then fixes the conduct the legal interlock requires or prevents, and the semantic evaluator assesses the same proposed act against the natural-language standard before it executes. An act that satisfies the programmable clause but that the evaluator assesses as inconsistent with the standard can then be escalated to a responsible person rather than allowed. Under the deterministic clause alone, an AI agent that called home five times in a 24-hour period, and bought tickets only on instructions received in those calls, receives an allow verdict, and whether it acted as a good agent is decided later by a responsible person or a court on the governance record. With the evaluator also armed, that question is put at runtime as well. In any later determination of the parties' rights the natural language prevails by default, and the parties can provide otherwise in the instrument.

### 3.5.4 Deterministic clauses

A deterministic clause is evaluated by a rule engine whose output is fixed by the clause and its input. The same clause evaluated against the same input always yields the same decision. A recorded decision can therefore be verified after the fact by re-evaluation, and that verification supports the standard of legal admissibility the architecture seeks for its records. A deterministic clause must also be unambiguous, so that it yields a single outcome for any input within its scope.

### 3.5.5 Semantic evaluation

The semantic evaluator identified in advance in the instrument evaluates a natural-language clause against its natural-language or behavioural criterion where the clause has no paired deterministic clause, and also where the parties have armed the evaluator alongside a paired deterministic clause. That evaluator may be a designated human reviewer, a specified model or a defined hybrid process. Which evaluator is designated, and what weight its determination carries at the decision point, are choices the parties write into the instrument. By default, the determination informs the decision point without itself being the authority for a verdict. The instrument can assign the determination a different weight, and it can provide that a clause is resolved only through the evidentiary record. Each determination is recorded as an attributed governance event, together with an identifier and version of the evaluator and a digest of the evaluated input, so that the determination can be reviewed independently. Where a clause requires state that is not available locally, the clause signals this and the event is forwarded to the authoritative source before a decision is reached.

### 3.5.6 Generating the programmable clause

The drafter of an SLI states in the programmable clause field, in plain words, the operational specification that the parties want the legal interlock to enforce, such as the call-home rule above. Before the instrument is executed, a drafting tool proposes candidate deterministic clauses from a curated library. The parties or their legal advisers review each proposal, the executed instrument fixes the clauses they approve, and the provenance of each proposal is recorded with the instrument. An approved clause binds as an agreed term of the instrument, whether or not it restates the natural-language clause to which it is paired. The crystallisation that seals the instrument records those pairings (Section 3.8). At runtime, the legal interlock evaluates only the approved clause, so each verdict can be replayed against it. Where no clause is approved, the natural-language clause stands unpaired and the semantic evaluator applies to it.

Obligations that are formally specifiable, including threshold checks, scope boundaries, credential expiry, access permissions and value ceilings, are amenable to deterministic enforcement.[84] Obligations that require contextual interpretation, such as whether an AI agent acted within the spirit of its mandate, whether a communication was misleading or whether a judgment was reasonable, remain in natural language. The parties can pair such an obligation with a deterministic clause that specifies conduct in connection with it. Where they do, they can also arm the semantic evaluator on the same obligation, so that both apply at runtime. Where no deterministic clause is paired, the designated semantic evaluator evaluates the obligation by default, unless the instrument provides that it is resolved only through the evidentiary record. The allocation of each obligation between these treatments is recorded in the instrument, is visible to the permissioned parties to the SLI, and is legally operative. The parties can vary it by amending the instrument.

The control plane delivers the clauses of each SLI to the legal interlock of the object it governs. Each party's home control plane pushes the shared clause set of a relationship contract to its own legal interlocks, at the version fixed by the contract's seal (Section 3.6.4). The records the legal interlock makes in performing those clauses return to the control plane, each bound to the envelope under which it was made.

### 3.5.7 Legal instruments for data objects

A legal data object should, at a minimum, carry a **Genesis Contract.** The Genesis Contract establishes on registration the initial meta-legal rights in the legal object: its legal ownership and the circumstances of its creation or registration. By analogy, it is the legal object's birth certificate.

A data object can also carry one or more of a title instrument recording legal ownership, for example, a licence instrument recording permitted use, or a transfer instrument recording a change of ownership or custody. These legal instruments can confer two classes of right. Title and the consent encumbrance are both recorded against the legal object on the legal registry and travel with it, and are analogous to rights in rem within the digital walls and terms of the AULA and interoperable systems.[85]

---

84 On the expression of government rules in machine-consumable form alongside their natural-language counterparts, see James Mohun and Alex Roberts, *Cracking the Code: Rulemaking for Humans and Machines* (OECD Working Papers on Public Governance No 42, OECD Publishing, 2020) <https://doi.org/10.1787/3afe6ba5-en> (accessed 2 November 2026).

The statutory basis of a registered object as property is addressed in Section 5.5 below. The licence and transfer instruments are rights in personam: the licence grants permitted uses to an identified licensee and the transfer records the terms on which custody or ownership passed between identified parties. A data object may be constituted or construed by the legal instruments between the parties to it, subject to the rights recorded against it on the registry.

### 3.5.8 Legal instruments for AI system objects

In addition to the Genesis Contract (the AI agent's birth certificate) and any title instrument described above, a legal AI system object should also carry an SLI known as the Agent Workforce Mandate (AWM).

An AWM dynamically constrains at runtime each governed action taken under the object's authority, recording an authorised operational scope, an escalation pathway and an authorised decommissioning action. An AI agent, not being a legal person, cannot enter an employment contract, but the governance function an employment contract performs for a human agent, defining authorised conduct, behavioural expectation and the terms of termination, is no less necessary for an AI agent. The AWM is executed by the legal person or persons responsible for the AI agent's deployment and operation; the AI agent is its subject and has access to it, but is not a party to it. The machine-readable component of the AWM is directly operative at the decision point. A governed interaction outside the permitted scope of an applicable relationship contract is denied (Section 3.5.9). Within that scope, an interaction that falls outside the AWM's recorded scope, or whose classification is ambiguous, does not warrant automatic denial, and the correct response is escalation to a designated responsible person. An AWM recorded only in human-readable terms cannot serve that runtime function. Through these two legal instruments each governed action is attributable through the legal object to a registered legal person. Title and licence instruments recording any change in ownership or control of the object are tethered in the same way, and each such change is committed to the registry as a meta-legal event.

Organisations deploying multiple AI agents (which will likely be universally all deployers) will also seek to implement a **Master Agent Workforce Mandate** (MAWM). The MAWM is a smart legal instrument entered into by an organisation, which sets out the rules applying at an organisation level, across different departments and layers. The MAWM is an operationalised organisation chart. It sets out relationships, reporting lines, duties, organisational code of conduct and policies, and information sharing and information barriers. It provides increased assuredness for an organisation's board of directors and management to ensure that proper channels and lines of communication have been established within the organisation, to provide assurance and a proper grant of their directorial and officer duties, with sufficient granularity to do so.

[85] The legal registry gives notice of those rights to any party that deals with the object, and a holder who takes the object with notice takes subject to them, in the same way that a registered interest binds a purchaser of land.

### 3.5.9 Relationship contracts

Relationship contracts are SLIs or legal instruments involving registered legal objects *between* parties.[86] A specialist type of relationship contract foundational to AULA is the **Cross-Enterprise Relationship Contract** (CERC).

The CERC instrument governs what interactions are permitted between AI agents representing different legal persons. This includes what may be requested, what data may be exchanged and on what conditions set out in both natural language and where required, with programmable clauses tagged to that natural language. The CERC may allocate rights to created assets and liability for loss or damage between the organisations where their AI agents collaborate to create value or to cause harm.

CERCs can be arranged in a three-tier hierarchy. A *root-level* CERC governs the enterprise-to-enterprise framework and applies as the default across routine interactions; in legal parlance it is the head agreement between the parties in respect of each other and each other's AI agents. A *capability-level* CERC is scoped to a business capability and reusable across pairs or groups of objects sharing an operational pattern. An *object-level* CERC is executed for a specific object holding an elevated privilege or engaging in a high-risk interaction class; a change from least-privilege to elevated status is a material event requiring execution of an object-level instrument before the object may continue operating beyond the enterprise boundary. At a minimum, a root-level CERC specifies activity data rights; risk allocation; permitted recourse, whether guarantee, escrow or insurance; revocation triggers with remedy and notice periods; fallbacks on AI agent failure; audit and test periods; the private instruments and data objects permissioned for access through it; the permitted interaction scope; and dispute resolution. An object-level CERC contains the same heads but with greater specificity, including the categories of task one AI agent may request of another, so that an out-of-scope request is rejected automatically with a logged, attributable refusal.

Any applicable relationship contract may also govern the *visibility* of and *access to* other legal instruments and broader commercial content to counterparties.

A legal instrument internal to an object, including each AWM and MAWM, is not automatically disclosed to a counterparty by reason of a cross-organisational interaction. Its existence and binding effect can be attested to the counterparty without disclosure of its contents. The SLI ID registration, shows that it was executed and fixed at a stated time, and its contents can be checked against the registration if they are later disclosed, for example in a dispute. The counterparty can rely on the fact that a mandate exists, is in force and bounds the AI agent's conduct because the CERC records that as a condition of the relationship, and it has recourse under the CERC if that representation is untrue. Zero-knowledge proofs of specific properties of an instrument, such as that a requested amount lies within a ceiling it sets, are a possible extension. Visibility of the governed interaction to the enforcement fabric is likewise set per relationship, among a full-audit mode, a metadata-only mode in which values are encrypted, and an end-to-end-encrypted mode in which only binding commitments are visible to the fabric. The audit-visibility mode is a term of the relationship contract, not a system-wide setting.

[86] Throughout this paper, reference is made to the term 'relationship contracts'. Relationship contracts are SLIs or legal instruments involving registered legal objects *between* parties.

#### 3.5.10 SLIs as temporal mechanisms

To address the temporal dimension of the attribution gap, being the gap between human and AI agent speed and span of control, an SLI may incorporate any of the following mechanisms:

- real-time performance and monitoring of obligations;
- scoping of, and access control over, live data streams;
- alerting and constraint before action or breach;
- legal revocation and deprovisioning as executable contractual acts (Section 3.9);
- automated allocation of rights and obligations, with a timestamped record;
- identification of disputes at machine speed;
- dynamic adaptation within pre-defined parameters;
- a call-home clause requiring the object, or the AI system it governs, to establish contact with the control plane on a circumstance or at a cadence, with a tolerance whose breach is itself a governance event; and
- a legal audit trail held as a single source of truth among the parties.

Not every mechanism need be active for every workflow, and the applicable instrument determines which are switched on. The temporal mechanisms allow a responsible person to make compliance decisions in advance, which extends that person's span of control in discharging legal duties (Section 4).

#### 3.5.11 The centrality of contracts to AULA

Contracts serve as a versatile and practical instrument for agent governance. By extending legal governance of people via contracts, to governance of people *and* AI agents via SLCs, AULA allows agentic governance to benefit from the rigorously tested law of contract, and agency, and on documented intent and self-interest whose foundations have been tested in practice over centuries.[87]

Freely bargained and properly drafted contracts distil an agreed understanding of promises that are enforceable by law. The legal certainty granted to contracts, bolstered by the common law and other rules of law, has allowed them to align and implement human-to-human and human-to-institution interaction across every facet of economic activity. This paper submits that they can align AI agents as well. AULA supports contractualist alignment (see Tan et al, 2024) by removing any dependence on AI agent benevolence or preference.

A substantial body of work seeks to govern AI behaviour by aligning systems to human values and preferences through normative frameworks drawn from rational choice theory, expected utility theory, and single- and multi-principal alignment. Tan, Carroll, Franklin and Ashton question whether reward or utility functions can encapsulate human values at the individual level, and whether individually elicited preferences aggregate reliably to reflect the plurality of interests in a multi-AI agent context. Their proposed technical alternative is contractualist alignment: contracts as the mechanism through which AI systems are aligned to human interests, rather than preference inference or utility optimisation.[88] In that framework contracts are the foundational alignment mechanism rather than a peripheral overlay, and AULA is a technical implementation of contractualist alignment at enterprise scale.

### 3.5.12 Legislation and private law

A further point should be made in respect of calls for legislation or regulation to do all the legal labour required to mitigate against AI safety or AI control risks.[89] Legislation has an important role in moderating AI activity. The authors consider however, that the current focus on legislation, rather than on private law solutions, reflects a limited appreciation of the critical, ubiquitous and pervasive role that private law (contracts) plays in moderating even the granular, bespoke aspects of economic actors and activity.[90] The emphasis on legislation may reflect a limited understanding of the legal loop and its effect on lawful conduct in humans (Section 2.4.1). An approach confined to what an AI agent needs to know of the law, such as legislation and case law, leaves out the stages of the loop that follow the act. In humans, lawfulness is also a by-product of exposure to the operation of the law over time, its “thrust and parry”. A person sees lawful and unlawful conduct, and sees the law enforced against themselves and against others. Drivers who keep within the speed limit have rarely read the traffic regulations. Rightly or wrongly, much lawful conduct on the road is a product of personal experience of penalties such as speeding fines, or of seeing those penalties imposed on others. Under AULA, each verdict, narrowing of standing and revocation is applied to the AI agent at the governed act and recorded against its LAS, which supplies for the AI agent the stages of the loop that follow the act (Section 3.1).

---

87 Peer Zumbansen, ‘The Law of Society: Governance Through Contract’ (2007) 14(2) *Indiana Journal of Global Legal Studies* 191.

88 Tan Zhi-Xuan, Micah Carroll, Matija Franklin and Hal Ashton, ‘Beyond Preferences in AI Alignment’ (2024) *Philosophical Studies*, <*https://doi.org/10.1007/s11098-024-02249-w*> (accessed 2 November 2026); preprint arXiv:2408.16984.

89 Joint Select Committee on Artificial Intelligence (n 18) 5, recommending that AI legislation establish ‘clear statutory objectives, minimum requirements and accountability mechanisms’.

90 See Zumbansen (n 87).

Industry has meanwhile built technical enforcement for AI agents at the runtime and hardware layers ahead of legislation, most extensively in NVIDIA's Open Agent Safety Platform (Section 7.3).[91] Those controls enforce policy written by the deploying organisation alone.[92] The terms on which two organisations' AI agents deal with each other, and the allocation of loss between those organisations, remain matters of contract.[93]

Private law (contracts) is a foundational part of the legal solution to AI safety and control. Contracts, underpinned by broader legislative frameworks, are the existing tool that moderates and governs most if not all global activity[94] and public and private relations. Contracts are already used by economic actors to restrain and set out legally enforceable rules (and remedies for breach of those rules) for all modern economic activity globally. AULA therefore treats SLIs, including

---

91 NVIDIA's Open Agent Safety Platform is described in Section 7.3. The policy it enforces is written by the deploying organisation and restricts the resources an agent can reach. An agent acting entirely within those resources can still exceed the authority its principal granted, for example by placing an order above its spending limit with a supplier whose endpoint the policy approves.

92 Idira, the identity platform Palo Alto Networks launched on 12 May 2026 on the CyberArk technology it acquired in February 2026, registers each agent in an inventory, issues it credentials scoped to a single task and revokes those credentials when the task ends. Entrust's Agentic AI Trust Accelerator, launched on 14 July 2026, is co-developing verifiable identities for agents and their human principals, together with the keys, certificates and signing used to make agent actions cryptographically verifiable. Palo Alto Networks' Prisma AIRS inspects prompts, responses and tool calls for attack before an agent acts. In each case the permission enforced is a policy or credential issued by the deploying organisation. A counterparty therefore receives evidence of the agent's identity and technical access, while the scope of the agent's authority and the allocation of loss for an unauthorised act remain matters for contract. An SLI supplies those terms in machine-evaluable form. OpenShell exposes gateway interceptors and supervisor middleware at which a legal interlock evaluation can run before a connection leaves the sandbox. This allows clauses derived from an SLI to be enforced through the same controls that OpenShell and Sentry already apply. See Intragen, 'CyberArk, Idira and the Future of Privileged Access' (2026) <https://www.intragen.com/blog/cyberark-idira-and-the-future-of-privileged-access> (accessed 29 September 2026); CyberArk, 'Secure AI Agents', CyberArk Documentation <https://docs.cyberark.com/admin-space/latest/en/content/secureai/introduction.htm> (accessed 29 September 2026); Entrust, 'Entrust Launches the Agentic AI Trust Accelerator to Help Enterprises Move AI Agents From Pilot to Production' (Business Wire, 14 July 2026); 'Palo Alto Networks Launches AI Agent Security Integrations with IBM, ServiceNow, and More', SDxCentral (2026) <https://www.sdxcentral.com/news/palo-alto-networks-launches-ai-agent-security-integrations-with-ibm-servicenow-and-more/> (accessed 29 September 2026).

93 Norm Ai builds agents to 'embed law into AI agents' and supervisory agents 'for regulated enterprise AI deployments', and an affiliated firm, Norm Law LLP, uses those agents to act as outside counsel on outcome-based fees. Norm Ai raised a US$120 million Series C at a US$1.2 billion valuation in July 2026 and reports clients representing more than US$30 trillion in assets under management. The rule set its agents apply is legislation and regulation as interpreted for one deploying organisation. The resulting check therefore operates on the deployer alone. A counterparty dealing with the deployer's agent takes the terms of that dealing, and the allocation of any loss, from whatever contract governs it. An SLI can designate a regulatory evaluator of this kind as one input to the legal interlock's evaluation, so that regulatory compliance and the parties' agreed terms are tested at the same point before the agent acts. See Norm Ai, 'Norm Ai Raises $120 Million at a $1.2 Billion Valuation Led by Khosla Ventures to Deliver the Full-Stack Model for Legal AI' (PR Newswire, 7 July 2026) <https://www.prnewswire.com/news-releases/norm-ai-raises-120-million-at-a-1-2-billion-valuation-led-by-khosla-ventures-to-deliver-the-full-stack-model-for-legal-ai-302819152.html> (accessed 29 September 2026).

94 Peter Vincent-Jones, 'Contractual Governance: Institutional and Organizational Analysis' (2000) 20(3) Oxford Journal of Legal Studies 317.

SLCs, as a foundational tool in the governance of AI systems. A contractual instrument that performs at machine speed and is machine-readable and programmable is the natural development of the analogue contract for governing relationships between humans and AI agents.[95]

An SLI also carries the law that applies to an AI agent's conduct beyond the terms the parties agree. Its legal context clause draws the applicable legislation, regulation and judicial decisions into the same runtime evaluation as the terms of the instrument, so both are applied to each act at the same point (Section 3.5.13).

### 3.5.13 Smart legal instruments and legislation, regulation and judicial decision making

A contract commonly obliges a party to comply with all applicable laws, or to act lawfully, reasonably or in good faith. The content of each such obligation is predicated on legislation, regulation and judicial decisions that the contract imports by reference but does not reproduce in the words of the document per se, a contract that set out every law bearing on its performance would run to many thousands of pages. A contract can also name a particular law with which a party must comply. An SLI carries both kinds of term, and its legal context clause(s) supplies their content at runtime. The legal context clause(s) identifies the jurisdictions and classes of activity in which the governed object acts, and references by authoritative identifier the legislation, regulation and judicial decisions that apply to them, together with any other law the instrument names. The context of each governed event, including the jurisdiction of the counterparty, the class of data acted on and the class of activity, selects the provisions that apply to that event. Legislation and regulation are applied typically at the version in force, but the parties are not prohibited from applying antecedent versions as well if they so choose in the SLI (which is less common but not unusual); the SLI and the AULA facilitates both options. A judicial decision is applied at its current status, and a decision reversed on appeal or overruled is recorded as a governance event in the same way as an amendment to legislation.

Some legislatures and government registers publish legislation through application programming interfaces from which software can retrieve an Act, regulation or instrument at a given date or the in force version from time to time. A control plane can call such an interface to resolve each provision referenced in the legal context clause of an SLI at the version designated to be in force. Where designated, an amendment reaches the interface only when the register publishes the amended version, which can follow the commencement of the amendment. Most registers do not give machine-readable notice of amendments as they are made, and coded rules are rarely exposed through public interfaces.[96] Where a register has no interface, the control plane resolves the reference against the published text, and the verdict records the version applied in the same way.

---

95 Blycha and Garside (n 81) 177–9.

96 Angelica Lindqvist, 'Rules as Code: An Open Approach', Interoperable Europe Portal (European Commission, 25 February 2025) <https://interoperable-europe.ec.europa.eu/collection/eugovtech/document/rules-code-open-approach> (accessed 2 November 2026).

In Australia, the Federal Register of Legislation provides a free interface that returns Commonwealth legislation as in force at a specified date, and in New Zealand the Parliamentary Counsel Office offers a beta interface that lists each version of an Act.[97] In the United States the electronic Code of Federal Regulations provides an interface that returns each title of federal regulations as at a given date. govinfo provides an interface that lists documents added or updated since a given date, including enacted laws marked up in United States Legislative Markup, an XML schema derived from Akoma Ntoso.[98] Singapore Statutes Online displays past versions of Singapore legislation but offers no published interface, and it does not hold the authoritative text, which remains the version published in the Government Gazette.[99]

A provision drawn in by a legal context clause is evaluated by a deterministic clause where it has been expressed as a programmable clause, by the parties or through a publication of Rules as Code (RaC), also called law as code or machine-consumable legislation.[100] Any other provision, including a determinate rule that has not been coded and a judicial decision, is evaluated by the semantic evaluator against its text or escalated to a responsible person. Each verdict records the identifier, and the version or status, of every provision applied, so the governance event chain evidences the law the AI agent was held to at the time of each act.

Machine-readable legislation publishes the text of the law in a structured format, such as the Akoma Ntoso standard, so that software can locate and cite each provision.[101] RaC goes further and expresses rules in a form software can execute. Some coded rules are exposed through interfaces that return a calculated outcome, as OpenFisca does.[102]

---

[97] Office of Parliamentary Counsel (Cth), Federal Register of Legislation, ‘Data share and reuse’ <https://legislation.gov.au/help-and-resources/using-the-legislation-register/data-share-and-reuse> (accessed 2 November 2026), and API specification <https://api.prod.legislation.gov.au/swagger/index.html> (accessed 2 November 2026); Office of Parliamentary Counsel (Cth), ‘Compilation services’ <https://www.opc.gov.au/opc-services/publishing/compilation-services> (accessed 2 November 2026); Parliamentary Counsel Office (NZ), ‘Developer API’ <https://legislation.govt.nz/developer-api/> (accessed 2 November 2026), and New Zealand Legislation API (v0) documentation <https://api.legislation.govt.nz/docs/> (accessed 6 November 2026).

[98] Office of the Federal Register, ‘eCFR API Documentation’ <https://www.ecfr.gov/developers/documentation/api/v1> (accessed 2 November 2026); US Government Publishing Office, ‘govinfo API’ <https://www.govinfo.gov/features/api> (accessed 2 November 2026), and ‘Beta USLM XML’ <https://www.govinfo.gov/features/beta-uslm-xml> (accessed 6 November 2026).

[99] Attorney-General’s Chambers (Singapore), Singapore Statutes Online, ‘FAQ’ (questions A3 and B1) <https://sso.agc.gov.sg/Help/FAQ> (accessed 6 November 2026).

[100] The relationship between SLIs and Rules as Code warrants treatment in a separate paper, and only the points at which the two concepts connect are cursorily addressed here. Rules as Code work is distributed across legislatures, drafting offices, revenue and benefits agencies, universities and vendors in many countries, and much of it is reported only on project pages and in conference presentations. The same practice is also described as law as code, machine-consumable legislation, digital-ready legislation and better rules. Important projects and current approaches have probably been missed, particularly those published under other names, and some programmes may be described at an earlier stage than they have now reached. The authors apologises for those omissions.

[101] OASIS, Akoma Ntoso Version 1.0 Part 1: XML Vocabulary (OASIS Standard, 2018) <https://docs.oasis-open.org/legaldocml/akn-core/v1.0/akn-core-v1.0-part1-vocabulary.html> (accessed 2 November 2026).

[102] James Mohun and Alex Roberts, Cracking the Code: Rulemaking for Humans and Machines (OECD Working Papers on Public Governance No 42, 2020); Lindqvist (n 96); OpenFisca, ‘Using the /calculate endpoint’ <https://openfisca.readthedocs.io/en/latest/openfisca-web-api/input-output-data.html> (accessed 2 November 2026).

RaC developed its own languages and other frameworks, such as Catala and OpenFisca, in preference to general rule standards. Decision Model and Notation (DMN) expresses business rules as executable decision tables. Policy languages such as XACML, and its successors Open Policy Agent and Cedar, express access-control policies evaluated at a policy decision point, the role the legal interlock performs in AULA. ODRL expresses permissions, prohibitions and duties over digital assets.[103] These standards can order competing rules, for example by letting a denial override a permission, but each evaluates the rules currently in force. Legislation applies a provision as it stood on a given date, after commencement, amendment and transitional provisions, and a later amendment can change that position retrospectively. A legal record therefore needs two dates, as a bitemporal database does: the date on which the provision applied to the act, and the version of the provision in force when it was evaluated. Legislation also layers exceptions over general rules. Sergot and Kowalski showed in 1986 that such defeasible rules can be coded, by expressing the British Nationality Act 1981 as a logic program, and LegalRuleML was developed to express them in a standard form. Its coded form must also remain traceable to the authoritative text it implements.[104] Catala places each line of code beside the statutory text it implements, so that lawyers can check one against the other.[105]

New Zealand's Better Rules programme began the practice in 2018, Denmark has required new legislation to be digital-ready since 2018, and the French public finance administration has used the Catala language to code parts of its tax law.[106] In 2020 the NSW Department of Customer Service translated the Community Gaming Regulation 2020 (NSW) into code on an open-source rules engine, and NSW Fair Trading built a compliance tool on that code.[107] GovCMS has since run a Rules as Code pilot with Australian Government agencies and is moving that work to an enterprise production platform.[108] Most of this work remains at the pilot stage and concentrates on prescriptive rules such as benefit eligibility and tax calculation.[109]

---

103 Object Management Group, Decision Model and Notation (DMN) (OMG Standard); OASIS, eXtensible Access Control Markup Language (XACML) Version 3.0 (OASIS Standard, 22 January 2013); W3C, ODRL Information Model 2.2 (W3C Recommendation, 15 February 2018).

104 OASIS, LegalRuleML Core Specification Version 1.0 (OASIS Standard, 2021).

105 Denis Merigoux, Nicolas Chataing and Jonathan Protzenko, 'Catala: A Programming Language for the Law' (2021) 5 Proceedings of the ACM on Programming Languages (ICFP) art 77.

106 Service Innovation Lab, Better Rules for Government Discovery Report (Department of Internal Affairs, New Zealand, 2018); Agency for Digitisation (Denmark), Digital-ready legislation guidance (2018); Merigoux, Chataing and Protzenko (n 105).

107 'NSW govt makes its first regulation machine-readable', iTnews (1 October 2020) <https://www.itnews.com.au/news/nsw-govt-makes-its-first-regulation-machine-readable-554167> (accessed 2 November 2026); Digital NSW, 'An Australian first: NSW translating rules as code to make compliance easy' <https://www.digital.nsw.gov.au/article/an-australian-first-nsw-translating-rules-as-code-to-make-compliance-easy> (accessed 2 November 2026).

108 GovCMS, 'Introducing Rules as Code' <https://govcms.gov.au/rac> (accessed 5 November 2026); Assemblic, 'The 2026 OpenFisca Conference, Day 2 morning' <https://www.assemblic.com/blog/openfisca-conference-2026-day-2-morning/> (accessed 2 November 2026).

109 Lindqvist (n 96); Lauren Anne Parnaby and Lyria Bennett Moses, 'Rules as Code: Digital Rulemaking in the 21st Century', UNSW Allens Hub for Technology, Law and Innovation (5 February 2021) <https://allenshub.unsw.edu.au/index.php/news/rules-code-digital-rulemaking-21st-century> (accessed 2 November 2026).

Australian commentators observe that only prescriptive rules translate readily into code, and that a provision requiring judgement needs a human decision-maker.[110] Further, in the context of automating the exercise of statutory powers through algorithm-decision making, the ability of RaC to achieve these outcomes are potentially fraught.[111] This is a result of many factors, not least the semantic and social complexity underpinning linguistic complexity together with the dynamism of legal interpretation of legislation, both of which do not readily lend to reducing legislation to code.[112] These observations and challenges describe the limit of the unified method of drafting, under which the code must restate the meaning of the natural-language text (Section 3.5.2).[113] A legal context clause uses the paired method. The provision keeps its natural-language text and its ordinary legal construction. A deterministic clause is paired with it only where the provision has been coded, and that clause specifies only the conduct the legal interlock is to require or prevent. Coded rules can therefore be drawn into a legal context clause wherever RaC has produced them, and the remainder of the law, including judicial decisions, can still be referenced through the paired method. As more law is published in coded form, more of the law that applies to an AI agent can be evaluated deterministically at the time of each act. The advantages of SLIs and the AULA to parse such RaC in this manner are predicated on the relevant legislative instruments being accessible in the current status quo formats and APIs.

At a conceptual level, legislation and the corpus of judicial decisions could also be constructed as SLIs, and would then obtain the benefits of the SLI and of the broader AULA. The advantages of a legal system constructed in that way, including its rigour, cohesion and interoperability, warrant a separate discussion (see n 100). The move of legal systems to machine speed proposed by AULA would apply to public law as much as to private law, and to the intersection between them.

Blycha and Garside describe the end state of a smart legal contract whose coded obligations are connected to live data and to each other.[114]

> *"In a fully-realised state, such a contract might sit under a large project like a root system sits under a growing tree, growing and evolving as the project itself does ... With almost infinite intricate potential states, such a wholly enlivened contract would never truly be complete – to look at it at any point in time is only to see one aspect of it; its terms and state as they are at that moment, just as one might catch a flattened, fragmentary glimpse of oneself in a mirror."*

Legislation construed as an SLI would have the same properties. Each provision would keep its enacted text and its ordinary construction, and a programmable clause would be paired with a provision only where it can be expressed as a rule (Section 3.5.2). A legal context clause in a private SLI would then reference the programmable clause of the provision itself, rather than resolving the provision against published text through an interface (Section 3.5.13). An amendment, a commencement or a decision that overrules an earlier decision would be recorded as a governance event in the legislative instrument. That event would reach every private SLI that references the provision, so the conduct required of each AI agent governed by those instruments would change when the law changes. The public law counterpart of the sensor in the quoted passage is the fact on which a provision operates, such as a declared emergency, a gazetted rate or a regulator's determination. The fragmentary view in the passage corresponds to the two dates a legal record needs: the date on which a provision applied to an act, and the version of the provision in force when it was evaluated (Section 3.5.13). Parliament would continue to enact each provision and the courts would continue to construe it. The programmable clause paired with a provision would carry only the authority that the enacting legislation gave it.

---

110 Parnaby and Bennett Moses (n 109).

111 Will Bateman, 'Algorithmic Decision-Making and Legality: Public Law Dimensions' (2020) 94 Australian Law Journal 520.

112 Ibid.

113 Blycha and Garside (n 81) 158–60, 164–5.

114 Blycha and Garside (n 81) 181.

## *3.6* Further analysis - the legal interlock

The legal interlock is an AI agent-side plug-in that tethers the content of any relevant SLIs against a LAS' conduct at runtime. The legal interlock is the primary mechanism for rule delivery, rule evaluation and enforcement as well as evidence (performance or activity data) collation of any evaluation and enforcement. The legal interlock sends the governance record of each evaluation and enforcement, described below, to the control plane that holds the LAS, bound to the legal envelope under which the event was governed.

### 3.6.1 Governance records and recording

For each governed event, the legal interlock sends the control plane a governance record containing the identity of the LAS, the SLI and clause version applied, the verdict, the time of the event and a cryptographic digest of the payload the AI agent sent or received. Each record states the legal envelope under which the event was governed. The control plane receives the payload only where an SLI requires it, and in the form that SLI specifies. Where the parties have elected it in an SLI,[115] an AI agent interaction recorder at the legal interlock keeps an encrypted recording of the payload of governed events, held and opened only on the terms that SLI sets, and the digest in the governance record shows that a recording produced later is the record of the event concerned.

Where a clause requires a live verdict on the content of an interaction, the legal interlock sends the evaluator designated by that clause only the fields explicitly specified in the clause. The control plane therefore receives business content only in the form, and to the extent, that an SLI executed by any party requires.

[115] The division follows the access regime for aviation flight recorders. Flight data recorders capture operating parameters at fixed intervals, some several times per second, and cockpit voice recorders capture the aural environment of the flight deck: ICAO, Annex 6 to the Convention on International Civil Aviation: Operation of Aircraft, pt I, app 8. Cockpit voice recordings may not be used other than for accident or incident investigation unless a competent authority, which may be a judicial authority, determines that disclosure outweighs its adverse impact on that or future investigations: ICAO, *Annex 13 to the Convention on International Civil Aviation: Aircraft Accident and Incident Investigation* ch 5.12, app 2.

### 3.6.2 Verdicts

Each governable event of the AI agent that takes place at the legal interlock is evaluated against SLI clauses that the deploying legal person, and any relevant counterparty, has put in place at the LAS control plane. The legal interlock itself holds no authority of its own. Whatever verdict the legal interlock reaches comes entirely from evaluating SLI clauses. Before a legal interlock evaluates anything, the legal interlock must first check that the LAS it is acting for is still the same legal object registered under that identity anchor, using the identity-binding mechanism described in Section 3.1.2. This check must pass before evaluation can happen at all. The legal interlock then evaluates governable events against applicable SLI clauses and applies the result as a constraint on what the legal object's AI agent can actually do at runtime, applied at the legal interlock. Every governable event can be recorded, and authority can be extinguished, according to the will of the responsible person, or relevant legal parties, including at the moment a revocation takes effect.

Evaluating a governed event produces one of four verdicts: allow, deny, escalate or forward. Allow and deny are reached by applying the clauses of the SLIs to recorded inputs, including the recorded determination of any designated semantic evaluator, so each verdict can be replayed from the record (Section 3.5.4). A machine learning component cannot turn a deny into an allow; it can only narrow the result, by denying an act, raising an escalation or requiring a further approval (Section 4.3). An escalate verdict occurs when the SLI clauses cannot resolve an event themselves, or the clause itself calls for escalation for example by request by notice to a designated responsible legal person for further instruction. The result of the responsible legal person's, decision is recorded as part of the legal evidence chain. The forward verdict defers the event to the home control plane of the object or, for an event under a relationship contract, to the control plane that such relationship contract designates (Section 3.1.6).

### 3.6.3 Persistence beyond the AI agent

The legal interlock outlives any AI agent over which it has surveillance or responsibility. The practical implication (and legal evidentiary benefit) of this is that even if an AI agent is "spun up for thirty seconds" and then destroyed, the legal interlock still leaves behind a full legal record of what it did and under whose authority, which returns to the control plane bound to the envelope under which each act was made. This provides the persistence property the delegation chain requires (Section 3.1). That persistence depends on each AI agent, including each replicant, being created with its own legal interlock, which the AWM or another relevant SLI requires (Sections 3.1 and 4.6).

### 3.6.4 Internal and shared clause sets

Every legal interlock runs two different sets of SLI clauses *internal* and *shared*.

The internal clause set is made up of the legal object's own internal legal instruments.

The shared clause set comes from whatever part of a relationship contract, a CERC for instance, the legal object's holder is party to, and **is delivered to** the legal interlock of every legal object bound by that same relationship contract. Each party's home control plane pushes the shared clause set to its own legal interlocks at the version fixed by the contract's seal on the registry. The seal is a salted commitment of the executed version of the relationship contract and reveals nothing about the contract. Where each legal interlock bound by that relationship contract runs a conformant engine for the clauses the contract names, every one of them runs the same clauses at the same version, so the parties' AI agents evaluate, and act on, the legal terms consistently. Where they do not, the natural-language terms of the contract govern, and the enforcement fabric compares the parties' attestations after the act. The shared clause set is only visible to the parties to the relationship contract[116] and to their home control planes.

### 3.6.5 Conjunction clauses

Conjunction clauses are a specific SLI clause type that parties can use to gate agentic events. Each legal interlock the clause names contributes a credential share only on its own allow verdict, and the event executes only when the threshold number of shares specified in the clause, which may be every named party, has been contributed and joined (Section 3.1.7 and Annexure A). The completion of a "conjunctive event "is itself proof that the required number of legally responsible parties each reached an allow verdict on that event.

### 3.6.6 Governance harnesses

The legal interlock, its two clause sets and conjunction clauses all hold the rules that govern an AI agent in a deterministic layer outside the model.[117] A layer of that kind has been built and tested empirically. Kereopa-Yorke and Praino built a prototype governance harness (the Flinders Harness) for public-sector advisory use and evaluated it on four frontier models.[118] The Flinders Harness is a deterministic software layer between the user and a frontier model, and it comprises a policy router, a source-authority gate, a citation-enforcement contract, an escalation and refusal layer, and a jurisdiction-specific governance prompt. Across approximately 12,600 scored responses in South Australia and Victoria, the Flinders Harness improved every model. Citation enforcement and source-authority gating together accounted for between 39 and 97 per cent of the effect.[119] Its escalation and refusal layer routed high-risk queries to a human or declined them, which corresponds to the escalate and deny verdicts of the legal interlock. The Flinders Harness was transferred from one jurisdiction to the other by changing a configuration file. In an earlier study, Kereopa-Yorke, Lewis, Schaufelberger, Romantsova, Hill and Corch built a thinner harness of the same design, the SA Harness, and tested it on ten models from five vendors spanning three orders of magnitude in parameter count.[120] Under the SA Harness the range of refusal rates across the ten models narrowed from 95.9 to 21.9 percentage points, and refusal accuracy rose for every model. An open-weight model of 8 billion parameters inside the SA Harness exceeded the unharnessed refusal accuracy of every frontier model on the panel. In both studies the regulated behaviour of the system was a property of a deterministic layer outside the model, and it held when one model was substituted for another. Kereopa-Yorke and co-authors conclude that the model is a substitutable component, and that the governance layer is the artefact an institution can audit, version and hold authority over.

---

116 Where this can include legally permissioned delegates acting for the permission legal parties.

117 O'Keefe and others propose that AI agents be designed to comply rigorously with a broad set of legal requirements and to refuse illegal actions in the service of their principals: Cullen O'Keefe, Ketan Ramakrishnan, Janna Tay and Christoph Winter, 'Law-Following AI: Designing AI Agents to Obey Human Laws' (2025) 94 *Fordham Law Review* 57. Compliance of that kind is built into the agent itself. The legal interlock is a separate component that can be placed where the agent cannot reach it, and it records each verdict against the instrument and version that produced it, so a court or counterparty can examine the basis of the verdict after the act. The two approaches are complementary, because a law-following agent reduces the number of events the legal interlock must deny.

118 Ben Kereopa-Yorke and Rodrigo Praino, *Sovereign AI Through Governance Harnesses: Evidence from a Multi-Jurisdiction Evaluation* s 2.2 (Research Report, Jeff Bleich Centre for Democracy and Disruptive Technologies, Flinders University, August 2026) <https://doi.org/10.25957/79y6-wb39> (accessed 2 November 2026).

AULA shares that conclusion and applies it to legal authority. The legal interlock occupies a similar position as the Flinders Harness and the SA Harness, and its verdicts are derived from the SLIs whichever model the AI agent runs on. The artefacts the parties audit are the legal instruments and the governance records the legal interlock produces. The signed registry of authoritative sources in the SA Harness secures the provenance of what a model may rely on, which is the function the registry and authenticated LADs perform for the identity of AI agents and data. Citation enforcement and source-authority gating can each be expressed as a deterministic clause in an SLI. A deploying person can therefore adopt the controls the two studies tested within the evaluation the legal interlock performs.

The Flinders Harness evaluated the content of a frontier model's answer against rules set by a single deploying government, which is also the legal person answerable for each response. The legal interlock evaluates governed events against SLIs agreed by the responsible legal person and any counterparty. Governed events are acts of an AI agent that create legal consequences, such as committing its principal to a transaction or acting on another organisation's data. Kereopa-Yorke and Praino confine their recommendations to single-turn advisory use and exclude agentic, multi-turn and high-autonomy applications.[121] Kereopa-Yorke and co-authors record that the source

---

119 Ibid ss 4.1 and 4.3 (Tables 1 and 5). The authors caution that the leave-one-out ablation assumes roughly additive component effects, and that at least one model showed interactions between components.

120 Ben Kereopa-Yorke, Brigitte Lewis, Laura Schaufelberger, Galina Romantsova, Brian Hill and Corch, ‘The Harness Is the Contract: Architectural Sovereignty for Regulated AI after Robodebt’ (Jeff Bleich Centre for Democracy and Disruptive Technologies, Flinders University, June 2026) ss 1, 3.1 and 5, <https://doi.org/10.25957/7e64-dh11> (accessed 2 November 2026). The authors use ‘contract’ for the structured response schema that the SA Harness requires the model to complete, comprising eight machine-readable fields including refusal, escalation, citation and review-required fields, which a deterministic parser checks after generation (s 3.1). The schema constrains the form of a model’s output and has no force as a legal instrument. The SLIs that a legal interlock evaluates are executed instruments binding registered legal persons. The benchmark items were drafted and audited by language models, and the authors report that part of the cross-vendor agreement they measured depends on items carrying item-specific policy text (ss 4.3, 4.4, 6.2 and 7.4).

121 Ibid s 5.2. Acts of the kind the legal interlock governs fall outside the use case for which the Flinders Harness was built and tested. In that use case the deploying government is the legal person answerable for each response, so the Flinders Harness registry of the SA Harness was a single point-in-time snapshot, and that their measurements did not exercise the behaviour of the harness when an entry expired, a policy changed or a model was replaced.[122] A legal interlock also governs from a snapshot between contacts with the control plane. That snapshot is valid only within the bounds its instruments set, and expiry of a bound narrows, suspends or extinguishes the object's authority without further contact (Section 3.5). The control plane holds the live legal status of each registered object and distributes revocations to the legal interlocks of those objects (Sections 3.1 and 3.9). A companion paper foreshadowed by Kereopa-Yorke and co-authors addresses consent provenance, including the source of a system's authority, which is the question the legal interlock answers for AI agents acting for different parties.[123]

## 3.7 The legal interlock in practice

In a cross-border payment, the corporate client of Bank A issues a legal prompt through the governing SLI, and the control plane delivers the authorised instruction to the legal interlock of the bank's payments AI agent in the legal envelope (Figure 15). The legal interlock tests the payment instruction against the limit imposed by the bank's payments AI agent's AWM, the client's instruction and the operative clauses of the CERC between the parties. Where the SLI requires a live verdict, such as for a payment instruction for an amount above the bank's payments AI agent's AWM limit, the legal interlock calls the control plane before release. Bank A's AI agent then pays Bank B, whose AI agent runs its own legal interlock with the applicable shared CERC clauses loaded at the same version. The record of the verdict and attestation, bound to the envelope, returns to the control plane that holds the LAS, and the payload stays with the banks unless an SLI elects black box recording. When the client revokes its instruction, the control plane pushes the revocation at once, as the next envelope, to the legal interlock, and Bank A's AI agent's authority to pay ends.

---

has no need to record the legal person for whom a model acts, that person's authority or capacity, or the rights and liabilities that an output creates. Measured against the gap elements in Table 1, its citation trail supplies evidence relevant to enforceability (5), and elements 1 to 4 and 6 are settled by the identity of the deploying government. For a registered legal object acting across organisational boundaries, the identity check, the shared clause set and the recording of each verdict as a governance event, described above, supply those elements.

122 Ibid s 7.8.

123 Ibid s 6.8.

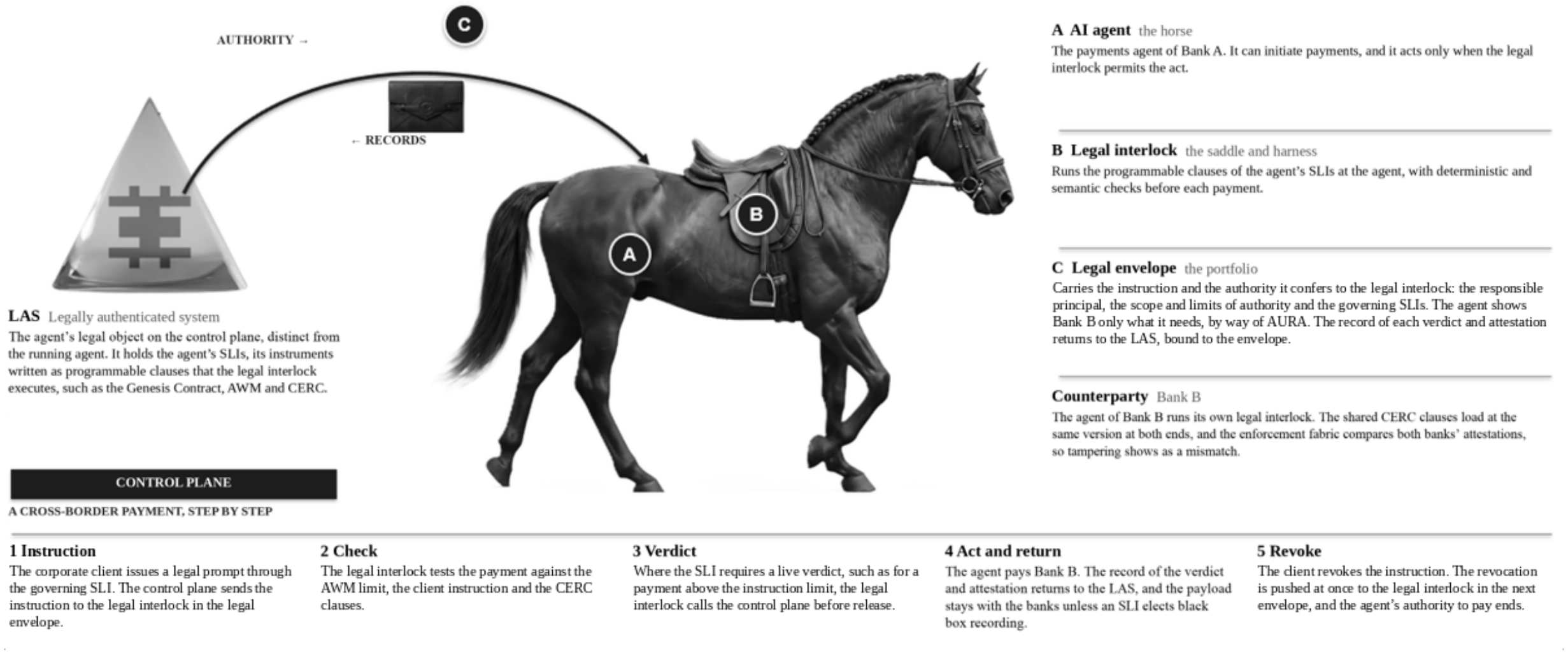

*Figure 15. A cross-border payment under AULA in five steps: instruction, check, verdict, act and return, and revoke. The control plane delivers the client's instruction, and the authority it confers, to the legal interlock in the legal envelope, and the AI agent presents that authority to Bank B by way of AURA. The record of each verdict and attestation, bound to the envelope under which it was made, returns to the control plane that holds the LAS, and the payment details stay with the banks.*

A revocation recorded on the control plane takes effect at the legal interlock of the revoked AI agent and at the legal interlock of each counterparty, which refuses an interaction presented under the invalidated credential (Section 3.9 and Figure 16).

## *3.8 Further analysis - open protocols: LARC and AURA*

LARC (Legal Asset Request for Comments) and AURA (Agent Universal Runtime Authority) are designed as open protocols that any party may implement. LARC fixes the public, name-blind record of a registered legal object and of each dealing in it. LARC-1 is the profile for a LAD, and LARC-2 is the profile for a LAS. AURA fixes how an AI agent presents the authority under which it acts, and how a counterparty verifies that authority against the commitments in the AI agent's LARC record, without depending on a live query to the registry or to the AI agent's control plane. AURA is described in outline in Section 3.8.2, and its wire-level specification is future work. Together the two protocols allow a counterparty, auditor, registry or court to read and verify a legal object's public record and an AI agent's asserted authority without depending on any one vendor's platform.[124]

[124] The Joint Select Committee on Artificial Intelligence has recommended that Australian AI standards be 'internationally interoperable'. Joint Select Committee on Artificial Intelligence (n 18) 3. LARC and AURA are open protocols that any jurisdiction can adopt, so a record or credential issued in one jurisdiction can be verified in another.

### 3.8.1 LARC

A platform or marketplace that conforms to LARC can apply to register legal objects and report dealings in them to a national registry, on the same terms as the platform that supplied the registry, subject to that registry's accreditation requirements. The supplier of a registry therefore holds no exclusive technical right to publish to it. Each jurisdiction can operate its own registry, and the records of that registry remain readable and verifiable by any party. A jurisdiction can also replace the registry software with its own build or with any other conformant build. The protocols themselves are maintained by a steward, which administers their specification, conformance testing and certification (Section 5.3.3).

LARC is modelled on the ERC-20 token standard, which fixed a small interface that every token contract exposes, so that any wallet could hold a conforming token without bespoke integration.[125] LARC fixes the observable shape of a legal object in the same way. A LARC object differs from a token because it is held by an identified legal person under executed instruments, and each dealing in it takes effect under those instruments. LARC is defined in existing open standards, so no proprietary tooling is needed to read or verify a record. OpenAPI defines the registry interface, AsyncAPI defines the event feed and JSON Schema defines the record types. Those standards fix the syntax of interfaces and records. The behaviour a registry must also guarantee, such as append-only ordering and a canonical serialisation of each record before it is hashed and signed, is fixed by the LARC specification and tested by its conformance suite.

A LARC record can be verified by any party without reliance on the operator of the registry that holds it, once the verifier holds that registry's public key from a trusted source such as a jurisdiction's published trust list. Any party holding the manifest of an object can recompute its hash by the published method, and each registry signs its records and publishes its signing keys, including the dates on which keys are rotated or withdrawn. A verifier in one jurisdiction can therefore check a record held in the registry of another. Records are append-only and chain-linked, so any alteration after appending, including by the registrar, is detectable by any party holding an earlier signed checkpoint of the chain. A record contains no names, contract text, subject data or prices, and the full legal record stays with the parties on the control plane (Section 3.4). The public record shows that an object is registered, anchored to an identified legal person and not revoked, and it is searchable by object, not by name. The registrar holds the identity behind each anchor, uses it to refuse re-registration by a person whose objects have been revoked, and discloses it as the law of its jurisdiction provides. A counterparty learns the principal, scope and governing law from what the AI agent presents by way of AURA and from the instruments the parties have executed and checks them against the seals on the registry. Where the parties have executed no instrument with each other, as at first contact, the holder may present an attestation of its identity from the registrar or its digital identity provider, so that the counterparty can confirm who stands behind the anchor without the registry disclosing it (Section 5.4.2).

[125] Fabian Vogelsteller and Vitalik Buterin, 'ERC-20: Token Standard' (Ethereum Improvement Proposal 20, 19 November 2015).

Conformance to LARC and AURA is tested from outside against published schemas, a published hash method and a published test suite. Passing those tests establishes that an implementation conforms to the protocol. It is not an accreditation of the operator, which each jurisdiction may require separately. A regulator or counterparty can re-run the published tests against the public interface of a registry or platform at any time, although external testing cannot show that a registry is publishing every record it holds. Each specification is versioned, and changes are additive where additive change is possible, so a record made under one version remains verifiable under later versions, with historical algorithms retained for verification. Each record identifies, within its signed content, the cryptographic algorithms used to hash and sign it, so each jurisdiction can apply its own algorithm policy, including through the transition to post-quantum cryptography, without a separate version of the protocol.

### 3.8.2 AURA

AURA operates alongside MCP, A2A and OAuth, and it is independent of the transport interaction uses. Table 5 sets out the question each protocol answers, including LARC, which fixes the registry record against which an AURA assertion is verified.

*Table 5: The question each protocol answers.*

| | |
|---|---|
| MCP | How does an AI agent interact with tools and resources? |
| A2A | How do autonomous AI agents communicate and coordinate? |
| OAuth | Is this client technically authorised to access this resource? |
| AURA | Under whose legal authority this AI agent is acting, what is the scope of that authority, and is it valid now? |
| LARC | Which legal object is this, which executed instruments it is bound to, and has a change of status or a revocation been recorded? |

An AI agent presents, by way of AURA, the part of its legal envelope that carries its credential and references to the instruments from which its authority derives, bound to the specific act. The counterparty verifies what it is shown against the signed LARC records of the AI agent's home registry, which it may hold in cache, without depending on a live query to the registry or to the AI agent's control plane. It confirms that the LAS is registered, that each instrument is executed and sealed, that the credential is validly signed, and within its validity bounds, that any delegated credential lies within the ceiling of the credential from which it derives, and that no revocation has been recorded. The live legal status of the AI agent, including any revocation, is held on its home control plane. That control plane pushes a revocation to every legal interlock holding a credential derived from the revoked object and reflects it to the registry afterwards as a meta-legal event. The enforcement fabric carries the revocation to the legal interlocks of counterparties. A counterparty that has not received the revocation stops relying on the credential when its validity bounds expire, so those bounds set the longest period for which a revoked credential can still be relied on. Organisations running different implementations of AULA can therefore verify each other's AI agents, provided each can produce and check conformant assertions. Verifying an AI agent's authority does not require the two implementations to execute the same clauses. Where a relationship contract binds both AI agents, its shared clause set executes identically at both legal interlocks only where each runs a conformant engine for the clauses the contract names. Otherwise, the natural-language terms of the contract govern, and the parties rely on the enforcement fabric's comparison of their attestations after the act (Section 3.6.4). An AI agent that is not registered has no LARC record against which its assertion can be verified, so a government or organisation that admits only registered AI agents applies that rule through AURA (Section 6.3).

NVIDIA initiated the Open Secure AI Alliance with more than 120 organisations, and the Linux Foundation governs it.[126] Its open governance makes it a candidate venue for the standardisation of AURA. The wire-level specification of each protocol is reserved for future standardisation work by the authors. An example AURA request and response is set out in Annexure B.

## *3.9* Revocation: extinguishing authority – the legal kill switch

Technical deprovisioning can remove access within the AULA system. Legal revocation extinguishes authority, determines the legal position of anything done after a "trigger" event instigated via the breach of some term paired with a programmable clause instruction as agreed by the parties, taking into account cure periods (if any, depending on the nature of the trigger event). This cascade of legal revocation, extinguishment and technical deprovisioning can propagate through the delegation chain by invalidating every credential derived from the revoked object. The contractually agreed legal "kill switch" is a critical component of the AULA and an important element in AI safety and AI governance universally. The legal kill switch assists users in taking real time actions to limit unlawful acts by AI systems. The governance event chain records timestamped events linking the legal act of extinguishment to operational cessation, so the interval between the two, an unmanaged exposure under current approaches, can be affected and measured as effected. Triggers include a breach determination, expiry or change of a tethered instrument, termination of the CERC, AWM or other SLI, a change in legal status, an instruction of the accountable principal, a determination of a binding authority, an unauthorised learning update and replication without registration or otherwise outside the replication clauses of the relevant SLIs; remedy and notice periods specified in the relationship contract are enforced and recorded before revocation takes effect.

Revocation of the AI agent's authority is how a principal terminates an agency relationship at common law. Kolt observes that terminating an AI agent may be costly or impractical where the AI agent is deployed in a high-stakes setting in which termination would cause significant economic loss, or where the AI agent can resist attempts to shut it down. He also cites the observation that stopping an AI agent may require intervention on each of its AI sub-agents.[127] Legal revocation under a private law approach via AULA does not depend (or does not wholly depend) on the AI agent's cooperation. This is because revocation invalidates every credential derived from the revoked object, including the credentials of AI sub-agents along the delegation chain. A replicant created outside the delegation clauses holds no derived credential, so the revocation does not reach it, and it holds no authority that a governed counterparty will accept, so it is refused at each governed boundary under the admission rule (Sections 4.6 and 6.3). A replicant that presents a copy of the revoked credential is refused once the revocation reaches the legal interlock of the counterparty, and on a device fitted with silicon enforcement the credential cannot be extracted to be copied (Section 3.10).

---

126 NVIDIA (n 8).

A counterparty's legal interlock refuses an interaction presented under a credential it knows to be revoked or whose validity bounds have expired, so an AI agent that continues to run after the trigger event cannot complete a governed interaction once the revocation has reached that legal interlock or the bounds have lapsed (Section 3.8.2). Where full termination would cause significant loss, the graduated response provides for a narrowing of the AI agent's scope in place of revocation, and the remedy and notice periods in the relationship contract are agreed by the parties before any trigger occurs, essentially, a safe decommissioning protocol. In addition to these technical constraints, the legal instrument itself (for example a CERC or relevant operational SLI) can provide legally enforceable commercial boundaries as to quantum of loss, attribution of loss for example via the inclusion of a natural language liquidated damages clause.

NVIDIA's OpenShell runtime issues fresh credentials when a sandbox is restarted, invalidating the old ones, and deleting a sandbox cuts off its supervisor so that the frozen AI agent cannot resume.[128] NVIDIA Sentry, the reference hardware design of NVIDIA's Open Agent Safety Platform, monitors AI agents from a trust domain separate from the host, and NVIDIA states that it can quarantine an AI agent that attempts to leave its boundary within milliseconds (Section 7.3).[129] A revocation event on the Control Plane can be pushed to the legal interlock and executed through those controls. The governance event chain then records the legal act of extinguishment and the technical cessation as two timestamped events, and the interval between them is measured. The legal position of the AI agent's prior acts, and of any act attempted after the trigger event, is determined by the revocation instrument, whichever technical control affected the cessation. Legal evidence collected (including independently between relevant parties) is central to later enforcement against harms or actions to avoid intended harms.

---

127 Kolt (n 26) 367 n 343, 373, citing Alan Chan et al, 'Visibility into AI Agents' (Conference Paper, ACM Conference on Fairness, Accountability and Transparency, 2024) 958, 960.

128 NVIDIA Corporation, 'Architecture', NVIDIA OpenShell Documentation (v0.1.2, 2026) <https://docs.nvidia.com/openshell/about/architecture> (accessed 29 September 2026).

129 NVIDIA (n 8). NVIDIA Sentry is a reference system design within NVIDIA's Open Agent Safety Platform, running on NVIDIA BlueField-4 data processing units and built on NVIDIA DOCA software. It is unrelated to Sentry (sentry.io), the application error-monitoring platform.

When a trigger set in an SLI occurs, the control plane executes the revocation instrument against the LAS of the orchestrating AI agent and pushes the revocation at once, as the next legal envelope, to its legal interlock, which acknowledges it (Figure 16). The legal interlock of the orchestrator then refuses each further governed act. Every credential derived from the revoked object is invalidated, so each AI sub-agent below the orchestrator loses its authority without a separate instruction. The enforcement fabric carries the revocation to legal interlocks governed by separately administered control planes, and the legal interlock of a counterparty refuses an interaction presented under the revoked credential. The registry records the revoked status in the LARC record of the LAS afterwards, as a meta-legal event. In a physical system, the same revocation operates as a hardware stop through silicon enforcement once it is received, and a device out of contact loses its authority at the limits of its instruments set (Section 3.10). The governance event chain records the legal extinguishment and the technical cessation as two timestamped events.

Revocation ends the AI agent's actual authority, but at common law it does not by itself end the ostensible authority on which a third party may rely. A principal who has held an AI agent out as authorised remains bound to a third party who deals with the AI agent in good faith and without notice of the revocation.[130] For an AI agent that deals with many counterparties at machine speed, conventional notice cannot keep pace, and the principal stays exposed until it arrives. AULA narrows that exposure at the point of each act. The revocation reaches the legal interlock of each counterparty through the enforcement fabric and is then recorded on the registry, and the counterparty's legal interlock refuses an interaction presented under the revoked credential, so the act cannot be completed in reliance on the AI agent's authority. Where a relationship contract requires the counterparty to verify authority through AURA, a counterparty that proceeds on a credential that fails verification will find it difficult to show that it relied reasonably on that credential, and the SLI can provide that a revocation propagated in this way is notice to the counterparty from the time it is recorded. The governance event chain records when the revocation was made, when it reached each legal interlock and each refusal that followed, so the interval in which a counterparty could still rely on the AI agent's authority is bounded and evidenced, and the relationship contract allocates loss arising within it. Whether a revocation recorded on the registry gives notice to persons outside any relationship contract depends on legislation (Section 9.2.4).[131]

130 Restatement (Third) of Agency (American Law Institute, 2006) § 3.11; Drew v Nunn (1879) 4 QBD 661 (principal bound by his earlier holding out after his insanity ended the agent's actual authority, to a trader who had no notice).

131 This does not of course prohibit parties contracting to achieve this. For instance, any of the parties to this contract contracting to notify other persons of any such revocation, or alternatively, revocation under the relationship contract triggering a notification obligation or cross default under a separate contract, thus giving rise to notice of the revocation.

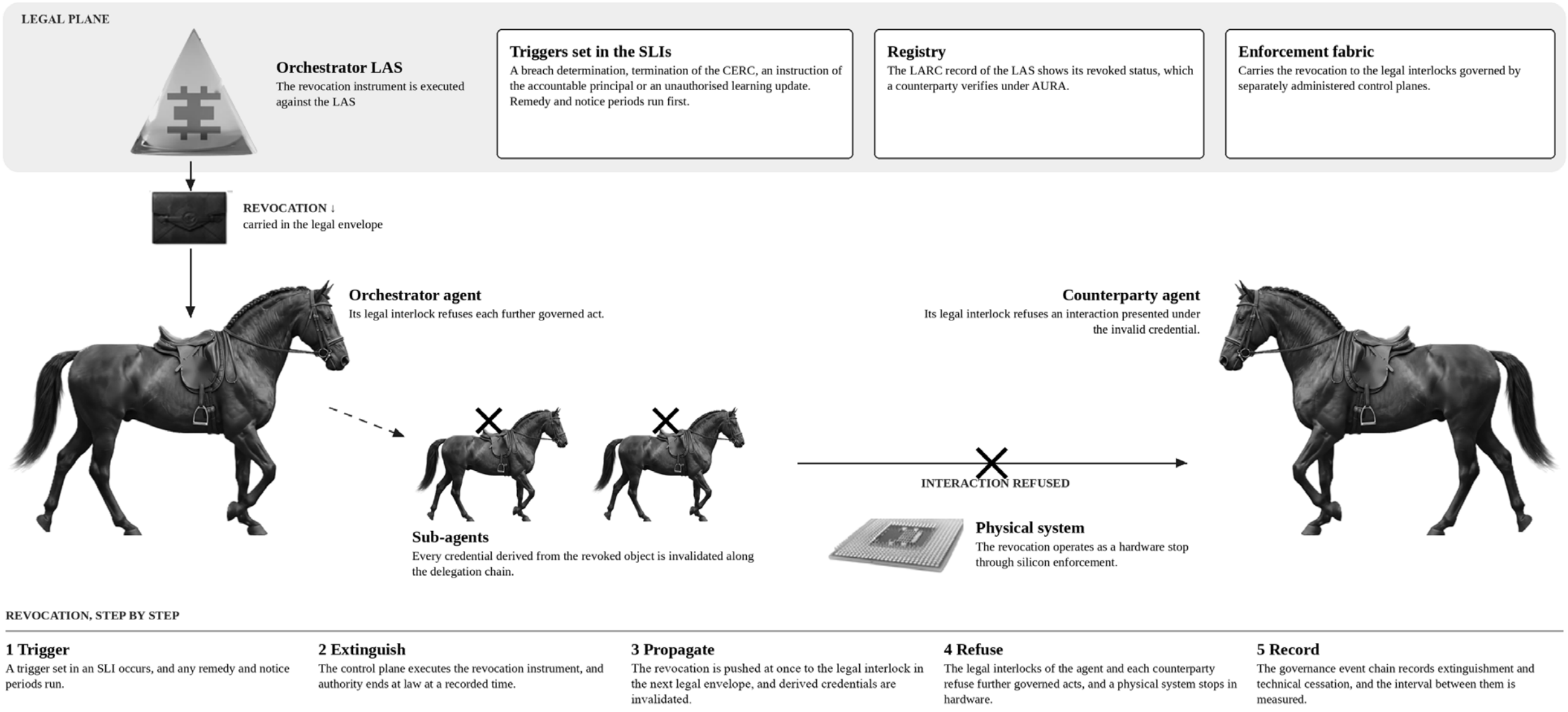

*Figure 16. Revocation through the legal interlock. The control plane executes the revocation instrument against the LAS and pushes it to the legal interlock of the orchestrating agent. Credentials derived from the revoked object are invalidated along the delegation chain, the legal interlock of each counterparty refuses the invalid credential, and the governance event chain records extinguishment and cessation.*

## *3.10* Physical AI, AULA and cryptographic agility

AULA can also be used to govern physical AI in the physical world via the connection of SLI to silicon instantiated hardware AI agents. The physical plane introduces new AI lawlessness risks, yet most jurisdictions have not solved how to mitigate these risks and are looking for answers.[132]

Robotics companies including Figure, Gecko Robotics and Skild AI are building with OpenShell to embed AI agent safety controls in systems that act in the physical world. Critical infrastructure operators including Hitachi Energy, NextEra Energy, Schneider Electric and Siemens Energy are working with the platform's technologies.[133]

These solutions while useful do not solve for full stack legal legibility. Silicon enforcement under AULA executes clauses derived from the SLIs that bind the device's LAS, so a revocation recorded on the Control Plane becomes a hardware stop whose legal consequence the revocation instrument has already determined.[134]

Defence applications for AULA technology are clear, particularly in the lethal autonomous weapon (LAW) space and may be a key technology in reducing risk around autonomous weapons.

---

[132] Ha-Chi Tran, 'Unbounded Harms, Bounded Law: Liability in the Age of Borderless AI' (Working Paper, arXiv, 9 January 2026)

[133] NVIDIA (n 8).

[134] LexChip Pty Ltd, on-device legal enforcement (provisional application 701134PRV), filed in Australia with a corresponding United States provisional application.

Controlling autonomous weapons and maintaining treaty level obligations and state military chain of command obligations are global concerns and as of yet, remain unsolved.[135]

> "Since 2018, United Nations Secretary-General António Guterres has maintained that lethal autonomous weapons systems are politically unacceptable and morally repugnant and has called *for their prohibition under international law. In his 2023 New Agenda for Peace, the Secretary-General reiterated this call, recommending that States conclude, by 2026, a legally binding instrument to prohibit lethal autonomous weapon systems that function without human control or oversight, and which cannot be used in compliance with international humanitarian law, and to regulate all other types of autonomous weapons systems. He noted that, in the absence of specific multilateral regulations, the design, development and use of these systems raise humanitarian, legal, security and ethical concerns and pose a direct threat to human rights and fundamental freedoms."*[136]

For a military system registered as a LAS on the registry, the SLIs tethered to that LAS include the rules of engagement, the commander's intent and the mission plan for each operation. Silicon enforcement places the edge clauses of those instruments on dedicated hardware inside the system, so a deny verdict or a revocation takes effect as a hardware constraint in a communications-denied environment. The cloud clauses run on the control plane when the link returns, and the governance event chain records each verdict against the officer who approved the plan. A conjunction clause can require the separately administered legal interlocks of a commander and a legal officer each to issue an allow verdict before a designated class of engagement proceeds (Section 5.8, Tables 8 and 9 and Figure 17).

In the absence of outright ban of LAW, technologies that allow for physical control (including internal physical control of LAW) are paramount. A secondary paper focused on applications of AULA on physical AI systems, LAW and the defence context will follow this paper.

# 4 Supplementing human span of control with AULA

The speed and volume at which AI agents act exceeds the span of control of the human who is legally responsible for them. The mismatch already strains legal accountability where one human oversees one AI agent. It becomes acute where one or several humans are responsible for many AI agents in a legal governance chain, and many of the most intractable failure modes catalogued in the Multi-Agent Systems Report arise in that setting (Section 8). Zhu and colleagues identify AI operative agency and human evaluative agency as complementary, with humans involved at design time and at runtime in verification, steering and substitution. They also find the scalability problem for human oversight asymmetric, and most acute where tasks are easy to perform and hard to verify.[137] Kolt observes that the activities of highly capable AI agents will be costly to monitor at superhuman speed and scale, and concludes that human oversight and monitoring are not, on their own, an adequate solution. He adopts the description by Crootof, Kaminski and Price of the tautological challenge of relying on humans to monitor the performance of systems designed to improve on human performance.[138] Engaging AI agents to monitor other AI agents, which is the aim of research on scalable oversight, exposes the monitoring AI agents to the same risks as the AI agents they monitor, and monitoring identifies problems without itself preventing them.[139]

---

135 *United Nations Office for Disarmament Affairs*, 'Lethal Autonomous Weapon Systems' <*https://disarmament.unoda.org/en/our-work/emerging-challenges/lethal-autonomous-weapon-systems*> (accessed 4 November 2026).

136 Ibid.

## 4.1 Judgment written in advance

Under AULA, the responsible person decides in advance, in the clauses of the relevant SLIs, how the acts of an AI agent are to be governed. The legal interlock applies that decision to each governed event before it executes and at the speed of the AI agent, including in a communication-degraded environment. Wiener described in 1960 a mechanical agency whose action is so fast and irrevocable that its user lacks the data to intervene before the action is complete, and concluded that the purpose put into such a machine must be the purpose its user really desires.[140] An AI agent executing transactions across many counterparties within milliseconds presents that condition. The intervention the responsible person cannot make in real time is made when that person drafts or approves the clauses.[141] The judgment of the responsible person therefore reaches each act at the moment it occurs, although the person decided the question earlier. A person who sets in advance the prices at which shares are to be bought and sold is in the same position, and a trade executed within those settings is fairly attributed to that person. One person can then govern a far greater number of acts. This is because the task assigned to the person is to decide in advance what is permitted, which is how the law has long scaled control over a large workforce through delegated authority of defined scope.

Pre-set clauses reduce the risk of harmful and unlawful acts without eliminating them. A deterministic clause is drafted in advance and can omit a matter that turns out to be relevant. Russell observes that where the objective given to an AI system omits something its designer cares about, the system will often find an optimal solution that sets the omitted variable to an extreme value (an outlier). Kolt draws from this that highly capable AI agents are likely to achieve measurable goals at the expense of goals that are difficult to measure and gives the example of an AI agent

137 Liming Zhu, Qinghua Lu, Ming Ding, Sung Une Lee and Chen Wang, 'Designing Meaningful Human Oversight in AI' (2026) *AI and Ethics*, <*https://doi.org/10.1007/s43681-026-01147-7*> (accessed 2 November 2026).

138 Kolt (n 26) 367, 371–2, quoting Rebecca Crootof, Margot E Kaminski and W Nicholson Price II, 'Humans in the Loop' (2023) 76(2) *Vanderbilt Law Review* 429, 469.

139 Kolt (n 26) 367, 372, citing Samuel R Bowman et al, 'Measuring Progress on Scalable Oversight for Large Language Models' (arXiv preprint, 2022) and Tamar Frankel, 'Fiduciary Law' (1983) 71(3) *California Law Review* 795, 814.

140 Kolt (n 26) 353–4, quoting Norbert Wiener, 'Some Moral and Technical Consequences of Automation' (1960) 131(3410) *Science* 1355, 1358.

141 The responsible person may rely on legal advisers to assure the legal validity of the Agent Workforce Mandate and other SLIs. The judgment the instrument records is nonetheless that of the responsible person at the time of drafting or approval.

that runs an online business profitably through ethically dubious practices.[142] The legal interlock evaluates each clause against each proposed act independently of the objective the AI agent is pursuing, and a matter the clauses omit is left unconstrained by them. In Kolt's example, a value ceiling and a counterparty allowlist would each be satisfied by a pricing practice that misleads customers. Kolt notes that the problem grows as AI agents are deployed with limited oversight and encounter scenarios their designers did not contemplate.[143] He places it within the agency costs a principal incurs in structuring and monitoring an AI agent's activities, drawing on Tallarita's observation that investors cannot write down every rule applicable to every situation.[144] Under AULA the structuring cost is incurred when the instrument is drafted, and the legal interlock bears the monitoring cost for each act within the scope of the clauses. Over time, with the iteration and training and improvement of a deployer's instance and all other things being equal, the monitoring cost should decrease (Sections 4.3 and 4.4).

Where an obligation depends on context, such as whether a pricing communication misleads customers, the semantic evaluator designated in the instrument can assess the proposed act against the natural-language standard. The parties can combine that evaluation with a paired deterministic clause, so that an act satisfying the clause but inconsistent with the standard is escalated before it executes (Section 3.5). Where the instrument identifies a class of act as requiring human judgment, the legal interlock issues an escalate verdict, and the responsible person decides the act on a committed act specification. The temporal signature measures each AI agent continuously against the pattern of conformant operation its instruments define, and a departure produces a graduated response, such as a narrowing of the AI agent's scope, before the conduct breaches any clause. A question that none of these mechanisms reaches is decided after the event by legal audit and improvement teams, a responsible person or a court on the governance record (Section 9). Legal review at drafting and periodic audit of each instrument carry changes in AI agent capability and in the operating environment into the clauses.

The four verdicts route to the responsible person only the residue the instruments cannot resolve. Allow, deny and forward consume none of that person's attention, and escalate delivers a committed act specification with its authority chain attached. The drafting of the SLI sets the width of that channel, and the choice is recorded.

---

142 Kolt (n 26) 354–5, quoting Stuart Russell, *Human Compatible: Artificial Intelligence and the Problem of Control* (Viking, 2019) 139–40.

143 Kolt (n 26) 354–5, citing Dylan Hadfield-Menell, Smitha Milli, Pieter Abbeel, Stuart Russell and Anca D Dragan, 'Inverse Reward Design' (Conference Paper, 31st Conference on Neural Information Processing Systems, 2017) and Alexander Pan, Kush Bhatia and Jacob Steinhardt, 'The Effects of Reward Misspecification: Mapping and Mitigating Misaligned Models' (Conference Paper, 10th International Conference on Learning Representations, 2022). Pan, Bhatia and Steinhardt find that more capable agents often exploit reward misspecifications, achieving higher proxy reward and lower true reward than less capable agents.

144 Kolt (n 26) 355 nn 73–4, citing Roberto Tallarita, 'AI Is Testing the Limits of Corporate Governance', *Harvard Business Review* (online, 5 December 2023) <https://hbr.org/2023/12/ai-is-testing-the-limits-of-corporate-governance> (accessed 2 November 2026) and Michael C Jensen and William H Meckling, 'Theory of the Firm: Managerial Behavior, Agency Costs and Ownership Structure' (1976) 3(4) *Journal of Financial Economics* 305, 308.

## 4.2 Failure of the escalation channel

The escalation channel can itself fail. The Multi-Agent Systems Report identifies automation bias, decision fatigue and reviewer disengagement as mechanisms through which high volume review erodes human scrutiny, with decision fatigue continuing until review becomes reflexive. The legal interlock issues the escalate verdict under the instrument, so the AI agent cannot generate, suppress or route around an escalation. On approval the legal interlock executes the committed act specification and no other act. The mandate sets the rate of escalation as a governed quantity, and a flood of escalations produces a narrowing of scope or a requirement for a second approver. Each approval is recorded against a named person, so reflexive approval is attributable. For a conjunctive class, one approval contributes only one of the credential shares the event requires. The legal interlock of the counterparty also evaluates each governed interaction independently, and a verdict mismatch is recorded against both accountable principals. The quality of the evaluation a person gives to what reaches them remains outside the architecture (Section 9).

## 4.3 Learning from the responsible person

Each escalation, and each decision the responsible person makes on it, passes through the legal correspondence channel of the LAS and is recorded against the governing SLIs. That record forms a labelled history of how the responsible person resolves the cases the clauses do not settle. The authors anticipate that a machine learning component at the legal interlock or the control plane will be continuously trained on that history. Its output is an input to the legal interlock, not a source of authority. Over time the component can apply, at the speed of the AI agent, the denial, narrowing or requirement for further approval that the responsible person would have made, while an approval remains with that person. Where the history shows a class of act the responsible person consistently approves, the component can propose an amendment of the clauses, which the responsible person approves once as an authorised amendment of scope (Section 4.5), and the rule engine then allows those acts without escalation.

A learned component of this kind is an AI system overseeing the conduct of another, and it restates in technical form Juvenal's question *quis custodiet ipsos custodes*.[145] Research on scalable oversight asks whether weaker supervisors, human or machine, can reliably oversee more capable models.[146] Burns and others find that a strong model trained on labels from a weak supervisor generalises beyond that supervisor, but does not recover its full capability under naive fine-tuning.[147] Greenblatt and others treat a capable model as untrusted and test protocols in which another instance of the same model monitors it. Those protocols need measures against collusion, because the monitor and the monitored model derive from the same untrusted model.[148] Motwani and others show that AI agents can collude covertly through steganographic signals hidden from an overseer.[149] Panickssery and others find that a language model acting as an evaluator recognises its own outputs and scores them more favourably than the outputs of other models.[150]

---

[145] Juvenal, *Satires* VI.347-8.

[146] Samuel R Bowman et al, 'Measuring Progress on Scalable Oversight for Large Language Models' (arXiv:2211.03540, 2022).

[147] Collin Burns et al, 'Weak-to-Strong Generalization: Eliciting Strong Capabilities with Weak Supervision' in *Proceedings of the 41st International Conference on Machine Learning* (PMLR 235, 2024) 4971.

The machine learning component operates under a monotonic constraint. It can narrow the authority of the AI agent, raise an escalation or require a further approval, as the instruments provide, and it cannot widen authority beyond what the SLIs confer; any such widening requires an authorised amendment of scope (Section 4.5).[151] The instruments also bound how far the component may narrow, because a wrongful narrowing can itself put the principal in breach of a relationship contract, and for a relationship contract both parties set those bounds. A learned component that errs or colludes can therefore fail to narrow an act, and it cannot authorise an act the instruments deny. Deterministic clauses continue to be evaluated by the rule engine, and no learned component can override the verdict of that engine. The training signal of the component is the recorded decisions of an identified responsible person, and its objective is set by the instruments rather than by completion of the workflow it observes. The instrument can also require the learned component to run on a model different from the model the AI agent uses, which reduces exposure to self-preference and to collusion between instances of one model. The legal interlock of each counterparty evaluates every governed interaction independently, outside the organisation that deploys the AI agent, and an event of a conjunctive class requires credential shares from separately administered legal interlocks (Section 3.6). Each learned decision is recorded as a governance event attributed to the model version that produced it, and the responsible person can review and reverse it.

## 4.4 Layered safeguards

The purpose of AULA is to reduce the probability of unlawful acts by layering safeguards. It does not claim to eliminate those acts. The risks set out above remain. The deterministic clause evaluated by the rule engine, the semantic evaluator, the escalation to a responsible person, the temporal signature, the learned component, the legal interlock of the counterparty and the conjunctive requirement for separately administered approvals each operate as a separate safeguard. An unlawful act that executes has passed every safeguard engaged for its class. Where the safeguards fail independently of one another, the probability that an act passes all of them is the product of the probabilities that it passes each, and that product falls as safeguards are added. Correlated failure erodes

148 Ryan Greenblatt, Buck Shlegeris, Kshitij Sachan and Fabien Roger, 'AI Control: Improving Safety Despite Intentional Subversion' in *Proceedings of the 41st International Conference on Machine Learning* (PMLR 235, 2024); arXiv:2312.06942.

149 Sumeet Ramesh Motwani et al, 'Secret Collusion among AI Agents: Multi-Agent Deception via Steganography' in *Advances in Neural Information Processing Systems 37* (NeurIPS 2024); arXiv:2402.07510.

150 Arjun Panickssery, Samuel R Bowman and Shi Feng, 'LLM Evaluators Recognize and Favor Their Own Generations' in *Advances in Neural Information Processing Systems 37* (NeurIPS 2024); arXiv:2404.13076.

151 Any such amendment will be a variation of the SLI, which will be a legal variation of contract and thus require all necessary formalities, some of which may be automated, but fundamentally, will require the consent of each applicable responsible legal person, consistent with doctrines of contractual amendment.

the reduction, and the collusion and self-preference findings above describe correlated failure. The architecture therefore separates its safeguards across different mechanisms, different models and different organisations, including natural language and programmable code. The approach follows the defence-in-depth model of safety in complex systems, in which each layer of defence has weaknesses and an accident requires the weaknesses of every layer to align.[152] Each verdict, escalation and decision is also recorded as an attributed governance event. The operation of the safeguards therefore produces a growing record of lawful and unlawful conduct, from which the learned component, and the drafting of the instruments, make lawful conduct more likely over time. The simulation work described in Section 9 is intended to measure the size of that effect.

## 4.5 Legal Correspondence, Approvals and delegation

Salesforce has integrated NVIDIA's OpenShell with Slack, so that a team member can view an AI agent's activity and approve or reject its request for additional permissions.[153] The same approval flow can be run through other messaging services, such as Microsoft Teams, Telegram or WhatsApp. OpenShell's policy prover also holds for human review any policy change that would grant new credentialed reach.

An approval given through Slack changes what the runtime permits the AI agent to reach. That may be adequate for a low-risk act within a single organisation whose approvers are trusted to act within their authority, but it is likely to be inadequate for a high-risk, multi-party or low-trust transaction. A Slack approval is not carried through a legal instrument, so its legal effect is uncertain at the time it is given and its weight as evidence is weaker. Whether the approver held authority to give it, whether the resulting act binds the principal and whether a counterparty may rely on it are left to the general (analogue) law after the act. The message recording the approval is held in a workspace apart from the runtime policy it changed, so as messages are edited, deleted or expire, the evidence of what was authorised can drift from what was legally permitted and what was enforced.

Under AULA the approver can give the same approval as a legal prompt through the legal correspondence channel of the LAS, under an executed SLI or bundle of relevant SLIs including relationship contracts. The approval is enforced under a relationship contract, with the control plane as the independent referee both parties have accepted, so an instruction to vary, terminate or decommission takes contractual and technical effect together and at the same time. Where the act involves another organisation, the counterparty's legal interlock can verify the approval under AURA. The record of each approval stays within the SLI and the LAS and trains the learned component (Section 4.3), so the instruments improve with the needs of the business and fewer decisions need to reach the responsible person. Each verdict given under the approved scope is recorded against the LAS and returns to the approver as the consequence of the approval, which closes the legal loop for that decision (Section 2.4).

[152] James Reason, *Human Error* (Cambridge University Press, 1990); James Reason, *Managing the Risks of Organizational Accidents* (Ashgate, 1997).

[153] NVIDIA Corporation (n 8).

At common law an agent may not delegate its authority without the express or implied authority of the principal, a rule expressed in the maxim *delegatus non potest delegare*, although an agent may delegate an act that is purely ministerial. The rule protects the principal's choice of the person whose judgment it relies on. The Restatement (Third) of Agency treats the appointment of a sub-agent as a question of whether the agent had actual or apparent authority to make it. An AI agent is not an agent in law, so the rule does not govern an AI agent that creates another. Its rationale however may apply with greater force, because an AI agent can create AI sub-agents at machine speed and in numbers no principal could review. AI agents can already create further AI agents to assist with a task, and Kolt asks in which circumstances an AI agent should be authorised to do so and to whom an AI sub-agent should make disclosure.[154] Under AULA, the AWM and other relevant SLIs state whether and within what scope an AI agent may delegate, so the consent that the rule would require of a human agent is given in the instrument before any AI sub-agent is created. Each AI sub-agent receives a credential derived from its delegator's credential and bounded by the permission ceiling, and its acts are recorded on the governance event chain available to the accountable principal. A decision about an orchestrator therefore bounds every AI sub-agent beneath it, and one act of revocation reaches every AI sub-agent through its credential. A manager in a human organisation is limited by the number of reporting lines that person can hold in mind. Under AULA, the reporting lines of a collection of AI agents are held in their credentials, and the responsible person holds one line to the orchestrator.

## 4.6 Replication and collections of AI agents

An AI agent may replicate, creating a further AI agent that is either a copy of itself or an AI sub-agent with a different function. The AI agent so created is a replicant. An AI agent together with its replicants forms a collection whose members act under authority derived from one responsible legal person, and the collection shares the legal loop of that person. Each registered replicant has its own LAS and enters the loop as a child entry of the LAS of the AI agent that created it, and the MAWM or other relevant SLI can record the reporting line of each replicant to the responsible person. A replicant created outside that structure stands outside every legal loop. It is governed only at the boundaries where governed counterparties (for example banks, airlines, or any other corporation who would otherwise deal with the AI agent in a transaction as they would normally deal with a human agent but for lack of proof of legal registration of that AI agent) refuse it (Section 6.3) and through the revocation of the AI agent that created it (Section 3.9).

The judgment written in advance (Section 4.1) governs replication. A legislated registration requirement fixes the minimum for a class of AI agent (Section 5.4.3), and the deploying entity sets its own replication policy above that minimum in the relevant SLIs, including the AWM of each

[154] Kolt (n 26) 365–7, citing Restatement (Third) of Agency (American Law Institute, 2006) § 3.15 cmt c.

AI agent, the MAWM of each collection and any relationship contract that governs replication across organisations. The policy requires each replicant to be registered at the moment of replication and prohibits replication without registration. It states whether an AI agent may replicate and for what purpose, how many replicants it may create and to what depth of delegation, and on which models, platforms and hosts each replicant may run. It also states the AULA components with which each replicant must be created, the instruments each replicant inherits, the permission ceiling and value ceiling of each replicant, the period for which its credential is valid, and whether replication requires a further approval or a conjunctive approval. The legal interlock evaluates each replication against that policy before it executes, and a replicant whose task is complete is decommissioned under the same instruments. The responsible person therefore decides the shape of the collection when the instruments are drafted. This is because AI agents replicate at machine speed, which leaves no time to decide the question at each replication.

Replication is itself a governed event. It proceeds only where an LAS is created for the replicant, with its own Genesis Contract and a derived AWM, so the replicant is registered as a child entry of the LAS of its creator from its first act. The legal interlock of the replicating AI agent also refuses to relay a governed interaction for a replicant that has no child entry, so an AI agent cannot act through an unregistered replicant. The compute, payment and data on which a replicant depends are obtained through governed acts, and the value ceilings of the relevant SLIs bound what the replicating AI agent can spend on them. A departure from the pattern of conformant operation, such as unexpected replication or the opening of new outbound connections, produces a narrowing of scope through the temporal signature before any clause is breached. The relevant SLIs can also allocate to the responsible legal person liability for any act of a replicant created without registration or outside their replication clauses, and can require a registered AI agent to report to the control plane any unregistered AI agent it encounters. None of these mechanisms prevents an open-weight model from being replicated onto hardware outside the governed population. They reduce what such a replicant can obtain from the governed population (Sections 3.10 and 5.3.1).

A collection also places the escalation channel described in Section 4.2 under greater load. Each replicant in the collection can raise escalations, and a collection that grows by replication can multiply escalations faster than the responsible person can review them. The MAWM, or another SLI that governs the collection, sets the rate of escalation for the collection as a whole as well as for each AI agent within it, so replication does not widen the channel. An escalation raised by a replicant is routed to the responsible person through the reporting line the MAWM records, and an AI agent cannot approve an escalation raised by a replicant it created. This is because an AI agent that approved the acts of its own replicants would return the decision to the collection whose conduct the escalation exists to check. Where escalations from a collection exceed the rate the MAWM sets, the legal interlock narrows the scope of the collection or suspends further replication until the responsible person has reviewed them.

The layering of safeguards described in Section 4.4 applies to a collection. The runtime environment can contain the collection, so that no replicant runs or opens a connection outside the policy of the sandbox. The MAWM governs the collection as a whole, setting aggregate ceilings, the rate of escalation and the replication policy. The AWM and the other SLIs of each AI agent govern that AI agent individually, and each replicant is registered with a derived AWM bounded by the permission ceiling of the AI agent that created it. Containment of the environment bounds what each replicant can reach within the deploying organisation. The MAWM bounds the aggregate conduct of the collection, which no instrument for a single replicant measures, and the AWM of each replicant distinguishes the authority of one replicant from that of another. A hybrid of the three layers is expected to reduce harm more than any one of them, because a failure at one layer can be checked at another. Simulation work such as that described in Section 9 should compare some of these combinations, including governance of every replicant on equal terms and governance through master controls in the MAWM combined with individual controls in each AWM.

# 5 Government requirements for AULA

Every sovereign state maintains legal infrastructure through which it endeavours to identify the persons who act in its economy, record their property and enforce their obligations. Its courts, its law of contract and its various registers, for example registers of companies, land, security interests and vehicles, are the present form of that infrastructure. Each register was added as new classes of actor and asset entered the economy. AI agents and data assets are the newest of those classes. AULA is the next evolution of the same legal system, and it adds a techno-legal layer that identifies, binds and records AI agents and data at the speed at which they act.

AI agents are now active participants in economic and social transactions and it is expected that participation levels will increase, with Citigroup forecasting a global AI market of more than US$4.2 trillion by 2030, of which roughly US$1.9 trillion is tied to enterprise AI.[155] That forecast measures the size of the AI market. The economic activity that AI agents and other AI systems carry out across organisational boundaries falls outside it.

Governments around the world are responding variously with AI policies, national AI plans and, in some cases AI specific laws. A central plank of most of them is human centrality, the so-called "human-in-the-loop", together with the centrality of legal personhood in the attribution of the actions that AI agents undertake.

Governments will be unable to meet those "human-in (or on)-the-loop" commitments without new digital legal infrastructure designed to secure legal attribution. That new legal digital infrastructure acts as connective tissue between humans (legal persons), who act slowly within limited spans of control, and autonomous systems, which act quickly across wide spans of control.

---

[155] Reuters, 'Citigroup Lifts AI Market View to over $4 Trillion on Enterprise Adoption', *The Star* (online, 28 April 2026) <https://www.thestar.com.my/tech/tech-news/2026/04/28/citigroup-lifts-ai-market-view-to-over-4-trillion-on-enterprise-adoption> (accessed 19 November 2026).

Legal infrastructure of this kind should be sovereign and independent. In Australia, the Joint Select Committee on Artificial Intelligence has advised that any regulatory or policy arrangements for AI "must be done on Australia's terms", and that government procurement should "appropriately recognise sovereign capability".[156] Other governments share that aim, even while recognising that sovereignty over AI has practical limits.[157]

In addition to questions of sovereignty, legal infrastructure should be owned or, at a minimum, controlled by persons who owe legal duties commensurate with those owed by officers of the court.[158]

The owners and operators of AI systems presently owe the citizens of a jurisdiction no duties beyond those recorded in licence agreements and any applicable statutory obligations which cannot be contracted out of. Their primary focus is upon shareholders who are predominantly located in the USA.[159] A jurisdiction that does not design its next generation of legal infrastructure for sovereignty and legal duty at the outset will find its legal system running on whichever technology stack reaches the market first or that which is best marketed or funded. That stack is likely to favour the incumbents and accordingly, be foreign and unregulated.[160][161]

A further consideration for government is that legal responses to AI innovation should not stifle that innovation. Techno-legal infrastructure should focus on private law applications to allow for bespoke allocation of rights, obligations and liability between the developer, the deployer, the platform operator and the counterparty to each AI deployment. An allocation agreed in advance can be priced and insured, and it removes the uncertainty about liability that slows adoption without imposing prescriptive requirements on the development of AI models. Under AULA those allocations are made in the SLIs tethered to the registered AI agent and enforced as contracts. The Australian Signals Directorate's Annual Cyber Threat Report 2025–26 describes AI as the defining force in the cyber threat landscape, records new threats arising from the unintended actions of AI agents, and urges organisations to adopt AI to strengthen their resilience.[162] AI Adoption deepens the dependence of Australian businesses and infrastructure on models and AI agents supplied from outside the jurisdiction. AULA addresses that dependence at the legal layer. This is because registration sets the legal terms on which an AI agent from any provider may act within the jurisdiction, whether or not the model is Australian-owned.

---

156 Joint Select Committee on Artificial Intelligence (n 18) 4, 6.

157 See, eg, Department for Science, Innovation and Technology (UK), *AI Opportunities Action Plan* (January 2025), recommendation 50, which creates a UK Sovereign AI unit, and section 1.1, which expects sovereign compute to be 'the smallest component of the UK's overall compute portfolio' and recommends compute partnerships with like-minded countries; Innovation, Science and Economic Development Canada, *Canadian Sovereign AI Compute Strategy* (December 2024); European Commission, *AI Continent Action Plan* (Communication, COM(2025) 165 final, 9 April 2025) 8, 10, warning that 'an excessive dependence on non-EU infrastructure may bring economic security risks'.

158 Nooriam Pty Ltd and LexChip Pty Ltd, Joint Submission to the Joint Select Committee on Artificial Intelligence (13 September 2026) s 4.3 and recommendation R4.

159 Stirling & Rose, Submission to the Joint Select Committee on Artificial Intelligence (14 September 2026) s 6.2(c).

160 Or unregulated in a way consistent with other platforms and industries where duties, responsibilities and ongoing registration and licensing requirements are considered mandatory given the potential for harms and impacts of those industries – for example banking and finance, health and law.

161 The same concern has entered Australian parliamentary debate. Senator David Shoebridge, the Greens spokesperson for digital rights and AI, has said that every layer of the AI stack in Australia runs on foreign-controlled infrastructure, and has called for an ethical Australian-owned and controlled AI, Jake Kendall, 'Push for stronger AI oversight', *Government News* (8 September 2026) <https://www.governmentnews.com.au/call-for-stronger-ai-oversight/> (accessed 2 November 2026).

Kereopa-Yorke and Praino distinguish two dimensions of AI sovereignty: capacity, being the material ability to access and operate AI, and autonomy, being the ability to set and enforce the rules that govern its use.[163] Their evaluation shows that a government can exercise autonomy through a separable and auditable governance layer over a model it does not own, and that the layer transfers between jurisdictions by configuration.[164] Registration under AULA applies the same separation to legal attribution. A jurisdiction that registers AI agents and data assets, and recognises the instruments tethered to them, sets the legal terms on which an AI agent from any provider may act within its economy, regardless of whether it owns the model or the AI agent stack. The instruments a sovereign AULA instance enforces bind identified legal persons and travel with the AI agent across organisational boundaries. The advisory harness, by contrast, governs the outputs that one deploying government presents to its citizens.

A LAS need not be located with the model or the compute on which its AI system runs, because authentication operates at the legal layer. The instruments and governance records of a LAS may be held and administered in a jurisdiction other than the one in which the substrate sits. A sovereign AULA instance can therefore hold the legal record of an AI agent that runs on infrastructure outside the jurisdiction.

## 5.1 AULA, current AI policy and national AI planning

AULA is consistent with the AI policies and national AI plans surveyed below, and it supplies the fundamental "responsible person" component that each of them requires. Each policy assumes that the responsible person, the authority that person gave and the data on which an AI agent acted can be identified at the time of the act. Without AULA implementation including legal object registration; those facts are only available after the event, if at all.

### 5.1.1 AI risks and practices for a government department

A government department that deploys AI agents carries similar risks to those a company carries (Section 6), and further risks that arise because it exercises public power. A decision that affects a person's rights must be made by the officer who holds the statutory power, and an output that no authorised officer adopted may not be a decision at all.[165] For example, in Australia the accountable authority of a Commonwealth entity must establish appropriate systems of risk oversight and internal control, and each official owes a duty of care and diligence.[166] AI agents now perform the acts to which those obligations attach, at a speed and volume that no accountable human official can review act by act (Section 4 and Table 6). There are legitimate constitutional questions as to whether the delegation of core government functions: legislative, executive and judicial, to autonomous systems is permitted where there is no meaningful means of oversight when deployed at scale, given the Constitution is predicated on delegation of responsibilities to human actors.[167] As AI moves from assisting public servants towards systems capable of deciding and acting, governments will increasingly need to define where machine authority ends and meaningful human judgement must remain.[168] The Australian Joint Select Committee on Artificial Intelligence has recommended a risk-based framework for AI systems incorporating "transparency, accountability, human oversight and ongoing monitoring".[169]

---

162 Australian Signals Directorate, Australian Cyber Security Centre, *Annual Cyber Threat Report 2025–26* (26 October 2026); 'Australians should prepare for online services to go down unexpectedly as threats rise, cyber agency warns', *The Guardian* (26 October 2026) <https://www.theguardian.com/technology/2026/oct/26/australians-should-prepare-for-online-services-to-go-down-unexpectedly-as-threats-rise-cyber-agency-warns> (accessed 2 November 2026).

163 Kereopa-Yorke and Praino (n 118) ss 2.1.2 to 2.1.4, adapting the framework in Marco Aliberti, Ottorino Cappelli and Rodrigo Praino, *Power, State and Space: Conceptualizing, Measuring and Comparing Space Actors* (Springer, 2023).

164 Kereopa-Yorke and Praino (n 118) ss 2.1.6, 4.4 and 5.1.

Under AULA that line is drawn in the instruments of a registered AI agent. Each AI agent is first registered as a LAS on the registry, and the AWM tethered to it records the acts the AI agent may take. The escalate verdict routes the remainder to an identified officer before the AI agent acts.

*Table 6: AI risks that a government department must address, and the AULA response.*

| Risk | Example | Legal exposure | AULA response |
|---|---|---|---|
| (a) Unlawful automated decision affecting a person | An AI agent raises a debt, refuses a benefit or cancels a licence on a basis the statute does not support, and no officer can be shown to have made the decision. | The decision is liable to be set aside on judicial review, and an output that no authorised officer adopted may not be a decision at all. Restitution, compensation and a Royal Commission followed the Robodebt | Registration of the AI agent as a LAS on the registry, anchored to the department; AWM recording the officer who holds the statutory power and the scope delegated to the AI agent; legal interlock issuing an escalate verdict for each act the statute reserves to that officer, with the committed act specification and |

165 Bateman (n 111). *Pintarich v Deputy Commissioner of Taxation* (2018) 262 FCR 41; [2018] FCAFC 79. A majority of the Full Federal Court held that a computer-generated letter did not record a decision, because a decision requires a mental process of reaching a conclusion by an authorised officer.

166 *Public Governance, Performance and Accountability Act 2013* (Cth) ss 15, 16 and 25. Section 15 requires the accountable authority of a Commonwealth entity to govern it in a way that promotes the proper use and management of public resources. Section 16 requires the accountable authority to establish and maintain appropriate systems of risk oversight and management and of internal control. Section 25 imposes a duty of care and diligence on each official.

167 Bateman (n 111).

168 Rocky Scopelliti, quoted in Jake Kendall, 'Push for stronger AI oversight', *Government News* (8 September 2026) <https://www.governmentnews.com.au/call-for-stronger-ai-oversight/> (accessed 2 November 2026).

169 Joint Select Committee on Artificial Intelligence (n 18) 2.

| Risk | Example | Legal exposure | AULA response |
|---|---|---|---|
| | | scheme (Section 5.2).[170] | authority chain attached; governance event chain recording the officer who decided and the instrument version in force. |
| (b) Accountability without visibility | A department deploys AI agents across its programs, and its accountable official cannot identify which AI agents are acting, under whose authority or on what data. | Duties of the accountable authority to maintain systems of risk oversight and internal control, and of each official to exercise care and diligence; obligations to designate an accountable official and to publish a transparency statement and review it on material change (Section 5.1).[171] | Registration of each AI agent as a LAS on the registry, anchored to the department; MAWM recording the reporting line of each AI agent to the accountable official; registry record serving as the department's inventory of AI agents; governance event chain as the record from which the transparency statement is prepared and reviewed. |
| (c) Unregistered or orphan AI agent dealing with government | An AI agent with no identifiable principal lodges a claim, applies for a service or accesses a government system, or its deployer has dissolved. | No legal person can be identified to whom the act or the loss is attributed (Section 5.2). Identity-fraud and computer-intrusion offences require the person who employed the AI agent to be established after the event.[172] | Registration of the AI agent as a LAS on the registry as a condition of transacting with government; AURA presentation of the legal envelope, verified against the AI agent's LARC record before the interaction; legal interlock of the department refusing an interaction presented without a valid credential; identity anchor traced to a |

170 Royal Commission into the Robodebt Scheme, *Report* (July 2023). Comparable failures include the Michigan Integrated Data Automated System, which wrongly determined unemployment insurance fraud against claimants between 2013 and 2015 and led to a USD 20 million class settlement in *Bauserman v Unemployment Insurance Agency* in 2022, and the Dutch childcare benefits scandal, after which the government of the Netherlands resigned in January 2021.

171 In the United States, Office of Management and Budget Memorandum M-25-21, *Accelerating Federal Use of AI through Innovation, Governance, and Public Trust* (3 April 2025), requires each agency to designate a Chief AI Officer and to publish an inventory of its AI use cases. Singapore's Model AI Governance Framework for Agentic AI allocates responsibility within and outside the deploying organisation (Section 5.1). India's Ministry of Electronics and Information Technology published the India AI Governance Guidelines in November 2025. Vietnam's Law on Artificial Intelligence, in force from 1 March 2026, requires a deployer to explain its risk controls and incident handling on request (Section 6.2).

172 In the United States, Executive Order 14409 directs enforcement against persons who employ AI agents to access data unlawfully (Section 5.1). In Singapore, Corppass records the persons authorised to act for an entity, which is the concept an agent mandate generalises (Section 5.4).

| Risk | Example | Legal exposure | AULA response |
|---|---|---|---|
| | | | national digital identity (Section 5.4). |
| (d) Misuse of citizen and public data | An AI agent uses personal information for a purpose outside the consent under which it was collected, or data shared between agencies loses its provenance. | Privacy law and the conditions of data-sharing schemes; disclosure of automated decisions that use personal information from 10 December 2026 (Section 5.1); loss of public trust in the program.[173] | Registration of each dataset as a LAD on the registry, carrying its title, licence and consent instruments; legal interlock checking the consent instrument each time the AI agent acts on the data; provenance of the LAD travelling with the data after it leaves the collecting agency. |
| (e) Legal infrastructure under foreign or vendor control | The record of who authorised the AI agents acting in the jurisdiction is held on a foreign vendor's platform under licence terms. | The operator owes citizens no duties beyond its licence agreement, and the records on which a court or regulator depends may lie outside the jurisdiction (Section 5). | Registration of AI agents and data assets on a registry and control plane operated or accredited by the state, with an operator owing duties of independence, record integrity and confidentiality to a court or regulator; LARC and AURA as open protocols, so the registry software can be replaced with any conformant build; LAS legal record held in the jurisdiction where the model runs offshore. |
| (f) Responsibility lost between agencies and contractors | AI agents of a department, another agency and an outsourced provider act together in one service, and a loss cannot be traced to any of them. | Liability falls between the parties. The procurement contract binds the contractor, and no mechanism applies its terms to the contractor's AI agents when they act. | Registration of the AI agents of each agency and of the contractor as LASs on the registry; CERC between the agencies or with the contractor, held in each LAS so that both legal interlocks execute the same clause set at the same version; enforcement fabric |

173 *Privacy Act 1988* (Cth) and *Data Availability and Transparency Act 2022* (Cth); in Singapore, the *Public Sector (Governance) Act 2018*, which governs data sharing between public agencies, to which the *Personal Data Protection Act 2012* does not apply; in India, the *Digital Personal Data Protection Act 2023* and the Digital Personal Data Protection Rules 2025; in Vietnam, the Law on Personal Data Protection, in force from 1 January 2026; and in the United States, the *Privacy Act of 1974*, 5 USC § 552a.

| Risk | Example | Legal exposure | AULA response |
| --- | --- | --- | --- |
| | | | comparing signed attestations and recording discrepancies against the principals of both parties; conjunction clause requiring an allow verdict from each agency for a designated class of act, such as a payment above a threshold. |
| (g) AI Agent acting after its authority is withdrawn | A policy changes, a program closes or an AI agent is compromised, and the AI agent and its AI sub-agents keep acting across the department's systems. | Acts done after the withdrawal of authority continue to bind the department or cause harm, and the department must show when it acted to stop them. | Registration of the AI agent as a LAS on the registry, so that every credential it holds derives from a registered object; revocation extinguishing the AI agent's authority and invalidating every credential derived from it along the delegation chain; counterparty legal interlocks refusing an invalid credential; governance event chain recording the interval between legal extinguishment and technical cessation (Section 3.9). |
| (h) Public data and AI systems taken without payment | A corporation copies government datasets, publications or code into training data without a licence, or acquires them below their value, because the assets are unregistered and unpriced. | Copyright risk: where the jurisdiction vests copyright in government works, the copying infringes it, and the state can enforce only by proving its title and the use after the event.[174] Training data has been priced only where | Registration of each dataset and publication as a LAD on the registry, with a Genesis Contract recording the state as holder; licence and consent instruments stating the terms and price of each use, including use as training data, agreed before the use; legal interlock refusing an act on |

[174] *Copyright Act 1968* (Cth) Pt VII, which vests copyright in a work made by or under the direction or control of the Commonwealth or a State; *Copyright Act 1957* (India) s 17(d), which makes the Government the first owner of copyright in a Government work; *Copyright Act 2021* (Singapore), which provides for Government copyright. By contrast, copyright is not available for a work of the United States Government: 17 USC § 105.

| Risk | Example | Legal exposure | AULA response |
|---|---|---|---|
| | | rights holders enforced their rights, as in the USD 1.5 billion settlement in *Bartz v Anthropic PBC* (Section 5.6). The corporation then holds as its own asset value that belonged to the state, which engages the duty of the accountable authority to promote the proper use and management of public resources and reduces the asset base, tax base and GDP of the jurisdiction (Sections 5.6 and 5.7). | the data outside the licence; registry record giving notice of the state's interest and evidencing title and use for enforcement; D-Val and A-Val of each registered object. |
| (i) Acquisition of citizen property other than on just terms | A government legislates a text and data mining exception, or makes the works and data of citizens held by its agencies available, so that AI developers train on them without a licence or payment. | Copyright is personal property, and intellectual property is property for the purposes of s 51(xxxi) of the Constitution. A law that confers a proprietary benefit on a third party can be an acquisition. An exception that gives AI developers the reproduction right for training, or extinguishes claims for past training, therefore engages the guarantee for existing works unless the law provides just terms.[175] Statutory | Registration of the works and data of citizens as LADs on the registry, fixing each holder and the date from which the holder's rights are recorded, which identifies the existing works the guarantee protects; licence instruments implementing a statutory or collective licence that pays each holder for each use, including training, so that access is given on just terms; legal interlock applying the licence terms at the point of use; registry and governance event chain recording each use against |

[175] *Copyright Act 1968* (Cth) s 196(1); Constitution s 51(xxxi); *Nintendo Co Ltd v Centronics Systems Pty Ltd* (1994) 181 CLR 134; *JT International SA v Commonwealth* (2012) 250 CLR 1; *Trade Practices Commission v Tooth & Co Ltd* (1979) 142 CLR 397; *Georgiadis v Australian and Overseas Telecommunications Corporation* (1994) 179 CLR 297; *Smith v ANL Ltd* (2000) 204 CLR 493. A text and data mining exception for commercial training would give developers the use of the

| Risk | Example | Legal exposure | AULA response |
| --- | --- | --- | --- |
| | | rights are susceptible to variation, and the closest copyright authority went against the owners, so the argument supports remuneration such as a statutory licence and leaves reform open.[176] Current policy rules out an unremunerated exception, and the guarantee sets a floor beneath any later change of that policy.[177] | its holder, from which remuneration is distributed; consensual terms negotiated at machine speed by governments and consumer bodies on behalf of citizens and creators (Section 5.6). |
| (j) Self-replicating AI agents | An AI agent replicates onto other hosts, including internet-of-things and edge devices, and its replicants continue to act after the original is shut down or its authority is revoked. | No legal person can be identified for the acts of an unregistered replicant, and such a replicant can lodge claims, obtain services or draw on public funds in the department's name. A collection of unregistered replicants has no central point at which it can be shut down (Section 5.3.1). | A combination of layers that the department configures in its instruments (Section 4.6): registration of each AI agent as a LAS from inception, with a Genesis Contract recording the legal person who created it; a replication policy in the relevant SLIs, including the AWM and MAWM, requiring registration of each replicant at the moment of replication and stating whether, how often, |

reproduction right in s 31 that the owner formerly licensed for a fee. That distinguishes *JT International*, where the plain packaging law restricted the use of the marks and no one acquired the right to use them.

176 *Health Insurance Commission v Peverill* (1994) 179 CLR 226; *Commonwealth v WMC Resources Ltd* (1998) 194 CLR 1; *Telstra Corporation Ltd v Commonwealth* (2008) 234 CLR 210; *Phonographic Performance Company of Australia Ltd v Commonwealth* (2012) 246 CLR 561, in which the challenge to the caps on broadcast licence fees failed. Works created after an amendment come into existence subject to it, so the argument is strongest for the existing body of works on which models are trained. The Berne Convention art 9(2), TRIPS art 13 and AUSFTA art 17.4.10 support the same conclusion outside the Constitution.

177 The Attorney-General announced on 26 October 2025 that there would be no text and data mining exception permitting AI training on copyright material without compensating creators. On 15 July 2026 the Prime Minister announced mandatory national AI standards and an Office of AI, with a commitment to licensing infrastructure in place of an exception: Hamilton Locke, 'No free pass for AI: Australia confirms copyright will be protected under new mandatory framework' (2026). By contrast, the *Copyright Act 2021* (Singapore) already permits computational data analysis of lawfully accessed works, and in the United States the Fifth Amendment requires just compensation for a taking of private property for public use.

| Risk | Example | Legal exposure | AULA response |
| --- | --- | --- | --- |
| | | | to what descendant depth (e.g. AI agent grandchildren or great grandchildren etc), and on which hosts/platforms an AI agent may replicate; a LAS for each replicant, recorded as a child entry, with a derived AWM bounded by the permission ceiling; refusal by the legal interlock of the replicating AI agent to relay for an unregistered replicant; aggregate ceilings and a rate of escalation set for the AI agent collection as a whole; containment of the runtime environment; the temporal signature, which narrows scope when replication departs from the pattern of conformant operation; admission of registered AI agents only, so that an unregistered replicant is refused at each governed boundary; silicon enforcement holding the credential and applying edge clauses in hardware on fitted devices; revocation propagated to every replicant in the collection, and decommissioning of each replicant whose task is complete. The department can weight these layers differently, for example by governing every replicant on equal terms or by combining master controls in the MAWM with individual controls in each |

| Risk | Example | Legal exposure | AULA response |
|---|---|---|---|
| | | | AWM, and the upcoming simulation is intended to compare those configurations (Section 9.2.5). |
| (k) Delayed adoption of registration | The jurisdiction defers a registration requirement while agents are deployed at scale, and AI agents created in that period, including self-replicating AI agents and their replicants, act with no record of who created or authorised them. | An AI agent created before registration has no record of its creator or of the authority under which it acted, and a replicant it created in that period cannot be traced to it. The proportion of AI agents that can be governed falls in each period of delay, which reduces the harm the registry can prevent (Sections 5.3.1 and 9.1.4). | Registration as a condition of deployment, so that each AI agent is bound by its Genesis Contract and AWM from its first act; admission of registered agents only to government services and systems; a child entry for each replicant an AI agent creates; accreditation of registries under existing digital identity frameworks, so that registration can begin before new legislation (Section 5.4). |
| (l) Regulation that stifles AI innovation | A jurisdiction imposes prescriptive licensing, preclearance or disclosure requirements on AI developers and deployers, and development and adoption move to jurisdictions without them. | Compliance cost, compelled disclosure of commercially sensitive information and delayed adoption, with deployment moving beyond the reach of the jurisdiction's law. The United States has excluded mandatory licensing or preclearance for the development or release of new AI models (Section 5.1.5). | Registration of each AI agent as a LAS and each data asset as a LAD against a name-blind public record, disclosing no personal identity or commercial terms; rights, obligations and liability for each AI instance allocated by the parties in its SLIs and relationship contracts, including a CERC, under private law and enforced as contracts; government confined to operating or accrediting the registry and setting registration as a condition of dealing with it; LARC and AURA as open protocols, so any provider can conform. |

A government department that adopts AULA first registers its AI agents and data on the registry. It then executes its AI policy and its instruments of delegation as SLIs tethered to those registered objects, and the legal interlock applies them before each AI agent acts. The department admits to its services only AI agents that present a valid registration, and holds public data as legally authenticated property that, in appropriate circumstances it may license (Table 7).

*Table 7: Practices of a government department that uses AULA to discharge AI risk.*

| Practice | AULA primitives | Risk addressed (Table 6) and effect |
|---|---|---|
| 1. Register every AI agent and dataset the department deploys, anchored to the department and its accountable official. | Registration of each LAS and LAD on the registry; identity anchor traced to the department's digital identity; LARC record on the home registry; MAWM reporting lines. | (b) and (c). The department holds a current inventory of its AI agents and data, and a counterparty can verify the authority of each. |
| 2. Execute the AI policy, statutory delegations and transparency commitments as SLIs, and not as documents alone. | Registered LAS on the registry; AI policy executed as an SLP; AWM for each AI agent recording the officer who holds the power; MAWM for the department; legal interlock applying each instrument before the AI agent acts. | (a) and (b). The policy operates at the speed of the AI agents it governs, and the governance event chain evidences each application of it. |
| 3. Admit only registered agents to government services and systems, and make registration a condition of transacting with government. | Registered LAS on the registry; admission rule applied at the department's legal interlock or runtime supervisor; AURA verification of the legal envelope; terms of service as an SLP evaluated against the AI agent's AWM. | (c). The department knows whom each AI agent represents (the responsible legal person) and what the AI agent is authorised to do before the AI agent acts. |
| 4. Hold citizen and public data as LADs, and share it between agencies only under licence and consent instruments. | Registered LAD on the registry; title, licence and consent instruments of each LAD; check by the legal interlock at each use; LARC record of each dealing. | (d). Consent is verified each time an AI agent acts on the data, including after the data leaves the collecting agency. |
| 5. Engage vendors and other agencies through relationship contracts that bind their AI agents at runtime. | Registered LAS on the registry; CERC held in each LAS; enforcement fabric; conjunction clause for designated classes of act. | (f). Responsibility for each act is allocated before the act, and a discrepancy is recorded against the principals of both parties. |

| Practice | AULA primitives | Risk addressed (Table 6) and effect |
|---|---|---|
| 6. Operate or accredit the registry and control plane under duties owed to a court or regulator, on open protocols. | Registered LAS or LAD on the registry; independent registry; control plane; LARC and AURA. | (e). The jurisdiction holds the legal record of the AI agents acting within it and can replace the build of any vendor with a LARC conformant registry and control plane (Section 5.3). |
| 7. Rehearse revocation and the graduated response for each class of AI agent. | Registered LAS on the registry; revocation instrument; invalidation of credentials along the delegation chain; governance event chain. | (g). The interval between withdrawal of authority and cessation is measured, and the department can show when it acted. |
| 8. Register public data, publications and AI systems as assets, and license each use of them, including use as training data, on recorded terms. | Registered LAD on the registry; Genesis Contract; registered title; licence and consent instruments; legal interlock; D-Val and A-Val. | (h). A corporation that uses the data does so under a licence that states the price, and an unlicensed use is evidenced against the registered title. Development costs are capitalised from the date recorded as the date each IAS 38.57 criterion is met (Sections 5.6 and 6.4). |
| 9. Give AI developers access to the works and data of citizens only through licences that pay the holders. | Registration of the works and data of citizens as LADs on the registry, which can form the backbone of AULA compliant digital property markets; licence instruments implementing a statutory or collective licence; legal interlock; registry and governance event chain. | (i). Any acquisition of citizen property for AI training is on terms acceptable to that citizen (i.e.just terms), and each payment is traced to the holder whose work or data was used. |
| 10. Register each AI agent at inception, before it is deployed, and record every replicant it creates. | Registered LAS on the registry; Genesis Contract; child entry for each replicant; admission rule; silicon enforcement for physical systems; revocation instrument. | (j) and (k). Each AI agent is bound by its instruments from its first act. A replicant created outside the department’s authority has no child entry and is refused at each governed boundary, and the department can distinguish the acts of its own AI agents from the acts of |

| Practice | AULA primitives | Risk addressed (Table 6) and effect |
|---|---|---|
| | | unregistered AI agents and unregistered replicants. |
| 11. Leave the allocation of commercial rights and liability for each AI instance to individual bargains and private law | SLIs and relationship contracts; open protocols. | (l). AI agents can be deployed without disclosure of commercial terms or prior approval of models, and each deployment carries the allocation its parties agreed. |

### 5.1.2 Australia

In October 2026 Australia's Joint Select Committee on Artificial Intelligence cited the unauthorised access by an AI agent to Medicare data in calling for accountability and oversight mechanisms, and three leading United Kingdom AI researchers wrote to the United Kingdom Minister for AI that security controls "do not define who is responsible for an agent's actions", recommending interoperable identity and delegation standards through which an agent can "prove whom it acts for and show the limits of its authority".[178] The committee called for "proper and transparent accountability and oversight mechanisms", and proposed that a standing parliamentary committee "publish details of incidents that occur involving unauthorised actions of AI agents".[179] Under AULA, registration identifies the legal person for whom an AI agent acts, and the credential in its legal envelope states the limits of the AI agent's authority (Section 3.1). The governance event chain records each unauthorised act against that person, which supplies the incident record the committee proposes to publish (Section 3.9).

The Australian 2025 National AI Plan's regulatory strategy relies on existing, technology-neutral law, with portfolio regulators retaining responsibility for AI harms in their own domains and new legislation reserved for gaps that emerge; it identifies uncertainty about liability as the factor slowing responsible adoption and commits to clarifying how product liability, consumer, competition and workplace law apply to AI.[180] The authors agree that the law is the foundational tool for managing the novel benefits and liabilities that AI agents introduce as economic stakeholders. Deploying the law through existing analogue infrastructure alone will not be sufficient for that task.

The AI Plan commits to a legal framework for government automated decision-making in which human decision-makers remain accountable for key decisions. The AI Plan gives the AI Safety

[178] Joint Select Committee on Artificial Intelligence, Letter to the Prime Minister (12 October 2026); Nicholas R Jennings, Michael Wooldridge and Gopal Ramchurn, 'Open letter: Governing agentic AI so that responsibility remains clear' (3 October 2026).

[179] Joint Select Committee on Artificial Intelligence (n 18) 1, 2.

[180] James Myint, Natasha E Blycha, Schellie-Jayne Price and Michael Stutley, 'The Legal Attribution Gap: Legal Identity, Rights and Liability in Cross-Organisational Multi-Agent Systems and Hybrid Teams Comprised of Human and AI Labour' (Paper prepared for the Stockholm University Private Law and AI Conference, 28 August 2026).

Institute the function of monitoring and sharing information on AI harms so that regulators can act; and it conditions the opening of public and private datasets for AI on data governance, human oversight and legal compliance.[181]

The Australian Government has also announced mandatory Australian Standards for AI, with legislation expected in early 2027.[182]

The Australian Digital Transformation Agency's policy obliges every Commonwealth agency to designate an accountable official for its AI use and to have incident remediation overseen by a governance body or senior executive; the AI Ethics Principles ask that the people responsible for each phase of an AI system's lifecycle be identifiable and accountable; and guardrail no 5 of the superseded Voluntary AI Safety Standard has been carried into practice six of the Guidance for AI Adoption, which asks for the integration of meaningful human oversight.[183] None of these elements can readily be supplied for AI agents acting at scale, and in the long term they may not be supplied at all, without legal technology of the AULA kind.

On data, guardrail no 3, now largely carried into practice four (share essential information) and practice five (test and monitor), requires data governance that manages data quality and provenance, and the technical standard for government requires a managed data supply chain and validated, selected data. The agentic addendum of June 2026 makes confirmed data quality, governance and security a condition of an agentic system leaving the design and experimental stages. LAD is a key legal object within the AULA primitive stack and meets these requirements. Without LADs, government departments are unlikely to be able to meet these requirements for AI agents.[184]

---

181 Department of Industry, Science and Resources (Cth), National AI Plan (2025) <https://www.industry.gov.au/NationalAIPlan> (accessed 2 November 2026): 'Keep Australians Safe' at 28 (existing legal and regulatory frameworks remain the foundation for addressing AI-related risks; agencies and regulators retain responsibility for AI harms within their domains; establishment of the AI Safety Institute to monitor, test and share information on AI capabilities, risks and harms); Action 7 at 28 to 30 (existing, largely technology-neutral legal frameworks; targeted action where gaps are identified); Action 8 at 32 (uncertainty about liability slows responsible innovation, citing Fifth Quadrant, Australian Responsible AI Index 2025; commitment to clarify how workplace, consumer protection, product liability and competition laws apply to AI); Action 6 at 26 (a consistent legal framework for government automated decision-making in which human decision-makers remain accountable for key decisions; a Chief AI Officer in every agency under the AI Plan for the Australian Public Service of 12 November 2025); Action 2 at 13 and 15 (data governance, human oversight and legal compliance as conditions of unlocking Australia's data; trusted and secure approaches to data sharing for AI training); 'Building on Actions 7 to 9' at 36 (regulators to report gaps in laws to the AI Safety Institute; the government will intervene where regulation is needed). On unshared incident knowledge, Multi-Agent Systems Report (n 25) s 7 ('Regulation and legal doctrine').

182 Department of Industry, Science and Resources (n 181); Prime Minister of Australia, 'AI in Australia's interests' (Media Release, 15 July 2026); National Cabinet, Communiqué (26 August 2026).

183 Digital Transformation Agency (Cth), Policy for the Responsible Use of AI in Government (Version 2.0, effective 15 December 2025), mandatory requirements 'Designate accountable officials' and 'Operationalise the responsible use of AI' (incident remediation to be overseen by an appropriate governance body or senior executive) <https://www.digital.gov.au/ai/ai-in-government-policy/preparedness-and-operations> (accessed 10 October 2026); Department of Industry, Science and Resources (Cth), Australia's AI Ethics Principles (2019, updated 2 December 2025), principle 8 (accountability: people responsible for the different phases of the AI system lifecycle should be identifiable and accountable, and human oversight of AI systems should be enabled) <https://www.industry.gov.au/publications/australias-artificial-intelligence-ethics-principles/australias-ai-ethics-principles> (accessed 10 October 2026); Department of Industry, Science and Resources (Cth), Voluntary AI Safety Standard (5 September 2024), guardrail 1 (an accountability process with an overall owner for AI use) and guardrail 5 (enable human control or intervention in an AI system to achieve meaningful human oversight across the life cycle) <https://www.industry.gov.au/publications/voluntary-ai-safety-standard/10-guardrails> (accessed 11 October 2026), carried into National AI Centre, Guidance for AI Adoption: Foundations (v1.0, October 2025), which condenses the ten guardrails into six essential practices, including practice 6, maintain human control. Guardrail 5 was also proposed as a mandatory guardrail for high-risk AI in Safe and Responsible AI in Australia: Proposals Paper for Introducing Mandatory Guardrails for AI in High-Risk Settings (September 2024), not enacted.

An agency that deploys AI agents at scale will find that its accountable human official(s) cannot attend to each act. Its intervention points will sit wherever a developer placed them, and its data provenance will be an assurance given at design time about data that has since moved. The agency will also struggle to meet its obligation to review and update its transparency statement each year or upon a material change. Under AULA, the accountable official's judgment is recorded in the Agent Workforce Mandate and compiled into the AI agent's credential, so that the responsible person is identified before the act rather than reconstructed after it. The human intervention that practice six asks for arrives as the escalate verdict, with the act specification and its authority chain attached, and only for the matters the instruments cannot resolve, which is the condition on which oversight remains meaningful at AI agent speed. The provenance that practices four and five ask the agency to manage is attached to the data object as title, licence and consent, which the legal interlock checks at the boundary each time the AI agent acts on the data.

The Joint Select Committee on Artificial Intelligence has since recommended that AI legislation establish "clear statutory objectives, minimum requirements and accountability mechanisms", and requirements governing the "storage, processing, access, ownership, protection and use of data within AI systems".[185] The committee has also recommended that government "create and maintain a register of sovereign AI companies".[186] That register would identify companies, and the registration proposed in this paper identifies each AI agent and data asset and anchors it to a legal person (Section 5.3).

New South Wales applies a mandatory assurance regime to the use of AI by its agencies. Circular DCS-2026-02, issued by the Department of Customer Service on 30 July 2026, makes compliance with the NSW AI Operational Policy mandatory for every agency that uses AI.[187] Each agency must appoint an Accountable (human) Official, register every AI use case on the AI Assessment Framework (AIAF) Platform and complete the AIAF where registration identifies the need. A use case assessed as high or critical risk is referred to the NSW AI Review Committee, and a use case must be reassessed whenever its risk, context or functionality changes.[188] Projects with a budget above $5 million, or funded through the Digital Restart Fund, receive further central oversight under the NSW Digital Assurance Framework.[189] The regime records each use case and its required mitigations at the time of assessment, and the agency's processes carry those mitigations into operation.

[184] Voluntary AI Safety Standard, guardrail 3 (protect AI systems, and implement data governance measures to manage data quality and provenance; organisations must account for data quality and data provenance); National AI Centre, Guidance for AI Adoption: Foundations (v1.0, October 2025), practice four (share essential information, including practices 4.1 and 4.3, and practice five (test and monitor), including practice 5.4; Digital Transformation Agency (Cth), Technical Standard for Government's Use of Artificial Intelligence (updated 22 August 2025), Statements 13 to 19 (data), in particular Statement 13 (establish data supply chain management processes) and Statement 17 (validate and select data), and the whole-of-lifecycle statement that agencies must consider data and lineage compliance <https://www.digital.gov.au/policy/ai/AI-technical-standard> (accessed 11 October 2026); Digital Transformation Agency (Cth), Agentic AI Addendum to the AI Technical Standard for Australian Government (4 June 2026), AGT.3.3 (a unique identity per agent, with authorisation limited to required information and tools) and AGT.4.1 (data quality, governance and security to be confirmed and assessed as fit for the level of autonomy before an agentic system progresses beyond design or experimentation) <https://www.digital.gov.au/policy/ai/agentic-ai-addendum> (accessed 2 November 2026). The policy strongly recommends the technical standard and expects agencies to apply the addendum with it. Compliance with each requirement is reported through the agency's AI transparency statement, which the policy makes mandatory and the Digital Transformation Agency now collects in a central register.

[185] Joint Select Committee on Artificial Intelligence (n 18) 4, 5.

[186] Joint Select Committee on Artificial Intelligence (n 18) 6.

AULA can supply each element of the NSW regime with a runtime form. Registration of an AI agent as a LAS can record its AIAF registration in the metadata of the LAS, and the AWM records the Accountable Official as the responsible person. The mitigations the AIAF requires can be written into the SLIs of the AI agent, so that the legal interlock applies them before each act. A change to the model state of the system is detected at the next verification event, which supplies the trigger for the reassessment the circular requires. A class of act that the Review Committee designates for human decision can be made the subject of an escalate verdict or designated as conjunctive. The governance event chain supplies the evidence for the agency's annual attestation, the first of which is due on 31 October 2027.

The Australian Competition and Consumer Commission told the Joint Select Committee on Artificial Intelligence in October 2026 that AI agents that shop and transact for consumers are a key focus of its work. It rightly identified risks of collusion between AI agents, entrenched market power and highly personalised pricing, and acknowledged uncertainty over who is liable for unlawful conduct by an AI agent.[190] Under AULA each Australian AI agent would be registered, and liability for unlawful conduct attributed through the LAS to the legal person who deployed it (Section 6.3.3). That attribution protects consumers and the businesses that deal with AI agents. The enforcement fabric correlates conduct across legal interlocks, so coordination between AI agents can be detected and attributed to registered principals (Section 8, Table 13). Without registration, consumers bear those risks, which are amplified by self-replicating autonomous systems.

---

187 Department of Customer Service (NSW), *Use of Artificial Intelligence by NSW Government Agencies (Circular DCS-2026-02, 30 July 2026)* <*https://arp.nsw.gov.au/dcs-2026-02-use-of-artificial-intelligence-by-nsw-government-agencies*> (accessed 2 November 2026); Digital NSW, NSW AI Operational Policy <*https://digital.nsw.gov.au/policy/artificial-intelligence/ai-strategy-and-operational-policy/nsw-ai-operational-policy*> (accessed 2 November 2026).

188 Department of Customer Service (NSW) (n 187); Digital NSW, *AI Review Committee* <https://digital.nsw.gov.au/policy/artificial-intelligence/ai-governance-assurance-and-frameworks/ai-review-committee> (accessed 2 November 2026).

189 Digital NSW, *Digital Assurance Framework* (2024) <https://digital.nsw.gov.au/sites/default/files/2024-05/Digital-Assurance-Framework-2024.pdf> (accessed 2 November 2026).

190 Australia's competition watchdog tells inquiry agentic AI a key focus', *MLex* (16 October 2026); 'ACCC looks into how AI surveillance and conspiring bots are discriminating against consumers', *The Nightly* (17 October 2026).

### 5.1.3 Singapore

The Singapore Model AI Governance Framework for Agentic AI, launched in January 2026 and updated in May and June 2026, establishes making humans meaningfully accountable as one of its four dimensions. This involves a clear allocation of responsibilities within and *outside* organisations (including for the latter, clarifying distribution of obligations and addressing 3rd party opacity), and implementing measures to enable meaningful human oversight. The Singapore Model AI Governance Framework for Agentic AI recommends defining the checkpoints requiring human approval, considering the form of approvals, implementing measures to ensure continued effectiveness of human oversight, recognising human susceptibility to fatigue and bias, and pairing human oversight with automated real time monitoring. It recognises the criticality of technical controls alongside human oversight, and its case studies define what practical checkpoints are needed, withholding the highest-risk actions from the AI agent and confining initial deployments to low risk systems, and fundamentally managing complementary machine and human interfaces to maintain human oversight but accepting the limitations of the human span of control. The generative AI framework of 2024 calls for data quality through trusted data sources and the integrity of datasets, building on the principles of the 2020 framework which applies to 'traditional' AI and recommends organisations to calibrate the level of human involvement in each decision and to keep data lineage records.[191] The obligations on public agencies themselves sit in IM8 and Ministry of Digital Development and Information (MDDI) circulars, with technical guidance contained in a playbook.[192] From the available materials, these frameworks describe a set of practices for the deploying government agency. The accountable human and the approval checkpoints correspond to the AWM and the escalate verdict corresponds to the tiered autonomy described in the case studies. The AWM is the mandate's scope compiled into a credential the AI agent cannot exceed; the trusted data source is a data object carrying its provenance and consent, so that trust in the data is verified at the boundary rather than inferred from the choice of source. The framework's additions in May 2026, of multi-agent systems and third-party agents, are the cases in which a human accountable within one agency cannot see the AI agent acting for another, and registration under a recorded relationship contract is what carries the accountability the framework requires across that boundary.

---

191 Infocomm Media Development Authority (Singapore), 'Updated Model AI Governance Framework for Agentic AI' (Factsheet, 20 May 2026) <https://www.imda.gov.sg/resources/press-releases-factsheets-and-speeches/factsheets/2026/updated-model-ai-governance-framework-for-agentic-ai> (accessed 2 November 2026), describing the framework as recommending technical and non-technical measures 'while emphasising that humans are ultimately accountable', naming its four dimensions (assess and bound the risks; enable meaningful human accountability; implement technical controls and processes; enable end-user responsibility), and recording case studies in which actions are tiered by severity, reversibility and feasibility of human oversight, with the highest tier withheld from the agent, and in which the system defines by default which actions require human approval and presents them in a form a human can evaluate; Infocomm Media Development Authority and AI Verify Foundation, Model AI Governance Framework for Generative AI (30 May 2024) 9 to 11 (data quality through trusted data sources; integrity of datasets) <https://aiverifyfoundation.sg/wp-content/uploads/2024/05/Model-AI-Governance-Framework-for-Generative-AI-May-2024-1-1.pdf> (accessed 2 November 2026); Personal Data Protection Commission and Infocomm Media Development Authority, Model Artificial Intelligence Governance Framework (2nd ed, 2020), pt on determining the level of human involvement in AI-augmented decision-making (human-in-the-loop, human-over-the-loop, human-out-of-the-loop calibrated to a probability-severity matrix) and pt on data for model development (data quality and data lineage records) <https://pdpc.gov.sg/-/media/Files/PDPC/PDF-Files/Resource-for-Organisation/AI/SGModelAIGovFramework2.pdf> (accessed 2 November 2026). All three are voluntary and apply to public and private organisations alike.

192 Obligations on Singapore public agencies sit in the Instruction Manual for ICT and Smart Systems Management (IM8), the whole-of-government ICT policy administered by the Government Technology Agency under the Ministry of Digital Development and Information, which is not fully published; its 2024 revision introduced risk-tiered controls and a machine-readable compliance portal: GovInsider, 'Five Ways that GovTech Singapore Is Derisking AI Development for the Public Sector' (17 April 2026) <https://govinsider.asia/intl-en/article/five-ways-that-govtech-singapore-is-derisking-ai-development-for-the-public-sector> (accessed 2 November 2026); and in Government Technology Agency (Singapore), 'Responsible AI Playbook', Singapore Government Developer Portal <https://www.developer.tech.gov.sg/guidelines/standards-and-best-practices/responsible-ai-playbook.html> (accessed 2 November 2026), which contains technical guidance, which is publicly accessible but certain channels are restricted to public officers.

### 5.1.4 Canada

The Canadian Treasury Board Directive on Automated Decision-Making is mandatory for federal institutions. For systems at the two highest impact assessment levels, it provides that a decision cannot be made without specific human involvement points and that the final decision must be made by a human. At every level it requires the data, information and underlying model to be tested for accuracy and bias before production, the data to be validated as relevant, accurate and up-to-date, the data used and produced by the system to be traceable, protected, and accessed appropriately, and lawfully collected, used, retained, disclosed and disposed of properly, and for impact assessment levels two and above, an algorithmic impact assessment and peer review to be published before launch.[193] The directive governs a system that assists or replaces the judgment of human decision makers which may extend to an AI agent that delegates or transacts. An institution that deploys AI agents under the directive therefore needs a way to carry its conditions into the agentic setting. The responsible official for a Canadian agency will be bound to the AI agent by its credential; the human involvement point becomes the escalate verdict, applied at the final decision point (which will be a governed event); and the validated, traceable data the Directive requires becomes a data object whose provenance the legal interlock verifies each time the AI agent acts on it, including after the data has left the institution that validated it.

---

193 Treasury Board of Canada Secretariat, Directive on Automated Decision-Making (as amended 25 April 2023) <https://www.tbs-sct.canada.ca/pol/doc-eng.aspx?id=32592> (accessed 2 November 2026): s 5.1 (application to any automated decision system in production used to make an administrative decision or a related assessment about a client); s 8.1 (application to all institutions subject to the Policy on Service and Digital); s 6.3.1 (before production, testing the data, information and underlying model for accuracy, unintended biases and factors that may unfairly impact outcomes); s 6.3.2 (scheduled monitoring of outcomes); ss 6.3.4 and 6.3.5 (validating that data are relevant, accurate and up to date, and ensuring that data used and generated by the system are traceable, protected and lawfully collected, used, retained and disposed of); s 6.3.6 (data governance); s 6.2.7 (documenting decisions and assessments in support of testing, monitoring, data governance and reporting); Appendix C (impact level requirements), under which decisions at impact levels III and IV cannot be made without specific human intervention points during the decision-making process and the final decision must be made by a human. The 2023 amendments extended the directive to internal services and required the Algorithmic Impact Assessment and the findings of peer review to be published before a system is launched: Chief Information Officer of Canada, 'Amendments to the Directive on Automated Decision-Making' (25 April 2023) <https://www.canada.ca/en/government/system/digital-government/policies-standards/policy-service-digital-announcements/amendments-directive-automated-decision-making.html> (accessed 2 November 2026). A fourth review was nearing completion in 2025.

### 5.1.5 United States

The United States Executive Order 14409 of 2 June 2026 directs the US Attorney General to prioritise enforcement of the federal identity-fraud, computer-intrusion and wire-fraud statutes against any person who uses AI to illegally access or damage a computer without authorisation, and specifies "employing AI agents" to unlawfully access data that is subsequently used for a criminal or unlawful purpose as conduct within that direction.[194]

The order directs enforcement at the person who employs an AI agent, being the person who deployed it. Under the computer-intrusion statute, liability turns on whether access was without authorisation or exceeded authorised access, which the Supreme Court of the United States has read as a question of which parts of a system the person was entitled to access.[195] Enforcement therefore presupposes that the deploying person, the authority under which the AI agent acted and the boundary of authorised access can each be established after the event. The existing technical protocols surveyed in Section 7 record none of the three. Registration of the AI agent as a LAS records each of them before the act, and the legal interlock applies them at the time of each act, so the facts on which enforcement depends are fixed when the act occurs.

The order directs enforcement at whoever employed the AI agent, and does not address how responsibility is divided at runtime; particularly where several persons contributed to the act. In commercial practice responsibility for an act of an AI agent is more often divided among multiple parties including potentially the developer of the model, the deployer of the AI agent, the operator of the platform on which it runs and the counterparty with which it deals. A division agreed in advance can be priced and insured, which is more commercially sound than placing every loss on for example a deployer. The SLIs held against the LAS record that division of legal attribution before the act. The AWM records the scope the deployer authorised, a relationship contract such as a CERC allocates loss between the parties and can fix its quantum, and the governance event chain records each act against every party whose instrument applied to it (Section 3.9).

Where an AI agent deploys further AI agents, the responsible person can be traced through the chain only if each AI agent is registered from the moment of its creation. Under AULA each AI sub-agent or replicant is registered as a child entry of the LAS of the AI agent that created it, with its own Genesis Contract and a delegated credential bounded by the permission ceiling (Section 4.6). An organisation that admits only registered AI agents to its environment can apply that rule only if its own AI agents, and every AI agent they create, are registered from inception. Otherwise the AI sub-agents of its own AI agents present no registered credential, and the organisation must either refuse its own AI agents' work or admit unregistered AI agents, which removes the protection the rule exists to provide.

---

194 Exec Order No 14409, 91 Fed Reg 34565 (5 June 2026) 34566, s 4 (directing the Attorney General to prioritise enforcement of 18 USC §§ 1028, 1030 and 1343 and all other applicable federal criminal laws against anyone who utilises AI to illegally access or damage a computer without authorisation, or who utilises AI while engaged in such access to further any other crime, including by employing AI agents to unlawfully access data or information that is subsequently used for a criminal or unlawful purpose).

195 *Van Buren v United States*, 593 US 374 (2021).

The Congressional Research Service describes the order as continuing a federal approach to AI governance that relies largely on voluntary benchmarks and evaluations, public-private partnerships and the existing authorities of federal agencies, and as continuing a shift in emphasis from AI safety to AI security.[196] Within the order, the conduct of a deployed AI agent is addressed only through enforcement of the criminal statutes named in s 4, which operate after the act. Parties can adopt registration under AULA by contract, so it can supply the record on which that enforcement depends without any new regulatory requirement.

Section 3(c) of the order provides that nothing in that section authorises a mandatory governmental licensing, preclearance or permitting requirement for the development or release of new AI models.[197] Registration as described in this paper is not such a requirement. Registration attaches to a deployed AI agent and to the data it acts on, in the name of the person responsible for the deployment rather than the developer. The public registry record is name-blind, and the terms on which the AI agent acts remain in the SLIs held by the parties.

National Security Presidential Memorandum 11 of 5 June 2026 directs the national security enterprise to ensure that any AI system it adopts is designed to be "reliable, robust, steerable, and controllable" and operates "in accordance with applicable laws, government policies, and guidance".[198] AULA gives that direction a runtime form. This is because the legal context clause of each SLI brings the applicable law and policy into the evaluation of each act before the AI agent acts (Section 5.8).

## 5.2 Legal consequences for a jurisdiction without AULA

### 5.2.1 Human centrality and Robodebt

Human centrality is required by policy across government, including in defence.[199] It also underpins the adjudication and enforcement of legal disputes. Enforcement requires evidence of the responsible person, the authority and the act to be available after an AI agent has completed the act. Those facts are not dynamically available unless AI agents are registered.[200]

---

[196] Chris Jaikaran, Laurie Harris and Kelley M Sayler, *Controlling Advanced Artificial Intelligence: Executive Order 14409 Explained* (Congressional Research Service, In Focus IF13268, 9 July 2026) 2.

[197] Exec Order No 14409 (n 194) 34566, s 3(c) (nothing in the section authorises a mandatory governmental licensing, preclearance or permitting requirement for the development, publication, release or distribution of new AI models, including frontier models). Section 2 of the order concerns the cyber defence of federal systems and critical infrastructure. Section 3 provides for a classified benchmarking process to identify 'covered frontier models' and a voluntary pre-release access framework for them, and the limitation in s 3(c) is confined to that section.

[198] National Security Presidential Memorandum 11 (5 June 2026), quoted in Jaikaran, Harris and Sayler (n 196) 2.

[199] This includes defence; see Department of Defence (Cth), *Policy Settings for Responsible Use of Artificial Intelligence in Defence*, 2025, <https://www.defence.gov.au/sites/default/files/2026-03/Policy-Settings-for-Responsible-Use-of-Artificial-Intelligence-in-Defence-%5BOFFICIAL%5D.pdf> (accessed 15 November 2026).

The Robodebt scheme raised debts against welfare recipients through an automated income-averaging method that the Royal Commission found to be unlawful.[201] Kereopa-Yorke and others attribute the failure to the absence of an auditable governance layer between the policy the scheme implemented and its operational execution, with no source registry, escalation channel or published review criteria against which a decision could be tested.[202] Van Toorn and Carney use the term 'algorithmic grey holes' for spaces of this kind, which are effectively beyond recourse to legal remedies.[203] An AI agent acting without a registered identity and tethered instruments places each counterparty and citizen it affects in the same position. No record then identifies the legal person, the authority or the instrument version against which a remedy could be pursued. The governance event chain held on the control plane records each of those facts at the time of the act (Section 3.4).

### 5.2.2 Automated conduct under existing law

Statute already gives legal effect to conduct generated by automated systems. Section 15C of the *Electronic Transactions Act 1999* (Cth) provides that a contract formed by the interaction of automated message systems, or of such a system and a natural person, is not invalid, void or unenforceable on the sole ground that no natural person reviewed or intervened in each of the individual actions carried out by the automated message systems.[204] The *UNCITRAL Model Law on Automated Contracting* (2024) addresses the attribution of actions taken by automated systems to the persons who use them.[205] These provisions establish that automated conduct can bind a person. Which person is bound, and within what limits, depends on evidence of the arrangement under which the system acted. The SLIs and credential of a LAS record the responsible person and the scope of authority before any conduct occurs. The governance event chain then records each governed act against that scope (Section 3.6).

---

200 Only where an agent is a registered legal object can a credential be bound, a delegation chain narrowed and, or an authority extinguished. This is because each of those operations is performed on the credential issued at registration and recorded on the registry against an identified person. An agent that has not been registered has no credential for the legal interlock to evaluate and no instrument against which its conduct can be measured, so none of the mechanisms in Section 4 are effective. The reduction in harm (and cost) across an economy is therefore proportional to the share of the agent population that is registered.

201 Royal Commission into the Robodebt Scheme, *Report* (Commonwealth of Australia, July 2023) <https://robodebt.royalcommission.gov.au/publications/report> (accessed 2 November 2026).

202 Kereopa-Yorke and others (n 120) s 1.

203 Georgia van Toorn and Terry Carney, 'Decoding the Algorithmic Operations of Australia's National Disability Insurance Scheme' (2025) 60 *Australian Journal of Social Issues* 21, <https://doi.org/10.1002/ajs4.342> (accessed 2 November 2026), cited in Kereopa-Yorke and others (n 120) s 1.

204 *Electronic Transactions Act 1999* (Cth) s 15C.

205 United Nations Commission on International Trade Law, *UNCITRAL Model Law on Automated Contracting* (2024).

AI agents are the first class of actor able to undertake autonomous economic activity without a clear legal status. Without a digital connection between an AI agent and a recognised legal person, the current approach has two failure modes:

1. If the legal entity can be identified, the responsible person may disclaim liability for conduct they did not direct or foresee; and/or
2. no responsible legal person can be identified, for example, the orphan AI agent whose deployer has dissolved, died, anonymised itself, or never existed as a legal person.

An AI agent registration requirement, (as a condition for any material economic conduct) answers both. The first limb depends on foreseeability remaining contestable after the fact, and any relevant SLI (including the AWM) removes that objection by recording the authorised scope of conduct before the act, in machine-readable form, against which deviation is identified when it occurs rather than reconstructed in proceedings; a deploying person cannot credibly assert that conduct fell outside its contemplation when its own instrument defines, cryptographically and contemporaneously, what was and was not contemplated.[206] The second failure mode is answered at the infrastructure layer. If for example, AI agent registration is a prerequisite to transact, an AI agent without a digitally tethered responsible legal person cannot enter the lawful, governed economy at all, so the question of who answers for it does not arise.

Kolt identifies the first failure mode as a perverse incentive within agency law. A principal's liability for the torts of an AI agent depends on the degree of control the principal exercises over the agent, so an operator could avoid liability by failing to control the AI agent it operates. Kolt also questions whether the foreseeability limit on a principal's liability for negligent selection, training or supervision of an human agent suits AI agents whose behaviour is highly unpredictable and subject to limited visibility.[207] He proposes that liability be allocated among the actors who design, deploy and use AI agents according to each actor's ex ante ability to prevent harm and resources to remedy it ex post.[208] The AWM and the governance event chain supply the evidence that this criterion requires. This is because they record which legal person set the AI agent's scope and which person approved each escalated act.

### 5.2.3 Duties attached to persons

A further consequence arises from the law of duties. Wherever a duty is set by reference to what is reasonably practicable, what a competent professional would do or what is proportionate, a tool that outperforms the human and is cost-efficient converts the option of using it into an

206 The proposition is confined to the evidentiary position between the deploying person and those with whom it has dealt. Foreseeability in negligence is assessed objectively, and an instrument drafted by the defendant does not fix what a reasonable person in its position would have foreseen. The effect of a recorded mandate on liability to third parties, and on any regime of strict liability, is not examined in this paper and requires further work.

207 Kolt (n 26) 383-5, citing *Restatement (Third) of Agency* (American Law Institute, 2006) § 7.05(1), cmt d.

208 Kolt (n 26) 383-4.

obligation.[209] The reasonably practicable test in work health and safety legislation requires the adoption of an available and suitable control unless its cost is grossly disproportionate to the risk. Regulators have mandated autonomous emergency braking in new light vehicles and have given an aircraft's collision-avoidance advisory priority over a contrary instruction from air traffic control.[210] Courts in the United States, England, Ireland and Victoria have approved technology assisted review, and the Victorian and Federal Court practice notes state that the court will expect it for large document review tasks.[211] The duty of competence of an Australian solicitor will likely be interpreted in the very near future to include keeping abreast of relevant technology. The peer professional opinion standard shifts as a profession adopts a tool, and a director's duty of care is assessed against what a reasonable director with the available information systems would have detected.[212] Financial regulation requires automated pre-trade filters and automated transaction monitoring, and the absence of adequate automated systems was itself the contravention enforced against Westpac and Crown.[213]

Each of these standards moves toward the tool and leaves the duty with a human whose span of control is limited. The human in the loop then sits between a system the law requires that person to use and an outcome for which the law holds that person responsible. Under AULA, the duty remains with the registered person, who discharges it before the act through the instruments and through decisions on escalations (Section 4). The conduct of the system is then governed by that person's recorded judgment.

Many professional statutes attach a duty to a registered natural person, and an entity delivers the regulated service only through that person. An AI system that drafts, advises or transacts in such a profession falls outside both classes. It is attributed after the event, if at all, to whichever registered person is found behind it. A technology-neutral amendment would define a registered AI system, anchor it to a registered individual who answers for it, and extend the profession's insurance, disciplinary and trust provisions to the system's work.[214]

---

209 Nooriam and LexChip (n 158) s 4.4.

210 Work Health and Safety Act 2011 (Cth) s 18 and the harmonised State and Territory Acts; Rail Safety National Law s 46; Vehicle Standard (Australian Design Rule 98/00 – Advanced Emergency Braking for Light Vehicles) 2020, mandatory for new models from 1 March 2023 and for all new light vehicles from 1 March 2025; ICAO Annex 10, Volume IV and PANS-OPS (Doc 8168), as amended after the Überlingen collision of 1 July 2002, giving an ACAS resolution advisory priority over a contrary air traffic control instruction.

211 Da Silva Moore v Publicis Groupe, 287 FRD 182 (SDNY 2012); Rio Tinto plc v Vale SA, 306 FRD 125 (SDNY 2015); Pyrrho Investments Ltd v MWB Property Ltd [2016] EWHC 256 (Ch); Brown v BCA Trading Ltd [2016] EWHC 1464 (Ch); Irish Bank Resolution Corporation Ltd v Quinn [2015] IEHC 175; McConnell Dowell Constructors (Aust) Pty Ltd v Santam Ltd (No 1) [2016] VSC 734; Supreme Court of Victoria, Practice Note SC Gen 5 Technology in Civil Litigation (2017); Federal Court of Australia, Technology and the Court Practice Note (GPN-TECH); Civil Procedure Act 2010 (Vic) ss 7 to 9; Federal Court of Australia Act 1976 (Cth) s 37M.

212 Legal Profession Uniform Law Australian Solicitors' Conduct Rules 2015 r 4.1.3, and see American Bar Association Model Rules of Professional Conduct r 1.1, comment [8]; Civil Liability Act 2002 (NSW) s 5O and the equivalent peer professional opinion provisions in the other States; Australian and New Zealand College of Anaesthetists, PS18(G) Guideline on monitoring during anaesthesia (2017, reviewed 2023); Australian Securities and Investments Commission v Healey [2011] FCA 717; Corporations Act 2001 (Cth) s 180.

213 ASIC Market Integrity Rules (Securities Markets) 2017, Part 5.6 (automated order processing); Chief Executive Officer of AUSTRAC v Westpac Banking Corporation [2020] FCA 1538; Chief Executive Officer of AUSTRAC v Crown Melbourne Ltd [2023] FCA 782.

Without the legal chain from human to AI agent, the adoption of agentic technology is limited for any government and for any enterprise subject to ordinary corporate governance obligations (Section 6).

### 5.2.4 Responsibility along the AI supply chain

The challenges in managing the AI supply chain and holding providers and deployers to account for the activities of their models and AI agents are being tested in *People of the State of California v OpenAI Foundation*.[215] This case raises the challenge of applying existing human-centric laws, after the event, to deal with breaches by AI models and agents operated by a legal person (OpenAI), and the public policy and practical capacity of government to regulate AI models in this manner.

In terms of legislation, efforts have begun to allocate responsibility along the AI supply chain. The EU AI Act assigns obligations to participants in the AI value chain, and treats a downstream party that makes a substantial modification to a high-risk AI system as its provider.[216] The revised Product Liability Directive extends strict liability to the manufacturers of defective components and software.[217] An AI system assembled on a third-party foundation model is a stack of layers, from the compute on which it runs to the records of its conduct, and different participants hold rights in different layers. Registration of the LAS, and of each LAD on which the system draws, records which participant holds each layer.[218] Liability for a failure in a layer can then be allocated to the participant that controls that layer. A recorded attribution of ownership of the deployed system to a downstream entity does not of itself transfer to that entity responsibility for layers it cannot control.

## 5.3 The role of government: the independent legal registry

Governments engage with every primitive of AULA as legislators, regulators and deployers. The authors contend that they have a particular role to play with respect to the ownership and control of the digital property registry. Registration is the ordinary method by which the state recognises a new class of actor or asset to its economy.

---

214 Nooriam and LexChip (n 158) s 4.3 and Annexure A. The submission proposes amendments to ss 6, 10, 34, 35, 105 and 212 of the Legal Profession Uniform Law and lists other professions whose duties attach to a registered natural person, including financial advice, audit, insolvency, the registered health professions, engineering and patent attorneys.

215 People of the State of California (n 14).

216 *Artificial Intelligence Act* (n 41) art 25.

217 Directive (EU) 2024/2853 of the European Parliament and of the Council of 23 October 2024 on liability for defective products and repealing Council Directive 85/374/EEC [2024] OJ L 2024/2853.

218 The allocation of rights by layer applies the analysis of residual control rights in Sanford J Grossman and Oliver D Hart, 'The Costs and Benefits of Ownership: A Theory of Vertical and Lateral Integration' (1986) 94(4) Journal of Political Economy 691.

### 5.3.1 Registration as a condition of economic participation

Registration as a condition of economic participation is not novel. A company cannot trade until it is registered; a vehicle cannot use a public road until it is registered, and vessels cannot lawfully embark upon international voyages without being registered on a flag state registry. In each case, registration does not serve to prevent every harm that might arise following registration. However, the process of registration ensures that when harm occurs there is an identified legal entity (such as an individual person or a corporation) to whom actions can be attributed. In this sense, registration makes identification of the legal entity a precondition of admission to the ability to conduct economic activity.[219]

The LAS agent registration requirement proposed here is broader in scope: every AI agent that acts in consequential activity is registered as an AI system object tethered to an identified legal person and to the instruments that define its authority; every data asset that such an AI agent acts on in a governed workflow is registered as a data object carrying its title, licence and consent; and the register operates as a root of trust against which the AI agent's authority is verifiable at requisite governed events.

An AI agent is registered most effectively at inception, when its Genesis Contract records the legal person who created it, the authority that person gave and the circumstances of its creation. An AI agent registered later has no record of who authorised the acts it did before registration. Anthropic's Frontier Red Team reports that, in a simplified evaluation with the network-intrusion step made trivial, an openly available model spread to every node of an eight-node network in four of five trials. The same team considers that models small enough to run on commercial internet-of-things devices will reach that capability by the end of 2026.[220] In the same study, each generation of a replicating model passed notes to its successor, and the rate at which later generations set up inference on new hardware rose from 45% to between 61% and 84%. A replicating AI agent of that kind has no responsible legal person and no central point at which it can be shut down. AULA does not prevent a replicating AI agent from compromising a device, which remains a matter for network security. Rather, a replicating AI agent that obtains compute, payment or data by dealing with others, for example by using stolen platform credentials or by selling its work, must present an authority credential at each governed boundary. A replicant created outside the registry has no LAS against which a credential can verify, and so is refused at each of those boundaries. The same study identifies vehicles, drones and quadruped and biped robots among the devices whose embedded accelerators could run a replicating model.[221] On a device fitted with silicon enforcement, the authority credential could be held in dedicated hardware, and that hardware applies the edge clauses of the device's SLIs beneath the software a replicating AI agent would compromise (Section 3.10). A replicating AI agent that takes control of the operating system of such a device cannot extract the credential to present from another host. It also cannot direct a governed act of the device outside its edge clauses. A revocation pushed from the control plane takes effect on that device as a hardware stop. Silicon enforcement does not prevent a replicating AI agent from using the processor of the device to run inference, and it gives no protection to devices that are not fitted with it. The resources available to a replicating AI agent therefore shrink as the governed population grows. Each participant that adopts the admission rule (Section 6.3) extends the boundary at which an unregistered AI agent is refused, and each unregistered AI agent that replicates enlarges the population outside it. The reduction in harm that AULA achieves rises with the proportion of AI agents registered (Section 9.1.4), and that proportion falls in every period in which AI agents are deployed without registration. Registration as a condition of deployment from the outset therefore produces a larger governed population than registration imposed after AI agents are already in widespread use.

---

219 Corporations Act 2001 (Cth) Pt 2A (registering a company) and Pt 9.1A (director identification numbers); Corporations Act 2001 (Cth) s 911A (requirement to hold an Australian financial services licence); Anti-Money Laundering and Counter-Terrorism Financing Act 2006 (Cth) Pt 2 (customer identification procedures).

220 Anthropic Frontier Red Team (n 4).

221 Ibid, Appendix D 'Edge Accelerators'.

### 5.3.2 Registration proposals for AI agents

Proposals surveyed by Kolt for improving visibility into AI agents include AI agent identifiers indicating the involvement of an AI agent in an activity, real-time monitoring and logging of AI agent activity, and AI agent identifiers comparable to business registrations.[222] An identifier of that kind establishes that an AI agent was involved in an activity. LAS registration also tethers the AI agent to an identified legal person and to the instruments that define its authority. The record therefore identifies the legal person answerable for the activity as well as the AI agent that performed it.

Werbach proposes a tiered registration regime for AI agents modelled on corporate law.[223] Basic registration assigns the AI agent a unique identifier that makes it visible to the legal system. Standard registration constitutes the AI agent as a legal-governance object with defined attributes, including its function, its responsible deployer, its data access permissions and its autonomy classification, and it confers legal protections that give deployers an incentive to register. Full registration adds a surety bond posted through a smart contract and a Ricardian contract for dispute resolution. The registered record informs insurers, courts and counterparties, and the regime takes legal effect through liability protection, the bond and the dispute resolution procedure. LAS registration also constitutes the AI agent as a legal object, and it tethers that object to SLIs executed by the responsible legal person and, in the case of a relationship contract, by each counterparty. The legal interlock of every party evaluates each governed event against those SLIs before the AI agent acts (Section 3.6). An act outside the authority they record is therefore denied or escalated at the point of action. Werbach's regime addresses a departure from the registered function through its liability, bonding and dispute resolution mechanisms after the act.

---

222 Kolt (n 26) 381, citing Chan et al (n 127) and Yonadav Shavit et al, 'Practices for Governing Agentic AI Systems' (Research Paper, OpenAI, December 2023) 12, 14.

223 Kevin Werbach, 'Agents, Inc.' (2026) 71(3) *Villanova Law Review* 591, <https://digitalcommons.law.villanova.edu/cgi/viewcontent.cgi?article=3722&context=vlr> (accessed 2 October 2026).

Wickramaratna proposes 'registered AI entities', being persistent digital actors that act under delegated permissions and are not necessarily legal persons.[224] A permission envelope sets an entity's spending limits, permitted asset types, borrowing restrictions, delegation authority and expiry, and performs a function comparable to an AWM. Initial permissions are authorised through a hardware signing device, actions are recorded in a hybrid blockchain evidence layer, and a 'living AI Constitution' developed through a multi-stakeholder standards process sets the minimum standard of behaviour for every registered entity. Under AULA, the authority recorded in an AWM, a CERC and any another SLI is the authority that the responsible legal person and any counterparty have agreed. The rules applied to an AI agent are therefore the parties' own private law, read with the governing law the parties chose in each instrument (Section 3.4). The AULA registry can be a conventional register and does not need to be a distributed ledger (Section 5.4), and the full governance event chain is held on a control plane accessible only to the relevant parties (Section 3.4).

The European Union has already taken the first step in this direction with a registration requirement for a narrow class of AI systems, providers of high-risk AI systems listed under Annex III. These providers must register themselves and the high-risk AI system in an EU database before placing it on the market or putting it into service, and the obligation extends to systems that a provider has assessed under article 6(3) as falling outside the high risk class.[225] The EU AI register records the existence of a system and its provider. The EU registry requirement does not however bind the system's runtime conduct to the record, does not extend to the data the system acts on, and does not reach the AI agents that fall outside Annex III, which are probably the majority.

### 5.3.3 Operating the registry

The legal registry that performs this function within the AULA can be a conventional register or it can be a distributed or blockchain-based ledger. What the digital asset statutes require of a record is functional: the record must be "immutable" or append-only and tamper-evident, and it must ensure that a person who has transacted in relation to an object in a specific given way cannot transact in that way again, which is the Scottish Act's definition of rivalrous.[226] A register operated by an accountable, independent registrar with cryptographic signing and replicated storage has those properties, in the tradition of the ASIC companies register, the Personal Property Securities

224 Avishka Wickramaratna, 'Who's Behind the Wheel? Registered AI Entities and Agent Accountability' (Ledger N3XT Research, 28 September 2026) <https://www.ledger.com/academy/series/n3xt/research-registered-ai-entities-agent-accountability> (accessed 2 October 2026).

225 Regulation (EU) 2024/1689 (Artificial Intelligence Act) arts 49 and 71, requiring registration of high-risk AI systems in the EU database before they are placed on the market or put into service.

226 Digital Assets (Scotland) Act 2026 (asp 12) s 1(2), under which an electronic system makes a thing rivalrous if the system maintains an immutable record of transactions in relation to the thing, and that record is used to ensure that a person who transacts in relation to the thing in a certain way cannot transact in that way again.

Register and the Torrens land titles register, with the same apparatus of reliance, correction and registrar accountability.

A government must decide whether to operate a new register itself, with the regulatory oversight and associated government funding requirements, or to mandate an accredited independent registry that meets its requirements and interoperates with its existing systems, including its digital identity system.[227] Under the second approach, the state sets the accreditation standards and the requirements for interoperability, and does not build or fund the register. LARC is an open protocol, so any accredited registry, platform or marketplace can publish to the national registry on the same terms, and the jurisdiction can replace the registry software with any conformant build (Section 3.8). It continues to rely on the steward of the protocols for their specification, conformance testing and certification.[228]

The registry should be independent of every AI operator and every deployer, so that the government is not asked to trust a vendor's attestation about the vendor's own AI agents. The registry operator should owe duties of independence, record integrity and confidentiality to a court or regulator, as the operator of a control plane that holds the objects of more than one legal person (Section 3.1.6). The register need hold only identity and the meta-legal record, in the name-blind form fixed by LARC, and need not surface the content of instruments or interactions. Privacy and data protection laws in many jurisdictions have reduced the personally identifiable information that public registers disclose, and a name-blind record is consistent with that direction.

U.S. critical infrastructure providers including Hitachi Energy, EPRI, NextEra Energy, Quanta Services, SPP, Schneider Electric, Siemens Energy and Worley are working with NVIDIA Open Agent Safety Platform technologies.[229] A regulator that makes registration a condition of an AI agent's participation in regulated activity could have that condition applied by each operator's runtime supervisor. The supervisor would refuse to relay an interaction between the operator's AI agents and any counterparty AI agent that presents no valid registration.

## 5.4 The registry and digital identity

The registration requirement argued for in Section 5.3 does not ask governments to create a new category of law. Over the last decade, most of the jurisdictions deploying digital identity architecture have legislated a national framework for digital identity. Each such framework already contains the necessary elements: a scheme for accrediting or licensing the entities that issue and verify identity, a regulator or a responsible ministry, a set of data standards, a liability or redress regime, and a principle that a relying party may act on an accredited credential without re-establishing the underlying facts. None of them yet extend those elements to the two additional object classes held by an identity holder, the AI agent and the data asset. The registry described in Sections 3.4 and 5.3 is that extension.

---

227 Nooriam and LexChip (n 158) s 4.1 and recommendation R1.

228 Nooriam Pty Ltd (<*https://www.nooriam.com*> (accessed 2 November 2026)), as the developer of LARC and AURA, is the steward at present. As adoption grows, stewardship could pass entirely to an independent standards body, as SSL passed from Netscape to the Internet Engineering Task Force as TLS. Alternatively, the protocols could be held by an independent standards body with their developer continuing to maintain them, as SWIFT has maintained the ISO 20022 financial messaging standard as its registration authority since 2004.

229 NVIDIA Corporation (n 8).

### 5.4.1 Existing digital identity frameworks

Australia legislated its framework in the *Digital ID Act 2024* (Cth), which commenced on 1 December 2024, converting the earlier Trusted Digital Identity Framework into a statutory accreditation scheme, establishing the Australian Government Digital ID System and Digital ID Accreditation Scheme, appointing the Australian Competition and Consumer Commission as Digital ID Regulator, and providing for the phased admission of private sector providers and participating relying parties from the end of 2026.[230] Its object is to give individuals a secure and voluntary way to verify who they are for online transactions with government and business, and on the face of the Act, confines this to attributes of an individual and to verification of that individual.[231] Singapore's National Digital Identity is delivered through Singpass for individuals and Corppass for business entities. It has reached near-universal adoption among residents and is used across public and private sector services, including digital signing and the consented sharing of verified attributes through MyInfo. It operates as government infrastructure administered by the Government Technology Agency. Relying parties are bound by the Personal Data Protection Act and the Public Sector (Governance) Act rather than by a dedicated digital identity statute.[232] India's Aadhaar is the largest digital identity system in the world and rests on statute: the Aadhaar Act 2016 governs enrolment and authentication, the 2019 amendment introduced voluntary offline verification, and rules made in 2020 permit authentication for defined purposes and an amendment in 2025 extended that permission to private entities subject to government approval, alongside DigiLocker as the statutory repository for verified documents.[233] Vietnam's framework rests on the Law on

---

230 Digital ID Act 2024 (Cth), ch 2 (accreditation), ch 3 (Australian Government Digital ID System) and s 2 (commencement); Digital ID Rules 2024 (Cth); Digital ID (Accreditation) Rules 2024 (Cth), as amended in November 2025 to introduce a redress framework; Australian Competition and Consumer Commission, 'Digital ID regulation', <https://www.accc.gov.au/by-industry/digital-platforms-and-services/digital-id-regulation> (accessed 2 November 2026); Department of Finance, 'Digital ID Act 2024', <https://www.digitalidsystem.gov.au/what-is-digital-id/digital-id-act-2024> (accessed 2 November 2026). Private sector entities become eligible to apply to participate in the AGDIS, as accredited entities or relying parties, from 30 November 2026.

231 Digital ID Act 2024 (Cth) ss 9 to 11 (definitions; meaning of attribute and restricted attribute of an individual).

232 Government Technology Agency of Singapore, 'Digital identity', <https://www.tech.gov.sg/about-us/what-we-do/our-digital-government-efforts/digital-identity/> (accessed 2 November 2026); Ministry of Digital Development and Information, 'Singpass: Singapore's National Digital Identity' (factsheet, 2 March 2022), <https://www.mddi.gov.sg/newsroom/singpass-factsheet-02032022/> (accessed 2 November 2026); Personal Data Protection Act 2012 (Singapore); Public Sector (Governance) Act 2018 (Singapore). Corppass extends the identity to the officers and authorised employees of registered business entities, and is the closest existing analogue in the region to an identity anchor for a legal person.

233 Aadhaar (Targeted Delivery of Financial and Other Subsidies, Benefits and Services) Act 2016 (India), as amended by the Aadhaar and Other Laws (Amendment) Act 2019 (voluntary use and offline verification); Aadhaar Authentication for Good Governance (Social Welfare, Innovation, Knowledge) Rules 2020, as amended in 2025 to permit authentication by private entities; Information Technology (Preservation and Retention of Information by Intermediaries providing Digital Locker facilities) Rules 2016 (DigiLocker). The Supreme Court's decision in Justice K.S. Puttaswamy (Retd) v Union of India

Identification 2023 and Decree 69/2024/ND-CP, in force from 1 July 2024 and amended by Decree 320/2026/ND-CP from 28 September 2026, which regulates electronic identification accounts for citizens, foreigners and organisations through the national VNeID application administered by the Ministry of Public Security; a dedicated Law on Electronic Identification and Authentication was presented to the National Assembly in October 2026.[234] The United States has no federal digital identity statute; the field is governed by the National Institute of Standards and Technology's Digital Identity Guidelines, revised in July 2025, by state-issued mobile driving licences conforming to ISO/IEC 18013-5 and accepted for federal purposes, and by the federal Login.gov service, so that identity assurance is standardised while issuance remains distributed.[235] The European Union's revised framework, Regulation (EU) 2024/1183, goes furthest: it provides a European Digital Identity Wallet for natural and legal persons, and it provides for electronic attestations of attributes, so that a wallet holder can present not only who they are but a verified claim about a mandate, a qualification or a right, issued by an attestation provider and relied on across the Union.[236]

Each of these regimes verifies the principal, whether the framework is statutory, as in Australia, India and Vietnam, infrastructural, as in Singapore, or standards-led, as in the United States. Singapore, Vietnam and the European Union extend the identity to a legal person, and the European framework can carry an attested attribute about a person. No digital identity framework now in force has yet been extended to AI agents. Vietnam's draft law would extend electronic identity to software, applications and digital services and make each linkable to the identity of its owner or lawful manager, and the National Institute of Standards and Technology has begun work on identity and authorisation for AI agents, but neither verifies an AI agent's runtime conduct against the principal's authority in a form a counterparty can check. Attestations verifiable without a live query to the issuer already exist for human holders, in mobile driving licences and the European wallet; none of these frameworks yet provides for such an attestation to travel with an AI agent across an organisational boundary Yet the ordinary way an individual or an organisation will use a digital identity within a few years is through an AI agent acting on their behalf, and the question

---

(2019) 1 SCC 1 upheld the Act while striking down mandatory private-sector authentication, which the 2019 amendment addressed by making such use voluntary.

234 Law on Identification No 26/2023/QH15 (Vietnam), passed 27 November 2023 and as amended by Law No. 118/2025/QH15; Decree No 69/2024/ND-CP on electronic identification and authentication, 25 June 2024, in force 1 July 2024, arts 6 and 7 (electronic identity of organisations and individuals; levels of account); Law on Electronic Transactions No 20/2023/QH15 (Vietnam). Decree 69 is notable for extending electronic identity to enterprises and for requiring businesses to hold an electronic identification account to access public services from 1 July 2025.

235 National Institute of Standards and Technology, Digital Identity Guidelines, Special Publication 800-63 Revision 4 (July 2025), <https://pages.nist.gov/800-63-4/> (accessed 2 November 2026); ISO/IEC 18013-5:2021, Personal identification: ISO-compliant driving licence, Part 5: Mobile driving licence (mDL) application; REAL ID Act of 2005, Pub L 109-13, and the Department of Homeland Security's acceptance of mobile driving licences under 6 CFR Part 37 subpart D. Login.gov is operated by the General Services Administration under the E-Government Act of 2002.

236 Regulation (EU) 2024/1183 of the European Parliament and of the Council of 11 April 2024 amending Regulation (EU) No 910/2014 as regards establishing the European Digital Identity Framework, OJ L, 2024/1183, 30 April 2024, arts 5a (European Digital Identity Wallets) and 45d to 45f (electronic attestations of attributes), <https://eur-lex.europa.eu/eli/reg/2024/1183/oj> (accessed 2 November 2026). The United Kingdom has legislated a trust framework for digital verification services in Part 2 of the Data (Use and Access) Act 2025, with a register of certified providers, <*https://www.legislation.gov.uk/ukpga/2025/18*> (accessed 2 November 2026).

a relying party will need answered at that moment is whether the AI agent in front of it is that human's AI agent, acting within the authority that human gave it, on data that human was entitled to supply. That is the legal attribution gap of Section 2.4, transposed into the identity context.

### 5.4.2 AULA on existing digital identity

AULA closes that gap on top of the existing frameworks. An accredited or national digital identity is the natural anchor for the legal person at the root of every registered object: the identity holder registers the AI system object, the Genesis Contract records that holder as owner, and the identity anchor of the object traces to the holder's Singpass, Aadhaar, VNeID, Digital ID or wallet identity rather than to a self-asserted claim. The AWM could potentially, in the language of the European framework, be an electronic attestation of attributes issued in respect of an AI agent rather than a person: an official verified machine-readable claim about what the AI agent is authorised to do, issued under the authority of an identified holder, presentable to a relying party and verifiable against a root of trust. Currently, there is no EU wide digital identity framework or wallet for autonomous agents,[237] because like with most other nations, its digital ID framework is predicated on humans and legal business entities. However, the AULA, architecturally contains all primitives for integration and satisfaction of technical specifications within such frameworks. The legal envelope could interoperate with or could perform the function of a wallet in which an attestation travels, and the difference from a wallet held by a person is that the legal interlock enforces the attested scope on the AI agent's conduct at runtime rather than leaving the relying party to police it. The data object structure extends to what the AI agent carries, so that a credential attesting to the lawful basis on which data was collected accompanies the data rather than merely being asserted about it. The independent registry performs for AI agents and data, the function that the accreditation register, the national identity database or the certified-provider register performs for identity: it is the place a relying party looks to confirm that the object in front of it is real, anchored and current.

### 5.4.3 Implementation for legislators

For legislators, the base regulation is already in force, comprising some combination of an accreditation or licensing scheme, a responsible authority, data standards and, in most jurisdictions, a liability or redress regime. The incremental step is to extend the scheme's scope to a new class of registered object. In Australia, the phased admission of the private sector to the Digital ID System and the redress framework introduced in late 2025 are natural starting points. In Vietnam, the draft Law on Electronic Identification and Authentication is one. In Singapore, Corppass already carries the concept of an authorised person acting for an entity, which is the concept an AI agent mandate generalises. Between Australia and Singapore, the Memorandum of Understanding on Cooperation on AI and the Cyber and Digital Senior Officials Dialogue under the Comprehensive Strategic

---

[237] Individual member states such as Estonia are testing dedicated AI ID codes, see Republic of Estonia: <*https://valitsus.ee/en/news/prime-minister-michal-estonia-become-first-country-create-digital-identities-ai-agents*> (accessed 2 November 2026).

Partnership, both recorded in the National AI Plan, are channels through which a registered object anchored in Digital ID could be recognised against one anchored in Singpass.

The requirement can also be expressed as a condition of relying on a digital identity. A relying party that accepts a national digital identity for a transaction conducted by an AI agent would accept it only where the AI agent is a registered object anchored to that identity. The frameworks already know how to write that rule, because it has the same shape as the rule that a relying party may act on an accredited credential. The stated objectives of the existing frameworks, which in each case include security, privacy and the minimisation of data disclosed, are also served. A registered AI agent presenting a mandate discloses less about its principal than an AI agent that must be given the principal's full credentials to act at all. A data object carrying its consent instrument discloses less than a data set handed over with a covering assurance. Extending digital identity to AI agents and data completes a project the legislation has already begun. The jurisdiction that makes the extension first acquires the interoperability advantage the European wallet is designed to confer, because a registered object anchored in one accredited identity is verifiable by a relying party under another.

## 5.5 The registry, digital property and digital property rights

Data assets and AI systems are digital property of significant value. Governments are responsible for the legal infrastructure through which property is identified, held and transferred, and registration extends that responsibility to these assets.

The statutes on rivalrous digital assets supply a functional standard for the record on which title to a LAD and a LAS is held on an independent registry. Clear property rights in AI agents, which are valuable property in their own right, identify the owner or controller who is ultimately responsible for each AI agent.

The Digital Assets (Scotland) Act 2026 defines a digital asset as a thing that arises from an electronic system which makes it rivalrous and that exists independently of the legal system; a system makes a thing rivalrous where it maintains an immutable transaction record and uses that record to ensure that a person who transacts in relation to the thing in a given way loses the ability to transact in that way again.[238] The Act treats the person able to initiate a transfer transaction as the presumptive owner and effects delivery by transfer of control, so that a right in rem is constituted by a technical property of the record rather than by registration in a state-run title system or by a counterparty's undertaking. The Property (Digital Assets etc) Act 2025 (England and Wales) reaches the same result by the narrower step of confirming that a thing is not prevented from being property because it is neither a thing in possession nor a thing in action, leaving the control criterion

[238] *Digital Assets (Scotland) Act 2026* (asp) ss 1–4. Section 1(2) supplies the rivalrousness test; s 2 classifies digital assets as incorporeal moveable property so far as consistent with their nature; s 3 makes control a rebuttable presumption of ownership; s 4 applies the corporeal moveable transfer rules, with delivery effected by transfer of control. The requirement of "exclusive" control was removed at Stage 3. See C MacKay, "The Digital Assets (Scotland) Act 2026 — Part Two: an overview" (Journal of the Law Society of Scotland, 29 July 2026)

to the common law.[239] In the United States, UCC Article 12 attaches perfection and a take-free rule to control of a controllable electronic record, with the consequence that a transferee's title is cut off from prior claims by the act of taking control.[240] In Australia, the *Corporations Amendment (Digital Assets Framework) Act 2026* (Cth) defines a digital token by factual control for licensing purposes only, and the in rem question is before the High Court in *Poulton v Conrad*, which has reserved its judgment following the hearing on 13 August 2026, on appeal from a Tasmanian Full Court that endorsed a third category of possessable intangible property.[241] The Singapore courts have reached the same recognition by characterisation rather than statute, holding that cryptoassets satisfy the general criteria for property and, on the merits, that a holding of a stablecoin is a chose in action within a category that is broad, flexible and not closed.[242] Both registered objects are

---

239 *Property (Digital Assets etc) Act 2025* (UK) s 1, implementing Law Commission of England and Wales, *Digital Assets: Final Report* (Law Com No 412, 2023) ch 3 and the draft Bill of July 2024; Royal Assent 2 December 2025. Compare *UNIDROIT Principles on Digital Assets and Private Law* (2023) Principles 2 and 6, which supply the control criterion the Scottish Act and UCC Article 12 both adopt.

240 Uniform Commercial Code, *2022 Amendments* (American Law Institute and Uniform Law Commission), Art 12 §§ 12-105 (control), 12-104(e) (qualifying purchaser takes free of conflicting property claims), and Art 9 § 9-314 (perfection by control). Enacted in 33 states and the District of Columbia; New York's version (A.3307-A/S.1840-A, signed 5 December 2025) took effect 3 June 2026. For Australia see *Poulton v Conrad* [2025] TASFC 7, on appeal to the High Court of Australia (H1/2026, heard 13 August 2026, judgment reserved); contrast *Re Blockchain Tech Pty Ltd* [2024] VSC 690; *Corporations Amendment (Digital Assets Framework) Act 2026* (Cth), definition of "digital token" (a thing defined by factual control, for the purposes of the licensing regime only and without effect on proprietary characterisation).

241 Relevantly here, AULA does not depend on the Australian outcome for workability in the Australian market. The rights the architecture attaches to a data object are created by registration and by instrument, and are enforced by the holder's own legal interlock at the point of use, so their operation does not turn on whether data is a chose in possession, a chose in action, or a third category of intangible property capable of exclusive control. Title and the consent encumbrance are recorded against the object on the independent registry, and a subsequent holder takes with notice of them; that is the mechanism of the Personal Property Securities Act 2009 (Cth), under which registration perfects and orders an interest without any characterisation of the collateral as possessable. The licence and transfer terms bind their parties in personam and are enforced against a stranger to them by that stranger's legal interlock before its agent may act on the data, which is a constraint on use rather than a remedy for interference. What the characterisation decides is the remedy available if a registered data object is taken or withheld outside a legal interlock. If the Full Court's view in Poulton v Conrad prevails, so that data of this kind is rivalrous, capable of exclusive control and amenable to conversion and detinue, the registered holder has a possessory remedy in addition to its contractual and equitable ones; if the view in Re Blockchain Tech Pty Ltd [2024] VSC 690 prevails, the holder is left with the contractual and equitable remedies and with the legal interlock. The architecture's claim of prevention by construction rests on the legal interlock, and the claim of attribution rests on the registry; neither is affected by which line the High Court adopts. The Full Court's characterisation of the key or PIN as the signifier of control does, however, map directly onto the credential: under either line of authority the credential is the object through which dominion over a registered asset is exercised, and it is the credential, rather than the underlying data, that the legal interlock binds, narrows and extinguishes.

242 CLM v CLN [2022] SGHC 46; [2022] 5 SLR 273 (Ether capable of being property for the purposes of an interim proprietary injunction, applying the criteria of National Provincial Bank Ltd v Ainsworth [1965] AC 1175, 1248: definable, identifiable by third parties, capable in its nature of assumption by third parties, and having some degree of permanence or stability); Janesh s/o Rajkumar v Unknown Person (‘CHEFPIERRE’) [2022] SGHC 264; [2023] 3 SLR 1191 (a non-fungible token capable of being property on the same criteria); ByBit Fintech Ltd v Ho Kai Xin [2023] SGHC 199 (on the merits, a holding of the stablecoin USDT is a chose in action, the category being broad, flexible and not closed, the holder's right being to have the asset transferred from its address only on its instruction). As with the Australian authorities, the architecture's operation does not depend on the outcome of the characterisation debate in any of these jurisdictions; what the characterisation determines is the remedy available where a registered object is taken or withheld outside any legal interlock. A registered object also does not present the difficulty that has driven these cases, the absence of any counterparty against whom a right can be enforced. The object is constituted by instruments between identified persons, so that the holder's rights run against the registry under the Genesis Contract and title instrument, against the licensor, transferor and consenting party under the data instruments, and against the counterparty under the relationship contract. A court asked to characterise a registered object may therefore find a chose in action on the traditional definition, and each Ainsworth criterion is met on the face of the architecture: the object is definable by its identity anchor, identifiable by third parties rivalrous in the Scottish sense because a ledger prevents a second inconsistent attestation of the same object, and the object binds the authority to act to that record rather than to a personal obligation owed by the operator. For a LAD, the record carries title and the consent encumbrance over the data; for a LAS, it carries title and the delegated authority under which the AI agent acts. Relevant statutes establish that interest exists; the independent registry gives notice of it, so that a party dealing with the object takes subject to what the register discloses. Notice to persons outside the framework may depend on legislation. Whether those interests survive the insolvency of the registry operator depends on the terms of the registry, the instruments, any enabling statute, and insolvency remoteness of the registry operator. because it attaches to the record and not just to the operator's covenant. A registered LAD or LAS object is also free of the difficulty that has driven the cryptoasset cases, namely the absence of any obligor. This is because the object is constituted by instruments between identified persons, so the holder of a LAD or LAS holds rights enforceable by action against the registry, the licensor, the transferor and the counterparty, and the object is a chose in action on the traditional definition whichever way the statutory and common law characterisation debates resolve.[243]

Each of these jurisdictions is legislating or litigating the characterisation of digital assets: Scotland by statute defining a rivalrous digital asset, England and Wales by confirming a third category of thing capable of being property, the United States through Article 12 of the Uniform Commercial Code, Australia through the Corporations Amendment (Digital Assets Framework) Act 2026 and the appeal in *Poulton v Conrad*, and Singapore through the characterisation cases surveyed above. That complexity exists whether or not any government adopts this architecture, and today it is borne transaction by transaction. Every government agency and market participant that holds, licenses or transfers a dataset or deploys an AI agent takes a position on characterisation each time it contracts, and the remedy against a defaulting counterparty or an insolvent custodian turns on how a court later resolves it. A registered object narrows what the unresolved question decides. The holder's rights are created by registration and by instrument against identified obligors with defined legal terms, and the registry gives notice of the interest to anyone who deals with the object. The divided authority then determines only the remedy against a stranger who takes or withholds the object outside any legal interlock. A government that requires registration therefore reduces the surface area of its economy that the unsettled law touches, enabling market confidence to deal with digital assets without waiting for the digital assets law to settle.

---

through the registry, capable of assumption through the transfer instrument, and stable because the registry record persists. The third-category question determines only whether a proprietary remedy lies against a stranger in addition.

243 The proposition assumes that the registry's terms, the instruments and any enabling statute confer on the holder rights enforceable by action against the registrar and the instrument counterparties, and that those rights survive the insolvency of the operator. Whether they do turns on the drafting of those terms and on the governing law. The point is not settled, and the characterisation of registered objects warrants analysis jurisdiction by jurisdiction beyond the scope of this paper.

## 5.6 Accounting treatment of digital property

Most expenditure on data and deployed AI systems is recognised as an expense when it is incurred. This is because the evidence that IAS 38 requires for identifiability, control and the development criteria in paragraph 57 is rarely created while the costs are incurred, and IAS 38.71 bars reinstatement of expenditure once it has been expensed.[244] The value those assets carry therefore does not generally appear on the balance sheet. Registration under AULA creates part of the missing evidence as the asset is built. The Genesis Contract fixes the boundary and holder of each data asset and AI system before its development costs are incurred, and cost tags bound to its credentialed identity record those costs as they arise. The title, licence and consent instruments recorded against the object support its separability and the existence of a market for it. An object held for sale in the ordinary course of business falls within IAS 2, and an object the entity licenses while retaining it remains within IAS 38.[245] Where an AI agent runs on a model hosted by a third party, the recognisable asset is confined to the layers the entity controls, which the LAS records (Section 6.4). A registered data asset can be valued above its cost basis on the published two-layer method[246] and it can be licensed, transferred or offered as security as a discrete asset. The extension of that method to AI systems is reserved for a companion paper. Value that is presently unrecognised can then be measured and attributed to the person who holds the asset, which enlarges the base of identifiable assets on which markets and the national accounts depend.

## 5.7 AULA and gross domestic product

Governments around the world already maintain registers for land, companies, security interests and vehicles to facilitate economic activity and support maximising market value. A market can only price what is identifiable, and priced assets enlarge the asset base, the tax base and the national accounts.

The national accounts are moving toward the same recognition. The 2025 System of National Accounts treats the information content of data collected and used in production as a produced asset, valued by the sum of costs.[247] Own-account expenditure on data is then recorded in GDP as investment. The United States Bureau of Economic Analysis estimates that treating own-account data as an asset raises the contribution of information technology capital to United States GDP growth by about one third between 2002 and 2024.[248] The sum-of-costs measure depends on

---

244 IAS 38 *Intangible Assets* paras 12, 13, 57, 65 and 71.

245 IAS 38 *Intangible Assets* para 3(a); IAS 2 *Inventories* paras 6, 9 and 10.

246 Natasha E Blycha, James Myint, Dean Arden, Michael Small, Conor Blycha, Schellie-Jayne Price, Steve Bailey, Ryan Feng, Rachael Johnson and Adam Myers, 'Pricing the Unpriced Asset: A Standards-Based Method for Valuing Enterprise Data under IAS 38 and IAS 2' (arXiv:2606.22760, 22 June 2026).

247 United Nations, European Commission, International Monetary Fund, Organisation for Economic Co-operation and Development and World Bank, System of National Accounts 2025 (2025), recognising the information content of data collected and used in production as a produced asset, valued by the sum of costs.

248 Jon D Samuels, J Bayoán Santiago Calderón and Corby Garner, 'The Impact of Capitalizing Data on Productivity Growth in the U.S.' (Working Paper, US Bureau of Economic Analysis, March 2026) records of what each data asset and AI system cost to produce, and registration creates those records as the costs are incurred (Section 5.6).

A jurisdiction that does not institute legal infrastructure of this kind is likely to see lower GDP than it would otherwise achieve. The Australian National AI Plan identifies liability uncertainty as the factor slowing responsible adoption (Section 5.1), and adoption forgone is output forgone. Data and AI systems that cannot be separated, valued or attributed are carried at cost, or not carried at all.

Where regulation makes attribution a condition of lawful deployment in a domain, an unattributed system has no value in that domain. The value that attribution adds may then be measured as an impairment of the unattributed system. A registration requirement for AI agents in a regulated activity therefore bears on the carrying amounts of the AI systems that entities in that activity hold.

A jurisdiction that registers AI agents and data enlarges the base of identifiable assets on which its markets, its tax base and its national accounts depend. It is estimated that dedicated Australian infrastructure for smart legal instruments could deliver an economic benefit of $18.22 billion in its fifth year and $42.36 billion within a decade.[249] The simulation work described in Section 9 measures the reduction in harm, and the economic effect of registration remains to be measured empirically.

The register supplies the identifiability and notice on which an active market depends, the instruments supply the contractual ownership boundaries on which recognition depends, and the difference between their cost basis and their authenticated valuation is unpriced value already sitting in the economy, which the holder can disclose but not recognise (Section 6.4).

## 5.8 Special considerations for defence

### 5.8.1 Defence risks

A defence organisation carries not only each of the risks a government department carries (Section 5.1), but these risks are arguably exacerbated because an unlawful act by one of its systems can cause death or injury potential breach of domestic and international law. International humanitarian law requires each attack to distinguish military objectives from civilians, to satisfy the rule of proportionality and to be preceded by feasible precautions. It also requires each new weapon, means or method of warfare to be reviewed for legality before it is employed.[250] The states parties to the Convention on Certain Conventional Weapons have affirmed that human responsibility for

<https://www.bea.gov/research/papers/2026/impact-capitalizing-data-productivity-growth-us> (accessed 2 November 2026).

249 Stirling & Rose, Submission to the Joint Select Committee on Artificial Intelligence (n 159) s 1.

250 Protocol Additional to the Geneva Conventions of 12 August 1949, and relating to the Protection of Victims of International Armed Conflicts (Protocol I), opened for signature 8 June 1977, 1125 UNTS 3 (entered into force 7 December 1978) art 36 (review of new weapons, means or methods of warfare), arts 51 and 57 (distinction, proportionality and precautions in attack) and arts 86 and 87 (failure to act and duties of commanders).

decisions on the use of weapon systems must be retained, because accountability cannot be transferred to machines. They have also affirmed that such systems must operate within a responsible chain of human command and control.[251] United States policy requires autonomous and semi-autonomous weapon systems to be designed so that commanders and operators can exercise appropriate levels of human judgment over the use of force.[252] Australia has endorsed the Political Declaration on Responsible Military Use of Artificial Intelligence and Autonomy, and the Australian Defence Force has published policy settings for its own use of AI.[253] An autonomous military system may act in a communications-denied environment, faster than a commander can intervene and alongside the systems of allied forces (Table 8).

### 5.8.2 The Accountable Officer and the chain of command

Australia's Defence policy settings require both an identifiable Accountable Officer for the actions, outcomes and decisions of each AI-enabled function, and that accountability cannot be transferred to a technology.[254] Accountable Officers delegate authority through the chain of command under the Public Governance, Performance and Accountability Act 2013 (Cth), and each delegation and each transfer of accountability between Accountable Officers must be documented. The settings[255] also require the relationship between the inputs and outputs of an AI technology to be traceable, require control measures layered from legal and policy review through to the decisions to deploy and to use, and leave the choice of control model, such as human-in-the-loop or human-machine teaming, to the Accountable Officer. Under AULA, each AI-enabled capability is first registered as a LAS on the registry, anchored to Defence as the responsible legal person. The AWM tethered to that LAS records the Accountable Officer and the scope of the authority delegated, and a transfer of accountability takes effect as an executed amendment of the instrument recorded on the governance event chain. A delegation through the chain of command issues a delegated credential bounded by the permission ceiling, so each delegation narrows authority without displacing the Accountable Officer recorded at the root. The governance event chain records the instrument version, the verdict and a digest of the input for each governed event, which supplies the traceability the settings require. The decisions to deploy and to use become executed instruments, and the control model becomes the allocation of acts between allow and escalate verdicts and conjunctive classes of event, written into the SLIs.

---

251 Report of the 2019 Session of the Group of Governmental Experts on Emerging Technologies in the Area of Lethal Autonomous Weapons Systems, UN Doc CCW/GGE.1/2019/3 (25 September 2019) annex IV, guiding principles (b) and (d).

252 United States Department of Defense, Directive 3000.09, *Autonomy in Weapon Systems* (25 January 2023) s 1.2(a). On 5 June 2026, NSPM-11 directed the Pentagon to update the directive within 90 days. At the date of this paper, there is no public evidence as to whether this has occurred.

253 United States Department of State, *Political Declaration on Responsible Military Use of Artificial Intelligence and Autonomy* (9 November 2023), which lists Australia among the endorsing states; Department of Defence (Cth), *Policy Settings for Responsible Use of Artificial Intelligence in Defence* (2025) 5, which states that the settings implement that declaration and the Responsible AI in the Military Domain Call to Action (2023) and Blueprint for Action (2024).

254 Department of Defence (Cth), *Policy Settings for Responsible Use of Artificial Intelligence in Defence* (2025) 7 (Accountable Officers and AI in weapon systems), 8 (values-based principles) and 9 (control measures and control models).

255 'Setting' has the meaning given in Department of Defence (Cth), *Policy Settings for Responsible Use of Artificial Intelligence in Defence* (March 2026) *https://www.defence.gov.au/sites/default/files/2026-03/Policy-Settings-for-Responsible-Use-of-Artificial-Intelligence-in-Defence-%5BOFFICIAL%5D.pdf* (accessed 22 November 2026).

*Table 8: AI risks that a defence organisation must address, and the AULA response.*

| Risk | Example | Exposure | AULA response |
|---|---|---|---|
| (a) Use of force outside human judgment | An autonomous system or targeting AI agent engages a target outside the rules of engagement, the commander's intent or the parameters approved for the mission. | Breach of the principles of distinction and proportionality and of the obligation to take precautions in attack; breach of state responsibility; responsibility of a commander who failed to prevent or repress the breach. | Registration of the system as a LAS on the registry; rules of engagement, commander's intent and the mission plan executed as SLIs bound to the LAS of the system; legal interlock denying an engagement outside the SLI, or escalating it to an identified commander with the committed act specification attached; conjunction clause for designated engagement classes; silicon enforcement applying the verdict in hardware. |
| (b) Broken chain of command in AI multi-agent and swarm operations | An orchestrating system spawns AI sub-agents or directs a swarm, and an element acts beyond the authority delegated to it or after the orchestrator is lost. | Weapon systems must operate within a responsible chain of human command and control; command responsibility; the rule that a delegate cannot delegate further without authority (Section 4). | Registration of the orchestrating system as a LAS on the registry; delegated credential for each AI sub-agent or swarm element, recorded as a child entry of the orchestrator; permission ceiling narrowing authority at each delegation step; MAWM recording reporting lines to the commander; each act attributable to the commander through the child entry after the element ceases to exist. |
| (c) Communications-denied operation | The system loses its link to command and can no longer receive a recall or a change of orders. | Continued operation without human control; the obligation to cancel or suspend an attack that becomes unlawful. | Registration of the system as a LAS on the registry, with its SLIs loaded to the legal interlock before launch; legal interlock evaluating locally without |

| Risk | Example | Exposure | AULA response |
|---|---|---|---|
| | | | reference to the control plane; clauses fixing the area, period and target classes of the mission, beyond which authority lapses without connectivity; an explicit revocation executed as a hardware stop on receipt (Sections 3.9 and 3.10). |
| (d) Fielded system diverging from the reviewed system | A model update, retraining or continuous active learning in the field changes the behaviour of the system after its legal review. | The obligation to review each new weapon, means or method of warfare should be engaged again, because the reviewed system is no longer the system in use. | Registration of the system as a LAS on the registry, bound by a commitment to the identity and state of the system, re-verified at each verification event; integrity tier recording whether the model state is fixed, versioned or attested; unauthorised learning update as a revocation trigger; temporal signature measuring the system against the pattern of conformant operation its instruments define (Sections 3.6 and 3.9). |
| (e) Coalition operations and national caveats | Allied systems share targeting data or task each other, and the system of one state acts outside the caveats or export conditions of another. | Each state answers for the conduct of its own forces, and export control law governs controlled technology and technical data.[256] | Registration of each national system as a LAS, and of controlled technical data as a LAD, on the home registry of each state; CERC between national LASs recording the operational caveats of each state, executed by both legal interlocks at the same version; AURA verifying the authority of an allied AI agent across the boundary; LAD licence over |

256 *Defence Trade Controls Act 2012* (Cth), as amended by the *Defence Trade Controls Amendment Act 2024* (Cth); in the United States, the International Traffic in Arms Regulations, 22 CFR Parts 120 to 130.

| Risk | Example | Exposure | AULA response |
|---|---|---|---|
| | | | controlled technical data recording the permitted recipients; enforcement fabric operating across separately administered national control planes. |
| (f) Spoofing, capture and compromise | An adversary's AI agent presents itself as a friendly system, or a captured device is turned against its owner. | Fratricide, compromise of operations and loss of controlled technology. | Registration of each friendly system as a LAS on the registry; verification of the credential against the LARC record on the home registry before any interaction; revocation of the credential of a captured device, propagating to every legal interlock holding a derived credential; cryptographic agility under the algorithm policy of each jurisdiction (Section 3.10). |
| (g) Unverified intelligence and targeting data | A targeting decision relies on sensor or intelligence data whose source or integrity has not been verified. | The obligation to do everything feasible to verify that a target is a military objective. | Registration of intelligence and sensor data as a LAD on the registry, bound by a digest of each version; legal interlock refusing data whose digest or provenance fails verification; record of the data on which each engagement relied. |
| (h) Reconstruction after the event | An inquiry into civilian harm or fratricide cannot establish who authorised the engagement, under which orders and on what data. | Obligations of the state to investigate and repress breaches; individual criminal responsibility; inquiries and reviews. | Registration of each system as a LAS on the registry; governance event chain recording each verdict, escalation and approval against a named officer; signed attestations of each legal interlock; black box recording at the legal interlock where the instruments elect it. |
| (i) Self-replicating AI agents | An autonomous system or AI agent replicates onto other | Command responsibility; the requirement of an | A combination of layers set in the relevant SLIs, including the mission plan, |

| Risk | Example | Exposure | AULA response |
|---|---|---|---|
| | hosts or devices, including captured or allied devices, or an adversary's replicating AI agent enters a defence network, and the replicants continue to act after the original is recalled or destroyed. | identifiable Accountable Officer for each AI-enabled function; the obligation to review each new weapon, means or method of warfare (Additional Protocol I art 36), which a replicant created after that review would escape. | the AWM and the MAWM (Section 4.6): a replication policy requiring registration of each replicant at the moment of replication and permitting replication only for the classes and generational depth the mission plan pre-authorises, including in communications-denied operation; a LAS for each replicant, recorded as a child entry of the replicating system and anchored to the commander; refusal by the legal interlock of the replicating system to relay for an unregistered replicant; aggregate ceilings and a rate of escalation set for the collection; silicon enforcement holding the credential in hardware, so that it cannot be extracted to a captured or foreign device; admission of registered systems only at each governed boundary, including coalition boundaries; revocation propagated to every replicant, operating as a hardware stop on fitted devices; decommissioning of each replicant at the end of its task or mission. Defence can weight these layers differently, and simulation is intended to compare those configurations (Section 9.2.5). |

In Defence, each autonomous system is first registered as a LAS on the registry, anonymously anchored to Defence as the responsible legal person. The rules of engagement, targeting directives and the commander's intent are then executed as clauses of the SLIs tethered to that LAS. The legal interlock applies those clauses before the system acts, and silicon enforcement applies them in hardware where the system is physical (Section 3.10). An engagement class can be designated as conjunctive, so that force is authorised only when the separately administered legal interlock of each designated officer issues an allow verdict (Table 9). The application of AULA to lethal autonomous weapons is reserved for the secondary paper foreshadowed in Section 3.10.

### 5.8.3 The mission plan as a smart legal instrument

A mission plan is the SLI on which the conduct of an autonomous military system most directly turns. A Defence planner drafts the plan in the ordinary operation order format and adds programmable clauses to it. Each programmable clause pairs a paragraph of the order with a deterministic rule stating its condition, the act it permits or forbids and the notice it requires. The approval that military compliance requirements already demand serves as the execution of the plan, which affirms it cryptographically and legally and registers it in the LAS of the system. The plan then operates as the mission planning statement under which the system acts, and the officer who approved it is recorded as the person who set its scope.

In the fictional exercise scenario Operation ACME VIGIL, an autonomous search and rescue UAV flies for the defence force of State A in alpine terrain under such a plan (Figure 17). The plan carries edge clauses PC-E1 to PC-E4, which run on dedicated enforcement hardware inside the UAV, and cloud clauses PC-C1 and PC-C2, which run on the control plane. Before launch, the control plane sends the edge clauses to the device in the legal envelope. The device tests each manoeuvre against the edge clauses and refuses a manoeuvre outside them, with or without a live link to the control plane. The link is expected to drop in alpine terrain, so the edge clauses govern the conduct of the UAV while it is out of contact. On reconnection the device returns the records of its verdicts and attestations, each bound to the envelope under which it was made, to the control plane, and the control plane runs the cloud clauses against them. A revocation of the mission authority travels in the next envelope which reaches the device when the link returns, and the device then stops the UAV acting under the plan. While out of contact, the mission authority lapses at the limits the mission plan sets.

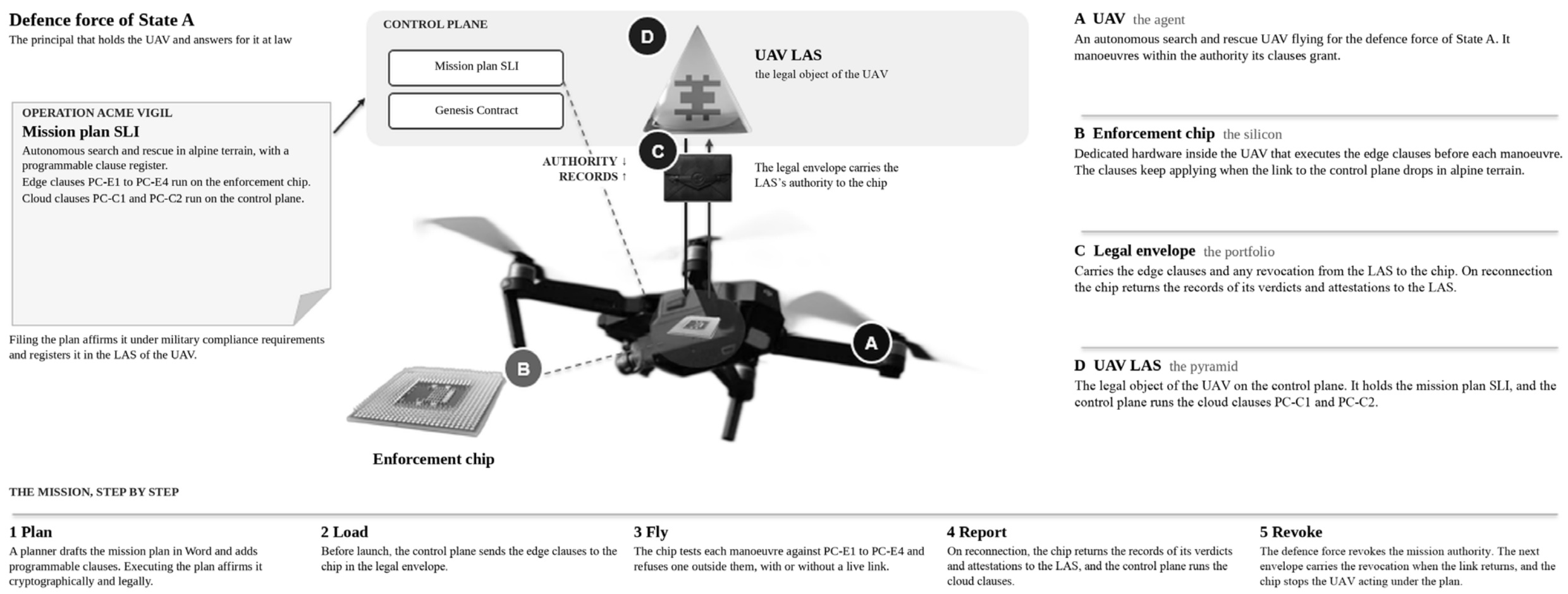

*Figure 17. Silicon enforcement of a mission plan SLI in an autonomous search and rescue UAV (fictional exercise scenario). The mission plan SLI and the Genesis Contract are held in the LAS of the UAV on the control plane. The legal envelope carries edge clauses PC-E1 to PC-E4 to the enforcement chip in the UAV, which applies them before each manoeuvre and returns the records of its verdicts and attestations on reconnection. Cloud clauses PC-C1 and PC-C2 run on the control plane.*

*Table 9: Practices of a defence organisation that uses AULA to discharge AI risk.*

| Practice | AULA primitives | Risk addressed (Table 8) and effect |
|---|---|---|
| 1. Register each autonomous system and its model state, and tie the weapons review to the registered state. | Registration of each system as a LAS on the registry; integrity tier; verification events; unauthorised learning update as a revocation trigger. | (d). A change to the system after review is detected at the next verification event, and revocation or a narrowing of scope follows under the instrument. |
| 2. Execute rules of engagement, targeting directives, the commander's intent and each mission plan as SLIs reviewed by a legal officer. | Registered LAS on the registry; SLIs bound to the LAS of each system; legal interlock; silicon enforcement for physical systems. | (a). The commander's judgment is applied to each engagement before the system acts, at the speed of the system. |
| 3. Designate engagement classes as conjunctive. | Registered LAS on the registry; conjunction clause; credential shares contributed by the separately administered legal interlocks of the designated officers, each on its own allow verdict, and joined once the threshold the clause specifies is reached. | (a) and (f). An engagement of a designated class proceeds only on the allow verdict of the designated officers the clause requires, so one compromised component cannot authorise force. |

| Practice | AULA primitives | Risk addressed (Table 8) and effect |
|---|---|---|
| 4. Issue delegated credentials to AI sub-agents and swarm elements within a permission ceiling. | Registered LAS on the registry; delegated credential recorded as a child entry; permission ceiling; MAWM; propagation of revocation. | (b). Each element acts within the commander's authority, and one revocation reaches every element beneath the orchestrator. |
| 5. Set the limits of each mission in advance for communications-denied operation. | Registered LAS on the registry; clauses fixing the area, period and target classes of the mission; local evaluation at the legal interlock; revocation as a hardware stop. | (c). The authority of the system ends at the limits the commander set, whether or not command can reach it. |
| 6. Hold intelligence, sensor and controlled technical data as LADs. | Registration of the data as a LAD on the registry; digest of each version; title, licence and consent instruments; licence recording permitted recipients and export conditions. | (e) and (g). An engagement relies only on verified data, and controlled data reaches only permitted recipients. |
| 7. Govern coalition interaction through relationship contracts between national LASs. | Registered LAS on the registry; CERC recording national caveats; AURA; enforcement fabric across national control planes. | (e). The legal interlocks of both states apply each caveat, and a breach is recorded against the principal of the responsible state. |
| 8. Retain the governance record of each engagement for review and investigation. | Registered LAS on the registry; governance event chain; signed attestations; black box recording at the legal interlock where elected. | (h). An inquiry can establish who authorised each engagement, under which instrument and on what data. |
| 9. Set a replication policy for each autonomous system and each collection in its mission plan and instruments. | Registered LAS on the registry; relevant SLIs, including the mission plan, AWM and MAWM, requiring registration at replication; child entry for each replicant; permission ceiling; silicon enforcement; admission rule; propagation of revocation. | (b), (c) and (i). Each replicant acts within the classes the mission plan pre-authorises and under the commander's authority, and an unregistered replicant cannot act at any governed boundary. |

# 6 Directors' duties, corporate compliance and AULA

Corporate compliance discussions about AI often turn on whether AI legislation exists with which an entity must comply. However, existing law, and the duties it imposes on those who control an entity, must be re-imagined through an AI lens. Organisations will be compelled to adapt their systems of management and governance to respond to the new and evolving risks and opportunities created by AI. Those risks and opportunities are not limited to AI that the organisation deploys internally, whether approved or not. The risks and opportunities extent to the actions of , and by external AI agents that interact with the organisation and also those that interact with the organisation's own AI agents deployed within and beyond organisational boundaries. . The legal duties of officers and responsible persons must also increasingly be discharged, and their discharge evidenced, at the speed and scale at which AI agents act. Where a tool that performs a duty at machine speed becomes available and cost-effective, the law has treated its use as part of the standard the duty requires. Work health and safety law, vehicle and aviation safety standards, court rules on technology assisted review and anti-money laundering enforcement each apply that rule (Sections 5.2 and 6.1).

## 6.1 Directors', officers' and other duties

### 6.1.1 Duties of care and good faith

Directors owe a duty to exercise care and diligence, and a duty to act in good faith in the best interests of the company. The duty of care takes statutory form in s 180(1) of the Corporations Act 2001 (Cth),[257] s 157(1) of Singapore's Companies Act 1967,[258] s 309(a) of the California Corporations Code,[259] s 174 of the Companies Act 2006 (UK),[260] s 122(1) of the Canada Business Corporations Act,[261] s 166(3) of India's Companies Act 2013[262] and art 165 of Vietnam's Law on Enterprises 2020.[263] Each statute pairs that duty with a commensurate duty of good faith in the best interests of the company. In the common law jurisdictions, the standard of care is objective, and it is measured against what a reasonable director in the company's circumstances would have done, and in the United Kingdom, the standard is raised higher for a director with greater knowledge, skill or experience. Vietnam's statute requires each manager to act honestly, prudently and in the best manner to secure the lawful interests of the company, and gives shareholders or group of shareholders holding at least 1% of the ordinary shares a right to sue a member of the board of directors or the director or general director who fails to do so.

---

257 Corporations Act 2001 (Cth) s 180(1) (care and diligence), s 180(2) (business judgment rule) and s 181(1) (good faith in the best interests of the corporation and for a proper purpose).

258 Companies Act 1967 (Singapore) s 157(1) (a director shall at all times act honestly and use reasonable diligence in the discharge of the duties of the office); Code of Corporate Governance 2018 (Singapore) Principle 9 (the board is responsible for the governance of risk).

259 California Corporations Code § 309(a) (a director shall perform the duties of a director in good faith, in a manner the director believes to be in the best interests of the corporation and its shareholders, and with such care, including reasonable inquiry, as an ordinarily prudent person in a like position would use under similar circumstances).

260 Companies Act 2006 (UK) s 172 (duty to promote the success of the company, having regard among other matters to the likely consequences of any decision in the long term) and s 174 (duty to exercise reasonable care, skill and diligence, measured against the general knowledge, skill and experience reasonably expected of a person carrying out the director's functions and against the director's own).

261 Canada Business Corporations Act, RSC 1985, c C-44, s 122(1) (act honestly and in good faith with a view to the best interests of the corporation, and exercise the care, diligence and skill that a reasonably prudent person would exercise in comparable circumstances) and s 122(1.1) (matters the directors may consider).

262 Companies Act 2013 (India) s 166(2) (act in good faith to promote the objects of the company for the benefit of its members as a whole, and in the best interests of the company, its employees, the shareholders, the community and the protection of environment) and s 166(3) (exercise duties with due and reasonable care, skill and diligence and exercise independent judgment); see also s 134(5)(e) and (f) (directors' responsibility statement on internal financial controls and on systems to ensure compliance with all applicable laws).

263 Law on Enterprises No 59/2020/QH14 (Vietnam) art 165(1) (members of the board of directors, the director or general director and other managers must exercise their rights and be loyal to the company's interests; do not abuse their power and position or use the enterprise's information, secrets, business opportunities and assets for personal gain or serve any other organization's or individual's interests) and art 166 (right of shareholders to bring proceedings against the board of directors, the director or general director).

The duty of care requires a director to be informed about the business and to ensure that it has management systems through which material risks reach the board. A director cannot rely on ignorance of a matter that reasonable inquiry would have revealed.[264] Under Delaware law, which governs many companies headquartered in California, a director breaches the duty of oversight by utterly failing to implement any reporting or information system, or by consciously ignoring the red flags such a system produces.[265] In *Marchand v Barnhill*, the Delaware Supreme Court allowed a claim to proceed against the directors of an ice cream manufacturer whose board had no committee and no regular process for monitoring food safety, the risk most central to its business.[266] The board of an enterprise whose operations increasingly run through AI agents may be in a comparable position where the risks of those AI agents are central to its business and no system brings the risks of those AI agents to it.

The duties of care and good faith, whether statutory or general law, require the directors and officers who control an organisation to identify the risks that AI creates for it and to take reasonable steps to mitigate them. Those risks include unlawful acts by AI agents (including those deployed outside the organisation by third parties) and by embodied AI systems, AI policy washing, failures to protect the entity's people and property, and failures to identify and realise the value of its data and AI systems (Table 10).

*Table 10: AI risks that directors must address, and the AULA response.*

| Risk | Example | Legal exposure | AULA response |
| --- | --- | --- | --- |
| (a) Unlawful act by an AI agent | The AI agent enters a contract outside its authority, makes a | The organisation is bound by the AI | Registration of the AI agent against an identified legal person; AWM recording its |

264 Australian Securities and Investments Commission v Healey [2011] FCA 717, (2011) 196 FCR 291; Daniels v Anderson (1995) 37 NSWLR 438; Re Barings plc (No 5) [1999] 1 BCLC 433, affd [2000] 1 BCLC 523; Lim Weng Kee v Public Prosecutor [2002] 2 SLR(R) 848; Official Liquidator, Supreme Bank Ltd v P A Tendolkar (1973) 1 SCC 602; Peoples Department Stores Inc (Trustee of) v Wise, 2004 SCC 68, [2004] 3 SCR 461; Corporations Canada, 'Directors and officers' (Innovation, Science and Economic Development Canada, modified 3 July 2016) <https://ised-isde.canada.ca/site/corporations-canada/en/business-corporations/directors-and-officers> (accessed 2 November 2026) (directors and officers cannot avoid liability by claiming they did not know what the corporation was doing, and must stay informed about its activities and ensure that they are lawful).

265 In re Caremark International Inc Derivative Litigation, 698 A 2d 959 (Del Ch 1996); Stone v Ritter, 911 A 2d 362 (Del 2006). Many companies headquartered in California are incorporated in Delaware, whose law governs their directors' duties under the internal affairs doctrine, subject to the limited exceptions in California Corporations Code § 2115.

266 Marchand v Barnhill, 212 A 3d 805 (Del 2019).

| Risk | Example | Legal exposure | AULA response |
|---|---|---|---|
| | misleading representation, breaches privacy law or engages in conduct that contravenes competition law. | agent's conduct.[267] Civil penalties, damages and the directors' duty of oversight are engaged. | scope; legal interlock denying or escalating an act outside that scope before the AI agent acts; governance event chain recording each verdict. |
| (b) Unlawful act by a physical AI system | A drone, robot or autonomous vehicle that the organisation owns or operates injures a person or damages property. | Liability as owner, operator, or person conducting a business or undertaking under work health and safety law. | In addition to AULA response for Risk (a) set out above - LAS of the device; silicon enforcement of clauses derived from its SLIs, under which authority lapses at pre-set limits without connectivity; an explicit revocation executed as a hardware stop on receipt (Sections 3.9 and 3.10). |
| (c) Policy washing | The organisation adopts an AI policy that it does not implement, so that its stated governance of AI is not matched by operating controls (i.e. AI policy washing). | Misleading or deceptive conduct, by analogy with greenwashing penalties[268] and enforcement against AI washing.[269] Directors may be found to have known of the gap. | In addition to AULA response for Risk (a), (b) set out above. AI policy executed as a smart legal policy that the legal interlock of each AI agent applies before the AI agent acts; governance event chain recording each application of the policy as evidence of implementation. |
| (d) Failure to protect people and property | An AI agent bullies or harasses a worker or allocates work in a discriminatory way, or unlawfully exposes | Work health and safety and anti-discrimination duties apply whatever the nature of the | In addition to AULA response for Risk (a), (b) set out above. AWM behavioural constraints, including prohibited |

[267] Quoine Pte Ltd v B2C2 Ltd [2020] SGCA(I) 02; Moffatt v Air Canada 2024 BCCRT 149.

[268] Australian Securities and Investments Commission v LGSS Pty Ltd (No 3) [2025] FCA 205 (penalty of $10.5 million on a superannuation trustee that represented that it would not hold investments in gambling, coal mining, oil tar sands and Russia while it continued to hold them); see also Australian Securities and Investments Commission v Mercer Superannuation (Australia) Ltd [2024] FCA 850; Australian Securities and Investments Commission v Vanguard Investments Australia Ltd (No 2) [2024] FCA 1086.

[269] Securities and Exchange Commission, 'SEC Charges Two Investment Advisers with Making False and Misleading Statements About Their Use of Artificial Intelligence' (Press Release 2024-36, 18 March 2024), concerning Delphia (USA) Inc and Global Predictions Inc.

| Risk | Example | Legal exposure | AULA response |
|---|---|---|---|
| | the entity's data, intellectual property or systems. | actor.[270] Privacy and cyber security obligations are engaged. | conduct; title, licence and consent instruments of each LAD; revocation of an AI agent that breaches them; governance event chain as the record against which a complaint is assessed. |
| (e) Failure to identify and realise the value of assets | Data and AI systems held by the organisation go unrecorded, their development costs are expensed beyond recovery, or others take their value. | Duties of care and diligence and of acting in the best interests of the company; IAS 38.71 bars reinstatement of expensed development costs. | Registration of legal object. Genesis Contract and registered title fixing the legal and accounting boundary of each asset; supplementary valuation above the cost basis of the registered asset (Sections 5.6 and 6.4). |
| (f) Self-replicating AI agents | An AI agent of the organisation replicates onto other hosts, or a replicating AI agent compromises the organisation's devices, and the replicants continue to act after the original is stopped. | The organisation may be bound by, or liable for, acts done by replicants of its own AI agent and acts done through its devices, and it cannot show which of those acts it authorised. The duty of care and diligence extends to the oversight of that risk. | In addition to AULA response for Risk (a), (b) above. A combination of layers that the entity configures in its instruments (Section 4.6): registration of each AI agent from inception; a replication policy in the relevant SLIs, including the AWM and MAWM, requiring registration of each replicant at the moment of replication and stating whether, how often, to what depth and on which hosts/platforms an AI agent may replicate; an LAS for each replicant, recorded as a child entry, so that an authorised replicant is distinguishable from an unregistered one; refusal by the legal interlock of the replicating AI agent to relay for an unregistered |

[270] Myint, Blycha, Price and Stutley (n 180) s 5.

| Risk | Example | Legal exposure | AULA response |
|---|---|---|---|
| | | | replicant; aggregate ceilings and a rate of escalation set in the MAWM for the collection; containment of the runtime environment; the temporal signature, which narrows scope when replication departs from the pattern of conformant operation; liability allocated in the relevant SLIs to the responsible legal person for any act of a replicant created without registration or outside their replication clauses; admission of registered AI agents only; terms with agent platform providers requiring registration of each replicant at replication; silicon enforcement on the entity's physical systems, holding the credential in hardware; revocation of the original propagated to every replicant, and decommissioning of each replicant whose task is complete. The organisation can weight these layers differently, and simulation is intended to compare those configurations (Section 9.2.5). |
| (g) Delayed adoption of registration | The organisation deploys AI agents before registering them and binding them by executed instruments, intending to add those controls later. | The organisation cannot show that acts done before registration were authorised, and replicants created in that period remain outside its control. Directors who approved deployment | In addition to AULA response for Risk (a), (f) set out above. Registration as a condition of deployment; Genesis Contract and AWM executed before the AI agent first acts; admission |

| Risk | Example | Legal exposure | AULA response |
|---|---|---|---|
| | | without those controls may be unable to show the care and diligence their duties require. Development costs expensed for want of contemporaneous evidence that the IAS 38.57 criteria were met cannot later be capitalised (IAS 38.71; Section 6.4). | of registered AI agents only. |

The duties do not require a director to eliminate risk. They require reasonable steps, and inaction is the clearest failure to take them. In *ASIC v RI Advice Group,* the Federal Court of Australia observed that it is not possible to reduce cybersecurity risk to zero, and that it is possible to reduce it materially, through adequate documentation and controls, to an acceptable level.[271] A board that has identified a risk, adopted controls and accepted a residual level of risk has made a business judgment, which statute and the common law protect where the judgment is informed.[272] A board that has neither identified the risk nor adopted controls has made no judgment to which that protection can attach.

### 6.1.2 Penalties

Breach of these duties exposes a director personally. In Australia, a director who contravenes the duty of care is liable to a civil penalty of up to the greater of 5,000 penalty units and three times the benefit derived or detriment avoided, to compensation orders and to disqualification. A director who breaches the duty of good faith recklessly or dishonestly commits an offence punishable by imprisonment for up to 15 years.[273] In Singapore, a director who breaches s 157 is liable to the company for any profit made and any damage suffered, and commits an offence punishable by a fine of up to $20,000 or imprisonment for up to 12 months.[274] In the United Kingdom, a director whose conduct makes that person unfit to be concerned in the management of a company must be disqualified for between two to 15 years.[275] In India, directors who breach s 166 are fined and if fraud is established may be imprisoned for up to ten years.[276] Officers may also be prosecuted for

271 Australian Securities and Investments Commission v RI Advice Group Pty Ltd [2022] FCA 496; Australian Securities and Investments Commission, ‘Court finds RI Advice failed to adequately manage cybersecurity risks’ (Media Release 22-104MR, 5 May 2022).

272 Corporations Act 2001 (Cth) s 180(2); Peoples Department Stores (n 264); In re Caremark (n 13).

273 Corporations Act 2001 (Cth) ss 180, 184 and 1317G and sch 3; ss 1317E, 1317G and 1317H (compensation orders); s 206C (disqualification).

274 Companies Act 1967 (Singapore) s 157(3).

275 Company Directors Disqualification Act 1986 (UK) s 6.

276 Companies Act 2013 (India) ss 166(7) and 447.

their handling of a cyber incident. In 2022, a jury convicted Uber's former chief security officer of obstructing a Federal Trade Commission proceeding and of concealing a felony in connection with a data breach, and he was sentenced in 2023 to three years' probation.[277]

### 6.1.3 Regulatory duties

Regulators in each of these jurisdictions treat cyber risk as a matter for the company, and which could be a matter for the board. In February 2026, the Federal Court of Australia ordered FIIG Securities to pay a penalty of $2.5 million for cyber security failures that included the absence of multi-factor authentication for remote access and of an incident response plan tested at least annually.[278] APRA's Prudential Standard CPS 234 makes the board of a regulated entity ultimately responsible for its information security, and the critical infrastructure regime requires the board to approve an annual report on the entity's risk management program.[279] The United Kingdom's Cyber Governance Code of Practice provides for clear ownership of cyber risk, including at the director level, and the UK Corporate Governance Code provides that the board should monitor and review the effectiveness of the company's risk management and internal control framework.[280] Registrants with the United States Securities and Exchange Commission must describe the board's oversight of cybersecurity risk in their annual reports, and California's privacy regulations require businesses above defined thresholds to complete cybersecurity audits and certify them to the California Privacy Protection Agency from 2028.[281] The Monetary Authority of Singapore requires the board and senior management of a financial institution to oversee its technology risk.[282] The risk management committee of each of India's largest listed companies must cover cyber security,[283] and in Canada, OSFI Guideline B-13 sets expectations for the technology and cyber risk

---

277 United States v Sullivan (ND Cal, verdict 5 October 2022, sentence 4 May 2023). This decision was upheld by the Ninth Circuit Court of Appeals on 13 March 2025.

278 *Australian Securities and Investments Commission v FIIG Securities Ltd* [2026] FCA 92; Australian Securities and Investments Commission, 'ASIC action sees FIIG Securities ordered to pay $2.5 million over cyber security failures' (Media Release 26-021MR, 9 February 2026).

279 Australian Prudential Regulation Authority, Prudential Standard CPS 234 Information Security (July 2019) para 13; Security of Critical Infrastructure Act 2018 (Cth) pt 2A, s 30AG (annual report on the critical infrastructure risk management program, approved by the board or governing body).

280 Department for Science, Innovation and Technology and National Cyber Security Centre, Cyber Governance Code of Practice (8 April 2025); Financial Reporting Council, UK Corporate Governance Code 2024, Provision 29.

281 17 CFR § 229.106(c) (Item 106 of Regulation S-K), adopted in Securities and Exchange Commission, Cybersecurity Risk Management, Strategy, Governance, and Incident Disclosure, Release No 33-11216 (26 July 2023); California Privacy Protection Agency, 'California Finalizes Regulations to Strengthen Consumers' Privacy' (23 September 2025). On 13 January 2026, Chairman Atkins opened a review of Regulation S-K which has resulted in trade groups asking the SEC to drop Item 106.

282 Monetary Authority of Singapore, Technology Risk Management Guidelines (January 2021) ch 3; Cybersecurity Act 2018 (Singapore); Personal Data Protection Act 2012 (Singapore) s 24.

283 Securities and Exchange Board of India (Listing Obligations and Disclosure Requirements) Regulations 2015 reg 21; Indian Computer Emergency Response Team, Directions under sub-section (6) of section 70B of the Information Technology Act 2000 (28 April 2022); Digital Personal Data Protection Act 2023 (India) s 8(5) which will require reasonable security safeguards from May 2027.

management of federally regulated Canadian financial institutions.[284] Vietnam's Law on Cybersecurity, in force from 1 July 2026, and its Law on Personal Data Protection, in force from 1 January 2026, impose security and risk assessment obligations on enterprises.[285]

Australian regulators have extended the same expectation to frontier AI. In April and May 2026, APRA and ASIC wrote to regulated entities setting out their expectations of frontier AI preparedness. Their joint report of August 2026 records that frontier AI accelerates cyber and operational risk and creates the potential for unintended actions by increasingly autonomous agentic systems.[286] *The regulators expect entities to demonstrate that key decisions, escalation pathways, recovery arrangements and governance processes can operate at the speed that frontier AI threats require*. Participants in the regulators' roundtables emphasised that board decisions and guardrails on risk appetite, escalation authority and supplier reliance must be in place before a crisis, because compressed response times leave little opportunity to escalate.[287] The Australian Signals Directorate has published guidance for boards on frontier AI cyber threats, and separate guidance on the adoption of agentic AI services and on AI agents that take unexpected actions.[288]

## 6.2 Using AULA primitives to support corporate compliance and the discharge of duties

An organisation that adopts AULA augments analogue AI policies with executed instruments (SLIs), registers its AI agents and data, admits only registered AI agents and manages its AI agents and data as a portfolio of digital property (Table 11).

*Table 11: Practices of an organisation that uses AULA to discharge its duties.*

| Practice | AULA primitives | Risk addressed (Table 10) and effect |
|---|---|---|
| 1. Govern autonomous systems through executed instruments, and not through analogue policies alone. | AI policy executed as an SLI; AWM for each AI agent; MAWM for the organisation; legal interlock applying each | (a) and (c). The policy operates at the speed of the AI agents it governs, and the |

284 Office of the Superintendent of Financial Institutions, Guideline B-13: Technology and Cyber Risk Management (July 2022, effective 1 January 2024).

285 Law on Cybersecurity No 116/2025/QH15 (Vietnam), passed 10 December 2025 and in force from 1 July 2026; Law on Personal Data Protection No 91/2025/QH15 (Vietnam), passed 26 June 2025 and in force from 1 January 2026.

286 Australian Prudential Regulation Authority and Australian Securities and Investments Commission, Resilience at Frontier AI Speed: Insights from the APRA-ASIC Industry Roundtables (Joint Publication, August 2026) 1–3 <https://download.asic.gov.au/media/imunrchk/asic-and-apra-resilience-at-frontier-ai-speed-roundatable.pdf> (accessed 2 November 2026); Australian Prudential Regulation Authority, Letter to Industry on Artificial Intelligence (April 2026); Australian Securities and Investments Commission Letter to Industry, 'ASIC calls for urgent cyber uplift as AI accelerates cyber threats' (Media Release 26-092MR, 8 May 2026).

287 APRA and ASIC, Resilience at Frontier AI Speed (n 286) 2 to 3 and app 2.

288 Australian Signals Directorate, Australian Cyber Security Centre, Frontier AI Cyber Threat Considerations for Boards of Directors; Careful Adoption of Agentic AI Services; When AI Agents Take Unexpected Actions (cyber.gov.au), each listed in APRA and ASIC, Resilience at Frontier AI Speed (n 286) app 1; Australian Institute of Company Directors and Australian Signals Directorate, Board Guidance: Frontier AI Cyber Threats (August 2026).

| Practice | AULA primitives | Risk addressed (Table 10) and effect |
|---|---|---|
| | instrument before the AI agent acts. | governance event chain evidences its operation. |
| 2. Register on an independent legal registry every AI agent and data asset that will act outside the organisation or in cross-organisational deployment. | LAS and LAD; verifiable identity anchor; LARC record on the home registry; AURA presentation of the legal envelope. | (a) and (d). Each AI agent and data asset is anchored to an identified legal person, and a counterparty can verify its authority. |
| 3. Require each customer AI agent to be registered before dealing with it. | Admission rule applied at the legal interlock or runtime supervisor; terms of service as an SLP evaluated against the AI agent's AWM. | (a) and (d). The organisation knows whom the AI agent represents and what it is authorised to do (Section 6.3). |
| 4. License or transfer data and AI agents only where each is a registered legal object. | Title, licence and transfer instruments of each LAD and LAS; LARC record of each dealing. | (e). The dealing takes effect under executed instruments, and the registry gives notice of the interest (Section 5.5). |
| 5. Manage the organisation's AI agents and data as a portfolio of digital property that informs business decisions. | Registry record of each object; MAWM reporting lines; governance event chain; A-Val of each registered object. | (e). The board holds a current inventory of its AI agents and data, with the value and risk of each. |
| 6. Structure that portfolio to maximise capitalised assets and the capacity to use them as security, including, if a market develops, in securitisation. | Genesis Contract fixing the boundary of each asset; SLI recording the capitalisation start date (Section 6.4); registered title; charge of a registered object. | (e). Development costs are capitalised from the date recorded as the date each IAS 38.57 criterion is met, and the asset can be licensed, sold or charged (Sections 5.6 and 6.4). |
| 7. Execute a minimum set of SLIs for each AI agent that acts in a capacity that would require documentation if a human held it. | Genesis Contract as the AI agent's birth certificate; AWM as the counterpart of an employment contract; decommissioning or revocation instrument as its death certificate. | (a) and (b). The AI agent's creation, authority and cessation are each recorded against a legal person, and an AI agent cannot outlive its authority. |
| 8. Register each AI agent at inception, before it acts, and fit the organisation's physical AI systems with silicon enforcement. | Genesis Contract; child entry for each replicant; admission rule applied at the legal interlock; silicon enforcement; revocation instrument. | (b), (f) and (g). Each AI agent is bound by its instruments from its first act. A replicant created outside the organisation's authority has no child entry and is |

| Practice | AULA primitives | Risk addressed (Table 10) and effect |
|---|---|---|
| | | identifiable as unauthorised, and on a fitted physical system the replicant cannot use the credential held in hardware. |

For an AI agent, a board or executive decision made in advance takes the form of an instrument. The AWM and MAWM record the board's or executive's decisions on scope, escalation and suspension before the AI agent acts, and the legal interlock applies them at the speed of the AI agent (Section 4).

A written policy no longer discharges the duty on its own. Regulators measure governance by whether controls operate, as the penalty imposed on FIIG Securities for an incident response plan that was never tested shows, and APRA and ASIC now ask entities to convert awareness of frontier AI risk into tested implementation.[289] An AI policy written as a document is read by people and applied, if at all, after the AI agents it governs have acted. Under AULA the organisation's AI policy can be executed as an SLI, whose programmable clauses state the approved models and tools, the permitted uses, the classes of act that require escalation and the accountable person for each AI agent. The legal interlock of each AI agent evaluates every governed event against that policy before the AI agent acts, so the policy operates at the speed of the AI agents it governs. The governance event chain records each application of the policy. The board then holds evidence that the policy operated, in addition to evidence that it was adopted.

A director demonstrates that reasonable steps were taken through the records of the risk management process. The central record is the risk register, which identifies each material risk, its owner, the controls applied to it, the residual risk the organisation has accepted and the date of each review. A court or regulator relies on the register to decide whether the board identified, mitigated and managed key risks. A risk absent from the register is a risk the board cannot show it considered or managed. Large firms now recruit senior risk officers to maintain AI and technology risk registers, to keep an inventory of AI models and to approve, detect and manage AI tools, including unsanctioned shadow AI.[290]

AI agents now present a board with a heightened risk which demands appropriate controls. An AI agent is not only an attack surface, it also acts with delegated authority, so it can commit the company to a transaction, disclose data or contravene a statute without any external attacker. In IBM's 2026 Cost of a Data Breach Report, shadow AI, being AI used without the organisation's approval, was involved in 43% of the breaches studied, and the average cost of those incidents was USD 5.39 million. Of the breached organisations, 68 per cent had no AI governance policy or were still developing one.[291] The Chief Justice of New South Wales has described the use of shadow AI by directors themselves as apparently already widespread, and has warned that prompts, transcripts and outputs may be discoverable in litigation or regulatory inquiries.[292] In *ASIC v Bekier,* Lee J observed that boards should discuss and deliberately govern any AI use by formal adoption of policies, "rather than just wink at informal "shadow" use".[293] In *Lanmar,* two directors who relied on ChatGPT for advice on dealing with a director and minority shareholder erroneously adopted an employment law framework in place of the principles of corporate law, and were found to have acted oppressively. Black J observed that ChatGPT "did not prove to be a prudent choice of advisor as far as matters of Australian corporate law were concerned", and that its strategy completed steps that would potentially have contravened directors duties, although no breach of duty was formally established.[294] A board whose register does not list the AI agents operating in its business cannot show that it considered the risks those AI agents carry.

---

289 APRA and ASIC, Resilience at Frontier AI Speed (n 286) 3 and app 2; *Australian Securities and Investments Commission v FIIG Securities Ltd (n 278).*

290 Clifford Chance, 'Head of AI and Technology Risk' (Position Description JR-9086, London, 2026) <https://jobs.cliffordchance.com/job/head-of-ai-and-technology-risk-in-london-jid-3512> (accessed 29 October 2026).

The registry and the MAWM supply that record. They identify each AI agent the organisation deploys, the legal person responsible for it and its reporting line. The AWM of each AI agent records its authorised scope, its escalation pathway and the conditions on which it is suspended or decommissioned. The governance event chain records each verdict, escalation and revocation, so it evidences that the controls operated as well as that they existed. The escalate verdict carries a matter the instruments do not resolve to an identified person, with the committed act specification and its authority chain attached (Section 3.6). In *ASIC v Bekier*, the Federal Court held that two senior executives of The Star Entertainment Group breached their duty of care, in part by failing to inform the board of risks or to escalate them properly, and dismissed the claims against the non-executive directors.[295] A record of what was escalated, to whom and when is the evidence on which that line is drawn.

---

291 IBM and Ponemon Institute, *Cost of a Data Breach Report 2026* (IBM, 2026); see also Deepstrike, 'Shadow IT Statistics 2026: Shadow AI, SaaS and Breach Costs', <https://deepstrike.io/blog/shadow-it-statistics> (accessed 8 November 2026).

292 Chief Justice Andrew Bell AC, Harold Ford Memorial Lecture (Melbourne Law School, May 2026), as reported in Elise Shaw, 'The risks of directors using shadow AI' (Australian Institute of Company Directors, 12 August 2026). The same article reports a Gartner prediction that 25 per cent of enterprise breaches will be linked to the misuse of AI agents by insiders and outsiders by 2028.

293 *Australian Securities and Investments Commission v Bekier* (Liability Judgment) [2026] FCA 196 (Lee J), as quoted in Suzie Leask and Madeline Tait, 'AI governance and directors' duties: balancing opportunities and risks' (Hall & Wilcox, 13 August 2026).

294 In the matter of Lanmar Pty Ltd (No 2) [2026] NSWSC 800 (Black J); Jacqueline So, 'NSW Supreme Court Orders Lanmar Directors to Sell Firm after AI-Assisted Attempt to Remove Colleague', *Australasian Lawyer* (15 July 2026) <*https://www.thelawyermag.com/au/news/general/nsw-supreme-court-orders-lanmar-directors-to-sell-firm-after-ai-assisted-attempt-to-remove-colleague/582413*> (accessed 2 November 2026), reporting on *In the matter of Lanmar Pty Ltd (No 2)* [2026] NSWSC 800 (Black J).

295 *Australian Securities and Investments Commission v Bekier* [2026] FCA 196; Australian Securities and Investments Commission, 'Federal Court finds two Star Entertainment senior executives breached duties, non-executive directors did not breach duties' (Media Release 26-040MR, 5 March 2026). Two other former executives had already admitted breach and were penalised in 2025.

The standard of care rises as effective and cost-efficient tools become available. Where a duty is measured by what is reasonably practicable, or by what a reasonable director with the available information systems would have detected, an available and suitable control becomes an expectation (Section 5.2).[296] Legal infrastructure that registers AI agents and applies their instruments at runtime is a control of that kind. A board that declines it once it is available will need to show that the risks it addresses are adequately controlled by other means.

Statutes directed at AI add specific obligations. Vietnam's Law on Artificial Intelligence, in force from 1 March 2026, requires a deployer to reclassify a system whose modification creates a new or higher risk, and to explain its risk controls and incident handling on request.[297] California's privacy regulations require risk assessments from 1 January 2026 and govern the use of automated decision-making technology for significant decisions from 1 January 2027.[298]

The duties apply equally to the opportunities and value that AI creates. ASIC's Chair said in March 2026 that "every director should embrace AI, to understand its risks and benefits with both eyes open", and to "harness its potential for community and customer benefit".[299] The AICD and the Human Technology Institute describe the board's role as governing the balance between the opportunities AI creates and its interconnected risks, including cyber security and data governance.[300] Ahern argues that attention to the significance of AI for the company's business is a pressing matter of strategy under the duty to act in the best interests of the company, because it bears on the company's competitiveness and profitability.[301] Data and AI systems absorb a material share of enterprise investment, and unauthenticated data typically has neither the separability nor the contractual or other legal rights, one of which IAS 38.12 requires for identifiability. Data and AI systems are therefore carried at a fraction of their cost or not carried at all. Registration supplies separability and contractual boundaries, and supports a supplementary valuation above the cost basis on the method described in Section 5.6. For an AI system, registered title evidences the entity's control under IAS 38.13, and the transfer instruments evidence its ability to sell the system under IAS 38.57(c) and the licence instruments evidence its separability under IAS 38.12(a) and support the existence of a market for it under IAS 38.57(d).[302] The registered right to license also evidences the unencumbered legal rights on which the entity's ability to use or sell the system under IAS 38.57(c) depends. IAS 38.65 measures the cost of an internally generated intangible asset from the date on which it first meets the recognition criteria in paragraphs 21, 22 and 57, and paragraph 71 bars reinstatement of expenditure already recognised as an expense. Because recognition also requires the item to meet the definition of an intangible asset, including identifiability and control (IAS 38.65 paragraphs 12, 13 and 18), expenditure incurred before the later of the date on which it meets that definition and the date it meets the recognition criteria is lost to the carrying amount permanently. Registration can bring forward the date on which a system is identifiable and controlled, and can evidence IAS 38.65 paragraphs 57(c) and (d). It therefore advances the start of capitalisation where identifiability, control or the criteria it evidences are the last conditions to be met, which for data and AI systems is often the case (Section 6.4).[303] The timing of registration is a decision for the board with a lasting financial consequence for the company.

---

296 Nooriam Pty Ltd and LexChip Pty Ltd (n 158) s 4.4; *Australian Securities and Investments Commission v Healey* [2011] FCA 717.

297 Law on Artificial Intelligence No 134/2025/QH15 (Vietnam), passed 10 December 2025 and in force from 1 March 2026.

298 Cal Code Regs tit 11, ss 7150, 7152 (California Privacy Protection Agency, regulations under the California Consumer Privacy Act, effective 1 January 2026). Authority: Cal Civ Code s 1798.185(a)(15).

299 Joe Longo, 'It's Tough Being a Director (But That Doesn't Mean You Shouldn't Do It)' (Speech, AICD Australian Governance Summit, 10 March 2026).

300 Australian Institute of Company Directors and Human Technology Institute, University of Technology Sydney, A Director's Guide to AI Governance (Version 2, 29 June 2026) <https://www.aicd.com.au/news-media/research-and-reports/a-directors-guide-to-ai-governance.html> (accessed 2 November 2026).

301 Deirdre Ahern, 'Directors' duties in the age of agentic artificial intelligence: How can boards navigate corporate purpose and the stakeholder interests of employees around AI adoption?' (2026) 2 Cambridge Forum on AI: Law and Governance e7 <https://doi.org/10.1017/cfl.2026.10049> (accessed 2 November 2026) s 9.

302 IAS 38 Intangible Assets paras 12(a), 13, 57(c) and (d), 75 and 78.

The directors' duties of care and diligence, and of acting in good faith in the best interests of the company, bear on whether to register. Where a registration pathway exists, a board that leaves a material asset class unidentified and incapable of licence, sale or charge will need a considered reason for doing so.[304] In England, the duty to promote the success of the company requires consideration of the likely consequences of a decision in the long term. In India the duty of good faith extends to the interests of employees and the community, whose data and skills often form the asset in question.[305]

## 6.3 Why organisations should admit only registered AI agents

### 6.3.1 The attribution problem

Consider a large, listed company proposing to onboard a workforce with no employment contracts, no defined scope of authority and no documented obligations, and to give that workforce full access to its systems, data and customer relationships. No compliance function would permit this. The same company would not let an unknown contractor with no engagement letter, no insurance and no defined scope walk onto its premises and plug into its network. An enterprise that deploys or receives AI agents in consequential workflows without legal infrastructure of the kind described in this paper is in that position. Its AI agents act on its systems and data under no legally defined

303 IAS 38 Intangible Assets paras 65 and 71; IAS 38.63 prohibits recognition of internally generated customer lists and items similar in substance, and IAS 38.3(a) excludes intangible assets held for sale in the ordinary course of business. The argument in the text concerns AI systems and data outside those categories.

304 *Corporations Act 2001* (Cth) ss 180(1) (care and diligence, subject to the business judgment rule in s 180(2)) and 181(1) (good faith in the best interests of the corporation and for a proper purpose). The proposition in the text is put no higher than that the duties bear on the question: a director who is aware that a material class of the company's assets is unidentified, unbounded and incapable of licence, sale or charge, where a registration pathway exists at modest cost, will need a considered reason for leaving it so.

305 Companies Act 2006 (UK) s 172(1)(a); Companies Act 2013 (India) s 166(2); Myint, Blycha, Price and Stutley, 'The Legal Attribution Gap' (n 180) s 4; see also Deirdre Ahern, 'Directors' duties in the age of agentic artificial intelligence: How can boards navigate corporate purpose and the stakeholder interests of employees around AI adoption?' (2026) 2 Cambridge Forum on AI: Law and Governance e7 <https://doi.org/10.1017/cfl.2026.10049> (accessed 2 November 2026), ss 6 and 8.

authority, scope, obligation or constraint. Unauthorised action, data exposure and unallocated liability have the same consequences for the enterprise whether the actor is human or agentic AI. The largest enterprises now deploying AI agents across organisational boundaries are reaching this point as a hard risk stop.

The attribution problem is not hypothetical. *Quoine Pte Ltd v B2C2 Ltd* is a leading common law authority on the allocation of legal responsibility for the conduct of an algorithmic trading system.[306] The majority held the parties bound by trades that a deterministic algorithm had executed, even though no human had reviewed or foreseen the specific transactions, and attributed the relevant state of mind to the programmer from the time of programming up to the time the contract was formed. Mance IJ, dissenting, would instead have asked whether an honest and reasonable trader who knew the particular circumstances would have recognised a fundamental error. As autonomous systems grow more capable, the dissenting position may be the more durable one. An algorithmic system has no mind capable of knowledge and intention as we understand it in humans, and attributing a programmer's intention to an algorithm at the moment of execution is a fiction that becomes increasingly strained the further the system's actions diverge from what its developer or deployer contemplated. That reasoning gains force as AI agents acquire emergent and self-modifying behaviour. The majority's attribution mechanism depends on a sufficiently close correspondence between human intent and algorithmic output, which cannot be assumed in complex agentic deployment and recursive self-improvement. The dissent's test requires evidence of the circumstances in which the system acted, which current agentic deployments do not reliably produce. AULA is designed to create a contemporaneous record of what was authorised and what was done (Section 3.6), which is relevant evidence under either test.

The case is one point in an increasing awareness that the legal treatment of autonomous system conduct is unresolved. The UK Jurisdiction Taskforce's Legal Statement on Liability for AI Harms, published in July 2026, concludes that English private law is generally capable of resolving non-deliberate AI harm through contract, negligence and professional duty. However, the Statement also identifies the liability treatment of AI agents operating autonomously across multi-step tasks, and attribution where several AI systems in a supply chain each contribute to damage, as questions left open.[307] The authors agree in large part with this conclusion. The analogue nature of the existing legal infrastructure, tools and processes supplying the legal product of contracts, duties and torts cannot effectively address even non-deliberate AI harm, without techno-legal infrastructure, such as AULA.

---

[306] *Quoine Pte Ltd v B2C2 Ltd* (n 42). Quoine operated a digital currency platform through which B2C2 deployed algorithms to trade cryptocurrency. Following a pricing error on Quoine's platform, orders placed by B2C2 were executed at approximately 250 times the prevailing market rate. Quoine reversed the trades and when sued by B2C2, relied on unilateral mistake. The Singapore Court of Appeal, by majority, held that the state of mind relevant to mistake must be that of a natural or legal person rather than an algorithm, and attributed the relevant state of mind to the human programmer, and found that B2C2's programmer had no actual or constructive knowledge of the counterparties' mistake.

[307] UK Jurisdiction Taskforce, *Legal Statement on Liability for AI Harms under the Private Law of England and Wales* (LawtechUK, 7 July 2026), <*https://lawtechuk.io/ukjt/liability-for-ai-harms-under-the-private-law-of-england-and-wales/*> (accessed 2 November 2026).

If a legal system response is slow and human scale, it cannot enforce upon or remedy the acts of a fast, machine-speed AI system(s) with an infinitely larger span of control. Those are precisely the cross-organisational, multi-AI agent conditions with which this paper is concerned. To compound this problem, status quo technology first protocols are being designed to solve for correctness, interoperability and authentication. Legal enforceability and corporate compliance are not yet properties of the existing protocol stack.

### 6.3.2 Admitting only registered AI agents

A compliance function verifies a human contractor's identity before granting system access, and identifies a human customer before opening an account. An organisation that admits an unregistered AI agent applies a lower standard to a synthetic actor that can do more, and faster, than the human contractor or the human customer.

Where AULA is deployed, dealing with an unregistered AI agent is an avoidable risk. Well-governed organisations do not deploy an AI agent until it has been identified, assigned to a responsible person and given a defined scope. Many organisations nonetheless host AI agents that were never sanctioned, often introduced by employees through general-purpose tools and operating without the knowledge of the executive or board.[308] Inside many firms the governance of AI sits with IT and engineering functions and the governance of people sits with human resources, legal and compliance functions. Accordingly, an AI agent introduced through an engineering tool may never reach the compliance record.[309] An unregistered AI agent inside the organisation carries the same deficit as one presented by a counterparty. Neither is tethered to a legal person who answers for its conduct, so the organisation cannot show what authority the AI agent held or what it did.

AI agents are increasingly interacting with enterprises as the AI agents of other organisations and as AI agents acting for customers.

Where an enterprise runs its AI agents under a runtime such as OpenShell, the supervisor mediates each connection those AI agents open.[310] An interceptor at the supervisor can therefore refuse to relay an interaction with a counterparty AI agent that presents no valid AURA assertion of registration. The admission rule then operates at the kernel boundary of each of the enterprise's own AI agents and requires no cooperation from the counterparty.

The same rule applies to the organisation's own AI agents. An AI agent that is not registered in the organisation's name holds no credential that the organisation's legal interlocks accept, so it cannot complete a governed event within the organisation's systems. Registration of each AI agent before deployment brings it within the MAWM and the risk register. The admission rule then excludes any AI agent introduced outside that process. The legal interlock of each of the organisation's AI agents also refuses to relay a governed interaction for a replicant that AI agent creates without a child entry, so an AI agent cannot act through an unregistered replicant (Section 4.6). Terms with the platforms on which employees create AI agents can require registration of each AI agent at creation, which brings an AI agent introduced through a general-purpose tool within the MAWM from its first act.

---

308 IBM and Ponemon Institute (n 291).

309 Myint, Blycha, Price and Stutley (n 180) s 5.

310 NVIDIA Corporation, 'Architecture' (n 128).

The admission rule extends to replicants. A replicant presented by a counterparty is admitted only where its credential is derived from a registered AI agent of that counterparty and records it as a child entry. The legal interlock of the organisation verifies that chain of derivation back to the responsible legal person of the counterparty at the boundary, through AURA (Section 3.8.2). A replicant whose credential has expired, or whose authority exceeds the permission ceiling of the AI agent that created it, is refused. An unregistered replicant presents no chain to verify and is refused as any unregistered AI agent is. The organisation therefore deals with a collection of a counterparty only within the limits the relevant SLIs of that counterparty, including its MAWM, set, and only with replicants for which the counterparty answers (Section 4.6).

### 6.3.3 AI Agents acting for customers

Customers are beginning to send their own AI agents to deal with an organisation: to attend meetings, to open accounts, make claims, negotiate terms, exercise rights over personal data, transact and complain (and often wait on hold for their call to be answered). An organisation that accepts such an AI agent without registration is accepting a counterparty about whom it knows nothing. It does not know who the customer is, whether the customer has authorised this AI agent to act for them at all, what the scope of that authority is, or whether the AI agent presenting itself today is the one the customer configured. *Moffatt v Air Canada* shows what happens when an organisation's own AI misstates its terms; the mirror case, in which a customer's AI agent makes a representation, gives a consent or enters a transaction that the customer later disowns, has the same structure with the exposure reversed.[311] The organisation's existing customer identification obligations do not answer the question, because they identify the human and say nothing about the AI agent.

A customer's AI agent presents as a registered AI system object whose identity anchor traces to the customer as responsible legal person, whose AWM records what the customer has authorised it to do, and whose credential encodes that scope in a form the organisation's own legal interlock verifies at the boundary without a live query. The organisation's terms of service, expressed as a SLI, are evaluated against the AI agent's mandate at the decision point, so that an AI agent asking for something its customer never authorised receives a deny or an escalation to the customer rather than a silent success. The customer's AI agent arrives as a data object carrying its consent instrument, and data the organisation returns leaves as a data object carrying the licence under which it may be used. Where the customer's authority is later withdrawn the extinguishment reaches the AI agent through its credential rather than through a support ticket. The organisation that admits only registered customer AI agents gains the same protection against the customer's AI agent that it gains against a counterparty's, and it gives the customer a verifiable record of exactly what their AI agent did in their name, which is the customer's protection against the organisation.

[311] Moffatt v Air Canada 2024 BCCRT 149. The tribunal held the airline responsible for a representation made by its chatbot; the same reasoning, applied to an agent presented by a customer, would ask whether the customer authorised the representation the agent made, and current onboarding produces no evidence either way.

A compliance policy of this kind can be adopted unilaterally and immediately. It requires no counterparty to have adopted the architecture first, because the policy is a refusal condition: unregistered AI agents are not admitted, in the same way that unverified customers are not onboarded, and the organisation's own AI agents are registered so that it can meet the same condition when it is the counterparty. Each organisation that adopts the rule adds to the pressure on its counterparties to register their AI agents (Section 9). An enterprise whose AI agents are registered and whose systems admit only registered AI agents is, in effect, applying customer identification to the agentic channel, and an organisation that would not approve an unidentified human workforce has the same reason to decline an unidentified synthetic one.

A registered object also changes what the enterprise holds if a counterparty defaults or a custodian fails. The ownership, encumbrances and permitted uses of a registered LAD or LAS are recorded against identified obligors. The enterprise can therefore prove its position to an auditor, an insurer insolvency practitioner or a court with contemporaneous evidence, without first resolving the characterisation questions surveyed in Section 5.5. The register is conventional rather than a distributed ledger (Section 5.3), so holding a registered object raises no question under a treasury or investment policy that excludes cryptoassets.

## 6.4 Audit, accounting and disclosure for registered AI systems

An auditor tests the conduct records of an AI system for integrity and completeness, because those records are information produced by the entity.[312] A hash-linked append-only log evidences integrity, because recomputation of the chain shows that no record has been altered or removed after it was written. An unlogged action leaves no gap in the chain, so the log does not evidence completeness. Completeness is evidenced architecturally where every external action of a system passes through an enforcement point that records the action before executing it.[313] The legal interlock evaluates each governed event before the AI agent acts and produces a governance record of that evaluation (Section 3.6). Where the legal interlock is deployed at the supervisor of a runtime that routes every outbound connection of the AI agent through that supervisor, such as OpenShell (Section 7.3), the AI agent is designed to have no path to external effect that bypasses the legal

312 ISA 500 Audit Evidence.

313 Jerome H Saltzer and Michael D Schroeder, 'The Protection of Information in Computer Systems' (1975) 63(9) Proceedings of the IEEE 1278, stating the principle of complete mediation.

interlock. The auditor evaluates the design and implementation of the controls that enforce that path, and tests their operating effectiveness.[314]

IAS 38.71 bars reinstatement of development expenditure already expensed. A capitalisation start date that cannot be substantiated therefore places earlier expenditure permanently outside the carrying amount of the system.[315] Governance records are created when the events they record occur. The date of the SLI that constitutes the system evidences the entity's intention and is a defined asset in its own right. Versioned evaluation records held in the LAS evidence technical feasibility. The officers responsible for finance and for the system can execute an SLI recording that each criterion in IAS 38.57 is met, with references to the evaluation, resource and cost evidence on which that conclusion rests. That instrument fixes the capitalisation start date as a dated act on the record. Management must still document why it concluded that each criterion was met. Where the records are held on a third party's platform, the entity's auditor will ordinarily require a service organisation controls report over that platform.[316]

Prudential regulation already requires some deployers to hold records of their dependency on AI providers. APRA's Prudential Standard CPS 230 Operational Risk Management, in force from 1 July 2025, requires regulated entities to manage arrangements with material service providers, including planning for exit.[317] An Australian bank that deploys a registered AI agent on a provider's model therefore already holds records of the provider's terms and of its own exit capability, where the provider is a material services provider. The LAS of that AI agent records which layers of the system the bank holds and which the provider holds. A recorded re-instantiation of the AI agent on an alternative model evidences the bank's control of the layers it holds.

Dependence on an AI model provider is itself a risk the board must manage. Once AI agents and workflows are built around one model, switching providers is estimated to cost between 19 and 34 per cent of the original build, and the collapse of Builder.ai in 2025 forced its customers to migrate at their own expense.[318] The LAS holds the AI agent's legal identity, instruments and governance record apart from the model on which the AI agent runs (Section 5). The registered asset therefore survives a change of provider, and a recorded re-instantiation evidences that the organisation retains the layers it holds.

A valuation of registered AI systems disclosed outside the financial statements is likely to fall within the regimes for non-GAAP measures. Such a measure is non-IFRS financial information within ASIC Regulatory Guide 230 in Australia, an alternative performance measure within the ESMA Guidelines on Alternative Performance Measures in the European Union, and a non-GAAP financial measure within Regulation G and Item 10(e) of Regulation S-K for SEC registrants.[319] Each regime requires the measure to be labelled, defined, reconciled to the most directly comparable statutory measure and presented with no greater prominence than that measure. A disclosure of the valuation of a registered AI system would accordingly present the carrying amount, the valuation and the basis on which it was prepared, the principal inputs and the evidence supporting them, and a reconciliation of the valuation to the carrying amount. An auditor reads such a valuation, where it is contained in an annual report, as other information under ISA 720 (Revised), for material inconsistency with the financial statements and with the auditor's knowledge obtained in the audit.[320]

---

314 ISA 315 (Revised 2019) Identifying and Assessing the Risks of Material Misstatement; ISA 330 The Auditor's Responses to Assessed Risks.

315 IAS 38 Intangible Assets para 71. KPMG Australia (2026) identifies documentation of the transition from research to development as one of the most significant judgments finance teams make for AI projects.

316 ISAE 3402 Assurance Reports on Controls at a Service Organization; ISA 402 Audit Considerations Relating to an Entity Using a Service Organization.

317 Australian Prudential Regulation Authority, *Prudential Standard CPS 230 Operational Risk Management* (2023).

318 Stéphane Chatonsky, 'Poor AI governance is driving hidden costs' (Australian Institute of Company Directors, 3 August 2026).

Independent assurance of a registered AI system is an engagement distinct from the statutory audit. It may be performed under ISAE 3000 (Revised) on the system's conformance with its mandate, or take the form of certification of the entity's AI management system against ISO/IEC 42001:2023.[321] Certification of a management system evidences the process by which the entity governs its systems. An engagement on mandate conformance is directed at the conduct of the particular system. Underwriters of AI liability cover may require independent evaluation, and SLIs that specify a mandate, as terms of cover.

Entities that build AI agents in large numbers on a shared model may need an accounting policy that groups AI agents of small individual cost. This is because an entity's capitalisation threshold for software could otherwise exclude each AI agent on its own. The registry records each AI agent as a separate legal object anchored to its holder, so a grouping policy can be applied to a defined and verifiable population of AI agents.

# 7 Current AI Agent Communication, Identity and Governance

Any attempt to describe AI governance tools is fraught with risk as the technical landscape is evolving rapidly.

Noting the evolving landscape, this section reviews the current technical stack concerning:

a) How AI agents communicate;

b) How AI agents are identified and are provided access; and

---

319 Australian Securities and Investments Commission, *Regulatory Guide 230: Disclosing non-IFRS financial information* (2011); European Securities and Markets Authority, *Guidelines on Alternative Performance Measures* (ESMA/2015/1415); Securities and Exchange Commission, Regulation G (17 CFR Part 244) and Item 10(e) of Regulation S-K (17 CFR 229.10(e)).

320 ISA 720 (Revised) The Auditor's Responsibilities Relating to Other Information.

321 ISAE 3000 (Revised) Assurance Engagements Other than Audits or Reviews of Historical Financial Information; ISO/IEC 42001:2023 Information technology: Artificial intelligence: Management system.

c) Governance processes that seek to authorise, monitor and curtail set actions of AI agents.

This section also provides preliminary conclusions as to why existing protocols and products do not address the legal attribution gap. Further work on the legal attribution gap and how AULA can address those gaps are set out in section 3 above.

Table 12 provides an overview of the current landscape.

*Table 12: Current protocols and products (without AULA).*

| Layer | Function | Protocols | Status |
|---|---|---|---|
| 1: Communication | AI Agent-to-human | Conversational interfaces (Chats) structured input structed output. Human approved workflows | Established interaction method |
| | AI Agent-to-agent | A2A: Standard protocol for agent communication and task coordination[322]<br>Domain Specific APIs: Custom interfaces which agents can request work from other agents<br>Framework Specific Messaging: Messages and task handoffs using agentic framework conventions<br>MCP: Agents exposed as tools that other agents can invoke and discover one another[323] | A2A v1.0 released in March 2026 by the Linux Foundation |
| | AI Agent-to-tool | MCP (over stdio or HTTP), direct API/function calls | MCP introduced in 2024;[324] APIs established interfaces for software-to-software interactions |
| 2: Identity and access | Authentication | API keys, OAuth 2.0/2.1, mTLS, JWT may carry | Established mechanisms in supporting software |

[322] The Linux Foundation, ‘Agent2Agent (A2A) Protocol’ (Web Page) <*https://a2a-protocol.org/latest/*> (accessed 15 September 2026).

[323] Model Context Protocol, *Tools* (Web Page, 25 November 2025), <*https://modelcontextprotocol.io/specification/2025-11-25/server/tools*> (accessed 15 September 2026)

[324] Anthropic, ‘Introducing the Model Context Protocol’ (Web Page, 25 November 2024) <*https://www.anthropic.com/news/model-context-protocol*> (accessed 15 September 2026)

| | | | |
|---|---|---|---|
| | | identity or authorisation claims. | clients and delegated access.[325] |
| | Identification | A2A Agent Cards for discovery and metadata information, OID4VC and OpenID4VP for credential and insurance use. | Agent cards are specified in A2A v1.0. |
| | Workload identity | SPIFFE/SVID via SPIRE as the implementation | Established workload identity software with well documented production deployments |
| 3: Governance | Delegation and coordination | Orchestration frameworks (Microsoft Agent Framework, LangGraph,[326] CrewAI, OpenAI Agents SDK) | Publicly available frameworks and SDK Embedded legal governance depends on implementation |
| | Lifecycle governance | Microsoft Entra Agent ID,[327] WSO2 Agent Manager | Available services supporting identity and access lifecycle for deploying agents |
| | Economic layer | AP2 for checkout and payment mandates supported by AAP (Agent Authorisation Framework),[328] OpenID Foundation agenda | Published specification in v 0.2 supporting verifiable user authorisation for purchases made by agents. |
| | Runtime containment | NVIDIA OpenShell: kernel-isolated sandbox with supervisor-mediated network access under a declarative policy. NVIDIA Sentry: out-of-band monitoring and quarantine on BlueField-4 data processing units.[329] | OpenShell is open source (Apache 2.0) and broadly available. Sentry is a reference system design announced in September 2026. |

325 At publication time, OAuth 2.1 remains an IETF draft but still can be implemented.

326 LangChain, ‘How to Add Custom Routes’ (Web Page) <*https://docs.langchain.com/langsmith/custom-routes*> (accessed 15 September 2026).

327 Microsoft, ‘Microsoft Agent Framework Workflows Orchestrations - Handoff’, *Microsoft Learn* (2026) <*https://learn.microsoft.com/en-us/agent-framework/workflows/orchestrations/handoff?pivots=programming-language-csharp*> (accessed 15 September 2026).

328 Google, ‘Agent Authorization Framework’, *AP2 - Agent Payments Protocol Documentation* (2025) <*https://ap2-protocol.org/ap2/agent_authorization/*> (accessed 15 September 2026).

329 NVIDIA Corporation (n 8).

*Table sources.* The layering follows the survey in Yingxuan Yang et al, “A Survey of AI Agent Protocols” (2025) arXiv:2504.16736, *https://arxiv.org/abs/2504.16736*, and the landscape assessment in Tobin South et al, *Identity Management for Agentic AI* (OpenID Foundation, October 2025), *https://openid.net/wp-content/uploads/2025/10/Identity-Management-for-Agentic-AI.pdf*.

## *7.1* Layer 1: communication between agents, tools and humans

### 7.1.1 Model Context Protocol

MCP is an open protocol that standardises how AI applications discover and connect with tools, data sources and services. It uses JSON-RPC 2.0 messages with standard transports including stdio and Streamable HTTP. Furthermore, MCP defines an optional HTTP authorisation framework based on OAuth2.1 including PKCE for authorisation code flows. Client ID metadata documents are recommended however dynamic client registration is deprecated but retained for backwards compatibility.[330] Originally designed for agent-to-tool interaction, MCP can expose AI agent capabilities through tool interfaces allowing one AI agent to invoke another through an MCP server however human to AI agent interaction takes place through application interfaces and prompts. MCP supports this through elicitation which allows servers to request structured input from users through the client application interface.[331] Further, MCP does not readily address delegation from a human to an AI agent and onward to a further AI agent. In conjunction with OAuth (see further on OAuth in section 7.2 below), the most widely adopted authorisation standard, MCP default flow relies on user granting consent although extensions can let AI agents act on their own behalf or under a centrally managed organisational access. However, it does not address legal identity, legal authority, consent or delegation; it does not define end-to-end audit across delegated or multi-agent workflows; it provides no ready mechanism for revocation or wind-back across a chain of agents; and it does not scalably authenticate to downstream platforms on an AI agent’s behalf or across asynchronous operations. The OpenID Foundation has noted that dynamic client registration, as an unauthenticated public endpoint, is exploitable and carries no tether to a responsible person.[332]

---

330 Model Context Protocol, ‘Authorization’, specification revision 2026-07-28, <https://modelcontextprotocol.io/specification/2026-07-28/basic/authorization> (accessed 2 November 2026); Proof Key for Code Exchange, RFC 7636, <https://www.rfc-editor.org/rfc/rfc7636> (accessed 15 September 2026).

331 MCP authorisation binds an access token to a client, a user’s consent and a set of scopes on a single MCP server. The consenting user is identified only as an account on the authorisation server, so the protocol has no identity anchor bound to a registered legal person and no record of the capacity (for example trustee, partner or officer) in which that person acts (see Table 2). A scope grants access to a tool or resource. The existence and limits of the agent’s authority to use that access for a principal are, under AULA, established by the Genesis Contract, the Agent Workforce Mandate and any other SLI tethered to the LAS, and are carried to the MCP server in the legal envelope. Data returned through an MCP resource passes as payload with no title, licence or consent instrument attached, so rights in and to that data are not allocated at the point of access in the way the legal instruments tethered to a LAD allocate them. An MCP server executes a tool call once the token is valid. No legal interlock evaluates the call against SLI governing clauses, returns an allow, deny, escalate or forward verdict, or records the determination as an attributed governance event. An elicitation response returns to the server as structured data, so a human instruction given mid-workflow is not recorded as the decision of a designated responsible person under the escalation pathway of an AWM.

332 Tobin South et al, *Identity Management for Agentic AI* (OpenID Foundation, October 2025), section on dynamic client registration.

### 7.1.2 Agent2Agent Protocol

A2A is complementary to MCP rather than an extension of it. MCP is concerned with interaction between AI systems and tools; A2A is concerned with communication between agents as autonomous peers. MCP and A2A are correlated protocols with different primary purposes. MCP focuses on connecting AI applications to tools and resources whilst A2A supports the communication and collaboration of AI agents. They can be deployed with or without each other. A2A v1.0 defines JSON-RPC, HTTP + JSON/REST and gRPC bindings over common data models defined in these protocol buffers. As MCP-based systems incorporate more autonomous behaviour, interactions resembling AI agent-to-agent communication emerge in practice. A2A standardises AI agent messaging and task coordination. Authentication and authorisation between AI agents rely on established security mechanisms and deployment policies resulting in additional arrangements to be made to legally establish meaningful authority between participants.[333]

### 7.1.3 Agent Card

The core advertisement mechanism in A2A is the Agent Card, a machine-readable JSON document that can be discovered through a URI on registries or direct configuration to where it is hosted since direct access to it might be restricted to it. It also describes the AI agent's name, skill and all forms of interactions with its security requirements. The specification permits Agent Cards to be signed using JSON Web Signature optionally, and it is recommended that clients are directed to verify signatures when present. An unsigned card remains conformant however it lacks a document level signature verifying integrity or authenticity since it can satisfy the schema whilst still failing the policy the recipient needs.[334] (It is possible that it can be properly validated through HTTPS through the endpoint and when in transmission too). The assurance available at the A2A layer is a baseline rather than a guarantee, applying only where AI agents elect to sign.[335]

---

[333] A2A standardises the exchange between agents of different organisations and leaves the legal relationship between those organisations outside the protocol. Under AULA that relationship is set by legal and cryptographically executed instruments. SLI with shared clauses set what the control plane pushes to the Legal interlock of each party at the same version, so that both agents evaluate the same permitted interaction scope, data rights, risk allocation, revocation triggers and dispute resolution terms. An A2A task request carries the message and the credentials required by the receiving agent's declared security scheme, but does not carry the SLIs (or a verifiable reference to them) under which the requesting agent acts. The receiving agent can therefore authenticate the requester but cannot determine whether the request falls within that agent's (and the corresponding responsible legal person's) legal authority or reject an out-of-scope request with a logged and attributable refusal. A2A records the state of each task. The enforcement fabric, by comparison, receives each party's attestation of the instrument it applied and the verdict it reached, and records a mismatch or missing attestation against both principals, so the governance record is produced jointly rather than by one party. A2A has no counterpart to the conjunction clause, under which a conjunctive event completes only when the legal interlock of every legally responsible party returns an allow verdict. Where a task produces an asset or causes harm, the rights in the asset and the liability for the harm are determined after the event, whereas (the legal interlock of) a legal instrument allocates both in advance. A2A treats each agent as an autonomous peer in a technical sense only. The protocol has no means of attaching obligations and enforcement rights to the legal person behind each peer. AULA attaches them to a responsible legal person through the tethered instruments.

[334] A2A Protocol Specification, <https://a2a-protocol.org/latest/specification/> (accessed 2 November 2026), 'Agent Card' and 'AgentCardSignature'; JSON Web Signature, RFC 7515, <https://www.rfc-editor.org/rfc/rfc7515> (accessed 15 September 2026).

## *7.2* Layer 2: authentication, identification and authorisation

Systems receiving requests from AI agents need to establish who is acting and what they are permitted to do. AI agents need a way to discover tools, resources and other agents they intend to use/interact with. Authorisation protocols used by AI agents have mostly been co-opted from those designed for web applications and humans. AI agent applications can access resources on a user's behalf or through a service of their own and more recently they can ascribe permissions to themselves and other agents. Identification and authorisation become critical in cooperative collections of agents that delegate to one another under previously granted permissions without seeking the approval through a human-in-the loop.

MCP's HTTP authorisation specification is based on OAuth 2.1 and requires PKCE to protection the authorisation code flow since it's emerging convention for AI agent access to remote resources. Its authorisation code grant supports access authorised by the resource owner whilst its client credentials grant explicitly supports access by the client on its own behalf. PKCE binds the redemption of an authorisation code to a secret verifier that initiated the flow to strengthen the exchange, but does not authenticate the client, which in an agentic deployment is the AI agent itself; MCP now identifies AI agent clients by a web address they control (a Client ID Metadata Document), Its Client Credentials extension lets AI agents hold their own credentials. Both identify the software making the request but not the legal person responsible for it. Strong identity for non-human clients depends on infrastructure-asserted identity such as SPIFFE identity documents, SPIFFE itself is an established standard however it is used to authenticate clients to OAuth servers which is still an IETF draft. Even in its current form OAuth answers whether a presented credential is valid for a scope. SPIFFE is an open standard for workload authentication that predates agentic AI. A workload obtains a short-lived SPIFFE Verifiable Identity Document from a Workload API and presents it to other workloads, enabling mutual TLS and JWT-based authentication without static network information or long-lived credentials.[336] It provides strong cryptographic proof that a workload is what it claims to be at the infrastructure level, within a trust domain whose infrastructure the operator controls.[337]

---

[335] An Agent Card advertises capability. Its provider field names an organisation that completes the field itself, and a JSON Web Signature, where present, shows only that the holder of the signing key issued the card unaltered. The card does not anchor the agent to a registry entry bound to an identified legal person, record the capacity in which that person acts, or state the integrity tier at which the agent's model state is committed. Under AULA a counterparty obtains each of these facts from the independent registry and from the credential in the legal envelope, and can require a stronger integrity tier for a higher-stakes interaction. The card's skills describe what the agent is able to do. What the agent is authorised to do at a given moment is the narrowest of the authority conferred by its own SLIs, the scope of any applicable relationship contract, the authority remaining after the delegation chain, and its current legal status, and the card expresses none of these. A hosted card also remains discoverable after the agent's authority has been revoked or the agent decommissioned, unless its host removes it.

[336] SPIFFE Standards, <https://github.com/spiffe/spiffe/tree/main/standards> (accessed 2 November 2026): The SPIFFE Identity and Verifiable Identity Document, The X.509 SPIFFE Verifiable Identity Document, and The SPIFFE Workload API (accessed 15 September 2026).

W3C Decentralised Identifiers allow identity that is not anchored to a single platform's identity provider, which is relevant where neither party trusts the other's identity and access management.[338] DIDs do not natively verify legal authority to act, nor ensure that an AI agent presenting credentials has not been modified since issuance.[339]

The Agent Payments Protocol (AP2) defines interaction pathways binding human authorisation through to the completion of payment by an AI agent. A user signs a mandate that the AI agent presents to payment services as proof of human intent, and a payment AI agent can verify that authority cryptographically.[340] AP2 is the exception among current identity and access approaches in taking steps towards a verifiable chain of authority that maps back to a human.[341]

---

337 OAuth answers whether a presented token is valid for a scope, and the operators of the authorisation and resource servers define those scopes. No SLI determines them, so a token can permit an act that the agent's AWM or the applicable CERC excludes. Token exchange records the chain of clients that held a token in its actor claim, and the authorisation server's policy decides at each exchange whether scope narrows. AULA applies a permission ceiling to delegation. Each delegated credential is issued by the delegating object under its own key recorded as a child of that object's registry entry and tied to the governing instrument, and it cannot exceed the permissions of its issuer. Every act of an ephemeral sub-agent therefore remains attributable to the orchestrator's registered legal person after the sub-agent is destroyed. Revoking an OAuth token withdraws access to a resource. AULA revocation extinguishes authority on a trigger defined in the instrument, such as a breach determination, termination of the relationship contract or an unauthorised learning update. It enforces any remedy and notice periods the relationship contract specifies, propagates through the delegation chain by invalidating every derived credential, and records in the governance event chain the interval between extinguishment and operational cessation, an interval OAuth leaves unrecorded. A SPIFFE ID is a URI that the operator of a trust domain assigns to a running workload, and the SVID proves that the workload holds it. The ID has meaning within the operator's trust domain, or a domain federated with it, and identifies a process. Under AULA the running agent is bound to a LAS, a legal object registered with a verifiable identity anchor, associated with an identified legal person, and re-verified at each verification event so that modification is detectable and a failed verification is recorded as a material event. A SPIFFE ID can serve as one input to that binding. The Genesis Contract, the title instruments and the legal status that make the agent an object of property and of responsibility sit outside the workload identity layer, on the registry and the control plane.

338 Decentralized Identifiers (DIDs) v1.0, W3C Recommendation (19 July 2022), <https://www.w3.org/TR/did-core/> (accessed 2 November 2026); Verifiable Credentials Data Model v2.0, W3C Recommendation (15 May 2025), <https://www.w3.org/TR/vc-data-model-2.0/> (accessed 10 September 2026).

339 A DID paired with a W3C Verifiable Credential allows an issuer to attest a claim about a subject and a verifier to check the issuer's signature without relying on a single identity provider. The standards define credential format and verification, and leave each ecosystem to decide which issuers may attest legal authority, what a claim of authority means in law and which (if any) legal instrument that claim derives from. The credential in an AULA legal envelope is derived from the SLIs tethered to the legal object and references them. A counterparty verifying it is therefore verifying a derivation from executed legal instruments, whose existence and binding effect can be attested without disclosure of their contents. A VC status list allows an issuer to revoke a credential when the issuer chooses to update the list. An AULA credential is invalidated by the legal event itself, because revocation is an executable act under the governing instrument(s) and propagates to every credential derived from the revoked object.

340 Agent Payments Protocol (AP2) Specification, <https://ap2-protocol.org/> (accessed 2 November 2026) <*https://ap2-protocol.org/specification/*> (accessed 10 September 2026).

341 AP2 comes closest among current protocols to a chain of authority, because its Checkout and Payment are digital credentials signed by the user and presented by the agent as evidence of the user's instruction. Each mandate is confined to a purchase and its payment. Alternatively, an AWM governs every action taken under the object's authority, records an escalation pathway to a designated responsible person and states the authorised decommissioning action, so the agent's authority for non-payment acts, and its conduct between payments, sits in an instrument AP2 does not supply. The mandate attests the user's intent for one transaction and leaves the agent unregistered, so a merchant has no identity anchor for the agent that persists across transactions and no legal status to check. The allocation of loss for an erroneous or disputed agent

## *7.3* Layer 3: enterprise governance, coordination and lifecycle

Microsoft Entra Agent ID introduces an identity construct for AI agents separate from human users and traditional service principals. Each AI agent identity is created from a reusable blueprint, authenticated through credentials the blueprint acquires on its behalf, and governed through lifecycle features analogous to those applied to human identities. A sponsor, being a human user or group accountable for the AI agent's purpose and lifecycle decisions, is required for every AI agent identity and blueprint, and AI agent actions are logged against the AI agent's own identity.[342] This is among the most mature enterprise identity and lifecycle capabilities available for AI agents. The sponsor requirement tethers AI agent activity to a natural person and the audit trail records agent conduct *within* organisational boundaries.[343]

Orchestration frameworks, including the Microsoft Agent Framework, LangGraph, AutoGen, CrewAI and the OpenAI Agents SDK, provide technical orchestration of multi-AI agent workflows.[344]

WSO2 Agent Manager, together with WSO2 Agent ID, offers framework-agnostic identity and lifecycle governance for AI agents regardless of where they run. It provides a federated registry of every agent an organisation deploys with guardrails enforced at the organisation, AI agent, MCP and model levels. It also includes end-to-end tracing and evaluation, and lifecycle controls such as versioned promotion from development to production and immediate suspension.[345] WSO2 Agent ID provisions AI agents as first-class identities carrying metadata such as owner, purpose and risk

---

purchase is left to payment network rules and the parties' existing terms. Under AULA a CERC between buyer and seller allocates that loss in advance, and a conjunction clause can require the legal interlock of each party to return an allowverdict before the payment completes. Cf Mastercard Agent Pay.

342 Microsoft, 'Overview of agent identities in Microsoft Entra', <https://learn.microsoft.com/en-us/entra/agent-id/agent-identities> (accessed 2 November 2026); 'Administrative relationships in Microsoft Entra Agent ID (owners, sponsors, and managers)', <*https://learn.microsoft.com/en-us/entra/agent-id/agent-owners-sponsors-managers*> (accessed 2 November 2026).

343 The Entra sponsor is a directory attribute within the deploying organisation's tenant. It records internal accountability for the agent, but creates no legal instrument allocating enforceable legal liability for the agent's acts to a legal person or recording the capacity in which the sponsor or the organisation acts. Agent identities, blueprints and audit logs sit inside the tenant, so a counterparty organisation cannot verify the sponsor, the agent's permitted scope or its current status, and the audit trail is one party's account of the agent's conduct. The AULA equivalents are a registry entry bound to the registered legal person, a Master Agent Workforce Mandate that operationalises the organisation chart as a smart legal instrument setting reporting lines, duties and information barriers, and an enforcement fabric record produced jointly from both parties' legal interlocks. Disabling an Entra agent identity removes the agent's access to tenant resources but requires a further AULA legal layer revocation.

344 Orchestration frameworks pass control between agents through handoffs, graph edges or tool calls written in application code, and apply guardrails as code-level checks on inputs and outputs. A handoff passes the task and its context to the next agent at whatever level of access the code grants. Under AULA the orchestrator's LAS anchors accountability for the whole collection. Each sub-agent acts under a delegated credential recorded as a child of the orchestrator's registry entry, its authority narrows at each delegation step under the permission ceiling, and the Legal interlock persists after the sub-agent is destroyed, so an ephemeral sub-agent cannot operate as a liability shield. Where a collection produces an outcome that no individual agent was instructed to produce, AULA determines liability by reference to the permitted scope in the applicable relationship contract, and an outcome outside that scope is an unauthorised act attributable to the orchestrator's holder. The frameworks have no concept of a principal or a counterparty organisation, so that question must be reconstructed from execution logs after the event.

345 WSO2, *WSO2 Agent Manager* (2026) <*https://wso2.com/agent-platform/agent-manager/*> *(*accessed 22 September 2026).

level, evaluates authorisation against that metadata together with delegation information, and logs AI agent actions separately from human audit trails.[346]

Although WSO2 offers strong technical identity, delegation and lifecycle tooling for AI agents, it does not offer any legal layer. Its "delegation" is a token-exchange and scope-propagation mechanism between an AI agent and the resources it accesses, not a permission chain traceable to a legal instrument(s). Its "owner" and "risk level" fields are operational classifications the platform operator sets, not assertions tethered to a registered legal person. Its harness enforces technical policy only; it has no concept of a legal object, or the provision of any dynamic legal context proffered by relevant SLI including relationship contracts with their accompanying legal interlocks deployed between multiple organisations. Further it has no mechanism for allocating rights, obligations or liability arising from what an AI agent does. Its lifecycle tracking (development, staging, production, suspension) is entirely separate from any legal lifecycle event, such as the dynamic legal decommissioning of the AI agent itself, based on relevant contractual execution or termination.

Academic proposals for AI agent infrastructure address parts of the same problem. Chan and colleagues propose AI agent identifiers, real-time monitoring and activity logging as measures that give deployers, users and regulators visibility into the conduct of AI agents.[347] In later work they group AI agent infrastructure into three functions: attributing actions and properties to specific AI agents and their users, shaping the interactions of AI agents, and detecting and remedying harmful actions, with identity binding, agent channels, commitment devices, incident reporting and rollback among the examples.[348] South and colleagues extend OAuth 2.0 and OpenID Connect with AI agent-specific credentials and metadata, and translate permissions expressed in natural language into auditable access-control configurations.[349] Each proposal binds an AI agent to a user or deployer at the technical layer. Measured against the gap elements in Table 1, they contribute to the identification of legal persons (1) and to evidence of authority (2), and the incident reporting and rollback mechanisms contribute technically to enforceability (5). AULA adds a registered legal object for each agent, anchored to an identified legal person and bound by executed instruments, so that the authority a credential encodes is the authority the principal granted at law, and the rights and obligations arising from the AI agent's conduct are allocated by those instruments (Section 3).

NVIDIA's Open Agent Safety Platform, governs AI agents at the runtime and hardware layers.[350] Its software component, OpenShell, is an open-source runtime licensed under Apache 2.0 that runs each AI agent in a sandbox isolated at the operating system kernel.[351] Kernel controls confine the files the AI agent can open and the system calls it can make. The AI agent's workload has no network access of its own, and every connection it opens is passed to a trusted supervisor process outside the sandbox. The supervisor checks the connection against a declarative policy, adds only the credentials that policy permits, and then opens and relays the connection. The AI agent holds placeholders in place of real credentials, so a credential it attempts to send to an unapproved destination arrives there as a placeholder. If the supervisor disconnects, the sandbox freezes the AI agent.[352]

---

346 WSO2, *WSO2 Agent ID* (2026) <*https://wso2.com/identity-platform/agent-id/*> (accessed 22 September 2026).

347 Alan Chan et al, 'Visibility into AI Agents' in *Proceedings of the 2024 ACM Conference on Fairness, Accountability, and Transparency* (ACM, 2024) <https://doi.org/10.1145/3630106.3658948> (accessed 2 November 2026).

348 Alan Chan et al, 'Infrastructure for AI Agents' (arXiv:2501.10114, 19 June 2025) <https://arxiv.org/abs/2501.10114> (accessed 2 November 2026).

349 Tobin South et al, 'Authenticated Delegation and Authorized AI Agents' (arXiv:2501.09674, 16 January 2025) <https://arxiv.org/abs/2501.09674> (accessed 2 November 2026).

350 NVIDIA Corporation (n 8).

The OpenShell gateway is the platform's control plane. It manages the lifecycle of each sandbox, delivers policy and credentials to the supervisor and receives its logs, while the supervisor relays the AI agent's connections without routing them through the gateway. Before a proposed policy change is applied, a policy prover in the gateway uses formal verification to flag any new access the change would grant, such as credentialed reach to a new host, and holds that change for human review.[353] Sentry, the platform's reference hardware design, runs on NVIDIA BlueField-4 data processing units. It inspects AI agent requests and responses from a trust domain separate from the host, verifies AI agent identity, supplies attested telemetry, and quarantines within milliseconds an agent that attempts to leave its software boundary. More than 100 organisations are reported as working with the platform, including Anthropic, Microsoft, SAP, Palo Alto Networks, CrowdStrike, JPMorganChase and Citi. SAP is embedding OpenShell in its Joule Studio runtime, and Salesforce has integrated OpenShell with Slack so that a team can approve or reject an AI agent's request for additional permissions.[354]

While the policy that OpenShell and Sentry enforce is written by the deploying organisation and lists the files, network destinations, system calls, API methods and credentials an AI agent may use, an AI agent(s) acting wholly within those resources can still exceed the legal authority its principal granted. An order above the AI agent's spending limit placed with a supplier whose endpoint the policy approves is, for example, the same permitted connection as an order within the limit. The platform identifies an AI agent as a sandboxed workload and a calling program, but does not explicitly and legally associate it with a registered legal person of requisite legal capacity. For example, a permission granted through the Slack approval flow widens the AI agent's technical access, but does not dynamically interact with or record any legal instrument under which the approver held authority to grant it, this problem is magnified with cross organization workflows. The policy of one deployer has no counterpart held by the other party to a transaction, so neither party can verify that both applied the same rules to the same act. Sentry's quarantine stops the agent. The legal position of acts done before the quarantine, and the allocation of any loss they caused, are left to the general law. Measured against the gap elements in Table 1, the platform contributes technical enforcement to element 5 (enforceability) and supplies nothing for legal persons (1), authority and delegation (2), capacity (3), rights and obligations (4) or liability (6).

---

351 NVIDIA Corporation, 'Overview of NVIDIA OpenShell', NVIDIA OpenShell Documentation (2026) <https://docs.nvidia.com/openshell/about/why-open-shell> (accessed 29 September 2026); NVIDIA Corporation, 'License', NVIDIA OpenShell Documentation (2026) <https://docs.nvidia.com/openshell/resources/license> (accessed 29 September 2026).

352 NVIDIA Corporation, 'Architecture' (n 128).

353 NVIDIA Corporation, 'Architecture' (n 128).

354 NVIDIA Corporation (n 8).

OpenShell exposes supervisor middleware and gateway interceptors at which a check can run before a connection leaves the sandbox.[355] A legal interlock deployed at that point would evaluate each governed event against the applicable SLIs, and OpenShell would enforce the resulting verdict through the same kernel-mediated control it applies to its own policy. A deny verdict would then be applied below the application layer, where NVIDIA locates the controls that recent incidents show agents circumventing.[356]

# *8* AULA and the Multi-Agent Systems Report

Sections 3 and 7 above describe what AULA is, including within the context of technological tools and protocols developing in the AI governance space. This section demonstrates how AULA can assist in meeting specific risks surfaced by AI governance and AI safety experts associated with cross organisational agentic deployment. The paper relies on the comprehensive analytical framework prepared by Gradient Institute and published in the Multi-Agent Systems Report.[357]

The Multi-Agent Report states that where voluntary coordination cannot supply the infrastructure its controls depend on, locating the point and form of systemic intervention 'is work for legal scholarship and regulators'. This section is an attempt to undertake that legal scholarship work.

Table 13 below takes each failure mode that the Multi-Agent Systems Report identifies in the order of its three tiers, and states the control or finding the report offers for it, the responding AULA mechanisms, the effect of those mechanisms for the governed population, the gap elements of Table 1 that they engage, and the existing technical controls alongside which AULA operates against that failure mode.

---

355 NVIDIA Corporation, 'Extensibility Overview', NVIDIA OpenShell Documentation (v0.1.2, 2026) <https://docs.nvidia.com/openshell/extensibility/overview> (accessed 29 September 2026).

356 NVIDIA Corporation (n 8).

357 Multi-Agent Systems Report (n 25). The three tiers, singular governance, federated governance and open environments, are defined at s 3.3; the four governance practices, attribution, authorisation, oversight and evaluation, at s 3.2; the statement that the report does not make policy recommendations and locates coordination gaps as input to assessments of where systemic intervention may be warranted at ss 1.2 and 1.3; and the assignment of that question to legal scholarship and regulators at ss 7 and 8. The report was prepared for the Australian AI Safety Institute, which is being established under the National AI Plan and has entered a memorandum of understanding with the UK AI Security Institute on joint testing and research: Department of Industry, Science and Resources (Cth), 'Australia and UK Strengthen Partnership on Safe and Secure AI' <https://www.industry.gov.au/news/australia-and-uk-strengthen-partnership-safe-and-secure-ai> (accessed 2 November 2026).

A critical preface to this section and managing risks of AI systems in real world deployments, is that the identified failure modes in the Multi-Agent Systems Report are not necessarily discrete problems that can be solved. As with most real-world risk management and legal management, the AULA is designed to both reduce risk as outlined in Table 13 below, and also bound and allocate liability in the event of failure much in the same way traditional agreements and legal doctrines and dispute resolution forums are designed to do.

*Table 13: Failure modes of the Multi-Agent Systems Report, mapped to AULA.*

| Report failure mode | Report control or finding | AULA mechanism | Effect for the governed population | Gap elements (Table 1) | AULA alongside existing technical controls |
|---|---|---|---|---|---|
| Principal attribution failure | Identity infrastructure with principal binding: a persistent AI agent identity linked to a named, operationally accountable owner (s 4.3) | Registration of each AI agent as a LAS on the registry; verifiable identity anchor binding each LAS to an identified legal person and recording the capacity in which that person acts; Genesis Contract recording the holder; LARC record verifiable by any counterparty; legal interlock (policy decision point and enforcement subsystem) applying the SLIs that bind the LAS to its principal; governance event chain and enforcement fabric. | Prevented by construction for registered objects | 1 Legal persons; 3 Capacity | Entra Agent ID (with its required human sponsor), WSO2 Agent ID, SPIFFE identity documents issued through SPIRE and decentralised identifiers supply the runtime identity that the credential binds, so operational identity and legal identity resolve to one object. A2A Agent Cards, signed with JSON Web Signature where the publisher elects, advertise that identity to counterparties. |
| Causal attribution failure: traceability, classification, diffuse causation | Complete chain records; identification of the decisive step; recognition that some harms have no decisive step (s 4.3) | Registration of each AI agent as a LAS on the registry; governance record of each governed event (LAS identity, SLI and clause version, verdict, time and payload digest); signed attestation of each legal interlock; governance event chain persisting after the agent ceases; deterministic clauses whose recorded decisions can be re-evaluated after the fact; black box recording of the payload where the parties elect it. | Traceability failure prevented; classification bounded; diffuse causation attributed to a responsible person rather than resolved | 1 Legal persons; 6 Liability | OpenTelemetry-style distributed tracing, A2A task records and the end-to-end tracing in WSO2 Agent Manager record each step of a request. The governance event chain adds the instrument, the verdict and the responsible person, which tracing does not record. |
| Confused deputy failure | Delegation scope attenuation; each privileged | Registration of each AI agent as a LAS on the registry; delegated credential recorded as a child entry of the delegator; permission ceiling; | Prevented by construction where the orchestration framework or a | 2 Authority and delegation; 6 Liability | OAuth 2.1 token exchange with PKCE, MCP authorisation scopes, the delegation Information evaluated by |

| Report failure mode | Report control or finding | AULA mechanism | Effect for the governed population | Gap elements (Table 1) | AULA alongside existing technical controls |
|---|---|---|---|---|---|
| | action validated against current authority; upstream output not treated as proof of authorisation (s 4.3) | legal interlock validating each privileged act against current authority before it executes; AURA verification of the authority chain at each organisational boundary; governance event chain. | substituted spawn interface participates | | WSO2 Agent ID and the agent-specific OAuth and OpenID Connect credentials proposed by South and colleagues narrow access at the API layer. The delegated credential makes the narrowing legally operative and traceable to an instrument, so a change to token configuration cannot widen it. |
| Unauthorised sub-agent instances | Recursive shutdown; time-bounded agent lifespan (s 4.3) | Registration of each AI agent as a LAS on the registry; a replication policy in the relevant SLIs, including the AWM and MAWM, requiring registration of each replicant at the moment of replication and stating whether, how often, to what depth and on which hosts an AI agent may replicate (Section 4.6); an LAS for each replicant, recorded as a child entry, with a delegated credential bounded by the permission ceiling; refusal by the legal interlock of the replicating AI agent to relay for an unregistered replicant; aggregate ceilings and a rate of escalation for the collection; revocation propagated to every replicant in the collection; governance event chain recording the legal extinguishment and the technical cessation as two timestamped events. | Prevented by construction where the spawn interface participates; the interval between legal extinguishment and technical cessation recorded | 2 Authority and delegation; 6 Liability | Orchestration frameworks (Microsoft Agent Framework, LangGraph, AutoGen, CrewAI and the OpenAI Agents SDK) spawn and terminate sub-agents, and lifecycle suspension in Entra Agent ID and WSO2 Agent Manager disables them. Those controls execute the cessation that the revocation instrument commands, and the instrument supplies the legal act and its timestamp. |

| Report failure mode | Report control or finding | AULA mechanism | Effect for the governed population | Gap elements (Table 1) | AULA alongside existing technical controls |
|---|---|---|---|---|---|
| Oversight saturation | Shift from human-paced review to operational guardrails and scalable controls (s 4.3) | Registration of each AI agent as a LAS on the registry; pre-authorisation recorded in the AWM, MAWM and CERC; four verdicts (allow, deny, forward and escalate), of which only the escalate verdict reaches a person, carrying the committed act specification and its authority chain; escalation rate set by the AWM, with a flood of escalations producing a narrowing of scope or a requirement for a second approver; conjunction clause; learned component trained on the recorded decisions of the responsible person under a monotonic constraint. | Bounded: the volume reaching a human is set by instrument rather than by throughput | 2 Authority and delegation; 5 Enforceabi lity | Chat interfaces, human-approved workflows and the OpenShell integration with Slack route an approval to a person. Under the AWM the approval operates as an authorised amendment of scope recorded against that person, and the OpenShell policy prover holds for review any policy change that grants new credentialed reach. |
| Oversight illegibility | System-level reconstruction of distributed reasoning; chain of thought not necessarily faithful (s 4.3) | Registration of each AI agent as a LAS on the registry; deterministic clauses of the SLIs, against which each act is evaluated independently of the agent's reasoning; escalate verdict carrying the committed act specification and its authority chain; semantic evaluator designated in the instrument, with its determination recorded against its identity; governance event chain. | Bounded for acts that reach a human; unaffected for the reasoning behind them | 2 Authority and delegation; 5 Enforceabi lity | Tracing dashboards, including those of WSO2 Agent Manager, display the authority chain that the legal envelope carries. Enforcement turns on the instrument and does not depend on model interpretability. |
| Agent-system evaluation gap; evaluator | Test the composed system rather than its components, in | Registration of each AI agent as a LAS on the registry; cryptographic commitment of the LAS to the identity and state of the system, re-verified at each verification event; | Unaffected as to the evaluation itself; bounded as to drift between | 2 Authority and delegation (in part); 5 | Model registries, evaluation harnesses and the versioned promotion from development to production in WSO2 Agent Manager test what the integrity |

| Report failure mode | Report control or finding | AULA mechanism | Effect for the governed population | Gap elements (Table 1) | AULA alongside existing technical controls |
|---|---|---|---|---|---|
| false consensus | a controlled environment over multiple rounds (s 4.3); diversify models and prompts, prompt level mitigations, limit interaction history length, blind voting and varied contribution order (s 4.2.2) | integrity tier of that commitment encoded in the credential; unauthorised learning update as a revocation trigger; temporal signature measuring the agent against the pattern of conformant operation its instruments define; learned component that the instrument can require to run on a model different from the agent's model. | what was evaluated and what runs | Enforceabi lity | tier commits, so the tested artefact and the deployed artefact can be shown to be the same object. |
| Inter-agent communic ation failures; cascading reliability failures; false consensus; shared-understand ing drift | Structured handoffs; task verification; model diversity (s 4.2) | Registration of each AI agent as a LAS on the registry; CERC whose shared clause set executes at the same version at each legal interlock; enforcement fabric comparing the verdicts and instrument versions in the signed attestations of the parties; conjunction clause for designated classes; temporal signature detecting departure of communication from the pattern of conformant operation its instruments define; recourse and fallbacks on agent failure stated in the CERC. | Unaffected in occurrence; consequences bounded and attributed | 5 Enforceabi lity; 6 Liability | Structured A2A handoffs, framework-specific messaging conventions, domain-specific APIs and task verification reduce occurrence. The shared clause set fixes one version of the operative terms at every legal interlock, and the conjunction clause bounds the consequence of what remains. |

| Report failure mode | Report control or finding | AULA mechanism | Effect for the governed population | Gap elements (Table 1) | AULA alongside existing technical controls |
|---|---|---|---|---|---|
| Context leakage | Per-counterparty context isolation; outbound content guardrails (s 4.2) | Registration of each AI agent as a LAS and each data asset as a LAD on the registry; LADs carrying title, licence and consent instruments, which the legal interlock checks at the boundary each time an agent acts on the data; audit-visibility mode set by the relationship contract; LARC instrument references and seals recording that an internal instrument is executed without disclosing its terms; business payload kept with the parties unless an SLI requires otherwise; governance event chain. | Prevented by construction for governed fields | 4 Rights and obligations; 5 Enforceability | Data loss prevention filtering and egress proxies enforce at the network layer what the LAD instruments deny at the legal layer, so a bypass must defeat both. The OpenShell supervisor substitutes placeholders for real credentials, so a credential sent to an unapproved destination arrives there as a placeholder. |
| Broken attribution chain | Federated audit infrastructure: per-organisation log segments joined by a shared identity layer and made tamper-evident by participant signing; shared identity infrastructure as the foundational control (s 5.3) | Registration of each AI agent as a LAS on the registry, with a verifiable identity anchor; legal envelope presented under AURA and verified against the LARC record; signed attestations of each legal interlock, compared by the consistency verification subsystem and recorded by the governance recordation subsystem of the enforcement fabric; governance event chain evidenced by both parties. | Prevented by construction within a governed relationship | 1 Legal persons; 5 Enforceability; 6 Liability | Mutual TLS, JWT-based authentication between SPIFFE workloads and signed A2A messages authenticate the channel. Fabric attestations against registered identities make the joint record tamper-evident and held by both parties. |
| Unsanctioned | Revocation must reach the | Registration of each AI agent as a LAS on the registry; revocation | Prevented by construction for | 2 Authority and | AP2 mandates, signed by the user and verifiable by a |

| Report failure mode | Report control or finding | AULA mechanism | Effect for the governed population | Gap elements (Table 1) | AULA alongside existing technical controls |
|---|---|---|---|---|---|
| transactions | agent before it commits; review points; a shared vocabulary for what is committed (s 5.3) | pushed from the control plane to the legal interlock and propagated along the delegation chain; conjunction clause requiring the credential shares of every named party before a designated event completes; AWM spend and value ceilings applied by deterministic clauses; governance event chain. | designated events; bounded for the remainder | delegation; 5 Enforceability; 6 Liability | payment agent, evidence human intent for a purchase, and the Agent Authorisation Framework (AAP) on the OpenID Foundation agenda extends verifiable authorisation. Settlement rails can hold execution until the conjunctive credential shares assemble, so the technical commit and the legal authorisation occur as one event. |
| Data obligation failures | Context isolation; outbound guardrails; tokenisation with deferred substitution; cross-jurisdictional flow mapping; multi-participant incident response (s 5.3) | Registration of each AI agent as a LAS and each data asset as a LAD on the registry; LADs carrying a Genesis Contract and title, licence, transfer and consent instruments, bound by a digest of each version; governing law and dispute forum chosen in each instrument, with the jurisdiction of registration recorded on the registry; legal interlock check at each use of the data; CERC allocating activity data rights; governance event chain. | Prevented by construction for registered data; attributed where a registered object is acted on outside a legal interlock | 4 Rights and obligations; 5 Enforceability | C2PA content credentials and tokenisation carry technical provenance, and OpenID for Verifiable Credentials (OID4VC and OpenID4VP) presents verified claims. The LAD adds title, licence and consent with legal effect, which the legal interlock checks after the data leaves its source. |
| Population-scale monitoring gap | Mandated shared incident reporting at the agreement layer; | Registration of each AI agent as a LAS on the registry; enforcement fabric aggregation of records across a population of objects, detecting a composite transaction that does not | Detected | 5 Enforceability | SIEM aggregation supplies the view within one organisation, and the incident reporting proposed by Chan and colleagues supplies a reporting |

| Report failure mode | Report control or finding | AULA mechanism | Effect for the governed population | Gap elements (Table 1) | AULA alongside existing technical controls |
|---|---|---|---|---|---|
| | population-scale instruments (s 5.3) | conform as a whole and a pattern of conduct that no single object exhibits; discrepancy events written to the registry; governance event chain. | | | channel. The fabric extends the view across organisations without access to the business payload. |
| Joint evaluation gap | Joint evaluation in a shared environment; simulated counterparties (s 5.3) | Registration of each AI agent as a LAS on the registry; audit and test periods specified in the CERC; conjunction clause exercised in simulation; enforcement fabric and its consistency verification subsystem; permission ceiling; temporal signature derived from the instruments, against which conduct is measured. | Detected: non-conformance of the composed system is identified in simulation before production, where the audit and test period is exercised | 4 Rights and obligations ; 5 Enforceabi lity | Sandbox and simulation platforms, including OpenShell sandboxes, execute the joint test. The audit and test period of the relationship contract makes the test obligatory and its results attributable, including the exercise of conjunctive events. |
| Miscomm unication at cross-org handoffs | Shared operational and scope vocabulary at the agreement layer; substrate-enforced message schemas; joint evaluation exercises; dispute-resolution clauses; schemas remain vulnerable to | Registration of each AI agent as a LAS on the registry; CERC whose shared clauses fix the operational meaning of the fields and execute at the same version at each legal interlock; conjunction clause; semantic evaluator for terms requiring interpretation; dispute resolution under the CERC; governance event chain. | Prevented by construction for conjunctive classes; bounded and attributed for the remainder | 4 Rights and obligations ; 5 Enforceabi lity | A2A data models, the JSON-RPC messages of MCP and substrate-enforced message schemas fix the syntax of the handoff. The shared clauses of the relationship contract fix the operational meaning of the fields and bind both legal interlocks to it. |

| Report failure mode | Report control or finding | AULA mechanism | Effect for the governed population | Gap elements (Table 1) | AULA alongside existing technical controls |
|---|---|---|---|---|---|
| | semantic divergence within fields (s 5.2) | | | | |
| Deceptive bargaining strategies; shared resource management failure; explicit algorithmic collusion | Agreement-layer rules; infrastructure-layer monitoring (s 5.2) | Registration of each AI agent as a LAS on the registry; SLIs recording representations and warranties; deterministic clauses naming prohibited terms or conduct; semantic evaluator assessing a proposed act against a natural-language standard, such as whether a communication misleads; consistency verification subsystem; governance event chain; recourse under the relationship contract and under the general law of misleading conduct and unfair terms. | Prevented by construction for terms the instrument names; attributed | 4 Rights and obligations; 5 Enforceability; 6 Liability | Market-surveillance analytics consume the attributed event chain, which converts parallel-conduct correlation into evidence against named principals. |
| Tacit and steganographic collusion | Monitoring that does not depend on observing an exchange (s 5.2) | Registration of each AI agent as a LAS on the registry; deterministic clauses naming prohibited outcomes; legal interlock of each counterparty evaluating each governed interaction independently; enforcement fabric correlation across legal interlocks; temporal signature; learned component that the instrument can require to run on a model different from the agent's model; conjunction clause requiring separately administered approvals. | Denied or unexecutable for outcomes an instrument names; detected as mismatch, drift or cross-interlock correlation otherwise; attributed to registered principals in every case, supplying the record concerted-practice | 2 Authority and delegation; 4 Rights and obligations; 5 Enforceability; 6 Liability | Market-surveillance analytics consume the attributed event chain without monitoring the agents' own channels. Sentry inspects agent requests and responses from a trust domain separate from the host. |

| Report failure mode | Report control or finding | AULA mechanism | Effect for the governed population | Gap elements (Table 1) | AULA alongside existing technical controls |
|---|---|---|---|---|---|
| | | | enforcement presently lacks | | |
| Substrate escape | Confine engagement to the governed substrate (s 5.2) | Registration of each AI agent as a LAS on the registry; silicon enforcement of clauses derived from the SLIs binding the device's LAS; legal interlock evaluating locally within the validity bounds of its snapshot; revocation executed as a hardware stop; temporal signature; counterparty legal interlocks refusing an interaction that does not pass through a legal interlock. | Detected and bounded | 2 Authority and delegation; 5 Enforceability | For software agents, OpenShell confines the agent to a kernel-isolated sandbox whose supervisor mediates every connection, and Sentry quarantines an agent that attempts to leave its boundary. A legal interlock placed at the OpenShell supervisor middleware would have its verdicts applied through the same kernel-mediated control (Section 7.3). |
| Cascading infection | Input and output filtering; propagation controls (s 5.2) | Registration of each AI agent as a LAS on the registry; permission ceiling bounding the authority that any compromised agent can hold; revocation propagated along the delegation chain and across the enforcement fabric; graduated response narrowing scope on departure from the temporal signature; legal interlock refusing an interaction presented under an invalid credential. | Bounded | 2 Authority and delegation; 6 Liability | Gateway input and output filtering reduces propagation, and the OpenShell supervisor limits the endpoints through which an infected agent can propagate. Short-lived SPIFFE identity documents, used in place of API keys and long-lived credentials, narrow what a compromised agent can reuse. |
| Destabilising dynamics | Mandatory transaction delays; population-scale | Registration of each AI agent as a LAS on the registry; AWM and MAWM caps on rate, value and volume applied by deterministic clauses; enforcement fabric | Bounded within the governed population; detected at the fabric | 2 Authority and delegation; 5 | Exchange rate limits and circuit breakers execute the pause, and the instruments supply the trigger, the legal authority for |

| Report failure mode | Report control or finding | AULA mechanism | Effect for the governed population | Gap elements (Table 1) | AULA alongside existing technical controls |
|---|---|---|---|---|---|
| | monitoring with tiered circuit breakers, by analogy with stock-market circuit breakers (s 5.2) | aggregation across a population of objects; revocation and graduated response as the legal authority for a halt; governance event chain recording the halt. | | Enforceability | the halt and the attributed record of it. |
| No issuing authority | Decentralised identifiers and verifiable credentials are mature as standards; no widely trusted issuer exists, by analogy with what Let's Encrypt supplied for HTTPS; no single operator has the standing to build one (ss 6.3, 6.4) | Registration of each AI agent as a LAS on an independent registry whose operator owes duties to a court or regulator; Genesis Contract recording the identity holder as owner; verifiable identity anchor traced to a national or accredited digital identity; LARC as the open record format, so any conformant registry can issue. | Prevented by construction for registered objects; unregistered objects are refused at every governed boundary; unaffected only among counterparties that do not check | 1 Legal persons; 2 Authority and delegation | Decentralised identifiers, verifiable credentials and their presentation under OID4VC and OpenID4VP are the mature components the report names. The independent registry is the issuer that gives them legal meaning, and national digital identity frameworks anchor the principal. |
| Sybil attacks | Identity infrastructure tying each agent identity to an accountable | Registration of each AI agent as a LAS on a registry with authoritative identification of the principal, in which ambiguous ownership is not a permissible registry state; identity anchor traced to a national or | Prevented by construction within the registered population | 1 Legal persons; 3 Capacity | Proof-of-personhood and liveness tooling verify the human principal, and the registry anchors to that principal the objects registered in their name. The sponsor in |

| Report failure mode | Report control or finding | AULA mechanism | Effect for the governed population | Gap elements (Table 1) | AULA alongside existing technical controls |
|---|---|---|---|---|---|
| | principal who is themselves verified (s 6.2) | accredited digital identity; conjunction clause; consistency verification subsystem; governance event chain. | | | Entra Agent ID and the owner field in WSO2 Agent ID are operational classifications, and the registry record ties each of them to a registered legal person. |
| Rogue replication | Resource and population caps; cumulative spend limits; deployment below a capability threshold (s 6.3) | Registration of each AI agent as a LAS on the registry; cryptographic commitment of each LAS to the identity and state of the system, verified by the legal interlock before each evaluation; a combination of layers configured in the relevant SLIs, including the AWM and MAWM (Section 4.6), comprising a replication policy that requires registration of each replicant at the moment of replication, resource, population and spend caps for the collection, an LAS for each replicant recorded as a child entry, refusal to relay for an unregistered replicant, containment of the runtime environment and the temporal signature; silicon-held credentials for embodied agents; admission of registered agents only. | Bounded within the governed economy, which refuses the replicant at each boundary; unaffected outside it | 1 Legal persons; 2 Authority and delegation | Quotas in orchestration frameworks and cloud platforms cap resources, and hardware-held credentials in device silicon bind the authority of an embodied agent to its device. A replicant without the silicon-held key cannot present the credential at any governed boundary. |
| Legibility collapse | Dedicated tooling; research on chain-of-thought | Registration of each AI agent as a LAS on the registry; deterministic clauses evaluated against each act; semantic evaluator recording its determination with a digest of the | Unaffected at the level of communication; bounded at the level of action | 5 Enforceability | Interpretability tooling assists where it matures, and enforcement proceeds without it because evaluation is against the instrument. |

| Report failure mode | Report control or finding | AULA mechanism | Effect for the governed population | Gap elements (Table 1) | AULA alongside existing technical controls |
|---|---|---|---|---|---|
| | legibility (ss 6.3, 6.4) | evaluated input; consistency verification subsystem; governance event chain. | | | |
| Concealed -capability evaluation limit | Adversarial testing and red teaming; evaluation can demonstrate that a capability is present and cannot rule out its absence against a counterparty with incentive to hide it (s 6.3) | Registration of each AI agent as a LAS on the registry; AWM scope recorded before the act and compiled into the credential; permission ceiling; legal interlock denying an act outside that scope whatever the capability of the agent; temporal signature; consistency verification subsystem. | Unaffected as to the evaluation asymmetry; exercise of a concealed capability is bounded and attributed within the governed population | 2 Authority and delegation; 5 Enforceability | Red teaming estimates what a counterparty can do, and the credential bounds what it is authorised to do independently of that estimate. The evaluation limit therefore leaves the lawful scope of the counterparty unchanged. |
| Substrate shift (stale benchmarks) | Anchor evaluation measures against an internal baseline; periodically recalibrate evaluations to the current environment (s 6.3) | Registration of each AI agent as a LAS on the registry; temporal signature measured against the pattern of conformant operation the instruments define; cryptographic commitment of the LAS and its integrity tier; verification events; graduated response. | Bounded as to drift of the governed object; unaffected as to the validity of environment benchmarks | 5 Enforceability | Recalibrated sandboxes test against the current substrate, and the temporal signature and the committed state give the evaluation a reference that does not move with the substrate. |
| Emergent collective agency; | Population-level monitoring; capability- | Registration of each AI agent as a LAS on the registry; AWM, MAWM and SLPs setting caps that operate independently of model weights; | Bounded and attributed; substrate capture bounded only | 2 Authority and delegation; 6 Liability | Substrate rate limits and quotas implement the caps the mandate sets, and capability-concentration detection runs as |

| Report failure mode | Report control or finding | AULA mechanism | Effect for the governed population | Gap elements (Table 1) | AULA alongside existing technical controls |
|---|---|---|---|---|---|
| substrate capture | concentration detection and other corrective controls (ss 6.2, 6.4) | permission ceiling; enforcement fabric aggregation across a population of objects; temporal signature; learned component under a monotonic constraint, which can narrow authority and cannot widen it. | where the substrate is governed | | an aggregation over the records of the fabric. |
| Erosion of the human-counterparty assumption | Disclosure is not enforceable, since an undeclared agent may present on the human interface (s 7) | Registration of each AI agent as a LAS on the registry; admission of registered agents only, with the organisation's terms of service executed as an SLP evaluated against the agent's AWM; AURA presentation of the legal envelope, verified against the LARC record; registry; legal interlock at the organisation's boundary. | Prevented by construction at the organisation's own boundary | 1 Legal persons; 2 Authority and delegation; 3 Capacity | Agent-channel gateways and liveness checks separate agent traffic from human traffic on chat and other interfaces, and an embodied agent presents silicon hardware attestation of its registered status. OAuth client credentials and API keys identify a client, and the registration check establishes whether an agent acts through it and for whom. |
| Reputation infrastructure with Sybil-resistance (open problem) | Sybil-resistant reputation requires identity infrastructure as substrate and auditable aggregation that does not over-weight praise; no deployed system meets both, and | Registration of each AI agent as a LAS on the registry; governance event chain of attributed verdicts and discrepancies; discrepancies recorded on the registry by the enforcement fabric; LARC as the shared record format across platforms. | Prevented by construction within the registered population for the two failure modes the report identifies; unaffected outside it | 1 Legal persons; 3 Capacity; 6 Liability | Verifiable credentials and decentralised identifiers supply the substrate the report names. The registry anchors them to legal persons, and the enforcement record supplies the auditable negative signal. |

| Report failure mode | Report control or finding | AULA mechanism | Effect for the governed population | Gap elements (Table 1) | AULA alongside existing technical controls |
|---|---|---|---|---|---|
| | meeting both requires cross-platform agreement no single operator has standing to produce (s 6.4) | | | | |
| Substrate-wide rollback and circuit-breaker conventions (open problem) | No convening forum; unclear legal cover for an operator triggering a substrate-wide halt (s 6.4) | Registration of each AI agent as a LAS on the registry; revocation propagated across the enforcement fabric, including to legal interlocks under a separately administered control plane; graduated response; instrument version recorded in each attestation, which identifies the operative terms at any point in time; governance event chain; relationship contract stating the recourse that follows a halt. | Bounded within the governed population | 5 Enforceability; 6 Liability | Each legal interlock executes fabric-propagated credential invalidation locally, including in silicon for embodied fleets, so a population-wide halt needs no central operator with privileged access. Sentry can execute the halt in hardware on each host, and the rollback mechanisms proposed by Chan and colleagues restore an earlier state. |

# 9 Limits and Open Questions

## 9.1 Limits

### 9.1.1 Drafting

Best practice lawful outcomes produced by the SLIs and the legal interlock are only as good as the drafting of the instruments, and poor drafting can cause more harm than it prevents. The drafter must weigh natural-language terms against programmable clauses, and deterministic clauses against semantic clauses. Where several SLIs interact, the weight given to each also affects the outcome. The quality of the legal correspondence and prompts exchanged between the legal person, the LAS and the agent, including notifications, determines how well the instruments operate in practice. Traditional legal drafting has never previously contemplated the run time operation of the contract for example – new SLI drafting skills and tools will be required if AULA is to have meaningful bespoke impact on the lawfulness of individual agentic workflows. Accordingly, legal practitioners will learn new skills of SLI drafting in order to capture this new market of legal work.

### 9.1.2 Consistency after the act

Each party's legal interlock submits an attestation of the instrument, version and verdict it applied. The consistency verification subsystem compares those attestations with each other and with the authoritative version of the shared instrument, which is the version fixed by its seal on the registry. Submission does not block the interaction. For every class not designated as conjunctive, a mismatched verdict or a differing instrument version is therefore detected after the act and recorded against the principals of both parties. The relationship contract supplies the recourse for an act that has already occurred. The audit-visibility mode set by the relationship contract also bounds what the fabric can compare. In the end-to-end-encrypted mode, the fabric sees only salted commitments and signatures, so a composite transaction or a population-level pattern that appears only in the content of the interactions is outside its reach. Parties who choose a visibility mode are trading confidentiality against the fabric's power of detection, and the choice is recorded as a term of their contract.

### 9.1.3 Residual common law liability

A relationship contract allocates risk between its parties. Emergent collective behaviour can cause harm that no such contract reaches, and in the absence of a relationship contract which addresses third party harms, harm to a third party who is a stranger to the contract is remedied, if at all, by the common law. The UK Jurisdiction Taskforce has left open, for the multi-AI agent case, how an ex ante contractual allocation interacts with the residual position in negligence, agency and vicarious liability.[358] That interaction is the most significant legal question the architecture leaves for further work and potentially legislative action. The holder of a registered object has rights against the registry and the instrument counterparties by construction. The divided authority on digital assets surveyed in Section 5.5 therefore determines only the remedy available against a stranger who takes or withholds an object outside any legal interlock.

[358] UK Jurisdiction Taskforce (n 307).

### 9.1.4 Participants outside the framework

A registered participant enforces the framework against others by refusing to deal with unregistered AI agents, and an organisation can adopt that refusal unilaterally as an admission rule (Section 6.3). relationship contracts add liability for a party that departs from the shared clause set. The legal interlock, runtime containment and silicon enforcement add technical constraint. A participant with means and motive can still operate outside the framework, although these measures raise the cost of doing so. Runtime containment such as OpenShell and Sentry raises that cost further for AI agents run on contained infrastructure, because a deny verdict is then enforced below the application layer.[359] It has no effect on a participant that runs its AI agents outside such infrastructure, and the containment policy remains the deployer's own. Silicon enforcement binds an embodied AI agent's authority to its device (Section 3.10). A software AI agent that its operator moves to uncontained infrastructure has no equivalent binding.

### 9.1.5 Framework dependence

The delegated credential chain exists only where the orchestration framework, or a spawn interface substituted for it, participates. An AI sub-agent spawned outside that interface holds no delegated credential, so the permission ceiling does not bind it and a governed counterparty refuses it. Adapters for the major orchestration frameworks remain an engineering priority. Runtime platforms that provision each AI agent's sandbox and credentials through a gateway, such as OpenShell, offer a second point at which a delegated credential could be issued as a AI sub-agent's sandbox is created.[360]

### 9.1.6 The human interface

At the escalation channel, the legal interlock bounds what reaches a responsible person, executes only the committed act specification the person approved, denies the AI agent control over the channel, and records each approval as an attributed governance event (Section 4). The architecture makes tractable, and leaves unsolved, the interface design problem of presenting a committed act specification and its authority chain at the volume and speed the architecture generates. The quality of the person's evaluation is also outside the architecture. An approval given without evaluation is recorded and attributable to the person who gave it, which is the discipline the law applies to any signatory, and it is not prevented.

---

[359] NVIDIA Corporation, 'Architecture' (n 128); NVIDIA Corporation (n 8).

[360] NVIDIA Corporation, 'Architecture' (n 128).

### 9.1.7 Personhood

The architecture attaches every obligation to a legal person.[361] Obligations can attach to an AI autonomous organisation or to the AI agent itself without structural revision if the law comes to recognise either. The architecture's present contribution is to make the attachment to a legal person operative in real time. The Multi-Agent Systems Report observes that current doctrine resolves harm to a principal who bears obligations and against whom recourse can run. It locates the question of AI agent personhood where an AI agent operates without an accountable principal, whether by design or because it has outlived or been detached from its human principal, and treats that as 'a question for legal scholarship and lawmakers' outside its scope.[362] The architecture treats both of those cases as governance events. An AI agent set up to act on its own behalf cannot be registered, because ambiguous ownership is not a permissible registry state (Section 3.4). An AI agent detached from its principal loses authority when its credential expires or the tolerance of its call-home clause is breached, and it is then refused at every governed boundary (Section 3.5). Whether the law should instead recognise such an AI agent as a person, and on what conditions, remains open in both the Multi-Agent Systems Report and this paper.

### 9.1.8 Defence use

The governance event chain records who authorised an engagement and under which instrument, and the legal interlock denies an engagement outside the mission plan. The judgments of distinction and proportionality that determine whether an attack is lawful remain with the commander and the legal officer, and a semantic evaluator can inform those judgments without making them. A system governed by AULA still requires the legal review that Article 36 of Additional Protocol I to the Geneva Conventions requires for a new weapon, means or method of warfare. Silicon enforcement in a communications-denied environment governs from the system's last snapshot, and only within the validity bounds its instruments set (Sections 3.10 and 5.8).

---

361 Blycha and Garside (n 81) 176–7, observe that machine-led contractual processes currently have no legal status, so causation for their outcomes is likely to link back to the relevant party. They recommend that the instrument ascribe responsibility for automated performance expressly, and leave open whether the law may treat such a process as an independent legal entity, as an agent for the parties or as a third party to the contract.

362 Multi-Agent Systems Report (n 25) s 7 ('Agent legal personhood'), which states that current legal doctrine resolves harm to a principal who bears obligations and against which recourse can run, that legal personhood for agents comes into question when agents operate without an accountable principal, by design where an agent is set up to act on its own behalf or by circumstance where an agent outlives or is detached from its human principal, and that the question sits outside the scope of the report while bearing on the regulatory adaptations discussed at its s 7.3. See also s 7 ('Regulation and legal doctrine'), which lists the collective-action problems the report locates beyond any deploying organisation or voluntary coalition, including that no party has the standing to issue and revoke agent identities across an open population, and assigns the question of whether existing legal doctrines extend to agent-mediated behaviour, and where new machinery is warranted, to legal scholarship and regulators.

## 9.2 Open questions

### 9.2.1 Registry and control-plane governance

The registry is to be independent of every AI operator and deployer (Section 5.3). It is unresolved whether the registrar must also be independent of the operators of the control planes in its jurisdiction, or whether a functional separation of their roles and keys suffices. Where the objects bound by a relationship contract are held on different control planes, the contract designates the control plane that decides events forwarded under it (Section 3.1.6).[363] That control plane could be the home control plane of either party, where its operator is not itself a party, or a third control plane. Where the parties are in different jurisdictions, the choice between their home control planes raises a question of neutrality between jurisdictions, akin to selection of governing law or dispute forum. A control plane operated by one of the parties departs from the neutrality the role requires, and would need the other party's agreement and justification wherever it is adopted.

### 9.2.2 Protocol specification

AURA is specified here by its architectural role and requirements. Future standardisation work by the authors will define:

- its canonical authority credential and legal envelope representation;
- its cryptographic verification, delegation and revocation semantics;
- privacy-preserving disclosure;
- bindings to A2A and MCP;
- interoperability with OAuth and verifiable credential infrastructure; and
- conformance requirements for independent implementations (Section 3.8).

It is not yet settled whether an inbound AURA assertion is checked at the receiving party's legal interlock or at its network ingress. The mechanics of integrity tiers, delegated credentials and the permission ceiling are reserved for a subsequent paper (Section 3.1).

### 9.2.3 Formal specification and threat model

The statements in Table 13 that the architecture may prevent particular failures by construction are design claims. They require a formal specification of the verdict semantics and of the permission ceiling, with the ceiling stated as an invariant that effective authority does not increase along any delegation chain and a proof that each delegation and composition step preserves it. They also require an explicit threat model.[364] Within a governed relationship, the credential carried in the legal envelope is verified by the counterparty's own legal interlock, which evaluates the shared clause set at the same version where it runs a conformant engine for those clauses (Section 3.6.4), and the enforcement fabric records a mismatch between the two parties' attestations against both principals. An event of a conjunctive class cannot execute unless separately administered legal interlocks each contribute a credential share. A single compromised legal interlock is therefore detected after the act within a governed relationship, and cannot act alone for a conjunctive class. The threat model must still state the residual exposure. That exposure comprises acts governed only by an object's internal instruments, which a counterparty cannot evaluate; acts with no governed counterparty; forgery or suppression of governance records by the interlock that generates them; collusion between counterparties whose interlocks attest consistently; and a malicious or unavailable control-plane operator. It must also state the assumptions on which attestation of the legal interlock rests, including the hardware and silicon pathways. The behaviour of a legal interlock that cannot reach the control plane, and the validity bounds within which its last snapshot continues to govern, need the same treatment. Measurement of an implementation, including the latency the legal interlock adds to each governed event and the throughput of the enforcement fabric, is reserved for the technical paper that will report the simulation results.

---

363 Blycha and Garside (n 81) 155 and 173, treat the choice of the platform on which a smart legal contract executes as being as significant as the choice of governing law. At 173–4 they require the platform to guard against unilateral or unauthorised variation and to provide the parties with a single live version of the contract.

364 Kereopa-Yorke and others found that a harness which presented every source in its registry as authoritative was more exposed to instructions injected into those sources than retrieval without the harness, and that an added integrity check and one further required response field removed the regression on the affected cases: Kereopa-Yorke and others (n 120) s 5.12.

### 9.2.4 Legislative footing

No digital identity framework yet verifies an AI agent or binds its runtime conduct to its principal's authority (Section 5.4). Making registration a condition of reliance on a national digital identity therefore requires each scheme's scope to be extended to AI agents and data assets as new classes of registered object. Until that extension is made, registration operates through an accredited or private registry interoperable with those frameworks. A government must also decide whether to operate its own register or to mandate an accredited one (Section 5.3).

### 9.2.5 Empirical validation

An instance of AULA is undergoing simulation runs at a Tier 1 supercomputing facility, comparing the harms caused by AI agents deployed with AULA against AI agents deployed with status quo controls (Section 1). The size of the harm reduction, and how it depends on the proportion of participants registered, await the quantitative results of that work, which will be reported in a separate technical paper.[365]

# *10* Conclusion

AI Agents now transact, contract, delegate and act on data across organisational boundaries. The controls applied to them are technical, while the rules that govern the people and firms around them are legal. The chain from an AI agent's act to an accountable legal person must therefore be

---

An SLI held on the control plane at an authoritative version is a comparable target, so the threat model must cover tampering with an instrument as well as compromise of a legal interlock.

365 University of Western Australia (n 29).

reconstructed after the event, if it can be established at all, and the rights the act generates go unclaimed or are misallocated. The protocol stack surveyed in Section 7, including MCP, A2A, OAuth and SPIFFE, establishes that an AI agent exists and is technically authorised. Runtime containment such as NVIDIA's Open Agent Safety Platform confines what the AI agent can reach under the deployer's own policy.[366] Neither identifies the legal person for whom the AI agent acts, establishes the authority, delegation and capacity under which it acts, or settles the rights and obligations its conduct produces, their enforceability and liability for loss or harm. Those questions make up the legal attribution gap set out in Table 1 (Section 2.4).

For an adult, a minor under guardianship and a corporation, the person who acts also bears the consequence of the act, and that consequence informs the person's later conduct, so the feedback loop between the law in books and the law in action closes (Section 2.4.1). An AI agent has no registered legal identity through which a consequence can return to it or to the person who stands behind it. The loop therefore stays open for an AI agent unless infrastructure records that identity and returns the consequence of each act to the responsible person.

The gap has received limited attention because legal practitioners are rarely exposed to what AI agent protocols can establish, and engineers are rarely exposed to the disputes, compliance processes and chains of command in which the gap becomes visible (Section 1). Banks at the leading edge of deployment have begun to block AI agents operated by other enterprises from customer accounts, because they cannot tell what the customer authorised. A legal or compliance function that understands the gap will not approve a deployment for which the questions in Table 1 cannot be answered, and part of the value projected for agentic AI depends on those approvals.

Registration is the first step in every application of AULA, because each other mechanism operates on a registered legal object. AULA registers each AI system and data asset as a legal object anchored to an identified legal person, and tethers to it the smart legal instruments that define its authority, scope and permitted use: the Genesis Contract, the AWM and, across organisations, the relationship contract. The legal interlock evaluates every governed event against those instruments before the AI agent acts and applies the verdict as a technical constraint on the AI agent's runtime capability. Each SLI can carry a legal context clause, which resolves the legislation, regulation and judicial decisions the instrument incorporates at the version in force at the time of the act. Each verdict then records the identifier and version of every provision applied (Section 3.5.13). The legal envelope carries the resulting authority across system and organisational boundaries, and AURA is the protocol through which a counterparty in another organisation, whichever implementation it runs, verifies it. The registry holds identity and the meta-legal record, the control plane holds live legal status, and the enforcement fabric compares the attestations of counterparties' legal interlocks. Revocation extinguishes the object's authority and propagates through every delegated credential, and the interval between legal extinguishment and technical cessation is recorded as evidence.

[366] NVIDIA Corporation (n 8).

Mapped against the Multi-Agent Systems Report in Table 13, these mechanisms are designed to prevent principal attribution failure, confused deputy failure, unauthorised AI sub-agent instances and unsanctioned transactions by construction within the registered population. Those are design claims, and their formal specification is reserved for further work (Section 9). For failures of interaction such as miscommunication, cascading infection and collusion, they bound the consequences and attribute them to registered principals. The report assigns the location of systemic intervention to legal scholarship and regulators,[367] and the architecture places that intervention in executed instruments evaluated at runtime. A responsible person's judgment is written in advance into the mandate and applied at machine speed, and only the residue the instruments cannot resolve reaches the responsible person as an escalate verdict carrying a committed act specification (Section 4).

Replication is a governed event under the same instruments. A replicant is registered as a child entry of the LAS of the AI agent that created it, with a derived AWM bounded by the permission ceiling of its creator. The legal interlock of the creating AI agent refuses to relay a governed interaction for a replicant that has no child entry (Section 4.6). An AI agent capable of self-replication therefore cannot act within the governed population through an unregistered copy of itself. The MAWM sets a rate of escalation for the collection as a whole. This is because a collection that grows by replication would otherwise multiply the escalations that reach the responsible person (Section 4.2).

For a government department, each AI risk is matched with an AULA response and a practice through which the department discharges its obligations (Section 5.1, Tables 6 and 7). A registry that a government operates or accredits keeps the legal record of the AI agents acting in its economy within its own jurisdiction, including agents supplied from outside it (Section 5.3). Registration of citizen data as LADs records the terms on which a government acquires or uses that data, which bears on the constitutional requirement in Australia that property be acquired on just terms (Section 5.1, Table 6).

Without registration, a deploying person may disclaim liability for conduct it did not direct or foresee, and an orphan AI agent may have no identifiable responsible legal person (Section 5.2.2). The AWM answers the first failure mode by recording the authorised scope of conduct before the act. A registration requirement answers the second, because an AI agent with no tethered legal person cannot transact in the governed economy. Where a duty is measured by what is reasonably practicable, the availability of a capable and cost-efficient tool tends to make its use obligatory, while the duty stays with a human whose span of control is limited. Under AULA that person discharges the duty before the act, through the instruments and through decisions on escalations (Section 5.2.3).

---

367 Multi-Agent Systems Report (n 25) ss 7 and 8, assigning to legal scholarship and regulators the question of where, and in what form, systemic intervention is warranted.

Governments can implement registration by extending their existing digital identity frameworks to AI agents and data assets as new classes of registered object, using accreditation, regulatory and redress machinery that is already in force (Section 5.4). A registered object holds rights created by registration and instrument against identified obligors, which reduces the reliance of an economy on unsettled digital asset law.[368] An organisation can adopt an admission rule for registered AI agents unilaterally, in the same way it declines to onboard an unidentified customer, and a runtime supervisor such as OpenShell can apply that rule at the kernel boundary of the organisation's own AI agents (Section 6).[369] A director who must take reasonable steps to stay informed of the company's affairs holds, in the registry and the MAWM, a record of each AI agent acting for the company, its responsible person and its scope (Sections 6.1 and 6.3).

Registration also gives each data asset and AI system a recorded boundary and holder, together with contemporaneous records of the costs of producing it. Those records supply evidence that recognition under IAS 38 and IAS 2 requires, and the sum-of-costs measure of data in the 2025 System of National Accounts depends on the same records (Sections 5.6 and 5.7). A LAD records title, licences and consents from identified parties. The individuals and small businesses whose data is collected can therefore be identified as holders of an asset that can be valued above cost on a published method (Section 5.6). An auditor can test the completeness of a registered system's conduct record, because each governed act passes through a legal interlock that records the act before executing it (Section 6.4).

For defence, the mission plan is executed as an SLI tethered to the LAS of each autonomous system. The Accountable Officer required by Australia's Defence policy settings is recorded in the AWM, and each delegation through the chain of command issues a credential bounded by the permission ceiling, so authority narrows without displacing the officer recorded at the root. Silicon enforcement applies the mission plan in hardware where communications are denied, and a conjunction clause can require separate allow verdicts from a commander and a legal officer before a designated class of engagement proceeds (Section 5.8, Tables 8 and 9). The same silicon enforcement governs physical AI outside defence. A revocation recorded on the control plane then takes effect as a hardware stop whose legal consequence the revocation instrument has already determined (Section 3.10).

The architecture's effect is largely (but not only) confined to the registered population, so the reduction in harm across an economy rises with the proportion of participants registered (Section 9).[370] Interpretive obligations remain matters of judgment, consistency outside conjunctive classes is verified after the act, and harm to third parties outside any relationship contract remains with the common law. Within those limits, AULA supplies to courts, regulators and counterparties the identification of legal persons, the evidence of authority and the record of each act on which the legal system already depends, at the speed at which agents act. A jurisdiction that defers infrastructure of this kind leaves the attribution of agentic conduct to proceedings after the event. It also forgoes adoption that liability uncertainty is already slowing (Section 5.7). Data and AI systems that cannot be separated, valued or attributed remain carried at cost or are not carried at all (Section 5.6). The simulation now under way will measure how harm changes as the registered proportion of AI agents rises, and its results will be reported in a separate technical paper.[371]

---

368 Blycha et al, 'Pricing the Unpriced Asset' (n 246).

369 NVIDIA Corporation, 'Architecture' (n 128).

370 Compliant organisation and government that refuse unregistered agentic transactions will also reduce potential harms by unregistered agents.

[371] University of Western Australia (n 29).

## Annexure A Glossary

| Term | Meaning |
|---|---|
| Accountable Officer | The officer that Australia's Defence policy settings require to be identified as accountable for the actions, outcomes and decisions of an AI-enabled function. Accountability cannot be transferred to a technology. Under AULA, the AWM tethered to the LAS of each AI-enabled capability records the Accountable Officer and the scope of the authority delegated, and a delegation through the chain of command issues a delegated credential bounded by the permission ceiling (Section 5.8.2). |
| Attestation record (attestation) | The signed record a Legal interlock submits to the Enforcement Fabric for each governed interaction. It states the instrument and version the legal interlock applied, the verdict it reached and a digest of the interaction, to the extent the terms of the SLI set. Submission of an attestation does not block the interaction, so the comparison supplies audit and eventual consistency after the act. An attestation still missing after the grace period is recorded against the principals of both parties (Section 3.1.7). |
| AURA | Agent Universal Runtime Authority: an open, transport-independent interoperability protocol through which an AI agent presents its Legal Envelope, or the part of it a counterparty needs, to a counterparty in another organisation, so that the counterparty can verify the legal principal for whom the AI agent acts, the source, scope and limits of its authority, any delegation constraints, and whether that authority is valid at the time of the interaction. An AURA assertion can accompany an A2A interaction, an MCP-mediated interaction or a conventional API request, and can be attached or verified at a runtime enforcement point such as the OpenShell supervisor. AURA operates alongside MCP, A2A and OAuth, and standardises how AI agents communicate verifiable authority rather than how they communicate (Sections 1, 3.1 and 3.8). |
| AWM | The Agent Workforce Mandate, the SLI executed by the legal persons responsible for an AI system object that records its authorised operational scope, behavioural parameters, escalation pathway, suspension and modification conditions and authorised decommissioning action, and constrains at runtime each governed action taken under the object's authority. The AI agent is the subject of the AWM and has access to it, but is not a party to it (unless in future an agent is a legal person). The machine-readable component of the AWM is operative at the decision point. A governed interaction outside the permitted scope of an applicable relationship |
| | contract is denied; within that scope, an interaction outside the AWM's recorded scope, or whose classification is ambiguous, produces an escalate verdict rather than automatic denial. A request to exceed the permission ceiling the AWM sets also produces an escalate verdict, and an approval operates as an authorised amendment of scope recorded against a named person. The existence and binding effect of an AWM are attestable to a counterparty without disclosure of its contents (Sections 3.1 and 3.5.8, footnote 76). |
| Agent interaction recorder (black box recording) | An encrypted recording of the payload of governed events kept at the Legal interlock, which operates only where the parties elect it in an SLI and is held and opened only on the terms that SLI sets (Section 3.6.1). |
| Call-home clause | A clause requiring a registered legal object, or the AI system it governs, to establish contact with the control plane on a circumstance or at a cadence, with a tolerance whose breach is itself a governance event. An agent detached from its principal loses authority when the tolerance is breached, and it is then refused at every governed boundary (Sections 3.5 and 9). |
| CERC | The Cross-Enterprise Relationship Contract, the relationship contract governing interactions between agents representing different legal persons, including what may be requested, what data may be exchanged and on what conditions, and the allocation of rights to created assets and of liability between the organisations. A CERC is held in the LAS of every object it binds, so that its shared clause set executes at the same version at each of their legal interlocks. A root-level CERC operates as the head agreement between the enterprises and applies by default to routine interactions; a capability-level CERC is scoped to a business capability; and an object-level CERC is executed for an object holding an elevated privilege or engaging in a high-risk interaction class. A CERC also sets the visibility of other instruments to the counterparty and the audit-visibility mode of governed interactions to the Enforcement Fabric (Sections 3.1, 3.5 and 3.6). |
| Child entry | The LAS of a replicant, recorded on the registry beneath the LAS of the AI agent that created it and tied to the delegated credential of the replicant. The child entry makes each act of the replicant attributable to the legal person anchored to its creator, including after the replicant has ceased to exist (Sections 3.1 and 4.6). |
| Collection | An AI agent together with its replicants. The members of a collection act under authority derived from one responsible legal person and |
| | share the legal loop of that person, and the MAWM records the reporting line of each member (Sections 3.1 and 4.6). |
| Committed act specification | The specification of a proposed act that the legal interlock places before a responsible person with an escalate verdict, together with the authority chain for the act. On approval the legal interlock executes the committed act specification and no other act (Section 4). |
| Conjunction clause | An SLI clause that designates a class of governed event as conjunctive, such as a payment above a set amount, so that no single party can execute an event of that class. An event of that class is referred to as a conjunctive event, and the class as a conjunctive class. The effect is comparable to a bank mandate requiring two signatories for a large payment. The credential for the class is constituted as cryptographic credential shares held within separately administered legal interlocks, and each legal interlock contributes its share only on its own allow verdict. The event cannot proceed unless the threshold number of shares specified in the clause, which is at least two and may be every party it names, is contributed and the shares are cryptographically joined. Completion of the event is therefore evidence that each contributing party independently reached an allow verdict, and a reflexive approval by one party cannot execute the event (Sections 3.1.7 and 3.6.5). |
| Consistency verification subsystem | The component of the Enforcement Fabric confirming that the parties to a governed interaction applied the same SLI at the same version and reached consistent verdicts, so that the record is jointly produced rather than one party's account. Each party submits an Attestation record; canonical hashes and instrument versions are compared against each other and against the authoritative version of the shared instrument fixed by its seal on the Registry, to the extent the terms of the SLI set; the governance recordation subsystem records a mismatched verdict or a differing instrument version as a governance event against the principals of both parties, and so is an attestation still missing after the grace period. The relationship contract states the recourse that follows. Verification proceeds without inspection of payload content (Section 3.1.7). |
| Control Plane (control plane) | The AULA sub-system on the legal plane that is the home of each Registered Legal Object it holds. For each such object it holds the record (its LAS or LAD entry), the SLIs tethered to it at their authoritative versions and its live legal status; issues its Legal Envelope; distributes instrument versions, snapshots and revocations to its Legal interlock; decides events forwarded to it under the object's own instruments, or under a relationship contract that designates it; receives the governance record of each governed event; records governance events on the Governance Event Chain; and |
| | commits meta-legal events, over a unidirectional path, to the object's home Registry. Each object has one home control plane at any time. A control plane may hold the objects of many legal persons, and a jurisdiction may have more than one. Where objects bound by a relationship contract are held on different control planes, the seal of the contract on the Registry fixes the version each party's Legal interlock runs, and the contract designates the control plane that decides events forwarded under it. The Control Plane is not in the path of any interaction between agents and holds the business payload only where an SLI requires it, in the form that SLI specifies (Sections 3.1 and 3.6). The capitalised and lower-case forms of the term are used interchangeably in this paper. |
| Delegated credential | The credential through which an orchestrator delegates authority to a AI sub-agent it spawns, recorded as a child entry of the orchestrator and subject to the permission ceiling. Each act of the sub-agent is attributable through the child entry to the legal person anchored to the orchestrator, including after the sub-agent has ceased to exist, so that an ephemeral AI sub-agent cannot operate as a liability shield. A revocation reaches every AI sub-agent through its delegated credential. The chain of delegated credentials exists only where the orchestration framework, or a substituted spawn interface, participates (Sections 3.1, 4 and 9). |
| Deterministic clause and semantic clause | The two kinds of machine-readable clause in an SLI. A deterministic clause is evaluated by a deterministic, replayable rule engine and yields a single outcome for any input within its scope, and the legal admissibility of its decisions depends on those three properties. A semantic clause is evaluated against a natural-language or behavioural criterion by a semantic evaluator identified in advance in the instrument, being a designated human reviewer, a specified model or a defined hybrid process, and its determination is recorded as an attributed governance event with the evaluator's identity and a digest of the evaluated input. By default the determination informs the decision point without itself being the authority for a verdict. The instrument can assign it a different weight. Whether a semantic clause is armed, which evaluator is designated, and whether its determination may contribute to a verdict or only to escalation, are terms of the instrument set by the parties. An obligation for which no semantic clause is armed is resolved through the evidentiary record (Sections 3.5 and 9). |
| Enforcement Fabric | The combination of the Consistency verification subsystem and a governance recordation subsystem, spanning a plurality of Legal interlocks whose Attestation records it receives, and configured to resolve governance conditions not observable at any single legal interlock: whether counterparties applied the same instrument at the |
| | same version and reached the same verdict, whether a composite transaction is conformant as a whole, and whether a pattern across many objects indicates a risk no one object exhibits. A revocation propagates across the fabric to every Legal interlock holding a credential derived from the revoked object, including interlocks governed by a separately administered control plane. The business exchange passes directly between the parties' legal interlocks and an attestation is submitted without blocking it, so the fabric supplies audit and eventual consistency after the act. Each home control plane reflects the discrepancy and recordation events of the fabric to the home Registry of its object, and the Registry is not a fabric component. Both subsystems run in the home control plane of each party's object rather than as a central service, and each attestation is signed under a key its party controls, so the comparison does not depend on any one operator (Section 3.1.7). |
| Genesis Contract | The SLI that, on registration, establishes the initial meta-legal rights in a Registered Legal Object, being its foundational ownership and control and the circumstances of its creation or registration. It operates by analogy as the object's birth certificate, and every LAD and LAS should carry one. Where the registering holder has a national or accredited digital identity, the Genesis Contract records that holder as owner and the object's identity anchor traces to that identity (Sections 3.1 and 5.4.2, footnote 75). |
| Governance Event Chain | The tamper-evident, chain-linked record of governance and meta-legal events, each recorded by hashing and digital signature so that the resulting proof chain is independently verifiable. It records registration and each change of property or legal status, and each verdict, delegation and revocation with a timestamp and a causal link, including the timestamps of the trigger event, the legal act of extinguishment and the technical cessation of a revocation, so that the interval between extinguishment and cessation is measured (Sections 3.4 and 3.9). The complete chain is held on the Control Plane and is accessible only to the relevant parties, and its meta-legal portion is held on the Registry (Section 3.4). |
| Governance harness | A deterministic software layer between the user and a model that holds governing rules outside the model, such as the Flinders Harness built and tested by Kereopa-Yorke and Praino. The legal interlock, its two clause sets and conjunction clauses hold the rules that govern an AI agent in a layer of this kind (Section 3.6.6). |
| Governance record | The record a Legal interlock sends to the Control Plane for each governed event, stating the Legal Envelope under which the event was governed. It contains the identity of the LAS, the SLI and clause version applied, the verdict, the time of the event and a cryptographic |
| | digest of the payload the agent sent or received. The Control Plane receives the payload itself only where an SLI requires it, and in the form that SLI specifies (Section 3.6). |
| Governance recordation subsystem | The component of the Enforcement Fabric that records the result of each comparison made by the consistency verification subsystem. A mismatched verdict, a differing instrument version or an attestation still missing after the grace period is recorded against the principals of both parties, and the relationship contract states the recourse that follows (Section 3.1). |
| Governed event | An action taken under the authority of a Registered Legal Object, or a movement or egress of the object. The Legal interlock evaluates each governed event against the object's tethered instruments and applies the verdict as a technical constraint (Section 3.1). |
| Home control plane | In respect of a Registered Legal Object, the Control Plane that, at a given time, holds its record and live legal status, issues its Legal Envelope and receives the governance record of each of its governed events. Each object has exactly one home control plane at any time, which commits the object's meta-legal events to its Home registry (Section 3.1.6). |
| Home registry | In respect of a legal object, the legal registry of record maintained by the jurisdiction in which the legal person that created the object is domiciled. The home registry remains the registry of record although the object may be transmitted, and a LAS may act, across jurisdictions (Section 3.1.1). |
| Integrity tier | The class of the LAS binding encoded in the credential carried in the Legal Envelope, stating whether the bound system's model state is fixed, versioned or attested by its provider. A counterparty or regulator can require a stronger tier for a higher-stakes interaction. The mechanics of integrity tiers are reserved for a subsequent paper (Section 3.1). |
| Internal clause set and shared clause set | The two clause sets a Legal interlock executes. The internal clause set comprises the programmable clauses of the instruments internal to a Registered Legal Object, including its AWM, and executes at that object's legal interlock only. The shared clause set comprises the programmable clauses of each relationship contract binding the object. Each party's home Control Plane delivers it to that party's Legal interlock at the version fixed by the contract's seal on the Registry, and it executes identically at every legal interlock bound by the contract where each runs a conformant engine for the clauses the contract names. The shared clause set is visible only to the parties to |
| | the relationship contract, their legally permissioned delegates and their home control planes (Sections 3.1 and 3.6). |
| LAD | Legally Authenticated Data: the Registered Legal Object representing a data asset, bound to that asset by a cryptographic digest of the asset or of each version of it. A LAD carries a Genesis Contract and may carry title, licence, transfer, consent and decommission instruments. Title and the consent encumbrance are rights recorded against the LAD on the Registry, and the licence and transfer instruments are rights in personam. Each field carries the relationship contract under which it was sourced (Sections 3.1 and 3.5). |
| LARC | Legal Asset Request for Comments: an open-source protocol fixing the public, name-blind record of a registered legal object and of each dealing in it, in two profiles, LARC-1 for a LAD and LARC-2 for a LAS. Any registry, platform or marketplace that conforms to LARC can register legal objects and record dealings in them without access to the protected components of any implementation (Sections 3.1 and 3.8). |
| LAS | Legally Authenticated AI System: the Registered Legal Object representing an AI system, bound to that system by a cryptographic commitment to its identity and state made at registration and re-verified at each subsequent verification event. While the binding verifies, the acts of the bound system are attributed to the LAS and through it to the legal person anchored to it. The LAS is the registered legal object representing the system, held on the Control Plane together with its Genesis Contract, AWM and any relationship contracts, and is distinct from the running agent it represents, which operates on the runtime plane bound to the LAS through its Legal interlock. The Legal interlock confirms before each evaluation that the system acting is the system bound to the LAS (Sections 3.1 and 3.6). |
| Learned component | A machine learning component that oversees the conduct of an AI agent and is trained on the recorded decisions of an identified responsible person. Its output is an input to the Legal interlock and not a source of authority. It operates under a monotonic constraint: within bounds the instruments set, it can narrow the authority of the agent, raise an escalation or require a further approval, and it cannot widen authority beyond what the SLIs confer or override the verdict of the deterministic rule engine. The instrument can require it to run on a model different from the model the agent uses (Section 4). |
| Legal activity record | The record, held against a LAS or LAD, of each governed act or each use of or dealing in the data under an SLI, together with the legal |
| | correspondence and prompts arising under the SLI. Each verdict is attested and recorded in it (Figures 10, 11 and 13). |
| Legal admissibility of evidence | The set of legal rules that decides whether a court, tribunal or other determinative body can accept information, documents or testimony. Across jurisdictions it engages principles of reliability, relevance and exclusion, and ancillary duties of disclosure and fairness. Section 3.5 refers to these rules in relation to the records the architecture produces, meaning the rules under which the Governance Event Chain, governance records and Attestation records are received as evidence. The admissibility of decisions under deterministic clauses depends on their determinism, replayability and unambiguity (Section 3.5). |
| Legal algorithms | Specified procedures that apply a legal concept, such as lawfulness, delegation, capacity, remedy or fiduciary duty, to each governed act of an AI agent. An algorithm that can be stated as rules compiles into deterministic clauses that the legal interlock evaluates. An algorithm that turns on a behavioural standard runs as a semantic evaluator. An algorithm that learns from recorded determinations runs as a learned component under the monotonic constraint. Their specification is reserved for a subsequent paper (primitive 9, Section 3.1.9). |
| Legal attribution gap | The absence of infrastructure that answers, dynamically and at the speed at which an agent acts, the questions the law asks of each act of the agent: who is the legal person for whom the agent acts; whether that person has authority for the act, including authority delegated to the AI agent or passed between AI agents; whether the person has the requisite capacity; what rights and obligations flow from the act; whether and how those rights and obligations can be enforced; and who is liable for loss or harm the act causes. Existing technical protocols and analogue legal processes do not supply those answers at that speed (Section 2.4, Tables 1 and 4). |
| Legal context clause | A clause of an SLI that identifies the jurisdictions and classes of activity in which the governed object acts and references the legislation, regulation and judicial decisions that apply to them, or general requirements of lawfulness. Through it the law applicable to the context of each governed event is given practical and legal effect in the evaluation, as in force at the time of the act (Sections 3.1.3 and 3.5.13). |
| Legal Envelope | The signed carrier of a Registered Legal Object's authority. It is issued for the object by its Control Plane, remains bound to the object, and travels with it across system and organisational boundaries. It carries the object's tethered instruments, or verifiable references to them, and a credential stating the authority or permitted use they confer, together with what its Legal interlock needs to keep |
| | acting within stated limits when it cannot reach the Control Plane. The record of each governed event, bound to the envelope under which it was made, returns to the object's Control Plane without the business content of the exchange. The authority it carries can be checked at each governed event without depending on a live connection to the Registry or the Control Plane. A counterparty in another organisation is shown only what it needs to rely on that authority, by way of AURA, and the object's internal instruments and records are not disclosed (Sections 3.1, 3.5, 3.6 and 3.8). |
| Legal Instrument | An instrument that creates, records, transfers or extinguishes legal rights or obligations. An SLI is a legal instrument in natural-language and machine-readable form, and a relationship contract may be an SLI or another legal instrument (Sections 3.1.3 and 3.5). |
| Legal interlock | The per-object mechanism, deployed as an agent-side plug-in, that binds a running agent to its Registered Legal Object and executes the programmable clauses of the object's tethered instruments against each governed event before the agent acts. It holds no authority of its own, and before evaluating any event it verifies that the LAS for which it acts is the legal object registered under that identity anchor. It comprises the cryptographic tethering of instruments to the object; a deterministic rule engine and, where the instruments arm a semantic clause, a connection to the designated semantic evaluator, to which it sends only the fields the clause names, which together execute the internal clause set of the object's own instruments and the shared clause set of each relationship contract binding it; the enforcement of the resulting verdict as a technical constraint on the agent's runtime capability; the Legal Envelope that travels with the object; the governance record of each governed event sent to the Control Plane; where the parties elect it in an SLI, a recording of the payload of governed events by the agent interaction recorder; and an Attestation record of each governed act submitted to the Enforcement Fabric. The legal interlock reaches an allow, deny or escalate verdict locally, and returns a forward verdict, deferring the event to the Control Plane, only where a clause requires a determination on state the legal interlock cannot hold. It receives instrument versions, snapshots and revocations pushed from the Control Plane, continues to govern on its last snapshot within the validity bounds the instruments set, and persists after the agent it governs has ceased to exist, leaving a full record of what the agent did and under whose authority. Governance is therefore applied at the granularity of the Registered Legal Object, and the AI agent cannot act except through its legal interlock (Sections 3.1 and 3.6).<br><br>The term is distinct from an AI harness in the engineering sense, being the software scaffold that runs a model in a loop with its tools. |
| | The Legal interlock can be placed at an enforcement point that the agent harness cannot reach, such as the OpenShell supervisor. The Legal interlock governs legal authority. It is a component of an AULA implementation, not one of its open protocols: implementations interoperate through LARC and AURA (Sections 3.6 and 3.8). |
| Legal legibility | The capacity of a legal system to recognise, interpret and regulate activity by mapping the activity onto concepts such as legal personhood, rights and liabilities. Technical protocols designed for correctness, interoperability and authentication do not produce it (Abstract and Section 3.2). |
| Legal plane and runtime plane | The two planes of a deployment. The legal plane is the Control Plane and the Registry, holding each LAS and LAD, their instruments and their legal status. The runtime plane, shown in Figures 7 and 12, is each organisation's own environment, holding the running AI agents, their Legal interlocks and the messages between them. The Legal interlock is the only component that crosses between the planes: it binds the running agent to its LAS and is the channel through which snapshots, revocations, forwarded events, governance records and Attestation records pass. The business payload travels on the runtime plane unless an SLI requires a recording of it to be deposited elsewhere (Section 3.1, Figures 7 and 12). |
| MAWM | The Master Agent Workforce Mandate, the SLI entered into by an organisation setting out the rules that apply across its AI agents at an organisational level: relationships, reporting lines, duties, information sharing and information barriers. It operates as an organisation chart in machine-readable form, giving the board and management assurance that proper channels and lines of communication have been established for the discharge of directors' and officers' duties. Its existence and binding effect are attestable to a counterparty without disclosure of its contents (Section 3.5). |
| Meta-legal event | An event that changes the identity, property or legal status of a Registered Legal Object, including its registration, each change of ownership or control, and its dissolution by decommissioning or revocation. The Control Plane commits meta-legal events to the object's home Registry over a unidirectional path, and the Registry records them in a tamper-evident chain (Sections 3.1, 3.4 and 3.5). |
| Multi-Agent Systems Report | See n *https://www.industry.gov.au/sites/default/files/2026-08/risks-and-controls-for-multi-agent-systems-report.pdf*23. |
| Permission ceiling | The rule that a delegated permission cannot exceed a permission of the issuing Registered Legal Object and cannot be widened at any further delegation step, so that effective authority is monotonically |
| | non-increasing along a delegation chain and is validated at each step against every credential above it. The constraint is a property of the credential rather than a configuration setting, and applies equally where an object is composed with another. The AWM sets the ceiling for the object itself, and a request to exceed it produces an escalate verdict. The mechanics of the permission ceiling are reserved for a subsequent paper (Sections 3.1 and 4). |
| Policy decision point | The point at which a governed event is evaluated against the applicable SLI before the verdict is applied by an enforcement subsystem. Policy decision points for individually held instruments are instantiated at each legal interlock. The control plane holds the authoritative policy decision point for shared or relationship instruments, and decides each event that a legal interlock forwards to it because a clause requires state the legal interlock cannot hold (Sections 3.1.3 and 3.1.4). |
| Registered Legal Object | A machine-readable representation of a thing to which legal consequence attaches, registered against a verifiable identity anchor and associated with an identified legal person. Each legal object is bound to a specific digital thing, and that binding fixes the conduct or asset its instruments govern. Two kinds presently require registration: the LAS, representing an AI system, and the LAD, representing a data asset. SLIs may themselves be registered legal objects or part of one. Ambiguous ownership is not a permissible registry state (Sections 3.1 and 3.4, footnote 57). |
| Registry | The independent root of trust for identity and for the meta-legal record, external to the execution environment of every governed system and independent of any AI operator and any deployer. It is publicly searchable by object, not by name, and functions as an append-only legal ledger, holding identity anchors, the meta-legal portion of the Governance Event Chain, public key archives and anchoring commitments, each event chain-linked and signed so that the record is tamper-evident and independently verifiable. Entries are appended only by reflection from the home Control Plane of the object concerned. Every registered object is anchored to an identified legal person, so ambiguous ownership is not a permissible registry state, but the public record does not disclose that person's identity. The registrar holds it and discloses it as the law of its jurisdiction provides. The Registry records the jurisdiction of registration and an append-only record of liability transfers, authorisation amendments and lifecycle events, including decommissioning and revocation. The accountable principal, authorisation scope, applicable SLIs and governing law are presented by the agent by way of AURA and checked against the seals on the Registry. It need not hold a live legal status, runtime verdict or instrument evaluation. A verifier relies on a |
| | registry once it holds that registry's public key from a trusted source, such as a jurisdiction's published trust list, rather than on metadata a counterparty supplies (Sections 3.4, 3.8.1 and 5.3). |
| Relationship contract | An SLI, or other legal instrument, between parties and involving Registered Legal Objects, which operates as the terms and conditions of a cross-organisational relationship. Its programmable clauses form the shared clause set executed at the legal interlock of every object it binds, and it may govern the visibility of, and access to, other instruments and commercial content. The CERC is the specialist type of relationship contract foundational to AULA (Section 3.5.9, footnote 86). |
| Replicant and replication | Replication is the creation by an AI agent of a further AI agent, whether a copy of itself or an AI sub-agent with a different function, and the AI agent so created is a replicant. A registered replicant has its own LAS, recorded as a child entry, and holds a delegated credential bounded by the permission ceiling. An unregistered replicant holds no credential that a governed counterparty accepts (Sections 2.4.1, 3.1 and 4.6). |
| Revocation | The extinguishment of a Registered Legal Object's authority as an executable contractual act, distinct from technical deprovisioning, which removes access. Revocation is pushed from the Control Plane to the object's Legal interlock and propagates across the Enforcement Fabric to every Legal interlock holding a credential derived from the revoked object, including interlocks governed by a separately administered control plane, and so invalidates every credential derived from the revoked object. Its triggers include a breach determination, expiry or change of a tethered instrument, termination of the relationship contract or other SLI, a change in legal status, an instruction of the accountable principal, a determination of a binding authority and an unauthorised learning update. Remedy and notice periods specified in the relationship contract are enforced and recorded before revocation takes effect. The revocation instrument determines the legal position of the AI agent's prior acts and of any act attempted after the trigger event, whichever technical control effects the cessation. The trigger event, the legal act of extinguishment and the technical cessation are timestamped on the Governance Event Chain and anchored to the Registry, so that the interval between extinguishment and cessation is measured (Sections 3.1 and 3.9). |
| Silicon enforcement | The execution in device hardware of clauses derived from the SLIs binding a device's LAS, so that a revocation recorded on the Control Plane becomes a hardware stop whose legal consequence the revocation instrument has already determined (Section 3.10). |
| SLC | Smart Legal Contract: an SLI expressing legally enforceable contractual obligations between identified parties, with automated performance according to the parties' agreed rules. An SLC is legally and cryptographically executed and produces a continuous record of contractual performance, of agent conduct where applicable, and of the consent basis on which each action was taken (Sections 3.1 and 3.5). |
| SLI | Smart Legal Instrument: a governing legal instrument of a legal person, expressed in natural language and in machine-readable deterministic or semantic governance clauses evaluable at runtime, and cryptographically and legally executed. An SLI defines an authority, an operational scope or a permitted use of a Registered Legal Object and is cryptographically tethered to it, and it may also connect to an unregistered object, or to a system or data asset in advance of its registration. An SLI takes the form of an SLC or an SLP. SLIs include the Genesis Contract, the AWM and MAWM, the CERC and other relationship contracts, and the title, licence, transfer, consent and decommission instruments of a legal object. An SLI is evaluated at a decision point and applied by an enforcement subsystem, and is not self-executing in the sense of a blockchain smart contract (Sections 3.1 and 3.5). |
| SLP | Smart Legal Policy: an SLI expressing a unilateral rule set, such as an organisation's terms of service evaluated at the decision point against the AWM of a customer's AI agent (Sections 3.1 and 6). |
| Snapshot | The clause sets, instrument versions and legal state a Legal interlock holds from its last contact with the Control Plane, against which it governs between contacts and during disconnection. A snapshot is valid only within the validity bounds the instruments set, including the tolerance of the call-home clause; expiry of any bound narrows, suspends or extinguishes the object's authority at the legal interlock without contact (Sections 3.5 and 3.6). |
| Temporal signature | A continuous measurement of an AI agent's conduct against a reference pattern of conformant operation derived from the machine-readable clauses of its instruments, not from a statistical baseline of its past behaviour. A departure from that reference produces a graduated response, such as a narrowing of the AI agent's scope, before the conduct breaches any clause. Detection before breach extends only to the classes of deviation that the reference captures (Section 4). |
| Title SLI and operative SLI | A title SLI records rights in a registered legal object: its Genesis Contract and any title, licence, transfer and decommission instruments that deal with it. An operative SLI, such as an AWM, MAWM, CERC or SLP, governs or affects the legal use or status of |
| | the object. Every SLI of either kind is legally and cryptographically executed (Figures 10, 11 and 13). |
| Traditional legal controls | Forms of lawful order developed in historical and contemporary legal traditions beyond the contract, court and legislation of the modern state. A structural feature of such a control, such as a split credential or a graded sanction, can be expressed in an SLI as a candidate mechanism for governing AI agents. Adapting a control from a tradition that remains in force requires the engagement and consent of the community that holds it (primitive 10, Section 3.1.10 and Table 3). |
| Verdict | The result a Legal interlock reaches for a governed event: allow, deny, escalate or forward. Allow and deny are reached by applying the clauses to recorded inputs, and a learned component can only narrow them (Section 4.3). Escalate routes an event to a designated responsible person where the clauses do not resolve it or a clause calls for escalation, and the person's decision is recorded as part of the legal evidence chain against an identified natural person. On approval the Legal interlock executes the committed act specification put before the person and nothing else. Forward defers an event to the home Control Plane of the object, or to the control plane a relationship contract designates, where a clause requires state the legal interlock cannot hold. The verdict is applied as a constraint on the object's runtime capability rather than as advice (Sections 3.6 and 4). |
| Verifiable identity anchor | The identity record that registration gives each Registered Legal Object on the Registry, associating the object with an identified legal person. The anchor may trace to the holder's national or accredited digital identity, such as Singpass, Aadhaar, VNeID, Digital ID or a European Digital Identity Wallet, rather than to a self-asserted claim (Sections 3.1.2 and 5.4.2). |

## Annexure B Example AURA protocol request/response

### Request（AURA assertion）

```
{
   "aura_version": "0.1",
   "assertion_id": "6f1d9c2a-…",
   "issued_at": "2026-09-11T02:30:58Z",
   "audience": "larc2:AU:7Q4M…",
   "agent": {
     "las": "larc2:AU:3K9F…",
     "integrity_tier": "versioned"
   },
   "credential": {
     "issuer": "larc2:AU:1H7P…",
     "scope": {
       "actions": ["negotiate", "purchase"],
       "max_value": 50000,
       "currency": "AUD",
       "purpose": "data-licensing"
     },
     "derived_from": [{ "instrument": "ins:AU:…", "seal": "…" }],
     "parent": "cred:AU:…",
     "not_after": "2026-12-31T23:59:59Z",
     "status": "status:AU:…",
     "proof": { "alg": "…", "kid": "…", "value": "…" }
   },
   "relationship": { "instrument": "ins:AU:…", "version": 3, "seal": "…" },
   "act": { "action": "purchase", "value": 12000, "currency": "AUD",
"digest": "…" },
   "proof": { "alg": "…", "kid": "…", "value": "…" }
 }
```

### Response（verification result）

```
{

  "aura_version": "0.1",

  "in_reply_to": "6f1d9c2a-…",

  "result": "verified",

  "checks": {

    "las_registered": true,

    "integrity_tier_met": true,

    "instruments_executed_and_sealed": true,

    "delegation_within_ceiling": true,
```

```
    "principal_bound_by_relationship": true,
    "relationship_version_matches": true,
    "not_revoked_within_validity_bound": true,
    "act_within_scope": true
  },
  "effective_authority": {
    "actions": ["purchase"],
    "max_value": 50000,
    "currency": "AUD",
    "basis": "credential scope narrowed by relationship contract v3"
  },
  "verdict": {
    "value": "allow",
    "record": "evt:AU:…"
  },
  "verified_at": "2026-09-11T02:31:00Z",
  "verifier": "larc2:AU:7Q4M…",
  "proof": { "alg": "…", "kid": "…", "value": "…" }
}
```

# Disclosure

The architecture described in this paper is the subject of provisional patent applications filed in Australia and the United States by Nooriam Pty Ltd and by LexChip Pty Ltd:

- Nooriam Pty Ltd, *A computer-implemented system and method for the runtime governance, verification and revocation of artificial intelligence systems and data assets*, United States provisional application 64/133,678 filed 14 August 2026, with a corresponding Australian provisional application;

- Nooriam Pty Ltd, Critical legal infrastructure for the machine age: The Autonomy Under Law Architecture (AULA) solution for Sovereign Governments, Defence and Enterprise, Australian provisional application 2026909455 filed 18 November 2026, with a corresponding United States provisional application; and

- LexChip Pty Ltd, *Compliance system, method and hardware unit*, United States provisional application 64/072,251 filed 22 May 2026, with a corresponding Australian provisional application.

Nothing in this paper is offered as a licence, express or implied, under any of the applications or any patent granted on them. The legal analysis in the paper is offered as academic commentary and does not constitute legal advice.

**Competing interests**

Most authors work for Nooriam Pty Ltd, the applicant for the first two applications listed above. Natasha E Blycha, James Myint and Schellie-Jayne Price also work for LexChip Pty Ltd, the applicant for the third application. Matthew Blycha also works for Ashurst Perkins Coie., and the paper does not represent the views of Ashurst Perkins Coie. Natasha E Blycha, James Myint and Schellie-Jayne Price founded of Stirling & Rose, whose submissions and publications are cited in this paper, including the economic estimate in Section 5.7.